\documentclass{article}
\usepackage{amsthm,amsmath,amsfonts,amssymb}
\usepackage[numbers]{natbib}
\usepackage[colorlinks,citecolor=blue,urlcolor=blue]{hyperref}
\usepackage{graphicx}

\usepackage{enumitem,epsfig}
\usepackage[dvipsnames]{xcolor}
\usepackage{multirow}
\usepackage{multicol}
\usepackage{cleveref}
\usepackage{subcaption}
\usepackage[hmargin={1.0in, 1.0in}, vmargin={1.2in, 1.2in}]{geometry}
\usepackage{booktabs}
\usepackage{array}
\newcolumntype{Z}{>{\setbox0=\hbox\bgroup}c<{\egroup}@{\hspace*{-\tabcolsep}}}
\usepackage[title]{appendix}

\usepackage{setspace}
\usepackage{diagbox}
\usepackage{authblk}

\allowdisplaybreaks[4]

\theoremstyle{plain}
\newtheorem{theorem}{Theorem}[section]
\newtheorem{lemma}{Lemma}[section]

\newtheorem{proposition}{Proposition}[section]
\usepackage[title]{appendix}

\theoremstyle{remark}
\newtheorem{definition}{Definition}[section]

\newtheorem{remark}{Remark}[section]
\newtheorem{assumpt}{Assumption}

\long\def\acks#1{\vskip 0.3in\noindent{\large\bf Acknowledgments and Disclosure of Funding}\vskip 0.2in
\noindent #1}
\newenvironment{Proof}{{\noindent \it Proof:\quad}}{\hfill $\fbox{}$ \vspace*{5mm}}

\numberwithin{equation}{section}
\numberwithin{figure}{section}
\numberwithin{table}{section}

\newcommand{\ca}{\mathcal{A}}
\newcommand{\cb}{\mathcal{B}}

\newcommand{\cf}{\mathcal{F}}

\newcommand{\ch}{\mathcal{H}}

\newcommand{\cn}{\mathcal{N}}

\newcommand{\cx}{\mathcal{X}}

\newcommand{\be}{\mathbb{E}}
\newcommand{\bn}{\mathbb{N}}

\newcommand{\br}{\mathbb{R}}
\newcommand{\bz}{\mathbb{Z}}
\newcommand{\bone}{\mathbb{I}}

\newcommand{\fN}{\mathbf{N}}

\newcommand{\fZ}{\mathbf{Z}}

\newcommand{\fone}{\mathbf{I}}

\newcommand{\1}[1]{\boldsymbol{1}\{#1\}}

\newcommand{\Var}{\mbox{Var}}
\newcommand{\var}{\mbox{var}}
\newcommand{\Cov}{\mbox{Cov}}

\newcommand{\tr}{\mbox{tr}}
\newcommand{\sgn}{\mbox{sgn}}
\newcommand{\cum}{\mbox{cum}}

\newcommand{\indept}{\perp\!\!\!\perp}
\newcommand{\stra}[1]{\stackrel{#1}{\longrightarrow}}

\newcommand{\bbe}[1]{\mathbb{E}\big[{#1}\big]}

\newcommand{\blre}[1]{\mathbb{E}\left[{#1}\right]}

\newcommand{\bbp}[1]{\mathbb{P}\big({#1}\big)}

\newcommand{\blrp}[1]{\mathbb{P}\left({#1}\right)}

\newcommand{\bigp}[1]{\big({#1}\big)}
\newcommand{\lrp}[1]{\left({#1}\right)}

\newcommand{\lrcp}[1]{\left\{{#1}\right\}}

\newcommand{\lrbk}[1]{\left[{#1}\right]}
\newcommand{\lrag}[1]{\left\langle{#1}\right\rangle}
\newcommand{\lrabs}[1]{\left|{#1}\right|}
\newcommand{\lrnorm}[1]{\left\|{#1}\right\|}
\newcommand{\lrfl}[1]{\lfloor{#1}\rfloor}
\newcommand{\lrceil}[1]{\lceil{#1}\rceil}

\newcommand{\BE}{\begin{equation}}
\newcommand{\EE}{\end{equation}}
\newcommand{\BEqn}{\begin{eqnarray*}}
\newcommand{\EEqn}{\end{eqnarray*}}
\DeclareMathOperator*{\argmax}{argmax}

\newcommand{\bin}[2]{
  \left(
        \begin{array}{@{}c@{}}
          #1 \\ #2
        \end{array}
  \right)}

\newcommand{\dmat}[2]{
  \left(
    \begin{array}{cc}
      #1 &    \\
        & #2  \\
    \end{array}
  \right)}

\makeatletter
\newcommand*{\addFileDependency}[1]{
  \typeout{(#1)}
  \@addtofilelist{#1}
  \IfFileExists{#1}{}{\typeout{No file #1.}}
}
\newcommand*\showfontsize{\f@size{} point}
\makeatother

\newcommand{\blind}{1}

\begin{document}
 
\if1\blind
{
    \title{Dimension-agnostic Change Point Detection}	
		
    \author[1]{Hanjia Gao}
    \author[2]{Runmin Wang}
    \author[1]{Xiaofeng Shao}
    \affil[1]{Department of Statistics, University of Illinois at Urbana-Champaign}
    \affil[2]{Department of Statistics, Texas A\&M University}		
	\date{}	
	\maketitle
} \fi


\begin{abstract}
Change point testing for high-dimensional data has attracted a lot of attention in statistics and machine learning owing to the emergence of high-dimensional data with structural breaks from many fields. In practice, when the dimension is less than the sample size but is not small, it is often unclear whether a method that is tailored to high-dimensional data or simply a classical method that is developed and justified for low-dimensional data is preferred. In addition, the methods designed for low-dimensional data may not work well in the high-dimensional environment and vice versa. In this paper, we propose a dimension-agnostic testing procedure targeting a single change point in the mean of a multivariate time series. Specifically, we can show that the limiting null distribution for our test statistic is the same regardless of the dimensionality and the magnitude of cross-sectional dependence. The power analysis is also conducted to understand the large sample behavior of the proposed test. Through Monte Carlo simulations and a real data illustration, we demonstrate that the finite sample results strongly corroborate the theory and suggest that the proposed test can be used as a benchmark for change-point detection of time series of low, medium, and high dimensions.
\end{abstract}

\noindent\textit{Keywords}: Sample Splitting, Self-normalization, High-dimensional Data, Time Series, Cross-sectional Dependence.


\section{Introduction}\label{Sec:Intro}

Given a multivariate time series  $\{X_t\}_{t = 1}^{n} \in \mathbb{R}^p$ with both temporal and cross-sectional dependence, we are interested in testing the existence of at-most-one change point in the mean. Let $\mu_t = \be[X_t]$, it is equivalent to  testing the hypothesis
\begin{equation*}
    \ch_0:\ \mu_1 = \cdots = \mu_n
    \quad\mbox{versus}\quad
    \ch_1:\ \mu_1 = \cdots = \mu_{k_0} \neq \mu_{k_0+1} = \cdots = \mu_n,
\end{equation*}
where $k_0 = \lfloor n\varepsilon_0\rfloor $ denotes the unknown location of the change point and $\varepsilon_0 \in (0,1)$. Change point testing is a classical statistical problem that dates back to \cite{page1954continuous,page1955test} and there is a vast amount of literature in the areas of econometrics, statistics and machine learning for fixed-dimensional or low-dimensional data. For low-dimensional time series, we refer the readers to  \cite{aue2013structural} and \cite{casini2019} for excellent reviews of the subject and the huge literature cited therein.

In the past decade, there has been a surge of interest in developing change point testing/estimation methods for high-dimensional data, as motivated by the increasing need for the analysis of high-dimensional data with change points from many scientific areas, such as genomics, neuroimaging, finance and economics. Here we shall mention recent contributions by \cite{horvath2012change}, \cite{chan2013darling}, \cite{jirak2015uniform}, \cite{cho2016change}, \cite{wang2018high}, \cite{enikeeva2019high}, \cite{dette2020likelihood}, \cite{wang2022inference}, \cite{yu2021finite}, \cite{zhang2021adaptive}, \cite{wang2023computationally}, among others. Note that some of the above-mentioned works specifically target high-dimensional independent data, and do not allow for temporal dependence. For change point testing/estimation of high-dimensional time series (i.e., with temporal dependence), see \cite{jirak2015uniform}, \cite{dette2020likelihood}, \cite{wang2022inference}, \cite{chen2021inference}, to name a few.

A common feature of the methods developed for high-dimensional time series is that they need to handle the estimation of a high-dimensional long-run variance (LRV) matrix or componentwise LRV. Consistent estimation of LRV is a thorny issue in practice, as the choice of bandwidth is a notoriously difficult one and becomes especially challenging with the presence of change points and high dimensionality. Recently, self-normalization (SN, hereafter) based inference, which avoids direct consistent estimation of LRV,   has been extended to high-dimensional change point detection problem in \cite{wang2022inference} and \cite{zhang2021adaptive}.  We shall refer to \cite{liu2021high} for a timely review of the recent literature on high-dimensional change point testing and estimation.

For change point testing of multivariate time series,  the literature is naturally divided into two categories: methods developed and justified for low-dimensional time series, and methods that can accommodate high dimension and allow (or require) the dimension $p$ to be comparable to or exceeds sample size $n$. Most of the existing methods are designed for a specific dimensional regime (i.e., either low/fixed dimension or high/growing dimension). On one hand, the methods developed for low/fixed-dimensional problems may not be theoretically justified or even applicable when $p > n$. For example, the SN-based test in \cite{shao2010testing} works quite well when $p$ is relatively small compared to $n$, but it is no longer applicable when $p >n$ due to the non-invertibility of self-normalizer and can exhibit serious size distortion when $p$ is moderate relative to $n$ (e.g., $p=10$ and $n=100$) in the presence of moderate/strong temporal dependence. On the other hand, the method developed in the high-dimensional setting may not work for the low-dimensional data,  as the approximation accuracy of some test statistics highly relies on the central limit effect from the high dimension and the size can be quite distorted for data of low or moderate dimension. In addition, for the same test statistic, the limiting null distributions under different asymptotic regimes are usually different, which results in practical difficulty in using these tests as the calibration usually depends on the regimes, which are unknown in practice. This naturally motivates the question of whether there is a change point testing procedure that can work for weakly dependent time series of low, medium, and high dimensions, that is, be dimension-agnostic. 

The dimension-agnostic property of an inference procedure has been paid attention to in \cite{paindaveine2016high}, \cite{wasserman2020universal} but it was not until \cite{kim2020dimensionagnostic} that formalized the notion of dimension-agnostic inference, and proposed a sample splitting approach to several nonparametric testing problems. As argued in their paper, many test statistics developed in the literature often have a different limiting distribution in a fixed-dimensional regime ($p$ is fixed as $n\rightarrow\infty$) or a high-dimensional regime ($\min\{n,p\}\rightarrow\infty$ together at some relative rate). This typically leads to different calibration thresholds (or critical values) and the overall rejection rule is different in two regimes. But in practice, suppose we are given a dataset with (say) 120 samples in $30$ dimensions, should we calibrate assuming $p$ is fixed, or $n \gg p$, or $p/n\rightarrow 0.25$? The goal of dimension-agnostic inference is to develop a test statistic whose limiting null distribution is the same regardless of how the dimensionality $p$ scales with respect to sample size $n$. A major benefit of the dimension-agnostic inference is being consistent against various dimensional regimes, and the selection of dimension-dependent calibration threshold can be avoided.

In this article, we advance the dimension-agnostic inference to change point testing in time series for the first time. Although our work builds on \cite{kim2020dimensionagnostic} and follows their sample splitting and projection approach, it differs from their work in several fundamental ways. First, we deal with the time series data, whereas independent data is the sole focus in \cite{kim2020dimensionagnostic}. Note that time series has a natural ordering which implies the relatively limited way of sample splitting. For the iid data, sample splitting is not unique and it remains to develop a way of combining different sample splits for the purpose of dimension-agnostic inference, whereas for time series, there is no ambiguity with sample splits (e.g., an equal-sized sample split with time series is unique) so the concern over practical replicability due to randomness of sample splits is minimal. It is thus natural to use sample splitting for the purpose of dimension-agnostic inference in the time series setting. Second, we need to make some methodological adjustments to accommodate temporal dependence and a broad range of dimensionality. In particular, we need to introduce a trimming parameter to control the bias incurred by weak temporal dependence. Also, the self-normalizer used in the mean inference problem of \cite{kim2020dimensionagnostic}
is the classical studentizer in the formation of $t$ statistic, whereas we adopt the self-normalizer for change point testing for time series as used in \cite{shao2010testing}. Third, the theoretical argument is very different. The argument in \cite{kim2020dimensionagnostic}, which relies on the pointwise Berry-Esseen bound conditional on the half of the sample, no longer applies to the time series setting. To alleviate the difficulty, we develop some new conditioning arguments to study the large sample behavior of our proposed test statistic. Fourth, an added benefit is that our proposed test is not only dimension-agnostic but also agnostic to the degree of panel (i.e., cross-sectional) dependence. The robustness to the dimension and degree of panel dependence does not come for free and there is indeed a price to pay. As we demonstrate in simulation studies, there is a certain degree of power loss in some settings, which is in part due to the use of sample splitting, but the efficiency loss seems moderate in many scenarios. These findings are in general agreement with and an important complement to those in \cite{kim2020dimensionagnostic}.

The rest of the article is organized as follows. In Section \ref{Sec:Methodology}, we introduce our dimension-agnostic test statistic for a single change point alternative and state the main assumptions for three data-generating processes that encompass both fixed/growing dimensional regimes and weak/strong cross-sectional dependence. Section~\ref{Sec:Theory-null} and Section~\ref{Sec:Theory-power} present the main theory for our single change point test under the null hypothesis and alternative hypothesis, respectively. Simulation results are gathered in Section~\ref{Sec:NumResults-simul} whereas a real data illustration is provided in Section~\ref{Sec:NumResults-real}. Lastly, Section~\ref{Sec:Discussion} concludes. The generalization to a single sparse change point testing and to multiple change points testing are presented in the appendix. All technical proofs and additional simulation results are included in the online
supplement; see \url{https://arxiv.org/abs/2303.10808}.

Throughout this paper, we use $\lrfl{x}$ to denote the largest integer not exceeding $x$ for any real-valued $x$ whereas $\lrceil{x}$ to denote the smallest integer no smaller than $x$. For any $x,y\in\br^p$, we use $\lrag{x,y} = x^{\top}y$ to denote the inner product in $\br^{p}$  and use $\|x\|_2=\lrag{x,x}^{1/2}$ to denote the $L_2$ norm of $x$. For any matrix $A\in\br^{p\times q}$, $\|A\|$ denotes the spectral norm of $A$ whereas $\|A\|_F$ denotes the Frobenious norm. For two real-valued sequences $a_n,b_n$, we say $a_n = O(b_n)$ or $a_n \lesssim b_n$ if there exist $M,C>0$, such that $a_n \le C b_n$ for $n>M$. If there exist $M,C_1,C_2>0$, such that $C_1 b_n \le a_n \le C_2 b_n$ for $n>M$, then we say $a_n = O_s(b_n)$. In addition, we say $a_n = o(b_n)$ or $a_n \prec b_n$ if $a_n/b_n \rightarrow 0$ as $n\rightarrow\infty$. The symbols $\leadsto$ and $\stra{d}$ denote the process convergence and the convergence in distribution of random variables respectively. We use $\stra{p}$ to represent the convergence in probability and use $=^d$ to denote the equality in distribution. Additionally, we use $\cum(x_1,\cdots,x_k)$ to denote the joint  cumulant of the random variables $x_1,\cdots,x_k$.


\section{Methodology and Data-Generating Processes}\label{Sec:Methodology}

As mentioned in the introduction, our work is inspired by the recent dimension-agnostic inference proposed in \cite{kim2020dimensionagnostic}. Here we briefly review their method for one sample mean testing. Given $n$ iid observations $X_1,\cdots,X_n$ in the $p$-dimensional space with mean $\mu$, the goal is to test  $H_0: \mu=0$ versus $H_1: \mu\neq0$. They propose to split the data into two parts,  $\cx_1=\{X_i\}_{i=1}^{n_1}$  and  $\cx_2=\{X_i\}_{i=n_1+1}^{n}$. Furthermore, they define a $\cx_2$-dependent random function $f_{\mbox{mean}}(x) := \frac{1}{n_2} \sum_{j=n_1+1}^{n} X_j^{\top} x$, where $n_2=n-n_1$. 
Based on $f_{\mbox{mean}}(\cdot)$, \cite{kim2020dimensionagnostic} define $\bar{f}_{\mbox{mean}} = \lrp{\frac{1}{n_1} \sum\limits_{i=1}^{n_1} X_i}^{\top} \lrp{\frac{1}{n_2} \sum\limits_{j=n_1+1}^{n} X_j}$
and propose the studentized test statistic 
\begin{equation*}
    T_{\mbox{mean}}
  = \frac{\sqrt{n_1} \bar{f}_{\mbox{mean}}}{\sqrt{n_1^{-1} \sum_{i=1}^{n_1} \lrp{f_{\mbox{mean}}(X_i) - \bar{f}_{\mbox{mean}}}^2}}.
\end{equation*}
The key ingredients in forming the above test statistic are sample splitting, projection (as used in $f_{\mbox{mean}}$) and studentization. 

To justify the dimension-agnostic property, \cite{kim2020dimensionagnostic} show that the studentized statistic is asymptotically normal with an unconditional uniform Berry-Esseen bound under the null and mild assumptions on the moment and Lyapunov ratio (see Assumption 2.2 therein). In particular, their assumption can be satisfied by a wide class of distributions with sub-Gaussian or sub-Exponential tails without any restriction on $p$.

\subsection{Methodology}\label{Sec:Methodology-dense}

Below, we introduce the ``SS-SN" (i.e., sample splitting and self-normalization) methodology for single change point testing in the mean of a multivariate time series. Our test statistic is constructed by following the steps below:

\begin{enumerate}[label=(\roman*)]
    \item {\bf Sample splitting with trimming:} Mimicking the sample-splitting procedure introduced in \cite{kim2020dimensionagnostic}, we separate the entire sample into three pieces of unequal sizes. Specifically, let $\varepsilon \in (0,1/2)$ be the splitting ratio and assume that $\varepsilon_0 \in \lrp{\varepsilon, 1-\varepsilon}$,  we split the observed data into three pieces:
    \begin{equation*}
        \cx_1 = \lrcp{X_1, \cdots, X_m}, \qquad
        \cx_2 = \lrcp{X_{m+1}, \cdots, X_{n-m}}, \qquad
        \cx_3 = \lrcp{X_{n-m+1}, \cdots, X_n},
    \end{equation*}
    where $m = \lfloor{ n\varepsilon \rfloor}$ denotes the size of the first/third block and $N = n-2m$ denotes the size of the middle block.
    
    Furthermore, we introduce a trimming parameter $\eta \in (0, \varepsilon)$. Define $m_1 = \lfloor{ n(\varepsilon-\eta) \rfloor}$ and $m_2 = m - m_1$, then $\cx_1$ and $\cx_3$ can be respectively partitioned into two smaller pieces, i.e. $\cx_1 = \cx_{11} \cup \cx_{12}$ with $\cx_{11} = \lrcp{X_1, \cdots, X_{m_1}}$ and $\cx_{12} = \lrcp{X_{m_1+1}, \cdots, X_{m}}$, whereas $\cx_3 = \cx_{31} \cup \cx_{32}$ with $\cx_{31} = \lrcp{X_{n-m+1}, \cdots, X_{n-m_1}}$ and $\cx_{32} = \lrcp{X_{n-m_1+1}, \cdots, X_{n}}$. We shall use the data in $\cx_{11}$ and $\cx_{32}$ to estimate $\mu_1-\mu_n$ and then project the resulting estimate to the middle block $\cx_2$. For convenience, we call $\varepsilon$ the splitting parameter and $\eta$ the trimming parameter. 

    \begin{remark}
    Initially, we plan to use $\cx_1$ and $\cx_3$ to form an estimate of $\mu_1-\mu_n$, say $m^{-1}\sum_{t=1}^{m}X_t-m^{-1}\sum_{t=n-m+1}^{n}X_t$, and then project the data in the middle block $\cx_2$ onto this direction. This would be a direct extension of the splitting and projection idea presented in \cite{kim2020dimensionagnostic}. However, as demonstrated in \cite{wang2020hypothesis} and \cite{wang2022inference}, temporal dependence of high-dimensional time series may result in a bias term for the U-statistic-based estimate. It turns out that to make the bias asymptotically negligible in our setting, it requires a very stringent assumption on the growth rate of $p$ as a function of $n$. In contrast, the trimming technique used in our proposed test can greatly weaken the required technical conditions and also lead to improved finite sample performance.
    \end{remark}

    \item  {\bf Projection:} 
    After sample splitting and trimming, we obtain two sample mean estimates, denoted by $\hat\mu_1 = \frac{1}{m_1} \sum_{i=1}^{m_1} X_i$ and $\hat\mu_n = \frac{1}{m_1} \sum_{i=1}^{m_1} X_{n+1-i}$ respectively. Then we project the observations in block $\cx_2$ onto the direction of $\hat\mu_1 - \hat\mu_n$. By doing so, we obtain a scalar sequence $\{Y_j\}_{j=1}^{N}$ with $Y_j = \lrag{\hat\mu_1 - \hat\mu_n, X_{j+m}}$, $j=1,\cdots,N$. Note that the data we use for dimension reduction, i.e., $\cx_{11}\cup \cx_{32}$ is separated from the data we project to, i.e., $\cx_2$, by a distance of $\lfloor n\eta\rfloor$, to alleviate the bias problem we mentioned earlier. Under the one change point alternative and the assumption $\varepsilon_0\in (\varepsilon,1-\varepsilon)$, $\hat\mu_1 - \hat\mu_n$ quantifies the amount of the mean shift in the original data sequence, which is well preserved in the scalar sequence $\{Y_j\}_{j=1}^{N}$ as the shift in the mean of  $\{Y_j\}_{j=1}^{N}$ is approximately equal to $\|\mu_1-\mu_n\|_2^2$. Therefore,  the original $p$-dimensional change point testing for $\mu_1=\mu_n$ is converted into an equivalent univariate change point testing problem for $\|\mu_1-\mu_n\|_2=0$. It is worth noting that the temporal dependence in the series $\{Y_j\}_{j=1}^{N}$ is fairly complex.

    \item {\bf Forming a studentized test statistic:} 
    After dimension reduction and projection, we end up with a univariate series $\{Y_t\}_{t=1}^{N}$. Following the insights provided in \cite{kim2020dimensionagnostic}, studentization is a key for the dimension-agnosticness of their test. For mean testing, the classical $t$ statistic was used in \cite{kim2020dimensionagnostic} for the projected data since the data are iid and the testing problem is one sample. By contrast, we are dealing with time series and a change point testing problem, so some modification needs to be made. Specifically, we shall apply the SN-based test statistic in \cite{shao2010testing} to the projected data $\{Y_t\}$. For $1\le a\le b\le N$, with the cumulative sum defined as $S_{a,b} = \sum_{j=a}^{b} Y_j$, we further define
    \begin{equation*}
        T_n(k)
      = N^{-1/2} \sum\limits_{t=1}^{k} \lrp{Y_t - \bar{Y}_N},
        \qquad k = 1,2,\cdots,N-1,
    \end{equation*}
    where $\bar{Y}_N = S_{1,N}/N$. For $k=1,2,\cdots,N-1$, the self-normalizer is defined as
    \begin{equation*}
        V_n(k) 
      = N^{-2} 
      \lrp{  \sum\limits_{t=1}^{k} \lrp{S_{1,t} - \frac{t}{k}S_{1,k}}^2 
           + \sum\limits_{t=k+1}^{N} \lrp{S_{t,N} - \frac{N-t+1}{N-k}S_{k+1,N}}^2}.
    \end{equation*}
    Finally, we define the test statistic as 
    \begin{equation*}
        G_n = \sup\limits_{k=1,\cdots,N-1} T_n(k) V_n^{-1/2}(k).
    \end{equation*}
\end{enumerate}

\begin{remark}
Self-normalization for time series \cite{shao2010self} is an inference technique that has been developed for low and fixed-dimensional parameters in a low dimensional time series, following some early developments by \cite{kvb2000} and \cite{lobato2001testing}. It uses an inconsistent variance estimator to yield an asymptotically pivotal statistic and does not involve any tuning parameter or involves less number of tuning parameters compared to traditional procedures. See \cite{shao2015self} for a comprehensive review for low dimensional time series. There have been two recent extensions to the high-dimensional setting: \cite{wang2020hypothesis} adopted a one-sample U-statistic with trimming and extended self-normalization to inference for the mean of high-dimensional time series; \cite{wang2022inference} used a two-sample U-statistic and extended the self-normalization (SN)-based change point test in \cite{shao2010testing} to high-dimensional independent and dependent data.  Note that the test in \cite{wang2022inference}  is not expected to work for low dimensional time series as the theory requires the dimension to grow and the approximation by the limiting null distribution may be inadequate when the dimension is low or moderate. Additionally, the theoretical applicability of their test is limited to time series with weak cross-sectional dependence. We shall present some simulation comparisons in Section~\ref{Sec:NumResults-simul}.
\end{remark}

\begin{remark}
Sample splitting has been widely used in statistics and machine learning but it seems that most of its use is for independent data; see  \cite{shafer2008tutorial}, \cite{wasserman2009high}, \cite{rinaldo2019bootstrapping}, \cite{wasserman2020universal}, among others. In the context of time series, we are only aware of  \cite{lunde2019sample} and \cite{chang2022testing}. Specifically, sample splitting was used for the post-selection inference in the time series setting in \cite{lunde2019sample} and for unit root testing in \cite{chang2022testing}. The scope and property of our proposed SS-SN inference are very different from these papers and have no overlap with the existing literature.
\end{remark}

\begin{remark}
From an implementational perspective,  the test statistic is fairly easy to code and fast to compute due to the dimension reduction step involved. In particular, the calculation of $\{Y_j\}_{j=1}^{N}$ can be done at the cost of $O(np)$ and the calculation of SN test statistic based on one-dimensional sequence $\{Y_j\}_{j=1}^{N}$ can be done at the cost of $O(n^2)$. So the overall computational complexity is of order $O(n(n+p))$. By contrast, the SN-based test in \cite{wang2022inference} has the complexity of order $O(n^2p)$, and can be much more computationally expensive than ours when $p$ and $n$ are large. 
\end{remark}

\subsection{Data Generating Processes}\label{Sec:Theory-structure}

Since our main goal is to show that our test statistic works for time series in both fixed-dimensional and growing-dimensional settings and for time series with either weak cross-sectional dependence or strong cross-sectional dependence, we shall consider three types of data-generating processes in this paper. As summarized in Table \ref{Tab:DataStructure}, we investigate three cases for $\{X_t-\mu_t\}_{t=1}^{n}$: (1) stationary weakly dependent time series with fixed $p$ and arbitrary cross-sectional dependence; (2) linear process with growing $p$ that allows for weak temporal and cross-sectional dependence. This is similar to the setting in \cite{wang2020hypothesis}, where a nonlinear casual process was assumed. The results are expected to hold under the more general nonlinear process but at the expense of more complicated technical arguments; (3) static factor model for high-dimensional time series which can accommodate strong cross-sectional dependence and weak temporal dependence. 

These three DGPs are quite representative in the sense that they contain multiple dimensional regimes and dependence settings, and allow us to thoroughly investigate the properties of the proposed test and show its dimension-agnostic property and robustness to the magnitude of cross-sectional dependence.

\begin{table}[h!]
    \centering
    \caption{Three types of data-generating processes} \medskip
    \label{Tab:DataStructure}
    \begin{tabular}{c|c|l}
    \hline\hline
        {\bf Dimensionality} & {\bf Data-Generating Process} & \multicolumn{1}{c}{{\bf Dependence}} \\ \hline
        \multirow{2}{*}{fixed $p$} & \multirow{2}{*}{stationary sequence (DGP1)} & weak temporal dependence \\
        & & arbitrary cross-sectional dependence \\ \hline
        \multirow{4}{*}{diverging $p$} & \multirow{2}{*}{linear process (DGP2)} & weak temporal dependence \\
        & & weak cross-sectional dependence \\ \cline{2-3}
        & \multirow{2}{*}{static factor model (DGP3)} & weak temporal dependence \\
        & & strong cross-sectional dependence \\
    \hline\hline        
    \end{tabular}
\end{table}

Below we shall present the exact requirements for each data-generating process. 

\begin{definition}[DGP1]\label{Def:Model_Fixedp}
Assume that $X_t-\mu_t \in \br^p, t\in \fZ$ is a stationary sequence with $\be[X_t] = \mu_t$ and that the long-run variance $\Omega^{(1)} = \sum\limits_{k=-\infty}^{\infty} \Cov(X_t,X_{t+k})$ is a positive definite matrix. Further assume that $p$ is fixed and  $\frac{1}{\sqrt{n}} \sum\limits_{t=1}^{\lrfl{nr}} (X_t-\mu_t) \leadsto (\Omega^{(1)})^{1/2} B_p(r)$ in $D^p[0,1]$ as $n\rightarrow\infty$.
\end{definition}

Throughout the paper, we use $D^p[0,1]$ (or more generally, $D^p[a,b]$) to denote the space of $R^{p}$-valued ($p\in\fN$) functions on $[0,1]$ (or $[a,b]$) which are right continuous with left limits, endowed with the Skorokhod topology \cite{billingsley2008probability}. The assumption that the long-run variance matrix $\Omega^{(1)}$ is positive definite is common in SN-based inference \cite{shao2010self, shao2015self} and is widely used in the literature of time series analysis. The functional central limit theorem (or invariance principle) is a high-level assumption that can be verified for weakly dependent time series that satisfy certain mixing or near epoch dependence assumptions; see  \cite{lobato2001testing} and \cite{shao2010self} for related discussions. Note that there is no particular restriction on the cross-sectional dependence other than the positive definiteness of $\Omega^{(1)}$. 

Next, we introduce the second data-generating process which requires the dimension $p$ to grow with sample size $n$. 

\begin{definition}[DGP2]\label{Def:Model_lp} 
Assume that $X_t = \mu_t +  \sum\limits_{j=0}^{\infty} a_j \varepsilon_{t-j}$ for $t=1,\cdots,n$, where $\{\varepsilon_t\}_{t\in\bz}$ is an iid $p$-dimensional innovation sequence with mean zero and covariance matrix $\Gamma^{(2)}$ and $\{a_j\}_{j=0}^{\infty}$ is a sequence of $p \times p$ coefficient matrices. Here $p=p(n)$ is assumed to depend on $n$ and it grows to $\infty$ as $n\rightarrow\infty$.
\end{definition}

The sequence of coefficient matrices $\{a_j\}_{j=0}^{\infty}$ determines the temporal dependence of the series, whereas the cross-sectional dependence of $X_t$ is jointly determined by the covariance matrix $\Gamma^{(2)}$ and the coefficient matrices $\{a_j\}_{j=0}^{\infty}$. Additional assumptions will be imposed later to make sure that both temporal dependence and cross-sectional dependence are weak.

Finally, in order to accommodate strong cross-sectional dependence, we shall introduce the static factor model for high-dimensional time series \cite{fan2013large}.

\begin{definition}[DGP3]\label{Def:Model_factor}
Assume that 
\begin{equation}\label{Equ:factor}
    X_t = \mu_t + \Lambda F_t + Z_t, \qquad 1\le t\le n,
\end{equation}
where $Z_t \in \br^{p}$ is the idiosyncratic error, $\Lambda \in \br^{p \times s}$ is the factor loading matrix and $F_t \in \br^{s}$ is the underlying factor series. Without loss of generality, we shall assume that  $s$ is fixed as $p\rightarrow\infty$.
Additionally, we assume that  $\{F_t\}_{t=1}^{n} \indept \{Z_t\}_{t=1}^{n}$, where the low-dimensional factor series $\{F_t\}_{t=1}^{n} \sim (0,\Omega^{(3)})$ have weak temporal dependence, and  $\{Z_t\}_{t=1}^{n} \sim (0, \Sigma^{(3)})$ is a linear process with weak cross-sectional and  temporal dependence.  Here $p=p(n)$ is assumed to depend on $n$ and it grows to $\infty$ as $n\rightarrow\infty$.
\end{definition}

The factor model is commonly used to fit large-dimensional time series from economics and finance, when strong cross-sectional dependence is expected. Mathematically, $\Cov(X_t)=\Lambda \Cov(F_t) \Lambda^T +\Cov(Z_t)$, thus the strong cross-sectional dependence of $X_t$ originates from the low-rank part $\Lambda \Cov(F_t) \Lambda^{\top}$. Since both $\{F_t\}$ and $\{Z_t\}$ are weakly dependent over time and they are mutually independent, the original series $\{X_t\}$ also has weak temporal dependence.


\section{Asymptotic Theory Under the Null}\label{Sec:Theory-null}

In this section, we present the asymptotic theory for our proposed test statistic under the null hypothesis. We shall treat the three DGPs separately below as the arguments we adopt to derive the limiting null distribution are very different. This is different from \cite{kim2020dimensionagnostic}
where a unified treatment of low/high-dimensional settings is possible for the mean inference of iid data. In particular, they employ the uniform Berry-Esseen bound for the $t$ statistic, which is not available for our self-normalized change point test statistic in the time series setting. To this end, we shall develop a new conditioning argument to show that the limiting null distributions are identical across three regimes. 

Specifically, we shall show that under the null of constant mean, 
\begin{equation*}
    G_n
\stra{d} \sup\limits_{r\in[0,1]} \lrp{B(r) - rB(1)}V^{-1/2}(r) =: G,
\end{equation*}
where $\{B(r)\}_{r\in[0,1]}$ denotes the standard one-dimensional Brownian motion and $V(r)$ is given by
\begin{equation*}
    V(r) 
  = \int_{0}^{r} \lrp{B(s) - \frac{s}{r}B(r)}^2 ds 
  + \int_{r}^{1} \lrp{B(1) - B(s) - \frac{1-s}{1-r}\lrp{B(1) - B(r)}}^2 ds.
\end{equation*}
In practice, we reject the null hypothesis if the realized value of the test statistic exceeds the $1-\alpha$ quantile of $G$ at the significance level $\alpha\in (0,1)$. 
It is easy to see that the limiting null distribution (i.e., the distribution of $G$) is pivotal, and it can be  simulated by approximating the Brownian motion with standardized partial sum of iid N(0,1) random variables. Here we simulate the distribution of $G$ based on sample size of 5000 and 50000 Monte Carlo replicates, and summarize the critical values in Table \ref{Tab:critval}.
\begin{table}[h!]
    \centering
    \caption{Simulated critical values for $G$} \medskip
    \label{Tab:critval}
    \begin{tabular}{c|cccccc}
    \hline\hline
        Quantile Level & 90\% & 95\% & 97.5\% & 99\% & 99.5\% & 99.9\% \\ \hline
        Critical Value & 4.32 & 5.39 & 6.38 & 7.58 & 8.49 & 10.40 \\ 
    \hline\hline
    \end{tabular}
\end{table}

\subsection{Fixed-dimensional Stationary Sequence (DGP1) }

By using a novel conditioning argument and the property of the $p$-dimensional Brownian motion, we derive the limiting theory under the first DGP, which is formally summarized in the following theorem.

\begin{theorem}\label{Thm:NullDist_Gn_Fixedp}
Assume that $\{X_t-\mu_t\}_{t=1}^{n}$ is a stationary sequence as defined in Definition \ref{Def:Model_Fixedp}. Under the null, it holds that $G_n \stra{d} G$ when $p$ is fixed and $n\rightarrow\infty$.
\end{theorem}

\subsection{Growing-dimensional Linear Process (DGP2)}

Now we investigate the limiting null distribution of $G_n$ when the data is generated from a growing-dimensional linear process. To facilitate the subsequent analysis, we start from the simpler situation where $\{X_t-\mu_t\}_{t=1}^{n} \in \br^{p}$ is an iid sequence with mean zero and covariance matrix $\Sigma^{(2)} = (\Sigma^{(2)}_{ij})_{p\times p}$, and $\min\{p,n\} \rightarrow\infty$.  Note that the result for the iid setting will be a building block for our more general result for the linear process.  The following assumptions on $\Sigma^{(2)}$ are required to control the cross-sectional dependence within the data.

\begin{assumpt}\label{Assumpt:NullDist_Wn_iid_Growingp}
Assume that
\begin{enumerate}[label=(\roman*)]
    \item \label{Assumpt:NullDist_Wn_iid_Growingp-1}
    $\sum\limits_{\ell_1,\ell_2,\ell_3,\ell_4=1}^{p} \cum^2\lrp{X_{1,\ell_1}, X_{1,\ell_2}, X_{1,\ell_3}, X_{1,\ell_4}} = O\lrp{\|\Sigma^{(2)}\|_F^4}$.
    \item \label{Assumpt:NullDist_Wn_iid_Growingp-2}
    $\|\Sigma^{(2)}\| = o(\|\Sigma^{(2)}\|_F)$.
\end{enumerate}
\end{assumpt}

\begin{remark}
Assumption \ref{Assumpt:NullDist_Wn_iid_Growingp}\ref{Assumpt:NullDist_Wn_iid_Growingp-1} is also imposed in  \cite{wang2022inference}, and it is satisfied when the componentwise dependence within $X_t$ is weak, such as when the components of $X_t$ has AR(1) or banded correlation.  Assumption \ref{Assumpt:NullDist_Wn_iid_Growingp}\ref{Assumpt:NullDist_Wn_iid_Growingp-2} is equivalent to $tr\lrp{(\Sigma^{(2)})^4}=o(\|\Sigma^{(2)}\|_F^4)$ (see Remark 3.2 in \cite{wang2022inference}), which has been assumed in high-dimensional two-sample testing and change point testing; see \cite{chen2010two} and \cite{wang2022inference}, among others.
\end{remark}

For $r \in [\varepsilon, 1-\varepsilon]$, define the process as
\begin{equation*}
    W_n(r)
  = m_1 \sum\limits_{j=1}^{\lfloor{nr}\rfloor - \lfloor{n\varepsilon}\rfloor} Y_j
  = \sum\limits_{i=1}^{m_1} \lrp{X_i - X_{n+1-i}}^{\top} \lrp{\sum\limits_{j=1}^{\lfloor{nr}\rfloor - \lfloor{n\varepsilon}\rfloor} X_{j+m}}.
\end{equation*}

It follows that with the normalizing coefficient as $N_n = \sqrt{2 n m_1} \|\Sigma^{(2)}\|_F$, the asymptotic behavior of the normalized process $\{W_n(r)/N_n\}_{r\in[\varepsilon, 1-\varepsilon]}$ can be explicitly derived.

\begin{proposition}\label{Prop:NullDist_Wn}
Suppose the data $\{X_t\}_{t=1}^{n} \in \br^{p}$ is an iid sequence with mean zero and covariance matrix $\Sigma^{(2)} = (\Sigma^{(2)}_{ij})_{p\times p}$ and  Assumption \ref{Assumpt:NullDist_Wn_iid_Growingp} holds. Then it holds under the null that 
\begin{equation*}
    \frac{W_n(r)}{N_n} \leadsto B(r)-B(\varepsilon)~\mbox{in}~D[\varepsilon,1-\varepsilon]
\end{equation*}
\end{proposition}

With the auxiliary result summarized in Proposition \ref{Prop:NullDist_Wn} for the iid sequence, we are ready to establish a counterpart for the linear process via the Beveridge-Nelson decomposition [\cite{phillips1992asymptotics}]. To be specific, we define $A^{(j)} = \sum\limits_{\ell=j}^{\infty} a_\ell, j\in\bn$ and rewrite $\{X_t\}_{t=1}^{n}$ as
\begin{equation*}
    X_t = D_t - R_t, 
\end{equation*}
where $D_t =  A^{(0)}\varepsilon_t$, $\widetilde{D}_t = \sum\limits_{j=0}^{\infty} A^{(j+1)} \varepsilon_{t-j}$ and $R_t = \widetilde{D}_t - \widetilde{D}_{t-1}$. That is, the original linear process $X_t$ can be decomposed as an iid sequence $D_t$ minus a remainder term $R_t$.

To control both the temporal dependence and the cross-sectional dependence within the linear process $X_t$, we add some constraints on $\Gamma^{(2)}$ and $\{a_j\}_{j=0}^{\infty}$ through the following assumption.

\begin{assumpt}\label{Assumpt:NullDist_Wn_lp}
Assume that
\begin{enumerate}[label=(\roman*)]
    \item \label{Assumpt:NullDist_Wn_Lp-1}
    $\sum\limits_{k_1,\cdots,k_h=1}^{p} \cum^2(\varepsilon_{0,k_1},\cdots,\varepsilon_{0,k_h}) = O\lrp{\|\Gamma^{(2)}\|_F^h}$ for $h=1,2,3,4$;
    \item \label{Assumpt:NullDist_Wn_Lp-2}
    $\sum\limits_{k_1,\cdots,k_h=1}^{p} \lrabs{\cum(\varepsilon_{0,k_1},\cdots,\varepsilon_{0,k_h})} = O\lrp{\|\Gamma^{(2)}\|_F^h}$ for $h=1,\cdots,8$;
    \item \label{Assumpt:NullDist_Wn_Lp-3}
    $\|A^{(0)} \Gamma^{(2)} (A^{(0)})^{\top}\|_F = O_s\lrp{\|\Gamma^{(2)}\|_F}$ as $p\rightarrow\infty$;
    \item \label{Assumpt:NullDist_Wn_Lp-4}
    $\|\Gamma^{(2)}\| = o(\|\Gamma^{(2)}\|_F)$ as $p\rightarrow\infty$;
    \item \label{Assumpt:NullDist_Wn_Lp-5}
    $\sqrt{p} = O\lrp{\|\Gamma^{(2)}\|_F}$;
    \item \label{Assumpt:NullDist_Wn_Lp-6}
    $\blre{\varepsilon_{0,i}^8} \le C$ for $i=1,\cdots,p$ and some constant $C$;
    \item \label{Assumpt:NullDist_Wn_Lp-7}
    there exists some constant $\rho\in(0,1)$ s.t. $\|a_j\| \lesssim \rho^j$ for $j\in\bn$.
\end{enumerate}
\end{assumpt}

\begin{remark}
Assumption \ref{Assumpt:NullDist_Wn_lp}\ref{Assumpt:NullDist_Wn_Lp-1}, Assumption \ref{Assumpt:NullDist_Wn_lp}\ref{Assumpt:NullDist_Wn_Lp-2} and Assumption \ref{Assumpt:NullDist_Wn_lp}\ref{Assumpt:NullDist_Wn_Lp-4} can be jointly viewed as the counterpart of Assumption \ref{Assumpt:NullDist_Wn_iid_Growingp}, which extend the constraints for the iid data to the linear process. Assumption \ref{Assumpt:NullDist_Wn_lp}\ref{Assumpt:NullDist_Wn_Lp-3} and Assumption \ref{Assumpt:NullDist_Wn_lp}\ref{Assumpt:NullDist_Wn_Lp-5} specify the requirements on the Frobenious norm of the covariance matrix of the error $\Gamma^{(2)}$, which put some implicit conditions on the coefficient matrix $A^{(0)}$. 
Note that $\|A^{(0)}\Gamma^{(2)}(A^{(0)})^{\top}\|_F\le \|A^{(0)}\|^2\|\Gamma^{(2)}\|_F $, so  $\|A^{(0)} \Gamma^{(2)} (A^{(0)})^{\top}\|_F = O\lrp{\|\Gamma^{(2)}\|_F}$ if $\|A^{(0)}\|=O(1)$.
Furthermore, 
\begin{equation*}
    \|A^{(0)} \Gamma^{(2)} (A^{(0)})^{\top}\|_F^2 = \sum\limits_{1\le i,j,k,\ell \le p} (A^{(0)}_{ij})^2 (\Gamma^{(2)}_{jk})^2 (A^{(0)}_{\ell k})^2 = \sum\limits_{1\le j,k\le p} \|A^{(0)}_{\cdot j}\|_2^2 \|A^{(0)}_{\cdot k}\|_2^2 (\Gamma^{(2)}_{jk})^2,
\end{equation*}
where $A_{\cdot k}$ denotes the $k$-th column of the matrix $A$ and $\|\Gamma^{(2)}\|_F^2 = \sum\limits_{1\le j,k\le p} (\Gamma^{(2)}_{jk})^2$. Thus $\|\Gamma^{(2)}\|_F = O\lrp{\|A^{(0)} \Gamma^{(2)} (A^{(0)})^{\top}\|_F}$ if the matrix $A^{(0)}$ is banded with fixed bandwidth. Assumption \ref{Assumpt:NullDist_Wn_lp}\ref{Assumpt:NullDist_Wn_Lp-6} adds a uniform bound on the eighth moment of the innovation, whereas Assumption \ref{Assumpt:NullDist_Wn_lp}\ref{Assumpt:NullDist_Wn_Lp-7} assumes that the coefficient matrix $a_j$ decays exponentially in its spectral norm. Note that under Assumption \ref{Assumpt:NullDist_Wn_lp}\ref{Assumpt:NullDist_Wn_Lp-7}, it is trivial that $\|A^{(j)}\| \lesssim \rho^j$ for any $j\in\bn$.
\end{remark}

Define $\widetilde{N}_n = \sqrt{2n m_1} \|A^{(0)} \Gamma^{(2)} (A^{(0)})^{\top}\|_F$ as the normalizer for the linear process, then we can derive the process limit of $\{\frac{W_n(r)}{\widetilde{N}_n}\}$ under the null.

\begin{proposition}\label{Prop:NullDist_Wn_lp}
Under Assumption \ref{Assumpt:NullDist_Wn_lp}, if $\rho^{m_2/4}\|\Gamma^{(2)}\|_F = o\lrp{\frac{n}{\log(n)}}$, then it holds under the null that 
\begin{equation*}
    \frac{W_n(r)}{\widetilde{N}_n} \leadsto B(r)-B(\varepsilon)~\mbox{in}~D[\varepsilon,1-\varepsilon].
\end{equation*}
\end{proposition}

We summarize the derived result for the linear process in the following theorem.

\begin{theorem}\label{Thm:NullDist_Gn_Growingp_lp}
Assume that $\{X_t-\mu_t\}_{t=1}^{n}$ is a linear process as  defined in Definition \ref{Def:Model_lp}. Under Assumption \ref{Assumpt:NullDist_Wn_lp}, if $\rho^{m_2/4}\|\Gamma^{(2)}\|_F = o\lrp{\frac{n}{\log(n)}}$, then it holds under the null that $G_n \stra{d} G$ as $\min\{n,p\}\rightarrow\infty$.
\end{theorem}

From Theorem \ref{Thm:NullDist_Gn_Growingp_lp}, the desired limiting null distribution can be derived with an additional constraint between the sample size $n$ and the error covariance matrix $\Gamma^{(2)}$, which impose an implicit restriction on the growing rate of $p$ as a function of $n$. For example, when $\Gamma^{(2)}=I_p$, then the constraint reduces to $\rho^{m_2/4}p^{1/2} = o\lrp{\frac{n}{\log(n)}}$, which can be satisfied by $\log(p)=O(n)$. Therefore our restriction on $p$ is very mild. It also shows that the trimming we introduced in our procedure helps to
allow a broad range of $p$, as if $\eta=0$ and $m_2=0$ (i.e., no trimming), then 
the constraint becomes $p^{1/2} = o\lrp{\frac{n}{\log(n)}}$ so the growth rate of $p$ is quite limited. In finite sample, we also expect the trimming to help reduce the size distortion due to the bias caused by the temporal dependence in the data.

\subsection{Factor Model (DGP3)}

In this subsection we consider the last DGP in Definition ~\ref{Def:Model_factor}, when the data $\{X_t-\mu_t\}_{t=1}^{n} \in \br^{p}$ admits a factor model and the dimension $p$ is allowed to diverge as $n\rightarrow\infty$. Under the null, we assume without the loss of generality that $\mu_t = 0$ for $1\le t\le n$. In addition, assume that $Z_t = \sum\limits_{j=0}^{\infty} a_j \varepsilon_{t-j}$, where $\{\varepsilon_t\}_{t\in\bz}$ is an iid sequence with mean zero and covariance $\Gamma^{(3)}$. Here we slightly abuse the notation and use the notations introduced for the linear process case in Definition~\ref{Def:Model_lp}. Recall that $\Sigma^{(3)}=\var(Z_t)$. 

To derive the limiting null distribution, we require the following technical assumptions.

\begin{assumpt}\label{Assumpt:NullDist_Wn_factor}
Assume that
\begin{enumerate}[label=(\roman*)]
    \item \label{Assumpt:NullDist_Wn_factor-1}
    $\Lambda^{\top} \Sigma^{(3)} \Lambda$ is a positive definite matrix and it holds that  as $n\rightarrow\infty$, 
    \begin{equation*}
        \frac{1}{\sqrt{n}} \sum\limits_{i=1}^{\lrfl{nr}} \bin{\Lambda^{\top} Z_i}{F_i} 
    \leadsto \dmat{(\Lambda^{\top} \Sigma^{(3)} \Lambda)^{1/2}}{(\Omega^{(3)})^{1/2}} B_{2s}(r) 
    =^d \bin{x(r)}{y(r)}~\mbox{in}~D^{2s}[0,1], 
    \end{equation*}
    where $x(r) = (\Lambda^{\top} \Sigma^{(3)} \Lambda)^{1/2} B_s(r)$ and $y(r) =  (\Omega^{(3)})^{1/2} \tilde{B}_s(r)$, and $\{B_s(r): 0 \le r \le 1\} \indept \{\tilde{B}_s(r):0 \le r \le 1\}$ represent two independent standard $s$-dimensional Brownian motions.
    \item \label{Assumpt:NullDist_Wn_factor-2}
    $\|\Sigma^{(3)}\| = o\lrp{\|\Gamma^{(3)}\|_F}$.
    \item \label{Assumpt:NullDist_Wn_factor-3}
    there exists some $L_0\in\br^{s\times s}$ that is independent of $p$, such that $\frac{\Lambda^{\top} \Lambda}{\|\Lambda^{\top} \Lambda\|_F} \rightarrow L_0$ as $p \rightarrow \infty$.
    \item \label{Assumpt:NullDist_Wn_factor-4}
    there exists some $L_1\in\br^{s\times s}$ that is independent of $p$, such that 
    \begin{equation*}
        \frac{(\Lambda^{\top} \Sigma^{(3)} \Lambda)^{1/2}}{\|(\Lambda^{\top} \Sigma^{(3)} \Lambda)^{1/2}\|_F}
      = \frac{(\Lambda^{\top} \Sigma^{(3)} \Lambda)^{1/2}}{\sqrt{\tr\lrp{\Lambda^{\top} \Sigma^{(3)} \Lambda}}}
      \rightarrow L_1
    \end{equation*}
    as $p \rightarrow \infty$.
\end{enumerate}
\end{assumpt}

\begin{remark}
Assumption \ref{Assumpt:NullDist_Wn_factor}\ref{Assumpt:NullDist_Wn_factor-1} basically requires that both $\Lambda^{\top} Z_i$ and $F_i$ satisfy the functional central limit theorem. The joint process convergence result stated in Assumption \ref{Assumpt:NullDist_Wn_factor}\ref{Assumpt:NullDist_Wn_factor-1} then follows since we assume the full independence between $\{Z_i\}$ and $\{F_i\}$. Assumption \ref{Assumpt:NullDist_Wn_factor}\ref{Assumpt:NullDist_Wn_factor-2} is similar to Assumption \ref{Assumpt:NullDist_Wn_lp}\ref{Assumpt:NullDist_Wn_Lp-4}, though negligibility is  for the covariance matrix of the linear process instead of that of the error terms. Assumption \ref{Assumpt:NullDist_Wn_factor}\ref{Assumpt:NullDist_Wn_factor-3}-\ref{Assumpt:NullDist_Wn_factor-4} imply that, as the dimension $p$ diverges to infinity, both $\Lambda^{\top} \Lambda$ and $(\Lambda^{\top} \Sigma^{(3)} \Lambda)^{1/2}$  converge to a fixed matrix when standardized by their Frobenius norms.
\end{remark}

Observe that the factor model consists of a fixed-dimensional low rank component $F_t$ and a growing-dimensional linear process error $Z_t$, both of which have been investigated separately in previous sections. Following a similar argument as used before, we can show that the limiting null  distribution of our test statistic under the factor model, as summarized in Theorem \ref{Thm:NullDist_Gn_factor}. 

\begin{theorem}\label{Thm:NullDist_Gn_factor}
Assume that $\{X_t-\mu_t\}_{t=1}^{n}$ is generated from  a factor model as defined in Definition \ref{Def:Model_factor}. Under Assumption \ref{Assumpt:NullDist_Wn_lp} (applied to $Z_t$) and Assumption \ref{Assumpt:NullDist_Wn_factor}, if $\rho^{m_2/4}\|\Gamma^{(3)}\|_F = o\lrp{\frac{n}{\log(n)}}$, it holds under the null that $G_n \stra{d} G$ as $\min\{n,p\}\rightarrow\infty$.
\end{theorem}

In summary, the theoretical results in this section 
show that our test statistic $G_n$ converges to the same limiting null distribution $G$ under the three DGPs considered, and our test is not only dimension-agnostic but also robust to both weak/strong cross-sectional dependence and weak temporal dependence.


\section{Asymptotic Theory Under the Alternative}\label{Sec:Theory-power}

In this section we investigate the asymptotic power of our proposed test when there exists a single change point in mean. Recall that $k_0=\lrfl{n \varepsilon_0}$ denotes the location of the change point satisfying that $\varepsilon < \varepsilon_0 < 1-\varepsilon$, we consider the scenario
\begin{equation*}
    \mu_1 = \cdots = \mu_{k_0} = \mu, \qquad
    \mu_{k_0+1} = \cdots = \mu_{n} = \mu+\delta,
\end{equation*}
where $\delta$ is the mean shift. In the following we let $\Delta = \Delta_n = \sqrt{n} \delta$, and denote $r_0=\lim_{n\rightarrow\infty} \frac{k_0-m}{N} =  \frac{\varepsilon_0-\varepsilon}{1-2\varepsilon}$.

To facilitate the subsequent analysis, we use $\{\tilde{X}_t\}_{t=1}^{n} = \{X_t-\mu_t\}_{t=1}^{n} = \{X_t - \mu - \delta \bone\{t>k_0\}\}_{t=1}^{n}$ to denote the centered version of $\{X_t\}_{t=1}^{n}$. Let $\hat{\nu}_1 = \frac{1}{m_1} \sum\limits_{i=1}^{m_1} \tilde{X}_i$ and $\hat{\nu}_n = \frac{1}{m_1} \sum\limits_{i=1}^{m_1} \tilde{X}_{n+1-i}$. Recall that $N=n-2m$, $m=\lfloor n\epsilon\rfloor=m_1+m_2$. 
In the following proposition, we can express the  statistics $T_n(k)$ and $V_n(k)$ in terms of $\{\tilde{X}_t\}_{t=1}^{n}$.

\begin{proposition}\label{Prop:PowerExpression}
Under the alternative, we have that for $k=1,\cdots,N-1$,  
\BEqn
    N^{1/2} T_n(k)
&=& \lrag{\hat\nu_1-\hat\nu_n-\delta, 
    \sum\limits_{j=1}^{k}\tilde{X}_{j+m} - \frac{k}{N} \sum\limits_{j=1}^{N}\tilde{X}_{j+m}} \\
& & - \frac{\lrp{(k_0-m)\wedge k} \lrp{(N-k_0+m)\wedge(N-k)}}{N} \lrag{\hat\nu_1-\hat\nu_n-\delta,\delta}, \\
    N^2 V_n(k)
&=& \sum\limits_{t=1}^{k} 
    \left(
    \lrag{\hat\nu_1-\hat\nu_n-\delta, \sum\limits_{j=1}^{t}\tilde{X}_{j+m} - \frac{t}{k}\sum\limits_{j=1}^{k}\tilde{X}_{j+m}}
    \right. \\
& & \hspace{-1em}
    \left.
    - \frac{\lrbk{\lrp{(k_0-m)\wedge t} \lrp{(k-k_0+m)\wedge(k-t)}} \vee 0}{k} \lrag{\hat\nu_1-\hat\nu_n-\delta,\delta}
    \right)^2 \\
&+& \sum\limits_{t=k+1}^{N}
    \left(
    \lrag{\hat\nu_1-\hat\nu_n-\delta, \sum\limits_{j=t}^{N}\tilde{X}_{j+m} - \frac{N-t+1}{N-k}\sum\limits_{j=k+1}^{N}\tilde{X}_{j+m}}
    \right. \\
& & \hspace{-1em}
    \left.
    + \frac{\lrbk{\lrp{(k_0-m-k)\wedge (t+1-k)} \lrp{(N-k_0+m)\wedge(N-t+1)}} \vee 0}{N-k} \lrag{\hat\nu_1-\hat\nu_n-\delta,\delta}
    \right)^2.
\EEqn
\end{proposition}

With Proposition \ref{Prop:PowerExpression}, we are ready to derive the asymptotic behavior of the proposed test for each data-generating process. 
To facilitate the analysis, we define
\BEqn
    T(r,d) 
&=& B(r) - rB(1) - \lrp{r_0\wedge r} \lrp{(1-r_0)\wedge(1-r)} d, \\
    V(r,d) 
&=& \int_{0}^{r} \lrp{ B(s)-\frac{s}{r}B(r) - \frac{\lrp{s \wedge r_0} \lrp{(r-s) \wedge (r-r_0)} \vee 0}{r} d }^2 ds \\
&+& \int_{r}^{1} \lrp{ B(1-s)-\frac{1-s}{1-r}B(1-r) + \frac{\lrp{(1-s) \wedge (1-r_0)} \lrp{(s-r) \wedge (r_0-r)} \vee 0}{1-r} d }^2 ds,
\EEqn
where $d$ denotes some quantity that may depend on other variables. We further define 
\begin{equation}\label{Equ:Power}
    M(d) = \sup\limits_{r\in[0,1]} T(r,d) V^{-1/2}(r,d).
\end{equation}

\subsection{Fixed-dimensional Stationary Sequence (DGP1)}

When $\{\tilde{X}_t\}_{t=1}^{n} \in \br^{p}$ is a stationary sequence as defined in Definition \ref{Def:Model_Fixedp}, we can show in Theorem \ref{Thm:DensePower_Fixedp} that the asymptotic power of the proposed test depends on the limit of $\|\Delta\|_2$ when $p$ is fixed.

\begin{theorem}\label{Thm:DensePower_Fixedp}
Suppose that $\{\tilde{X}_t\}_{t=1}^{n}$ is a stationary sequence as defined in Definition \ref{Def:Model_Fixedp}, then it holds that,
\begin{enumerate}[label=(\roman*)]
    \item \label{Thm:DensePower_Fixedp-1}
    if $\|\Delta\|_2 \rightarrow 0$ as $n\rightarrow\infty$, we have that $\bbp{G_n>G_{1-\alpha}} \rightarrow \alpha$.
    \item \label{Thm:DensePower_Fixedp-2}
    if $\|\Delta\|_2 \rightarrow \infty$ as $n\rightarrow\infty$, we have that $\bbp{G_n>G_{1-\alpha}} \rightarrow 1$.
    \item \label{Thm:DensePower_Fixedp-3}
    if $\|\Delta\|_2 \rightarrow c \in (0,\infty)$ as $n\rightarrow\infty$, and assume that $\frac{\Delta}{\|\Delta\|_2} \rightarrow \Delta_0$, where $\Delta_0$ is a $p$-dimensional vector independent of $n$, then it holds that
    \begin{equation*}
        \blrp{G_n > G_{1-\alpha}}
    \rightarrow \int_{b_0\in\br^{p}} \lrp{\frac{4\pi}{\varepsilon-\eta}}^{-p/2} \exp\lrp{-\frac{1}{4}(\varepsilon-\eta) b_0^{\top}b_0} \blrp{M(d) > G_{1-\alpha}} db_0,
    \end{equation*}
    where 
    $M(d)$ is defined as Equation (\ref{Equ:Power}) with $d = d(\Delta,b_0) = c \sqrt{\frac{1-2\varepsilon}{u^{\top}\Omega^{(1)} u}} u^{\top} \Delta_0$ and $u = u(b_0,\Delta) = (\Omega^{(1)})^{1/2}b_0 - c\Delta_0$.  
\end{enumerate}
\end{theorem}

According to Theorem \ref{Thm:DensePower_Fixedp}, there are three regimes in the asymptotic power analysis. When $\Delta$ converges to zero or diverges to infinity in $L_2$ norm,  we  obtain asymptotically power $\alpha$ and power one respectively. In the intermediate case when the limit of $\|\Delta\|_2$ is a strictly positive constant,  the explicit formula of the asymptotic power is given. As we can see that the power depends on $\Omega^{(1)}$, $p$, $c$, $\Delta_0$, $\varepsilon$ and $\eta$. Note that the explicit form for the local asymptotic power was not derived for the SN test statistic in \cite{shao2010testing}. Nevertheless we can approximate their local asymptotic powers and compare them through simulations.

\begin{figure}[h!]
    \centering
    \caption{Power curves of SS-SN and SN test in \cite{shao2010testing} under DGP1} \medskip
    \label{Fig:PowerPlot-DGP1}
    \includegraphics[width=0.8\textwidth]{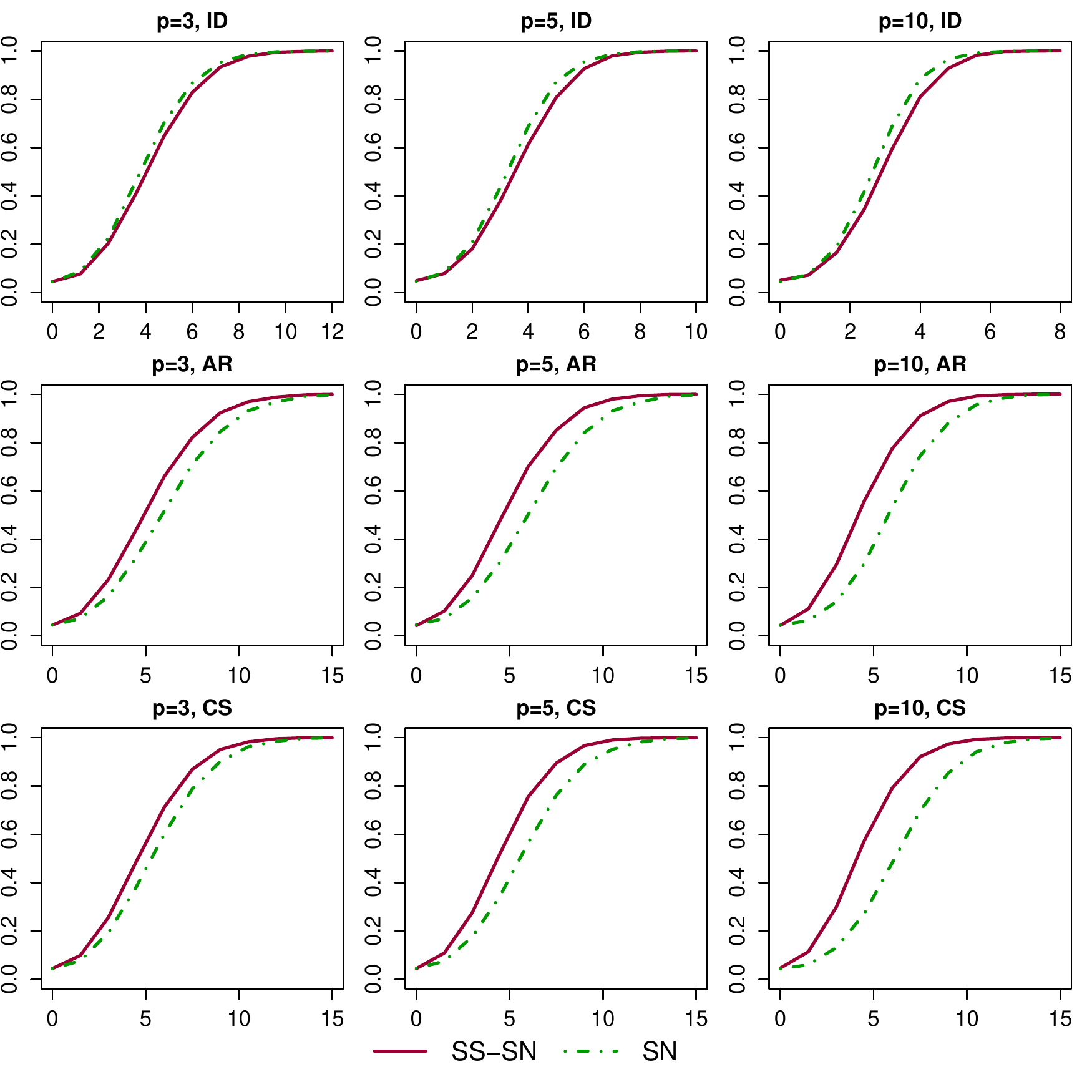}
\end{figure}

To approximate the local asymptotic powers, we generate a sample of $n=1000$ iid observations from the $p$-dimensional normal distribution with covariance matrix $\Omega^{(1)}$. We focus on a single mean shift at the location $k_0=\lrfl{n/2}$, that is, the first $\lrfl{n/2}$ observations in this sample have mean zero whereas the remaining observations have mean $\delta$. We consider $p\in\{3,5,10\}$ and set $\delta = c(1,\cdots,1)^{\top}/\sqrt{n}$ for a sequence of $c$'s. As for the covariance matrix, we consider three different types, namely, (1) ID: $\Omega^{(1)} = I_p$; (2) AR: $\Omega^{(1)} = (\Omega^{(1)}_{ij})_{p\times p}$ with $\Omega^{(1)}_{ij}=0.8^{|i-j|}$; (3) CS: $\Omega^{(1)} = (\Omega^{(1)}_{ij})_{p\times p}$ with $\Omega^{(1)}_{ij}=0.5+0.5\bone\{i=j\}$. We fix the splitting ratio $\varepsilon=0.1$ and the trimming ratio $\eta=0.02$ for the proposed method, and compare its power with that of \cite{shao2010testing}. The power curves against the value of $c$ are plotted in Figure \ref{Fig:PowerPlot-DGP1}, which are based on 5000 Monte Carlo replicates.

The three rows of Figure \ref{Fig:PowerPlot-DGP1} correspond to the three types of $\Omega^{(1)}$ whereas the three columns refer to $p\in\{3,5,10\}$. When compared to \texttt{SN}, our proposed method has some mild power loss when there is no componentwise dependence within the data, i.e., when $\Omega^{(1)}=I_p.$ 
Somewhat surprisingly, our method gains some advantage in the AR and compound symmetric cases and the power again seems to grow with respect to dimension. 
Therefore, contrary to the phenomenon discovered by \cite{kim2020dimensionagnostic}, our dimension agnostic test statistic does not always incur power loss, and when it does lose/gain 
power and by how much very much depend on the dimension and the  dependence structure in the data.

\subsection{Growing-dimensional Linear Process (DGP2)}

Now we consider the case when $\{\tilde{X}_t\}_{t=1}^{n}$ is a linear process as defined in Definition \ref{Def:Model_lp}. Again, based on the results in Proposition~\ref{Prop:PowerExpression}, we can  derive the limiting distributions of $T_n(k)$ and $V_n(k)$, and   present the asymptotic power of the proposed test below. As expected, there are  three different regimes based on the limit of $\frac{\|\Delta\|_2}{\|A^{(0)} \Gamma^{(2)} (A^{(0)})^{\top}\|_F^{1/2}}$.

\begin{theorem}\label{Thm:DensePower_lp}
Suppose that $\{\tilde{X}_t\}_{t=1}^{n}$ is a linear process as defined in Definition \ref{Def:Model_lp} and Assumption \ref{Assumpt:NullDist_Wn_lp} holds. If $\rho^{m_2/4}\|\Gamma^{(2)}\|_F = o\lrp{\frac{n}{\log(n)}}$, then it holds that,
\begin{enumerate}[label=(\roman*)]
    \item \label{Thm:DensePower_lp-1}
    if $\displaystyle\frac{\|\Delta\|_2^2}{\|A^{(0)} \Gamma^{(2)} (A^{(0)})^{\top}\|_F} \rightarrow 0$ as $\min\{n,p\}\rightarrow\infty$, we have that $\blrp{G_n > G_{1-\alpha}} \rightarrow \alpha$.
    \item \label{Thm:DensePower_lp-2}
    if $\displaystyle\frac{\|\Delta\|_2^2}{\|A^{(0)} \Gamma^{(2)} (A^{(0)})^{\top}\|_F} \rightarrow \infty$ as $\min\{n,p\}\rightarrow\infty$, we have that $\blrp{G_n > G_{1-\alpha}} \rightarrow 1$.
    \item \label{Thm:DensePower_lp-3}
    if $\displaystyle\frac{\|\Delta\|_2^2}{\|A^{(0)} \Gamma^{(2)} (A^{(0)})^{\top}\|_F} \rightarrow c \in (0,\infty)$ as $\min\{n,p\}\rightarrow\infty$,  we have that $\blrp{G_n > G_{1-\alpha}} \rightarrow \blrp{M(d) > G_{1-\alpha}}$, where 
    $M(d)$ is defined as Equation (\ref{Equ:Power}) with $d = -c\sqrt{\frac{(1-2\varepsilon)(\varepsilon-\eta)}{2}}$.    
\end{enumerate}
\end{theorem}

From the above results, we can see that the power depends on $c$, $\varepsilon$, $\eta$ and $r_0$. Compared to the local asymptotic power results in \cite{wang2022inference}, it is interesting to note that their test's asymptotic power is also discussed according to the limit of $\frac{\|\Delta\|_2}{\|A^{(0)} \Gamma^{(2)} (A^{(0)})^{\top}\|_F^{1/2}}$, which indicates the signal-noise-ratio. 

\begin{figure}[h!]
    \centering
    \caption{Power curves of SS-SN and $T(\eta_0)$ in \cite{wang2022inference} under DGP2} \medskip
    \label{Fig:PowerPlot-DGP2}
    \includegraphics[width=0.6\textwidth]{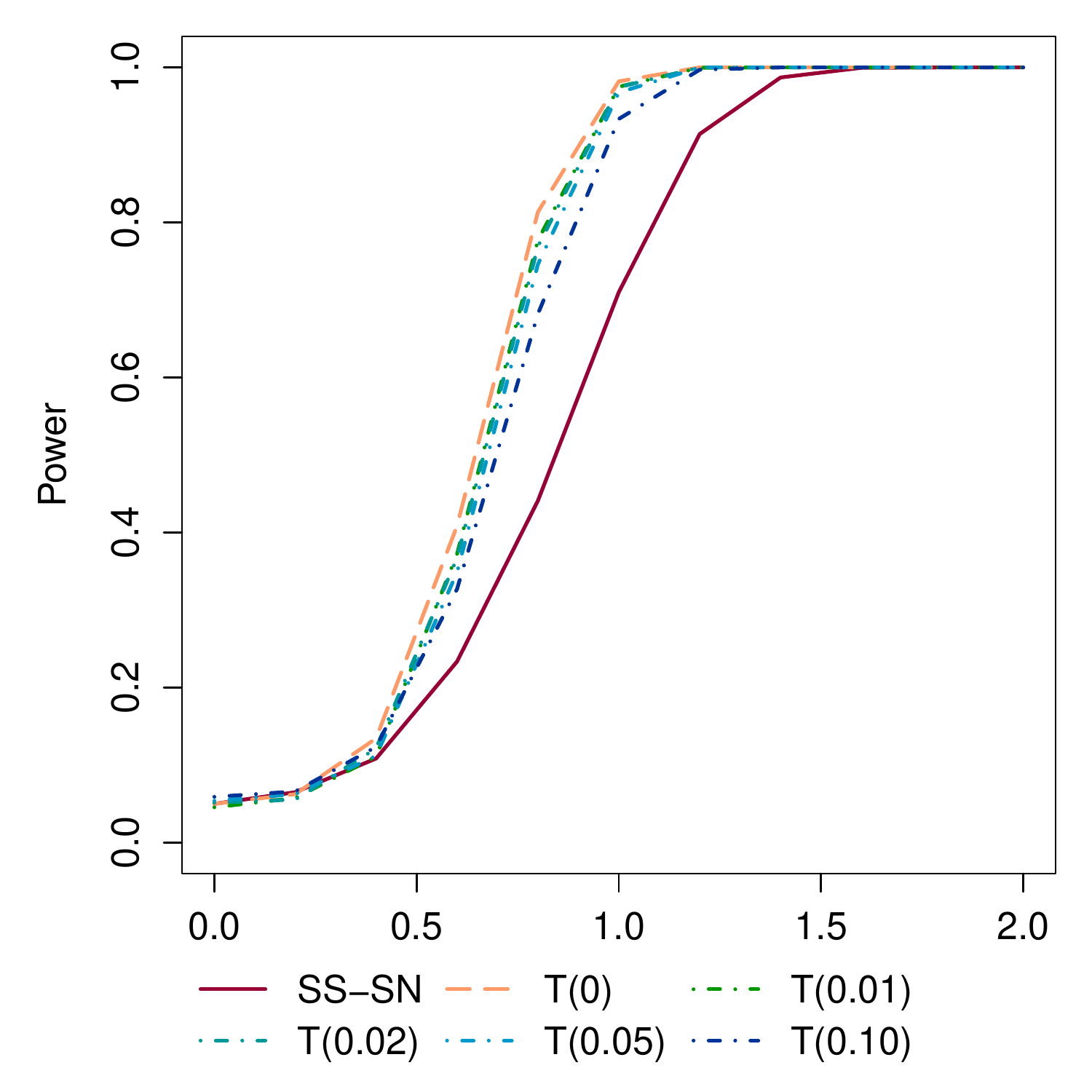}
\end{figure}

To compare the power curve in the intermediate case, we perform the simulations by generating a sample of $n$ iid observations from the $p$-dimensional normal distribution with covariance matrix $I_{p}$. Here we 
fix $n=p=1000$ and generate a single mean shift $\delta = c(1,\cdots,1)^{\top}/\sqrt{n}$ at the location $k_0=\lrfl{n/2}$. For the proposed method, we set the splitting ratio $\varepsilon=0.1$ and the trimming ratio $\eta=0.02$. As for the comparison, we adopt the U-statistic-based trimming method $T(\eta_0)$ proposed in \cite{wang2022inference} with their
trimming parameter $\eta_0 \in \{0.01, 0.02, 0.05, 0.10\}$ as well as the counterpart without trimming. The power curves against the value of $c$ are plotted in Figure \ref{Fig:PowerPlot-DGP2}, which are based on 2000 Monte Carlo replicates due to the expensive computation cost of \cite{wang2022inference}.

According to Figure \ref{Fig:PowerPlot-DGP2}, the method with no trimming, i.e., $T(0)$ beats all the others in power. The trimmed U statistic based 
tests in \cite{wang2022inference} have comparable performance and we observe an increasing power loss as the trimming parameter $\eta_0$ grows. Intuitively, 
with a larger trimming parameter, less pairs 
of observations are used in the trimmed 
U-statistics of \cite{wang2022inference}, so it is expected 
 to lose some efficiency. 
When compared to \cite{wang2022inference}, the proposed SS-SN method in this article has the least power and the power loss is moderate. This is consistent with the finding in \cite{kim2020dimensionagnostic}, who showed the power loss of their cross U-statistic based test 
in the high-dimensional setting. This is a reasonable
price to pay for the size accuracy across a wide range of dimensions.

\subsection{Factor Model (DGP3)}

The last case to consider is when the data admits a factor model as defined in Definition \ref{Def:Model_factor}. In this case, we need some additional assumptions to ensure the convergence of the test statistic.

\begin{assumpt}\label{Assumpt:DensePower_factor}
Assume that there exists some $\lambda_0\in\br^s$ that is independent of $p$, such that $\frac{\Lambda^{\top} \Delta}{\|\Lambda^{\top} \Delta\|_2} \rightarrow \lambda_0$ as $p \rightarrow \infty$.
\end{assumpt}

\begin{theorem}\label{Thm:DensePower_factor}
Suppose that $\{\tilde{X}_t\}_{t=1}^{n}$ is generated from the factor model as defined in Definition \ref{Def:Model_factor} and Assumption \ref{Assumpt:NullDist_Wn_lp} (applied to $Z_t$), 
Assumption \ref{Assumpt:NullDist_Wn_factor}, and Assumption \ref{Assumpt:DensePower_factor} hold. If $\rho^{m_2/4}\|\Gamma^{(3)}\|_F = o\lrp{\frac{n}{\log(n)}}$, then it holds that,
\begin{enumerate}[label=(\roman*)]
    \item \label{Thm:DensePower_factor-1}
    if $\|\Delta\|_2 = o\lrp{\max\{\|\Lambda\|,\|\Gamma^{(3)}\|_F^{1/2}\}}$, we have that $\bbp{G_n>G_{1-\alpha}} \rightarrow \alpha$ as $\min\{n,p\}\rightarrow\infty$.
    \item \label{Thm:DensePower_factor-2}
    if $\max\{\|\Lambda\|,\|\Gamma^{(3)}\|_F^{1/2}\} = o\lrp{\|\Delta\|_2}$, we have that $\bbp{G_n>G_{1-\alpha}} \rightarrow 1$ as $\min\{n,p\}\rightarrow\infty$.
    
    \item \label{Thm:DensePower_factor-3}
    if $\|\Delta\|_2 \sim \max\{\|\Lambda\|,\|\Gamma^{(3)}\|_F^{1/2}\}$,
    \begin{enumerate}[label=(\arabic*)]
        \item \label{Thm:DensePower_factor-3-1}
        when $\|\Lambda\| = o\lrp{\|\Gamma^{(3)}\|_F^{1/2}}$, and we additionally assume that 
        \begin{equation*}
            \frac{\|\Delta\|_2^2}{\sqrt{2(\varepsilon-\eta)} \|A^{(0)} \Gamma^{(3)} (A^{(0)})^{\top}\|_F}
        \rightarrow c:=c(\Delta),
        \end{equation*}
        then we have that $\blrp{G_n > G_{1-\alpha}} \rightarrow \blrp{M(d) > G_{1-\alpha}}$, 
        where $M(d)$ is defined as Equation (\ref{Equ:Power}) with $d = d(\Delta) = -c \sqrt{\frac{(1-2\varepsilon)(\varepsilon-\eta)}{2}}$.
        
        \item \label{Thm:DensePower_factor-3-2}
        when $\|\Gamma^{(3)}\|_F^{1/2} = O\lrp{\|\Lambda\|}$, we assume that
        \begin{equation*}
            \frac{\sqrt{2(\varepsilon-\eta)}\|A^{(0)} \Gamma^{(3)} (A^{(0)})^{\top}\|_F}{\|\Lambda^{\top}\Lambda\|} 
        \rightarrow c_1:=c_1(\Delta),
        \end{equation*}
        and $\displaystyle\frac{\|\Lambda^{\top} \Delta\|_2}{\|\Lambda^{\top}\Lambda\|} \rightarrow c_2:=c_2(\Delta)$ as well as $\displaystyle\frac{\|\Delta\|_2^2}{\|\Lambda^{\top}\Lambda\|} \rightarrow c_3:=c_3(\Delta)$. Then it holds that
        \begin{equation*}
            \blrp{G_n > G_{1-\alpha}}
        \rightarrow \int_{b_0\in\br^{p}} \lrp{\frac{4\pi}{\varepsilon-\eta}}^{-p/2} \exp\lrp{-\frac{1}{4}(\varepsilon-\eta) b_0^{\top}b_0} \blrp{M(d) > G_{1-\alpha}} db_0,
        \end{equation*}
        where $M(d)$ is defined as Equation (\ref{Equ:Power}) with $d = \omega_3\sqrt{\frac{1-2\varepsilon}{\omega_1^2+\omega_2^2}}$, where \begin{equation*}
            \omega_1 
          = \sqrt{\lrp{b_0^{\top} ((\Omega^{(3)})^{1/2})^{\top} L_0 - c_2 \lambda_0^{\top}} \Omega^{(3)} \lrp{b_0^{\top} ((\Omega^{(3)})^{1/2})^{\top} L_0 - c_2 \lambda_0^{\top}}^{\top}},
        \end{equation*}
        $\displaystyle\omega_2 = \frac{c_1}{\varepsilon-\eta}$, and $\omega_3 = c_2 \lambda_0^{\top} (\Omega^{(3)})^{1/2}$.     
    \end{enumerate}    
\end{enumerate}
\end{theorem}

Again, the asymptotic behavior of $G_n$ can be divided into three cases depending on the relationship between $\|\Delta\|_2$ and $\max\{\|\Lambda\|,\|\Gamma^{(3)}\|_F^{1/2}\}$. The most involved case is when $\|\Delta\|_2$ and $\max\{\|\Lambda\|,\|\Gamma^{(3)}\|_F^{1/2}\}$ have exactly the same order, which leads to nontrivial asymptotic power. In this intermediate case, the expression of local asymptotic power can be obtained in two separate scenarios: 
(1) When $\|\Gamma^{(3)}\|_F^{1/2}$ dominates $\|\Lambda\|$, the noise component $Z_t$ becomes the leading term, so it is not surprising to find that the limiting distribution of $G_n$ in this case matches that in Theorem \ref{Thm:DensePower_lp}\ref{Thm:DensePower_lp-3}. 
(2) When $\|\Gamma^{(3)}\|_F^{1/2} = O\lrp{\|\Lambda\|}$, this corresponds to the case the noise component is not the leading term, the expression is more complicated since both the low rank part $\Lambda F_t$ and the noise part $Z_t$ can be non-negligible. 

It is worth noting that the validity of the SN-based test in \cite{wang2022inference} requires weak cross-sectional dependence, and there seems few tests for high-dimensional time series that allows for strong cross-sectional dependence in the literature, with the exception of \cite{horvath2022}. In the latter paper, the authors developed a mean change point test tailored to time series generated from a factor model. The test is built on the basis of random centering applied to the original CUMSUM process and utilizes a bootstrap procedure to approximate the non-pivotal limiting null. By contrast, our test is much faster to implement due to the use of a simulated critical value. Also our test aims to be robust to both weak/strong cross-sectional dependence and the dimensionality (fixed and growing). 

To conclude this section, we summarize the power analysis results in Table \ref{Tab:PoweAnalysis}, which is consistent with our intuition. Specifically, the power is dependent on the signal-noise-ratio, which is proportional to $\|\Delta\|_2$, $\displaystyle \frac{\|\Delta\|_2}{\|A^{(0)} \Gamma^{(2)} (A^{(0)})^{\top}\|_F^{1/2}}$, and  $\displaystyle \frac{\|\Delta\|_2}{\max\{\|\Lambda\|,\|\Gamma^{(3)}\|_F^{1/2}\}}$ for DGP1, DGP2 and DGP3, respectively. The asymptotic powers are $\alpha$, $\beta\in (\alpha,1)$ and $1$, when the signal-noise-ratio goes to $0$, $c>0$ and $\infty$, respectively.

\begin{table}[h!]
    \centering
    \caption{Summary of asymptotic power analysis for three DGPs} \medskip
    \label{Tab:PoweAnalysis}
    \renewcommand{\arraystretch}{1.15} 
    \begin{tabular}{c|c|c|c}
    \hline\hline
        & DGP1  & DGP2 & DGP3 \\ \hline
        \multirow{2}{*}{$\blrp{G_n>G_{1-\alpha}} \rightarrow \alpha$} & \multirow{2}{*}{$\|\Delta\|_2\rightarrow0$} &
        \multirow{2}{*}{$\frac{\|\Delta\|_2}{\|A^{(0)} \Gamma^{(2)} (A^{(0)})^{\top}\|_F^{1/2}} \rightarrow 0$} & \multirow{2}{*}{$\frac{\|\Delta\|_2}{\max\{\|\Lambda\|,\|\Gamma^{(3)}\|_F^{1/2}\}} \rightarrow 0$} \\
        & & & \\ \hline
        \multirow{2}{*}{$\blrp{G_n>G_{1-\alpha}} \rightarrow \beta$} & \multirow{2}{*}{$\|\Delta\|_2\rightarrow c_1$} & \multirow{2}{*}{$\frac{\|\Delta\|_2}{\|A^{(0)} \Gamma^{(2)} (A^{(0)})^{\top}\|_F^{1/2}} \rightarrow c_2$} & \multirow{2}{*}{$\frac{\|\Delta\|_2}{\max\{\|\Lambda\|,\|\Gamma^{(3)}\|_F^{1/2}\}} \rightarrow c_3$} \\ 
        & & & \\
        $\beta\in(\alpha,1)$ & $c_1\in(0,\infty)$ & $c_2\in (0,\infty)$ & $c_3\in (0,\infty)$ \\ \hline
        \multirow{2}{*}{$\blrp{G_n>G_{1-\alpha}} \rightarrow 1$} & \multirow{2}{*}{$\|\Delta\|_2\rightarrow \infty$} & \multirow{2}{*}{$\frac{\|\Delta\|_2}{\|A^{(0)} \Gamma^{(2)} (A^{(0)})^{\top}\|_F^{1/2}} \rightarrow \infty$} & \multirow{2}{*}{$\frac{\|\Delta\|_2}{\max\{\|\Lambda\|,\|\Gamma^{(3)}\|_F^{1/2}\}} \rightarrow \infty$} \\ 
        & & & \\
    \hline\hline
    \end{tabular}
\end{table}


According to Table \ref{Tab:PoweAnalysis}, the power is dependent on the signal-noise-ratio, which is proportional to $\|\Delta\|_2$, $\displaystyle \frac{\|\Delta\|_2}{\|A^{(0)} \Gamma^{(2)} (A^{(0)})^{\top}\|_F^{1/2}}$, and  $\displaystyle \frac{\|\Delta\|_2}{\max\{\|\Lambda\|,\|\Gamma^{(3)}\|_F^{1/2}\}}$ for DGP1, DGP2 and DGP3, respectively. The asymptotic powers are $\alpha$, $\beta\in (\alpha,1)$ and $1$, when the signal-noise-ratio goes to $0$, $c>0$ and $\infty$, respectively. It is worth noting that our proposed test targets the dense alternative in the mean change of a multivariate time series. This can be well motivated by real data and is often the type of alternative we are interested in. For example, the financial crisis is expected to have an impact on a large number of sectors and their stock returns, so a dense change is expected if we study the stock returns time series for many sectors; see Section~\ref{Sec:NumResults-real} for an illustration. In genomic data analysis, detecting change-points in copy number variations in cancer cells is of great importance, and  change-points occurring at the same positions across many related data sequences corresponding to cancer samples are of particular interests as these change points can indicate cancer-related genetic loci; see \cite{fanMackey2017}. In the context of mean change testing for high-dimensional time series (i.e., with temporal dependence), our proposed test seems to be the first one that are able to capture dense mean change and are asymptotically valid for time series with either weak or strong cross-sectional dependence. As the amount of cross-sectional and temporal dependence is often unknown in practice, this robustness is desirable.


\section{Numerical Results}\label{Sec:NumResults-simul}

In this section, we examine the finite sample performance of the proposed methods in comparison with some existing SN-based ones in simulated studies. We only include SN-based ones into the comparison since our test statistic uses the self-normalizer proposed in \cite{shao2010testing} to the projected data, and restricting to SN-based tests  helps to make the comparison more interpretable. 
Under various dimensional settings and for several data-generating processes, Section \ref{Sec:NumResults-simul-size-singleCP} reports the empirical size accuracy whereas Section \ref{Sec:NumResults-simul-power-singleCP} investigates the power behavior. Some additional simulation results are reported in the supplement.

Throughout, the simulated data is generated from a $p$-dimensional AR(1) process, that is,
$    X_t - \mu_t = \kappa (X_{t-1} - \mu_{t-1}) + \epsilon_{t},~ 1 \le t \le n,$
where $\{\epsilon_t\}_{t=1}^{n}$ are iid $p$-dimensional multivariate normal random vectors with mean zero and variance $\Sigma$. Three structures for $\Sigma$ are considered, namely, (1) AR ($\Sigma_{i,j} = \rho^{|i-j|}$); (2) CS ($\Sigma_{i,j} = 0.5 + 0.5\fone\{i=j\}$); and (3) ID ($\Sigma_{i,j} = \fone\{i=j\}$). These three models correspond to weak cross-sectional dependence, strong cross-sectional dependence and independence across components, respectively. 

We denote the proposed test statistic as SS-SN and set the splitting parameter $\varepsilon=0.1$ and the trimming parameter $\eta=0.04$. As for the comparison, we consider the SN-based trimming test statistic proposed in Section 4 of \cite{wang2022inference}, which is denoted by $T(\eta_0)$ with $\eta_0$ being the trimming parameter. We consider the trimming parameter $\eta_0\in\{0,0.01,0.02,0.05,0.1\}$. Note that the theory in \cite{wang2022inference} requires $p$ to grow to infinity thus it is tailored to high-dimensional data, and its performance in the low-dimensional setting is unknown. By contrast, our proposed test is supposed to be dimension agnostic.

\subsection{Empirical Size}\label{Sec:NumResults-simul-size-singleCP}

\begin{figure}[h!]
    \centering
    \caption{Empirical size curves versus the logarithm of $p$} \medskip
    \label{Fig:SizePlot-1}
    \includegraphics[width=0.95\textwidth]{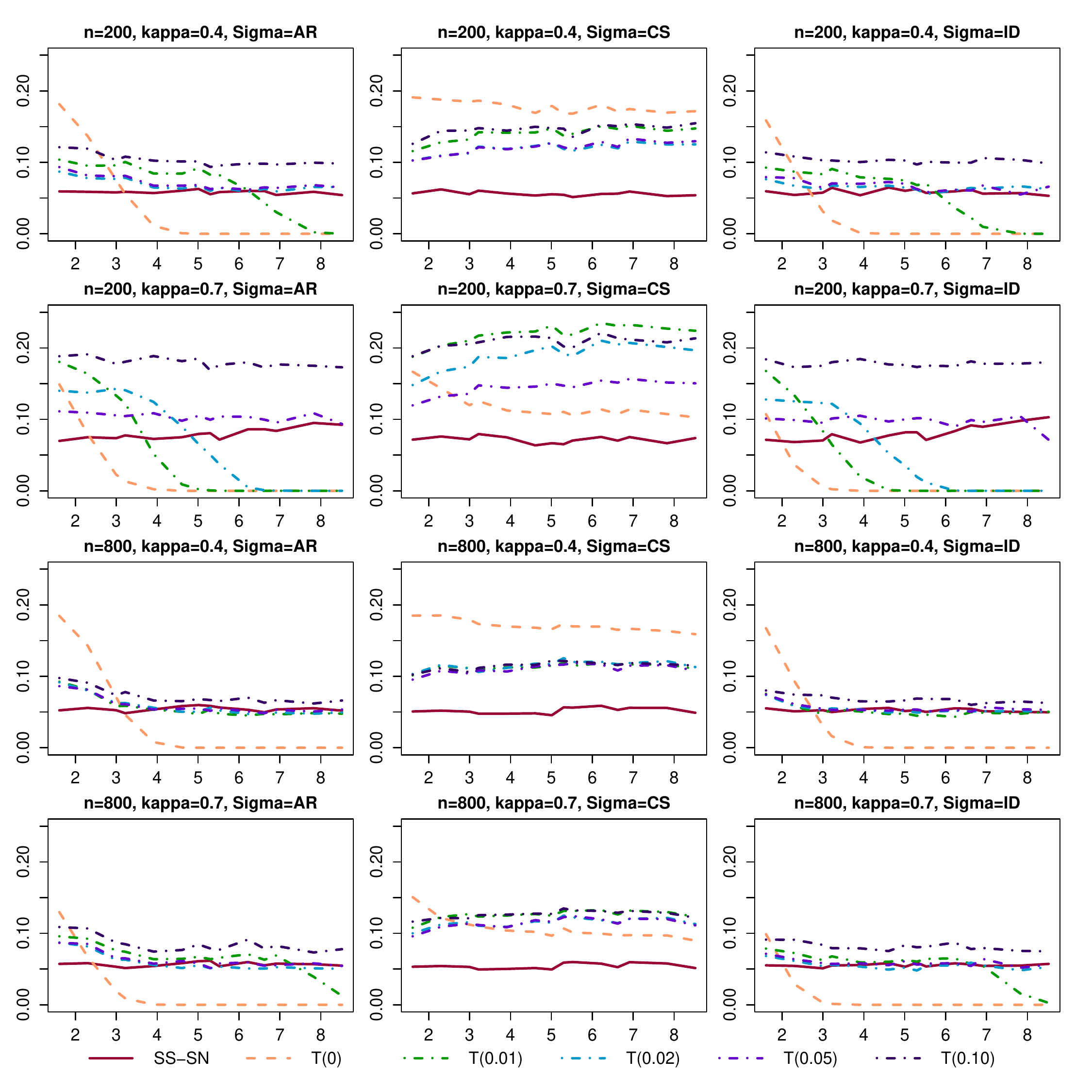}
\end{figure}

To examine the stability of size accuracy with respect to $p$, we plot the empirical size against the logarithm of $p$; see Figure \ref{Fig:SizePlot-1}. We consider $n\in\{200,800\}$, $p\in\{5,10,20,25,50,100,150,\\
200,250,500,750,1000,2500,5000\}$ and $\kappa\in\{0.4,0.7\}$. When $\Sigma$ takes the AR(1) form, we set $\rho=0.5$. We conduct 5000 Monte Carlo replicates for each setting.

In each plot, the solid curves in red correspond to SS-SN, whereas the dotted lines in other colors represent $T(\eta_0)$ with different $\eta_0$. It is apparent that the SS-SN test has a stable empirical size close to the nominal level $\alpha=0.05$ regardless of $p$, and the accuracy significantly improves as $n$ increases from $200$ to $800$. By contrast, the SN test $T(\eta_0)$ in \cite{wang2022inference} exhibits quite a bit of size distortion, especially in the compound symmetric case. This is not surprising as the theory in \cite{wang2022inference} suggests that their test only works for time series with weak cross-sectional dependence. Also, the trimming parameter $\eta_0$ in the test of \cite{wang2022inference} plays an important role in the size accuracy. Overall our SS-SN test has a great advantage in size. 

It is worth mentioning that when the trimming parameter is set as $\eta=0.02$, we see no obvious difference in the size accuracy from that when $\eta=0.04$. Related additional numerical results for the empirical sizes are reported in the supplement.

\subsection{Empirical Power}\label{Sec:NumResults-simul-power-singleCP}

Next, we investigate the power behavior of the proposed test against a single change point. In this case, we fix $n=200$, $\kappa=0.7$, $\rho=0.8$ for $\Sigma$ of AR(1) type, and consider $p\in\{3,10,100,500\}$. The location of the change point is set as $k=\lfloor{n/2}\rfloor$. We generate the mean vector $\mu_t$ by $\mu_t = \frac{c}{\sqrt{p}}(1,\cdots,1)^{\top} \bone\{t>k\}$, where $c$ is a parameter used to quantify the signal-noise-ratio.

\begin{figure}[h!]
    \centering
    \caption{Power curves (size-adjusted) against a single dense change point when $n=200$} \medskip
    \label{Fig:DensePowerCurve}
    \includegraphics[width=0.95\textwidth]{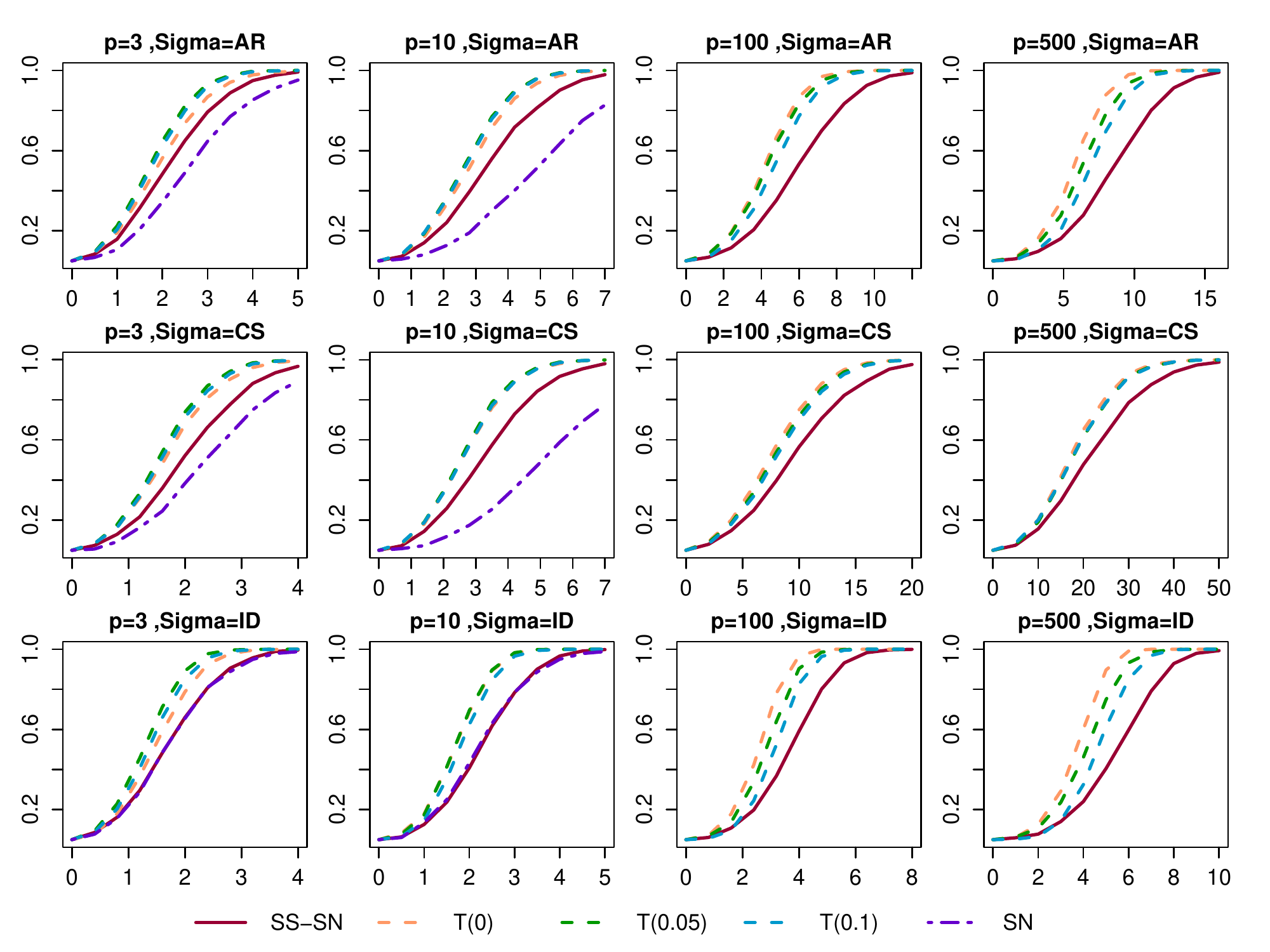}
\end{figure}

Apart from the trimming method $T(\eta_0)$ proposed by \cite{wang2022inference}, we also compare the results of our method with those of $\texttt{SN}$ introduced in \cite{shao2010testing} when $p$ is no larger than 10. Under each parameter setting, we plot the size-adjusted power of each method against $c$, see Figure \ref{Fig:DensePowerCurve}. In each figure, the three rows correspond to three structures of $\Sigma$ (i.e., AR, CS and ID) and the four columns stand for different values of $p$. All the simulation results are averaged over 5000 Monte-Carlo replicates.

As shown in Figure \ref{Fig:DensePowerCurve}, SS-SN has some moderate power loss when compared to the high-dimensional SN test $T(\eta_0)$ in \cite{wang2022inference}, which attains the most power under all settings, even when $p$ is small. In most cases, $T(\eta_0)$ is more powerful with smaller $\eta_0$, but the impact of the trimming parameter $\eta_0$ on the power seems not much in the case of cross-sectional dependence (that is, AR(1) and CS). In the ID case, the fixed-dimensional SN method of \cite{shao2010testing} has better power behavior than the proposed SS-SN methods, but it suffers quite a bit of power loss when the data exhibits some componentwise dependence (i.e., AR or CS), which matches our discussion in Section \ref{Sec:Theory-power}; also see Figure \ref{Fig:PowerPlot-DGP1}.


\section{Real Data Illustration}\label{Sec:NumResults-real}

Following  \cite{wang2023computationally}, we analyze monthly returns of $p=375$ securities in the S\&P 500 index collected over $n=165$ time points (Jan 2005-Nov 2018), which form a multivariate time series $\{r_{ij}\}_{1\le i\le n,~1\le j\le p}$ of size $n=165$  with dimension $p=375$. Let $\{r_{fi}\}_{i=1}^{n}$ and $\{r_{mi}\}_{i=1}^{n}$ denote the user-specified risk-free rates and market returns, respectively. \cite{wang2023computationally} modeled the data in the form of
\begin{equation*}
    r_{ij} - r_{fi} = \alpha_{ij} + \beta_j(r_{mi} - r_{fi}) + \varepsilon_{ij},
\end{equation*}
where $\{\varepsilon_{ij}\}_{1\le i\le n,~1\le j\le p}$ are errors with mean zero. The goal is to test for a single mean shift in $\{\alpha_i\}_{i=1}^{n}$ with $\alpha_i=(\alpha_{i1},\cdots,\alpha_{ip})^{\top}$, that is, to test $H_0: \alpha_1=\cdots=\alpha_n$ versus $H_1: \alpha_1=\cdots=\alpha_k\neq\alpha_{k+1}=\cdots=\alpha_n$ for some $1\le k<n$. After running the least square regression based on the above model, it is equivalent to testing for the single mean change based on the residual data $\{e_{ij}\}_{1\le i\le n, 1\le j\le p}$, as done in \cite{wang2023computationally}. 


\subsection{Data Analysis}

 An important part that is often ignored in applying the change-point testing procedures is to understand the temporal dependence and cross-sectional dependence within the time series being analyzed. 
As we mentioned earlier, many existing change-point detection methods require specific assumptions such as temporal independence or cross-sectional weak dependence in order for the method to work. To explore the cross-sectional and temporal dependence, one complication is that some component time series might contain a mean shift and some do not. To this end, we shall apply the change-point test in \cite{shao2010testing} to each component time series in $\{e_{ij}\}$, and estimate the change point location if the $p$-value is smaller than 0.05, and then separate the time series into two pieces. We further subtract the corresponding sample mean of each piece to obtain the mean-centered residual time series  $\{\tilde{e}_{ij}\}_{i=1}^{n}$. The consistency of the change-point location estimator based on \cite{shao2010testing} has been shown in \cite{zhaojiangshao2022}.


First, we use the Durbin-Watson test and the Breusch-Godfrey test to test for auto-correlation in the mean-centered residual time series $\{\tilde{e}_{ij}\}$. In particular, the Durbin-Watson test targets auto-correlation at lag 1 whereas the Breusch-Godfrey test is able to test for auto-correlation up to a specified lag $h$. Both methods can be implemented using \texttt{R} package \texttt{lmtest}. We apply both testing procedures to each component time series, that is,  $\{\tilde{e}_{ij}\}_{1\le i\le n}$ for $j=1,\cdots,p=375$. When targeting the auto-correlation at lag 1, the DW test and BG test respectively detect 59 and 62 significant components prior to the FDR control. When considering the auto-correlation at a larger lag, the BG test detects 90 components with significant auto-correlation up to lag 3 and 107 significant components with significant auto-correlation up to lag 5. 
To take into account multiple testing, we follow the procedures in \cite{holm1979simple}, \cite{benjamini1995controlling}, and \cite{benjamini2001control} to perform the FDR control, and the exact number of significant components before and after FDR control are reported in Table \ref{Tab:DW-BG-Pval}.

\begin{table}[!h]
    \centering
    \caption{Number of significant components before and after FDR control}\medskip
    \label{Tab:DW-BG-Pval}
    \begin{tabular}{c|c|c|c|c|c}
    \hline\hline
        Test & Lag & No FDR & Holm & BH & BY \\ \hline
        Durbin-Watson test & 1 & 59 & 1 & 10 & 1 \\ \hline
        \multirow{3}{*}{Breusch-Godfrey test} & 1 & 62 & 1 & 11 & 1 \\ \cline{2-6}
        & 3 & 90 & 9 & 28 & 11 \\ \cline{2-6}
        & 5 & 107 & 14 & 42 & 16 \\  
    \hline\hline
    \end{tabular}
\end{table}




Next, we investigate the cross-sectional dependence among the demeaned residual time series. In this case, we view each component as a vector of length $n=165$ and carry out pairwise association tests with the built-in \texttt{R} function \texttt{cor.test}. Specifically, we consider the Pearson correlation test, the Kendall's $\tau$ test, and the Spearman rank correlation test. Since the individual tests are dependent on each other, we select the procedure in \cite{benjamini2001control} to achieve the FDR control as it has weaker assumptions on the dependence between individual tests. Numerically, after the FDR control, the number of pairs with significantly associated securities detected by all three tests are 2144, 1457, and 1391 (out of 70125 pairs in total), respectively. Hence there is some degree of cross-sectional dependence in the residual time series, and this might be related to the fact that the least squares regression is only removing the market factor, but not the sector-specific factor, which implies the dependence for the returns of securities in the same sector.

\subsection{Change Point Testing Results}

In this section, we apply the SS-SN method to the residual time series $\{e_{ij}\}$ with $\epsilon=0.1$ and $\eta\in\{0.02,0.04\}$, and report the corresponding $p$-values in Table \ref{Tab:CompWangFeng}. For the comparison, we adopt the sum-$L_2$-type statistic Sum proposed in \cite{wang2019multiple}, the sum-$L_\infty$-type statistics Max(0) and Max(0.5) proposed in \cite{wang2023computationally}, and the adaptive test statistics DMS(0) and DMS(0.5) proposed in the same article. In particular, no tuning parameters are involved in Sum, Max(0), and DMS(0), whereas the tuning parameter $\lambda_n=\lrfl{0.2n}$ is used by Max(0.5), and DMS(0.5). The $p$-values of all the competing methods are also reported in Table 5 of \cite{wang2023computationally}.

\begin{table}[!h]
    \centering
    \caption{Comparison of $p$-values based on the residual time series $\{e_{ij}\}_{1\le i\le n;1\le j\le p}$} \medskip
    \label{Tab:CompWangFeng}
    \scalebox{1.0}{
    \begin{tabular}{c|c||c|c|c|c|c}
    \hline\hline
        SS-SN(0.02) & SS-SN(0.04) & Sum & Max(0) & Max(0.5) & DMS(0) & DMS(0.5) \\ \hline
        1.58e-03 & 7.32e-03 & 1.92e-01 & 3.92e-03 & 9.31e-04 & 6.17e-03 & 1.72e-03 \\
    \hline\hline
    \end{tabular}}
\end{table}

Both Max(0) and Max(0.5) are applicable to time series with both temporal and cross-sectional dependence and both methods target the sparse change point. The significant $p$-values of both methods suggest the existence of a sparse change pattern. The validity of Sum is shown only under the strong temporal independence assumption, and its applicability to the temporal-dependent time series, such as the one being analyzed, seems questionable. Although the $p$-value of Sum is non-significant, it does not necessarily negate the potential existence of a dense mean shift. Note that the adaptive test DMS(0) is obtained by performing Fisher's $p$-value combination for Max(0) and Sum, and DMS(0.5) is obtained from Max(0.5) and Sum, thus the validity of both DMS(0) and DMS(0.5) may be limited to the temporally independent data due to the use of Sum. In contrast, our proposed method accommodates both the temporal and cross-sectional dependence within the time series. The significant $p$-values for both choices of $\eta$ indicate the existence of a dense change pattern, which complements the findings in \cite{wang2023computationally}.




\section{Discussion}\label{Sec:Discussion}

There is a vast literature on change point testing for both low-dimensional and high-dimensional data with or without temporal dependence. Almost all existing test statistics only work in pre-determined dimensional regimes and tests developed for the low-dimensional setting may not work for high-dimensional data and vice versa. In this paper, we advance the dimension-agnostic inference first proposed in \cite{kim2020dimensionagnostic} for iid data to the mean change point problem in the multivariate time series setting. Specifically, We adopt the sample splitting with trimming, projection, and self-normalization ideas to develop new test statistics for a single change point alternative. On the theory front, we derive the limiting null distributions for the proposed test statistics under three data-generating processes, which encompass a broad range of dimensionality and arbitrary cross-sectional dependence. The limiting null distribution is pivotal and stays the same across both fixed and growing dimensional regimes, hence the selection of dimension-dependent calibration threshold can be avoided. Additionally, we provide a rigorous analysis of the asymptotic power behavior of the proposed tests. 
Monte Carlo simulation results strongly corroborate the theoretical phenomenon we discovered and suggest that these dimension-agnostic test statistics maintain very accurate size across a wide range of dimensions, albeit with a moderate amount of power loss in certain settings. The real data example also illustrates the versatility of the proposed test in real-world applications.


\acks{Wang's research is partially supported by NSF-DMS 2210007; Shao's research is partially supported by NSF-DMS 2014018 and NSF-DMS 2210002. The authors would like to thank Guanghui Wang for providing the code and stock return data used in \cite{wang2023computationally}.}


\appendix
\section{Generalization of the Proposed Test}\label{Appendix:Extension}

In this appendix, we investigate the possible generalizations to a single sparse change point testing and multiple change points testing. Additional simulation studies are presented in the online supplement; see \url{https://arxiv.org/abs/2303.10808}.

\subsection{A Single Sparse Change Point Alternative}\label{Appendix:Extension-sparseCP}

In our article, we limit the study to testing a single dense change point with a sufficiently large $L_2$ norm. It is natural to generalize the proposed method to accommodate the scenario where the single change point is sparse in mean. 

For this case, we keep all the procedures of the sample splitting, but instead of projecting $\cx_2$ along the direction of $\hat\mu_1 - \hat\mu_n$, we construct a sparse projection direction. Specifically, define 
\begin{equation*}
    k^{\ast} := \argmax_{k = 1,\cdots,p} \lrabs{\hat\mu_{1,k} - \hat\mu_{n,k}}.
\end{equation*}
Let $e_k = (0,\cdots,0,1,0,\cdots,0)^{\top} \in \br^p$ denote a vector where only the $k$-th component is $1$ and all other components are zero. Then $\hat\nu = \sgn(\hat\mu_{1,k^{\ast}} - \hat\mu_{n,k^{\ast}}) e_{k^{\ast}}$ will be used as the direction along which the projection is conducted, where $\sgn(\cdot)$ is the sign function. That is, $\hat\nu$ corresponds to the dimension in which the largest mean shift is observed. By projecting $\cx_2$ along this direction, we restrict the multivariate data to a single component and test for the change point in that specific coordinate. We denote the scalar sequence obtained in this case as $\{Y_j^{\ast}\}_{j=1}^{N}$, where $Y_j^{\ast} = \lrag{\hat\nu, X_{j+m}}$. The test statistic can be constructed in a similar way as in Section \ref{Sec:Methodology-dense}, with $Y_j$ replaced by $Y_j^{\ast}$. We denote the test statistic by $G_n^{\ast}$.

We conjecture that the limiting null distribution of $G_n^{\ast}$ is same as $G_n$, that is, $G_n^{\ast}$ converges to $G$ under the null for the three DGPs mentioned earlier. However, a rigorous theoretical justification of the test statistic $G_n^{\ast}$ under the null and the sparse alternative seems highly involved and is beyond the scope of this article. 

\begin{remark}
Note that the test statistics proposed in Section \ref{Sec:Methodology} and in this section target the dense change point and the sparse change point respectively. In practice, we do not have prior knowledge about the sparsity of the change. This motivates us to aggregate these two test statistics to achieve adaptive power. A common method to aggregate the two tests is to construct a Bonferroni test combining both the dense test and the sparse test.

Specifically, given the data, we can compute the dense test statistic $G_n$ as well as the sparse test statistic $G_n^{\ast}$. At the significance level $\alpha$, we reject the null hypothesis if the $p$-value of either test drops below the $\alpha/2$ threshold. As will be shown by the simulation studies, the Bonferroni test achieves encouraging performance across multiple scenarios, regardless of the sparsity of the mean shift.  
\end{remark}

\subsection{Multiple Change Points Alternative}\label{Appendix:Extension-multipleCP}

Another natural generalization is to accommodate the multiple change points alternative. In particular, we incorporate the proposed SS-SN methodology with the scanning test statistic of \cite{zhang2018unsupervised} and develop a new test statistic to test for multiple change points. The asymptotic theory is also presented in this section.

Mathematically, we aim to test
\begin{equation*}
    H_0: \mu_1=\cdots=\mu_n
    \quad\mbox{versus}\quad
    H_1^M: \mu_i \neq \mu_{i+1} \mbox{ for } i\in\{k_1,\cdots,k_M\} \mbox{ and } \mu_i=\mu_{i+1} \mbox{ otherwise.}
\end{equation*}
Under the alternative, $\{k_i\}_{i=1}^{M}$ denote the $M$ unknown  change points. Following the convention in the literature, we assume that $k_i=\lrfl{n\xi_i}$ and $\varepsilon_0 < \xi_1 < \cdots < \xi_M < 1-\varepsilon_0$ {are fixed unknown constants}. Same as the single change point setting, we assume that all the (relative locations of) change points are more than $\varepsilon_0$ away from the boundary. 

To facilitate the subsequent analysis, we denote all the mean shifts as $\{\delta_i\}_{i=1}^{M}$ with $\delta_i = \mu_{k_i+1} - \mu_{k_i}$ and define $\delta = \sum\limits_{i=1}^{M}\delta_i = \mu_{k_{M}+1}-\mu_{k_1}$ as the cumulative mean shift of the entire sequence. For convenience, we further define $k_0=\lrfl{n\xi_0}=m$ with $\xi_0=\varepsilon$, $k_{M+1}=\lrfl{n\xi_{M+1}}=n-m$ with $\xi_{M+1}=1-\varepsilon$, and $\delta_0=\mu_{k_1}$.

With a splitting parameter $\varepsilon$ and a trimming parameter $\eta$ satisfying that $0 < \eta < \varepsilon < \varepsilon_0$, we can repeat the same procedures described in Section \ref{Sec:Methodology} to obtain the scalar sequence $\{Y_j\}_{j=1}^{N}$. Recall that validity of the single change point testing is based on the fact that the mean shift is well preserved in the projected data, thus the original testing for $\mu_1=\mu_n$ is equivalent to the univariate change point testing problem for $\|\mu_1-\mu_n\|_2=0$. However, when it comes to the multiple change point testing problem, complication arises. Intuitively, the mean shifts in the projected sequence are approximately $\{\delta^{\top}\delta_i\}_{i=1}^{M}$. If $\delta\neq0$, then it is impossible that $\delta^{\top}\delta_i$'s are all zeros, hence at least some of the mean shifts are expected to be preserved in the projected data. Therefore, we can still convert the multivariate change point testing to the one-dimensional counterpart. However, if $\delta=0$, then $\{\delta^{\top}\delta_i\}_{i=1}^{k}$ are all zero, so  it no longer makes sense to apply the univariate testing procedure to the projected data as  only trivial power is expected. For simplicity, we exclude the case that $\delta=0$ in this article, as formulated in the following assumption. 

\begin{assumpt}\label{Assumpt:multiple}
Assume that under $H_1^{M}$, it holds that $\varepsilon_0 < \xi_1<\cdots<\xi_M < 1-\varepsilon_0$ are fixed unknown constants and  $\mu_1\neq\mu_n$.
\end{assumpt}

Under Assumption \ref{Assumpt:multiple}, testing for multiple change points in $\{X_t\}_{t=1}^{n}$ can be achieved by conducting the univariate test on $\{Y_j\}_{j=1}^{N}$. To this end, we apply the one-dimensional testing procedure proposed in \cite{zhang2018unsupervised}, which is an extension of the SN test in \cite{shao2010testing} to capture multiple change-points alternative. We first introduce the following notations for the forward process,
\begin{equation*}
    T_n^{f}(j_1,j_2,j_3) = \frac{1}{\sqrt{j_3-j_1+1}} \lrp{\sum\limits_{i=j_1}^{j_2} Y_i - \frac{j_2-j_1+1}{j_3-j_1+1} \sum\limits_{i=j_1}^{j_3} Y_i},
\end{equation*}
and $V_n^{f}(j_1,j_2,j_3) = L_n^{f}(j_1,j_2,j_3) + R_n^{f}(j_1,j_2,j_3)$, where
\BEqn
& & L_n^{f}(j_1,j_2,j_3) = \frac{1}{(j_3-j_1+1)^2} \sum\limits_{i=j_1}^{j_2} \lrp{\sum\limits_{t=j_1}^{i} Y_t - \frac{i-j_1+1}{j_2-j_1+1} \sum\limits_{t=j_1}^{j_2} Y_t}^2, \\
& & R_n^{f}(j_1,j_2,j_3) = \frac{1}{(j_3-j_1+1)^2} \sum\limits_{i=j_2+1}^{j_3} \lrp{\sum\limits_{t=i}^{j_3} Y_t - \frac{j_3-i+1}{j_3-j_2} \sum\limits_{t=j_2+1}^{j_3} Y_t}^2.
\EEqn
Similarly, we define the counterparts in the backward direction, that is
\begin{equation*}
    T_n^{b}(j_1,j_2,j_3) = \frac{1}{\sqrt{j_3-j_1+1}} \lrp{\sum\limits_{i=j_2}^{j_3} Y_i - \frac{j_3-j_2+1}{j_3-j_1+1} \sum\limits_{i=j_1}^{j_3} Y_i}, 
\end{equation*}
and $V_n^{b}(j_1,j_2,j_3) = L_n^{b}(j_1,j_2,j_3) + R_n^{b}(j_1,j_2,j_3)$ with
\BEqn
& & L_n^{b}(j_1,j_2,j_3) = \frac{1}{(j_3-j_1+1)^2} \sum\limits_{i=j_1}^{j_2-1} \lrp{\sum\limits_{t=j_1}^{i} Y_t - \frac{i-j_1+1}{j_2-j_1} \sum\limits_{t=j_1}^{j_2-1} Y_t}^2, \\
& & R_n^{b}(j_1,j_2,j_3) = \frac{1}{(j_3-j_1+1)^2} \sum\limits_{i=j_2}^{j_3} \lrp{\sum\limits_{t=i}^{j_3} Y_t - \frac{j_3-i+1}{j_3-j_2+1} \sum\limits_{t=j_2}^{j_3} Y_t}^2.
\EEqn
To formulate the test statistic, we mimic the notations used in \cite{zhang2018unsupervised} and define
\BEqn
& & \Xi = \{(r_1,r_2): 0 \le r_1 < r_2 \le 1\}, \\
& & \Xi(\varepsilon) = \{(r_1,r_2): \varepsilon \le r_1 < r_2 \le 1-\varepsilon\}, \\
& & \Xi_n(\varepsilon) = \{(\ell_1,\ell_2): 1<\ell_1<\ell_2<N=\lrfl{(1-2\varepsilon)n}, \ell_2-\ell_1>1\}.
\EEqn

Then the test statistic is defined as
\begin{equation*}
    G_n^{M}
  = \max\limits_{(\ell_1,\ell_2)\in\Xi_n(\varepsilon)} \lrabs{\frac{T_n^{f}(1,\ell_1,\ell_2)}{(V_n^{f}(1,\ell_1,\ell_2))^{1/2}}}
  + \max\limits_{(\ell_1,\ell_2)\in\Xi_n(\varepsilon)} \lrabs{\frac{T_n^{b}(\ell_1,\ell_2,N)}{(V_n^{b}(\ell_1,\ell_2,N))^{1/2}}}.
\end{equation*}

Note that in the projected sequence, the mean shifts are quantified as $\delta^{\top}\delta_i$ for $i=1,\cdots,M$. When there is a single change point, the only mean shift $\|\delta\|_2^2$ is always positive, hence we simply reject the null hypothesis if the self-normalized test statistic is too large. When there are more than one change point, the mean shifts $\delta^{\top}\delta_i$ can be either positive or negative. Therefore both positive and negative self-normalized statistics with large absolute value are strong evidence against the alternative. Hence when testing for multiple change points, we consider the maximum of absolute value of the self-normalized test statistic over all possible intervals $(\ell_1,\ell_2)\in\Xi_n(\varepsilon)$, which differs from the test statistic used against a single change point alternative.

Note that the computational complexity of $G_n^M$ is at the order of $O(n(n^2+p))$. To ease the computational burden for large $n$, one can opt for  discretised approximation proposed in \cite{zhang2018unsupervised}  by finding the maximum over a set of the cardinality $O(n)$, for which we spare the details. Also it is worth noting that when applying the SN test in \cite{zhang2018unsupervised} to the projected data $\{Y_j\}_{j=1}^{N}$, we removed the trimming parameter that is required in \cite{zhang2018unsupervised} and also in \cite{wang2022inference}.

Under suitable conditions, we can derive the limiting null distribution of $G_n^M$ for the three DGPs discussed early, which is summarized in Theorem \ref{Thm:NullDist_multiple} below.

\begin{theorem}\label{Thm:NullDist_multiple}
If Assumption \ref{Assumpt:multiple} holds and the sequence $\{X_t-\mu_t\}_{t=1}^{n}$ satisfies the assumptions of either Theorem \ref{Thm:NullDist_Gn_Fixedp} for DGP1, Theorem \ref{Thm:NullDist_Gn_Growingp_lp} for DGP2 or Theorem \ref{Thm:NullDist_Gn_factor} for DGP3, then it holds under the null that
\BEqn
    G_n^{M} 
\stra{d} G^{M}
&:=& \sup\limits_{(r_1,r_2)\in\Xi} \lrabs{\sqrt{r_2} T^{f}(r_1,r_2) (V^{f}(r_1,r_2))^{-1/2}} \\
& & + \sup\limits_{(r_1,r_2)\in\Xi} \lrabs{\sqrt{1-r_1} T^{b}(r_1,r_2) (V^{b}(r_1,r_2))^{-1/2}},
\EEqn
where $T^{f}(r_1,r_2) = B(r_1) - \frac{r_1}{r_2}B(r_2)$, $T^{b}(r_1,r_2) = B(1) - B(r_2) - \frac{1-r_2}{1-r_1}(B(1) - B(r_1))$, and
\BEqn
    V^{f}(r_1,r_2) 
&=& \int_{0}^{r_1} \lrp{B(s) - \frac{s}{r_1} B(r_1) }^2 ds \\
& & + \int_{r_1}^{r_2} \lrp{B(r_2) - B(s) - \frac{r_2-s}{r_2-r_1} (B(r_2) - B(r_1)) }^2 ds, \\[2mm]
    V^{b}(r_1,r_2) 
&=& \int_{r_1}^{r_2} \lrp{B(s) - B(r_1) - \frac{s-r_1}{r_2-r_1} (B(r_2) - B(r_1)) }^2 ds \\
& &  + \int_{r_2}^{1} \lrp{B(1) - B(s) - \frac{1-s}{1-r_2} (B(1) - B(r_2)) }^2 ds.
\EEqn
\end{theorem}

In practice, we reject the null hypothesis at the significance level $\alpha$ if the value of the test statistic $G_n^{M}$ exceeds the $1-\alpha$ quantile of $G^M$. Since the limiting null distribution $G^{M}$ is pivotal, we can repeat the simulation steps used in Section \ref{Sec:Theory-null} to simulate its distribution. The critical values based on 20000 Monte Carlo replicates and a sample size of 2000 are presented in Table \ref{Tab:critval-M}. 

\begin{table}[h!]
    \centering
    \caption{Simulated critical values for $G^M$} \medskip
    \label{Tab:critval-M}
    \begin{tabular}{c|cccccc}
    \hline\hline
        Quantile Level & 90\% & 95\% & 97.5\% & 99\% & 99.5\% & 99.9\% \\ \hline
        Critical Value & 20.71 & 23.16 & 26.75 & 35.50 & 45.74 & 102.97 \\
    \hline\hline
    \end{tabular}
\end{table}

Next, we look into the power behavior of $G_n^M$ and derive the regime where the power approaches one asymptotically.
We follow the notations defined in Section \ref{Sec:Theory-power} to denote the centered version of $\{X_t\}_{t=1}^{n}$ as $\{\tilde{X}_t\}_{t=1}^{n}$ with $\tilde{X}_t = X_t - \mu_t$. Additionally, we continue to use the notations of $\{W_n(r)\}_{\varepsilon\le r\le 1-\varepsilon}$ and $\{\tilde{W}_n(r)\}_{\varepsilon\le r\le 1-\varepsilon}$, where
$   W_n(r)
  = m_1 \sum\limits_{j=1}^{\lrfl{nr}-\lrfl{n\varepsilon}} Y_j
  = \sum\limits_{i=1}^{m_1} \lrp{X_i - X_{n+1-i}}^{\top} \lrp{\sum\limits_{j=1}^{\lrfl{nr}-\lrfl{n\varepsilon}} X_{j+m}}$, 
and $\tilde{W}_n(r)
  = \sum\limits_{i=1}^{m_1} \lrp{\tilde{X}_i - \tilde{X}_{n+1-i}}^{\top} \lrp{\sum\limits_{j=1}^{\lrfl{nr}-\lrfl{n\varepsilon}} \tilde{X}_{j+m}}.$

The following assumption is required to establish the power result.

\begin{assumpt}\label{Assumpt:MultiplePower}
Assume that there exists some positive normalizer $N_n$ (determined by specific DGP later on, see Theorem \ref{Thm:MultiplePower}) and $1\le i\le M$, such that $\frac{n^2}{N_n} \lrabs{\delta^{\top} \delta_{i}} \rightarrow \infty$ as $n\rightarrow\infty$ and $\|\delta\|_2^2 = O(\lrabs{\delta^{\top}\delta_i})$. Additionally if $1 < i < M$, either of the following conditions is satisfied:
\begin{enumerate}[label=(\roman*)]
    \item \label{Assumpt:MultiplePower-1}
    $\max\limits_{1\le j\le i} \|\delta_j\|_2^2 = O(\lrabs{\delta^{\top} \delta_{i}})$ and $\max\limits_{1\le j<i}\{|\delta^{\top} \delta_j|\} = o(\lrabs{\delta^{\top} \delta_{i}})$.
    \item \label{Assumpt:MultiplePower-2}
    $\max\limits_{M\geq j\geq i} \|\delta_j\|_2^2 = O(\lrabs{\delta^{\top} \delta_{i}})$ and $\max\limits_{M\geq j>i}\{|\delta^{\top} \delta_j|\} = o(\lrabs{\delta^{\top} \delta_{i}})$.
\end{enumerate}
\end{assumpt}

Intuitively with high probability, the forward scanning statistic $\displaystyle\lrabs{\frac{T_n^{f}(1,\ell_1,\ell_2)}{(V_n^{f}(1,\ell_1,\ell_2))^{1/2}}}$ should be at least able to detect the first $\delta_i$ which satisfies Assumption \ref{Assumpt:MultiplePower}\ref{Assumpt:MultiplePower-1} , whereas the backward scanning statistic should be able to at least detect the last $\delta_i$ that satisfies Assumption \ref{Assumpt:MultiplePower}\ref{Assumpt:MultiplePower-2}. In either case, we expect the value of $G_n^{M}$ to diverge to infinity, and the power goes to one as $n$ goes to infinity.

\begin{remark}
Assumption 6 describes the regime when the power of the proposed test goes to one. Basically, the null hypothesis will be rejected, if at least one mean shift $\delta_i$ has a large enough signal-to-noise ratio $\frac{n^2}{N_n} \lrabs{\delta^{\top} \delta_{i}}$. This indicates that not only both $\|\delta\|_2$ and $\|\delta_i\|_2$ need to be large, but also $|\sin(\theta)|$ needs to be bounded away from one, where $\theta$ is the angle between $\delta$ and $\delta_i$. 

To see this, note that Assumption \ref{Assumpt:MultiplePower} implies that $|\delta^{\top}\delta_i|\leq \|\delta\|_2\|\delta_i\|_2 = O(|\delta^{\top}\delta_i|)$, which is equivalent to $|\delta^{\top}\delta_i| = \|\delta\|_2\|\delta_i\|_2|\cos(\theta)| \asymp \|\delta\|_2\|\delta_i\|_2$. This indicates $|\cos(\theta)| \asymp 1$. For fixed $p$ case, this can always be satisfied unless $\delta_i$ is orthogonal to $\delta$. However for growing $p$ scenario, as $\theta$ could change as $p$ increases (both $\delta$ and $\delta_i$ change as well), $\delta_i$ is detectable only if $|\cos(\theta)|$ does not converge to zero. Equivalently, $|\sin(\theta)| \leq c < 1$ for all $p$, for some positive constant $c$.
\end{remark}

\begin{remark}
Two special cases of Assumption \ref{Assumpt:MultiplePower} are when either $\frac{n^2}{N_n} \lrabs{\delta^{\top}\delta_1} \rightarrow \infty$ or $\frac{n^2}{N_n} \lrabs{\delta^{\top}\delta_M} \rightarrow \infty$. In the first case, Assumption \ref{Assumpt:MultiplePower}\ref{Assumpt:MultiplePower-1} reduces to $\max\{\|\delta\|_2^2,\|\delta_1\|_2^2\} = O(\lrabs{\delta^{\top}\delta_1})$. This indicates that our test is powerful if $|\delta^{\top}\delta_1|$ is large enough, regardless of the magnitude of $\delta^{\top}\delta_2$, ..., $\delta^{\top}\delta_M$, since the first change is significant enough to be detected (by the forward statistic). And in the second case, the assumption can be simplified as $\max\{\|\delta\|_2^2,\|\delta_M\|_2^2\} = O(\lrabs{\delta^{\top}\delta_M})$, which indicates that the last change is always detectable (by the backward statistic).

There is another special case where only one change point presents, i.e., $M=1$. In this case, Assumption \ref{Assumpt:MultiplePower} reduces to $\frac{n^2}{N_n} \|\delta\|_2^2 \rightarrow \infty$, which is equivalent to the corresponding assumptions used in Theorem \ref{Thm:DensePower_Fixedp}\ref{Thm:DensePower_Fixedp-2}, Theorem \ref{Thm:DensePower_lp}\ref{Thm:DensePower_lp-2} and Theorem \ref{Thm:DensePower_factor}\ref{Thm:DensePower_factor-2}.
\end{remark}

Under Assumption \ref{Assumpt:MultiplePower}, we establish the asymptotic power results for all the three DGPs in Theorem \ref{Thm:MultiplePower}.

\begin{theorem}\label{Thm:MultiplePower}
Under Assumption \ref{Assumpt:multiple}, we have that,
\begin{enumerate}[label=(\roman*)]
    \item \label{Thm:MultiplePower-1}
    if $\{\tilde{X}_t\}_{t=1}^{n} \in \br^{p}$ is a stationary sequence as defined in Definition \ref{Def:Model_Fixedp}, and Assumption \ref{Assumpt:MultiplePower} is satisfied with $N_n=n$, then it holds that $\blrp{G_n^{M} > G^{M}_{1-\alpha}} \rightarrow 1$ as $n \rightarrow \infty$.
    
    \item \label{Thm:MultiplePower-2}
    if $\{\tilde{X}_t\}_{t=1}^{n} \in \br^{p}$ is a linear process as defined in Definition \ref{Def:Model_lp} and Assumption \ref{Assumpt:NullDist_Wn_lp} holds. Suppose that Assumption \ref{Assumpt:MultiplePower} is satisfied with $N_n=\sqrt{2n m_1} \|A^{(0)} \Gamma^{(2)} (A^{(0)})^{\top}\|_F$ and $\rho^{m_2/4}\|\Gamma^{(2)}\|_F = o\lrp{\frac{n}{\log(n)}}$, then it holds that $\blrp{G_n^{M} > G^{M}_{1-\alpha}} \rightarrow 1$ as $\min\{n,p\} \rightarrow \infty$.
    
    \item \label{Thm:MultiplePower-3}
    if $\{\tilde{X}_t\}_{t=1}^{n} \in \br^{p}$ admits a static factor model as defined in Definition \ref{Def:Model_factor} and Assumption \ref{Assumpt:NullDist_Wn_lp} (applied to $Z_t$), Assumption \ref{Assumpt:NullDist_Wn_factor}, and Assumption \ref{Assumpt:DensePower_factor} hold. Suppose that Assumption \ref{Assumpt:MultiplePower} is satisfied with $N_n=\max\{n\|\Lambda^{\top}\Lambda\|, \sqrt{2n m_1} \|A^{(0)}\Gamma^{(3)}(A^{(0)})^{\top}\|_F\}$ and $\rho^{m_2/4}\|\Gamma^{(3)}\|_F = o\lrp{\frac{n}{\log(n)}}$, then it holds that $\blrp{G_n^{M} > G^{M}_{1-\alpha}} \rightarrow 1$ as $\min\{n,p\} \rightarrow \infty$.
\end{enumerate}
\end{theorem}

Theorem \ref{Thm:MultiplePower} shows that each DGP attains power one asymptotically under Assumption \ref{Assumpt:multiple} and \ref{Assumpt:MultiplePower}. It can be viewed as a counterpart of the power results established in Section \ref{Sec:Theory-power}, but a subtle difference is that Assumption \ref{Assumpt:MultiplePower}\ref{Assumpt:MultiplePower-1} or Assumption \ref{Assumpt:MultiplePower}\ref{Assumpt:MultiplePower-2} are required when considering the multiple change points alternative. It is trivial that when $M=1$, the results of Theorem \ref{Thm:MultiplePower} coincide with the corresponding theorems established against a single local alternative in Section \ref{Sec:Theory-power}.

\clearpage

\begin{appendices}
\appendixpage
The supplementary material contains some additional numerical results and all the proofs for the theoretical results established in this article. In particular, Appendix \ref{Appdix:Additional_simul} includes some complementary simulation results to those in Section \ref{Sec:NumResults-simul} and some simulation studies regarding the generalizations presented in the appendix. Furthermore, we provide in Appendix \ref{Appdix:MainThm} the proofs for all the propositions and theorems stated in the main article and its appendix. The lemmas directly used by these proofs are presented in Appendix \ref{Appdix:AuxLemma-1} and some other auxiliary lemmas are included in Appendix \ref{Appdix:AuxLemma-2}. 
\vspace{10pt}


\section{Additional Simulation Results}\label{Appdix:Additional_simul}

In this section, we take the generalizations into account and present some additional simulation results. Throughout this section, to distinguish the SS-SN tests targeting a single dense change point, a single sparse change point, and multiple change points, we use $G_{n,2}$, $G_{n,\infty}$ to represent the SS-SN test against a single dense mean shift and a single sparse mean shift and use $\texttt{Bonf}$ to denote the Bonferroni test based on $G_{n,2}$ and $G_{n,\infty}$. Similarly, we use $G_{n,2}^M$, $G_{n,\infty}^M$ and \texttt{Bonf}$^M$ to represent the counterparts against multiple change points.

Specifically, Appendix \ref{Appendix:AddNumResults-size-singleCP} includes some additional simulation results regarding the empirical size for a single change point testing and Appendix \ref{Appendix:AddNumResults-size-multiCP} includes those for multiple change points testing. To investigate the power behavior of the generalized tests against a single sparse change point and against multiple change points, we perform additional simulation studies and present the numerical results in Appendix \ref{Appendix:AddNumResults-power-singleCP-sparse} and Appendix \ref{Appendix:AddNumResults-power-multiCP}.

\subsection{Empirical Size for Single Change Point Testing}\label{Appendix:AddNumResults-size-singleCP}

We have plotted the empirical sizes against the logarithm of $p$ in Section \ref{Sec:NumResults-simul-size-singleCP} based on a $p$-dimensional AR(1) process generated by
\begin{equation*}
    X_t - \mu_t = \kappa (X_{t-1} - \mu_{t-1}) + \epsilon_{t},
    \qquad 1 \le t \le n,
\end{equation*}
where $\{\epsilon_t\}_{t=1}^{n}$ are iid $p$-dimensional multivariate normal random vectors with mean zero and variance $\Sigma$. The variance $\Sigma$ takes three different structures, (1) AR ($\Sigma_{i,j} = \rho^{|i-j|}$); (2) CS ($\Sigma_{i,j} = 0.5 + 0.5\1{i=j}$); and (3) ID ($\Sigma_{i,j} = \1{i=j}$).

Using the same data-generating process, we report some additional results on the size accuracy. In particular, we set $\mu_t=(0,\cdots,0)^{\top}$ to be the zero vector for $t=1,\cdots,n$, set $\rho=0.8$ for AR(1) type $\Sigma$, and consider $n\in\{200,800\}$, $p\in\{3,10,100,500\}$ and $\kappa\in\{0,0.4,0.7\}$. The empirical sizes when $n=200$ and $n=800$ averaged over 5000 Monte-Carlo replicates are reported in Table \ref{Tab:AR1-Size-1} and Table \ref{Tab:AR1-Size-2} respectively.
	
\begin{table}[h!]
    \centering
    \scalebox{0.9}{
    \begin{tabular}{c|c|c|c|ccc|ccc|ccc}
    \hline\hline
        \multirow{3}{*}{$n$} & \multirow{3}{*}{$p$} & \multirow{3}{*}{$\kappa$} & \multirow{3}{*}{$\Sigma$} & \multicolumn{6}{c|}{Proposed} & \multicolumn{3}{c}{$T(\eta_0)$} \\ \cline{5-13}
       & & & & \multicolumn{3}{c|}{$\eta=0.02$} & \multicolumn{3}{c|}{$\eta=0.04$} & \multirow{2}{*}{$\eta_0=0$} & \multirow{2}{*}{$\eta_0=0.05$} & \multirow{2}{*}{$\eta_0=0.1$} \\ \cline{5-10}
       & & & & $G_{n,2}$ & $G_{n,\infty}$ & Bonf & $G_{n,2}$ & $G_{n,\infty}$ & Bonf & & & \\ \hline
       \multirow{36}{*}{200} & \multirow{9}{*}{3} & \multirow{3}{*}{0} & AR & 0.043 & 0.047 & 0.035 & 0.047 & 0.047 & 0.035 & 0.118 & 0.105 & 0.102 \\
       & & & CS & 0.047 & 0.047 & 0.037 & 0.049 & 0.050 & 0.038 & 0.107 & 0.096 & 0.099 \\
       & & & ID & 0.052 & 0.049 & 0.040 & 0.051 & 0.050 & 0.043 & 0.091 & 0.083 & 0.083 \\ \cline{3-13}
       & & \multirow{3}{*}{0.4} & AR & 0.054 & 0.057 & 0.042 & 0.051 & 0.056 & 0.042 & 0.209 & 0.121 & 0.141 \\
       & & & CS & 0.058 & 0.056 & 0.048 & 0.058 & 0.058 & 0.046 & 0.196 & 0.101 & 0.131 \\
       & & & ID & 0.056 & 0.054 & 0.047 & 0.054 & 0.053 & 0.047 & 0.190 & 0.090 & 0.119 \\ \cline{3-13}
       & & \multirow{3}{*}{0.7} & AR & 0.074 & 0.073 & 0.057 & 0.067 & 0.068 & 0.056 & 0.226 & 0.140 & 0.205 \\
       & & & CS & 0.065 & 0.069 & 0.061 & 0.066 & 0.071 & 0.060 & 0.196 & 0.128 & 0.197 \\
       & & & ID & 0.068 & 0.071 & 0.063 & 0.068 & 0.067 & 0.062 & 0.168 & 0.109 & 0.182 \\ \cline{2-13}
       & \multirow{9}{*}{10} & \multirow{3}{*}{0} & AR & 0.048 & 0.044 & 0.043 & 0.051 & 0.046 & 0.042 & 0.111 & 0.100 & 0.103 \\
       & & & CS & 0.055 & 0.050 & 0.046 & 0.052 & 0.053 & 0.048 & 0.117 & 0.108 & 0.113 \\
       & & & ID & 0.045 & 0.055 & 0.047 & 0.047 & 0.050 & 0.047 & 0.066 & 0.057 & 0.068 \\ \cline{3-13}
       & & \multirow{3}{*}{0.4} & AR & 0.053 & 0.057 & 0.049 & 0.052 & 0.054 & 0.048 & 0.193 & 0.100 & 0.126 \\
       & & & CS & 0.057 & 0.054 & 0.048 & 0.061 & 0.054 & 0.055 & 0.198 & 0.107 & 0.135 \\
       & & & ID & 0.053 & 0.057 & 0.055 & 0.050 & 0.053 & 0.055 & 0.093 & 0.070 & 0.107 \\ \cline{3-13}
       & & \multirow{3}{*}{0.7} & AR & 0.071 & 0.072 & 0.069 & 0.069 & 0.069 & 0.066 & 0.162 & 0.130 & 0.196 \\
       & & & CS & 0.080 & 0.066 & 0.069 & 0.077 & 0.068 & 0.070 & 0.144 & 0.138 & 0.196 \\
       & & & ID & 0.079 & 0.072 & 0.074 & 0.075 & 0.069 & 0.073 & 0.039 & 0.106 & 0.183 \\ \cline{2-13}
       & \multirow{9}{*}{100} & \multirow{3}{*}{0} & AR & 0.048 & 0.042 & 0.046 & 0.049 & 0.048 & 0.046 & 0.063 & 0.058 & 0.064 \\
       & & & CS & 0.051 & 0.051 & 0.049 & 0.051 & 0.050 & 0.050 & 0.124 & 0.121 & 0.117 \\
       & & & ID & 0.046 & 0.049 & 0.050 & 0.049 & 0.047 & 0.051 & 0.049 & 0.052 & 0.065 \\ \cline{3-13}
       & & \multirow{3}{*}{0.4} & AR & 0.055 & 0.056 & 0.057 & 0.056 & 0.054 & 0.055 & 0.022 & 0.070 & 0.102 \\
       & & & CS & 0.060 & 0.056 & 0.055 & 0.058 & 0.055 & 0.059 & 0.183 & 0.125 & 0.148 \\
       & & & ID & 0.055 & 0.057 & 0.060 & 0.054 & 0.057 & 0.055 & 0.000 & 0.065 & 0.101 \\ \cline{3-13}
       & & \multirow{3}{*}{0.7} & AR & 0.072 & 0.066 & 0.072 & 0.071 & 0.067 & 0.075 & 0.003 & 0.097 & 0.180 \\
       & & & CS & 0.073 & 0.072 & 0.075 & 0.068 & 0.069 & 0.068 & 0.113 & 0.149 & 0.204 \\
       & & & ID & 0.087 & 0.078 & 0.093 & 0.076 & 0.075 & 0.083 & 0.000 & 0.093 & 0.171 \\ \cline{2-13}
       & \multirow{9}{*}{500} & \multirow{3}{*}{0} & AR & 0.045 & 0.048 & 0.049 & 0.051 & 0.042 & 0.046 & 0.059 & 0.054 & 0.064 \\
       & & & CS & 0.050 & 0.048 & 0.043 & 0.049 & 0.048 & 0.045 & 0.119 & 0.110 & 0.104 \\
       & & & ID & 0.049 & 0.043 & 0.045 & 0.047 & 0.046 & 0.047 & 0.051 & 0.049 & 0.061 \\ \cline{3-13}
       & & \multirow{3}{*}{0.4} & AR & 0.058 & 0.056 & 0.060 & 0.055 & 0.054 & 0.061 & 0.000 & 0.069 & 0.102 \\
       & & & CS & 0.054 & 0.058 & 0.052 & 0.056 & 0.058 & 0.053 & 0.166 & 0.122 & 0.149 \\
       & & & ID & 0.056 & 0.049 & 0.056 & 0.057 & 0.048 & 0.057 & 0.000 & 0.063 & 0.099 \\ \cline{3-13}
       & & \multirow{3}{*}{0.7} & AR & 0.093 & 0.072 & 0.097 & 0.079 & 0.068 & 0.081 & 0.000 & 0.103 & 0.174 \\
       & & & CS & 0.078 & 0.069 & 0.079 & 0.074 & 0.070 & 0.077 & 0.103 & 0.151 & 0.211 \\
       & & & ID & 0.104 & 0.070 & 0.102 & 0.074 & 0.064 & 0.083 & 0.000 & 0.099 & 0.177 \\ \hline\hline
    \end{tabular}}
    \caption{Empirical size of single change point testing method when $n=200$}\label{Tab:AR1-Size-1}
\end{table}

\begin{table}[h!]
    \centering
    \scalebox{0.9}{
    \begin{tabular}{c|c|c|c|ccc|ccc|ccc}
    \hline\hline
        \multirow{3}{*}{$n$} & \multirow{3}{*}{$p$} & \multirow{3}{*}{$\kappa$} & \multirow{3}{*}{$\Sigma$} & \multicolumn{6}{c|}{Proposed} & \multicolumn{3}{c}{$T(\eta_0)$} \\ \cline{5-13}
       & & & & \multicolumn{3}{c|}{$\eta=0.02$} & \multicolumn{3}{c|}{$\eta=0.04$} & \multirow{2}{*}{$\eta_0=0$} & \multirow{2}{*}{$\eta_0=0.05$} & \multirow{2}{*}{$\eta_0=0.1$} \\ \cline{5-10}
       & & & & $G_{n,2}$ & $G_{n,\infty}$ & Bonf & $G_{n,2}$ & $G_{n,\infty}$ & Bonf & & & \\ \hline       \multirow{36}{*}{800} & \multirow{9}{*}{3} & \multirow{3}{*}{0} & AR & 0.054 & 0.051 & 0.037 & 0.051 & 0.047 & 0.035 & 0.124 & 0.111 & 0.110 \\
       & & & CS & 0.049 & 0.046 & 0.037 & 0.047 & 0.048 & 0.035 & 0.112 & 0.101 & 0.098 \\
       & & & ID & 0.049 & 0.051 & 0.042 & 0.050 & 0.052 & 0.038 & 0.099 & 0.081 & 0.081 \\ \cline{3-13}
       & & \multirow{3}{*}{0.4} & AR & 0.048 & 0.047 & 0.035 & 0.052 & 0.047 & 0.036 & 0.202 & 0.107 & 0.111 \\
       & & & CS & 0.048 & 0.050 & 0.040 & 0.049 & 0.049 & 0.037 & 0.198 & 0.099 & 0.101 \\
       & & & ID & 0.048 & 0.047 & 0.044 & 0.049 & 0.049 & 0.043 & 0.186 & 0.081 & 0.091 \\ \cline{3-13}
       & & \multirow{3}{*}{0.7} & AR & 0.053 & 0.055 & 0.042 & 0.055 & 0.058 & 0.043 & 0.200 & 0.114 & 0.132 \\
       & & & CS & 0.055 & 0.051 & 0.044 & 0.054 & 0.054 & 0.046 & 0.176 & 0.099 & 0.119 \\
       & & & ID & 0.051 & 0.050 & 0.042 & 0.048 & 0.053 & 0.043 & 0.154 & 0.075 & 0.096 \\ \cline{2-13}
       & \multirow{9}{*}{10} & \multirow{3}{*}{0} & AR & 0.050 & 0.043 & 0.042 & 0.050 & 0.041 & 0.039 & 0.109 & 0.098 & 0.091 \\
       & & & CS & 0.049 & 0.051 & 0.046 & 0.049 & 0.049 & 0.044 & 0.115 & 0.101 & 0.095 \\
       & & & ID & 0.049 & 0.048 & 0.043 & 0.047 & 0.048 & 0.042 & 0.075 & 0.062 & 0.071 \\ \cline{3-13}
       & & \multirow{3}{*}{0.4} & AR & 0.050 & 0.052 & 0.046 & 0.049 & 0.049 & 0.045 & 0.186 & 0.094 & 0.096 \\
       & & & CS & 0.047 & 0.051 & 0.040 & 0.049 & 0.050 & 0.040 & 0.176 & 0.087 & 0.103 \\
       & & & ID & 0.047 & 0.050 & 0.046 & 0.050 & 0.052 & 0.049 & 0.088 & 0.065 & 0.075 \\ \cline{3-13}
       & & \multirow{3}{*}{0.7} & AR & 0.055 & 0.059 & 0.052 & 0.055 & 0.058 & 0.047 & 0.146 & 0.103 & 0.118 \\
       & & & CS & 0.053 & 0.051 & 0.051 & 0.056 & 0.057 & 0.054 & 0.133 & 0.115 & 0.126 \\
       & & & ID & 0.056 & 0.053 & 0.051 & 0.054 & 0.053 & 0.049 & 0.025 & 0.061 & 0.083 \\ \cline{2-13}
       & \multirow{9}{*}{100} & \multirow{3}{*}{0} & AR & 0.051 & 0.056 & 0.053 & 0.052 & 0.049 & 0.046 & 0.072 & 0.057 & 0.065 \\
       & & & CS & 0.049 & 0.047 & 0.046 & 0.046 & 0.048 & 0.044 & 0.120 & 0.116 & 0.117 \\
       & & & ID & 0.051 & 0.051 & 0.056 & 0.051 & 0.051 & 0.055 & 0.059 & 0.054 & 0.060 \\ \cline{3-13}
       & & \multirow{3}{*}{0.4} & AR & 0.050 & 0.053 & 0.052 & 0.053 & 0.049 & 0.053 & 0.022 & 0.057 & 0.076 \\
       & & & CS & 0.047 & 0.046 & 0.042 & 0.048 & 0.047 & 0.041 & 0.167 & 0.108 & 0.118 \\
       & & & ID & 0.055 & 0.044 & 0.051 & 0.046 & 0.053 & 0.050 & 0.000 & 0.054 & 0.066 \\ \cline{3-13}
       & & \multirow{3}{*}{0.7} & AR & 0.057 & 0.058 & 0.059 & 0.057 & 0.060 & 0.056 & 0.002 & 0.064 & 0.085 \\
       & & & CS & 0.052 & 0.051 & 0.047 & 0.053 & 0.053 & 0.046 & 0.101 & 0.119 & 0.131 \\
       & & & ID & 0.054 & 0.052 & 0.056 & 0.050 & 0.056 & 0.057 & 0.000 & 0.053 & 0.079 \\ \cline{2-13}
       & \multirow{9}{*}{500} & \multirow{3}{*}{0} & AR & 0.046 & 0.049 & 0.049 & 0.047 & 0.051 & 0.051 & 0.063 & 0.054 & 0.063 \\
       & & & CS & 0.044 & 0.046 & 0.044 & 0.045 & 0.049 & 0.045 & 0.111 & 0.106 & 0.101 \\
       & & & ID & 0.049 & 0.050 & 0.054 & 0.053 & 0.054 & 0.057 & 0.056 & 0.051 & 0.061 \\ \cline{3-13}
       & & \multirow{3}{*}{0.4} & AR & 0.049 & 0.050 & 0.052 & 0.050 & 0.050 & 0.049 & 0.000 & 0.056 & 0.066 \\
       & & & CS & 0.046 & 0.047 & 0.043 & 0.049 & 0.048 & 0.041 & 0.162 & 0.113 & 0.119 \\
       & & & ID & 0.045 & 0.051 & 0.054 & 0.046 & 0.052 & 0.046 & 0.000 & 0.056 & 0.066 \\ \cline{3-13}
       & & \multirow{3}{*}{0.7} & AR & 0.057 & 0.053 & 0.059 & 0.056 & 0.050 & 0.057 & 0.000 & 0.058 & 0.086 \\
       & & & CS & 0.055 & 0.052 & 0.055 & 0.058 & 0.055 & 0.056 & 0.094 & 0.117 & 0.122 \\
       & & & ID & 0.056 & 0.053 & 0.058 & 0.053 & 0.053 & 0.059 & 0.000 & 0.061 & 0.083 \\ \hline\hline
    \end{tabular}}
    \caption{Empirical size of single change point testing when $n=800$}\label{Tab:AR1-Size-2}
\end{table}

In general, the numerical results exactly match the plots in Section \ref{Sec:NumResults-simul-size-singleCP} and there is no major difference between the size accuracy of $G_{n,2}$, $G_{n,\infty}$ and \texttt{Bonf}. Specifically, with $n=200$, though there is some slight size distortion when $\kappa=0.7$, our method has already achieved accurate empirical size in most situations. The size accuracy of our method significantly improves as the sample size increases to $n=800$, where our empirical size is stable and accurate across the table. By contrast, the trimming-based SN test by \cite{wang2022inference} has noticeable size distortion especially when the dimension $p$ is low and such distortion doesn't seem to improve with larger sample size. This is not surprising since the methodology in \cite{wang2022inference} is tailored to high-dimensional data. Also it appears that the results for $\eta=0.02$ and $0.04$ are similar, showing the insensitivity of the results to the choice of $\eta$.

\subsection{Empirical Size for Multiple Change Point Testing}\label{Appendix:AddNumResults-size-multiCP}


Next, we look into the size accuracy of the generalized test targeting multiple change points based on the same data-generating process as used in Section \ref{Sec:NumResults-simul-size-singleCP} and in Appendix \ref{Appendix:AddNumResults-size-singleCP}. In this case, we consider $n\in\{100,300\}$, $p\in\{5,10,20,25,50,100,150,200,250,500,750,1000,2500,5000\}$. When $\Sigma$ takes the form of \texttt{AR}, we consider $\rho\in\{0.4,0.7\}$.

For the generalized SS-SN method, we set the splitting ratio $\varepsilon=0.1$ and consider the trimming ratio $\eta=0.02$.

To compare the performance of the multiple change points testing method, we adopt the test statistic $T_n^{\Diamond}$ formulated in Section 2.2 of \cite{wang2022inference}. Note that $T_n^{\Diamond}$ takes a trimming parameter $\eta_0$, then we follow the recommendations of \cite{wang2022inference} to set $\eta_0=0.1$ in our simulated studies and denote it as $T^M(0.1)$ to distinguish from $T(\eta_0)$ used in the previous section. 
Note that the multiple change-point test in \cite{wang2022inference} only works for high-dimensional independent data and does not provide good size in the presence of temporal dependence (results not shown), we shall just focus on the comparison for independent data sequence. 

\begin{figure}[h!]
    \centering
    \includegraphics[width=0.9\textwidth]{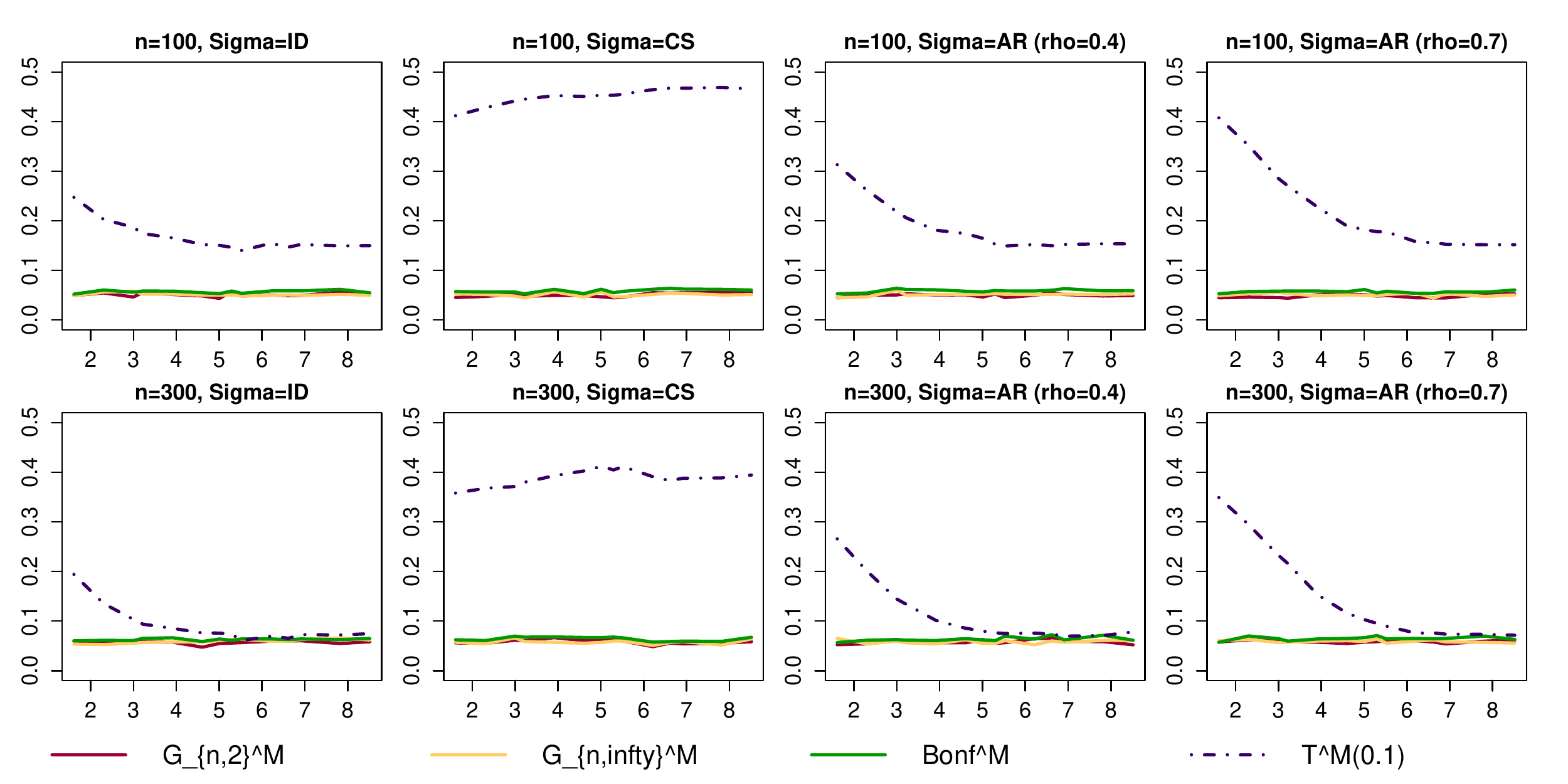}
    \caption{Empirical size curves versus the logarithm of $p$ against the multiple change points alternative}
    \label{Fig:SizePlot-M-2+3+4}
\end{figure}

All the empirical sizes are averaged over $T=5000$ Monte Carlo replicates, and we plot the empirical size against the logarithm of $p$ in Figure \ref{Fig:SizePlot-M-2+3+4}. 

The two panels of Figure \ref{Fig:SizePlot-M-2+3+4} correspond to $n=100$ and $n=300$, and the four columns correspond to four types of $\Sigma$, namely, ID, CS, AR with $\rho=0.4$ and $\rho=0.7$. In each plot of Figure \ref{Fig:SizePlot-M-2+3+4}, we use the solid lines in red and orange to represent the results of $G_{n,2}^{M}$ and $G_{n,\infty}^{M}$. The solid green curve stands for $\texttt{Bonf}^M$ based on both dense and sparse tests. The dotted purple curve corresponds to $T^M(0.1)$ proposed in \cite{wang2022inference}. 

According to Figure \ref{Fig:SizePlot-M-2+3+4}, all of $G_{n,2}^{M}$, $G_{n,\infty}^{M}$ and $\texttt{Bonf}^{M}$ achieve stable size around the nominal level under all settings, and the sizes for SS-SN methods are already accurate when $n=100$. 
By contrast, the high-dimensional SN method $T^M(0.1)$ has noticeable size distortion in most cases. In particular, the size distortion worsens as the cross-sectional dependence strengthens, and we can see increasing distortion when $\rho$ increases from 0.4 to 0.7 in the AR case and can notice the severest size distortion in the CS case. Note that the test $T^M(0.1)$ is not expected to work in the CS case as the strong cross-sectional dependence is expected to alter the limiting null distribution derived in \cite{wang2022inference}. 
In general, when the componentwise dependence is weak, $T^{M}(0.1)$ achieves a more accurate empirical size either as the sample size $n$ grows or $p$ increases. Overall, the size accuracy and stability of our SS-SN tests across all dimensional settings is impressive and they are far superior to those for $T^M(0.1)$.

As a supplement to Figure \ref{Fig:SizePlot-M-2+3+4}, we present some numerical results in Table \ref{Tab:Multi-Size}, Here, we consider $n\in\{100,300\}$, $p\in\{5,10,100,500\}$. For the proposed method, we fix the splitting parameter $\epsilon=0.1$ but additionally consider the trimming parameter $\eta=0.04$. Again, the empirical sizes averaged over 5000 Monte-Carlo replicates are reported in Table \ref{Tab:Multi-Size}.
	
\begin{table}[h!]
    \centering
    \scalebox{0.9}{
    \begin{tabular}{c|c|cc|ccc|ccc|c}
    \hline\hline
        \multirow{3}{*}{$n$} & \multirow{3}{*}{$p$} & \multicolumn{2}{c|}{\multirow{3}{*}{$\Sigma$}} & \multicolumn{6}{c|}{Proposed} & \multirow{3}{*}{$T^{M}(0.1)$} \\ \cline{5-10}
       & & & & \multicolumn{3}{c|}{$\eta=0.02$} & \multicolumn{3}{c|}{$\eta=0.04$} & \\ \cline{5-10}
       & & & & $G_{n,2}^{M}$ & $G_{n,\infty}^{M}$ & $\mbox{Bonf}^{M}$ & $G_{n,2}^{M}$ & $G_{n,\infty}^{M}$ & $\mbox{Bonf}^{M}$ & \\ \hline 
      \multirow{16}{*}{100} & \multirow{4}{*}{5} & \multicolumn{2}{c|}{ID} & 0.049 & 0.049 & 0.052 & 0.049 & 0.047 & 0.054 & 0.247 \\ \cline{3-11}
      & & \multicolumn{1}{c|}{\multirow{2}{*}{AR}} & $\rho=0.4$ & 0.045 & 0.045 & 0.053 & 0.045 & 0.048 & 0.053 & 0.313 \\ \cline{4-11}
      & & \multicolumn{1}{c|}{} & $\rho=0.7$ & 0.045 & 0.049 & 0.053 & 0.046 & 0.049 & 0.050 & 0.408 \\ \cline{3-11}
      & & \multicolumn{2}{c|}{CS} & 0.045 & 0.053 & 0.057 & 0.047 & 0.052 & 0.054 & 0.412 \\ \cline{2-11}
      & \multirow{4}{*}{10} & \multicolumn{2}{c|}{ID} & 0.055 & 0.056 & 0.060 & 0.055 & 0.054 & 0.060 & 0.203 \\ \cline{3-11}
      & & \multicolumn{1}{c|}{\multirow{2}{*}{AR}} & $\rho=0.4$ & 0.050 & 0.046 & 0.054 & 0.053 & 0.052 & 0.060 & 0.261 \\ \cline{4-11}
      & & \multicolumn{1}{c|}{} & $\rho=0.7$ & 0.046 & 0.053 & 0.057 & 0.047 & 0.053 & 0.057 & 0.352 \\ \cline{3-11}
      & & \multicolumn{2}{c|}{CS} & 0.048 & 0.051 & 0.057 & 0.049 & 0.048 & 0.058 & 0.428 \\ \cline{2-11}
      & \multirow{4}{*}{10} & \multicolumn{2}{c|}{ID} & 0.049 & 0.050 & 0.055 & 0.048 & 0.051 & 0.058 & 0.152 \\ \cline{3-11}
      & & \multicolumn{1}{c|}{\multirow{2}{*}{AR}} & $\rho=0.4$ & 0.051 & 0.051 & 0.058 & 0.055 & 0.053 & 0.060 & 0.174 \\ \cline{4-11}
      & & \multicolumn{1}{c|}{} & $\rho=0.7$ & 0.053 & 0.051 & 0.057 & 0.054 & 0.048 & 0.060 & 0.189 \\ \cline{3-11}
      & & \multicolumn{2}{c|}{CS} & 0.049 & 0.046 & 0.053 & 0.048 & 0.049 & 0.055 & 0.451 \\ \cline{2-11}
      & \multirow{4}{*}{10} & \multicolumn{2}{c|}{ID} & 0.051 & 0.050 & 0.059 & 0.052 & 0.055 & 0.062 & 0.155 \\ \cline{3-11}
      & & \multicolumn{1}{c|}{\multirow{2}{*}{AR}} & $\rho=0.4$ & 0.050 & 0.051 & 0.058 & 0.050 & 0.047 & 0.057 & 0.152 \\ \cline{4-11}
      & & \multicolumn{1}{c|}{} & $\rho=0.7$ & 0.045 & 0.054 & 0.054 & 0.051 & 0.048 & 0.060 & 0.158 \\ \cline{3-11}
      & & \multicolumn{2}{c|}{CS} & 0.055 & 0.052 & 0.062 & 0.050 & 0.055 & 0.060 & 0.464 \\ \hline
      \multirow{16}{*}{300} & \multirow{4}{*}{5} & \multicolumn{2}{c|}{ID} & 0.059 & 0.053 & 0.060 & 0.051 & 0.055 & 0.061 & 0.194 \\ \cline{3-11}
      & & \multicolumn{1}{c|}{\multirow{2}{*}{AR}} & $\rho=0.4$ & 0.052 & 0.065 & 0.057 & 0.051 & 0.066 & 0.062 & 0.266 \\ \cline{4-11}
      & & \multicolumn{1}{c|}{} & $\rho=0.7$ & 0.058 & 0.060 & 0.057 & 0.057 & 0.057 & 0.058 & 0.349 \\ \cline{3-11}
      & & \multicolumn{2}{c|}{CS} & 0.056 & 0.057 & 0.062 & 0.056 & 0.054 & 0.058 & 0.358 \\ \cline{2-11}
      & \multirow{4}{*}{10} & \multicolumn{2}{c|}{ID} & 0.058 & 0.052 & 0.061 & 0.057 & 0.056 & 0.061 & 0.135 \\ \cline{3-11}
      & & \multicolumn{1}{c|}{\multirow{2}{*}{AR}} & $\rho=0.4$ & 0.054 & 0.054 & 0.061 & 0.058 & 0.054 & 0.061 & 0.201 \\ \cline{4-11}
      & & \multicolumn{1}{c|}{} & $\rho=0.7$ & 0.062 & 0.063 & 0.070 & 0.049 & 0.059 & 0.060 & 0.294 \\ \cline{3-11}
      & & \multicolumn{2}{c|}{CS} & 0.055 & 0.054 & 0.061 & 0.056 & 0.055 & 0.058 & 0.368 \\ \cline{2-11}
      & \multirow{4}{*}{100} & \multicolumn{2}{c|}{ID} & 0.047 & 0.058 & 0.059 & 0.054 & 0.060 & 0.063 & 0.076 \\ \cline{3-11}
      & & \multicolumn{1}{c|}{\multirow{2}{*}{AR}} & $\rho=0.4$ & 0.056 & 0.062 & 0.064 & 0.056 & 0.057 & 0.062 & 0.085 \\ \cline{4-11}
      & & \multicolumn{1}{c|}{} & $\rho=0.7$ & 0.055 & 0.060 & 0.065 & 0.055 & 0.057 & 0.059 & 0.116 \\ \cline{3-11}
      & & \multicolumn{2}{c|}{CS} & 0.061 & 0.055 & 0.067 & 0.060 & 0.056 & 0.064 & 0.403 \\ \cline{2-11}
      & \multirow{4}{*}{100} & \multicolumn{2}{c|}{ID} & 0.059 & 0.059 & 0.064 & 0.057 & 0.053 & 0.055 & 0.069 \\ \cline{3-11}
      & & \multicolumn{1}{c|}{\multirow{2}{*}{AR}} & $\rho=0.4$ & 0.064 & 0.052 & 0.063 & 0.060 & 0.058 & 0.067 & 0.076 \\ \cline{4-11}
      & & \multicolumn{1}{c|}{} & $\rho=0.7$ & 0.060 & 0.059 & 0.065 & 0.062 & 0.057 & 0.065 & 0.075 \\ \cline{3-11}
      & & \multicolumn{2}{c|}{CS} & 0.048 & 0.051 & 0.057 & 0.053 & 0.058 & 0.062 & 0.391 \\ \hline\hline
      \end{tabular}}
    \caption{Empirical size of multiple change points testing}\label{Tab:Multi-Size}
\end{table}

As shown in Table \ref{Tab:Multi-Size}, the proposed multiple change points testing method achieves stable size accuracy regardless of $p$ and the overall empirical size is already accurate when $n=100$. Overall, the finding is quantitatively similar to Figure \ref{Fig:SizePlot-M-2+3+4}.

\subsection{Power Analysis for Single Sparse Change Point Testing}\label{Appendix:AddNumResults-power-singleCP-sparse}

Next, we investigate the power behavior of the proposed test against a single sparse change point. The data is generated from the same AR(1) model as in previous sections. In this case, we fix $n=200$, $\kappa=0.7$, $\rho=0.8$ for AR(1) type of $\Sigma$, and consider $p\in\{3,10,100,500\}$. The location of the change point is set as $k=\lfloor{n/2}\rfloor$. For the sparse change point, the mean vector $\mu_t$ is generated by
\begin{equation*}
    \mu_t
  = \left\{
    \begin{array}{ll}
        (0,\cdots,0)^{\top},              & 1 \le t \le k \\
        c(0,0,1,0,\cdots, 0)^{\top}, & k+1 \le t \le n
    \end{array}
    \right.
\end{equation*}
Here, we use $c$ to quantify the signal-noise-ratio, which ranges over a respective set of values under each pair of $(p,\Sigma)$.

For the proposed SS-SN tests, we set the splitting parameter $\varepsilon=0.1$ and the trimming parameter $\eta=0.04$. The same competing methods are adopted as in Section \ref{Sec:NumResults-simul-power-singleCP}.

\begin{figure}[h!]
    \centering
    \includegraphics[width=0.9\textwidth]{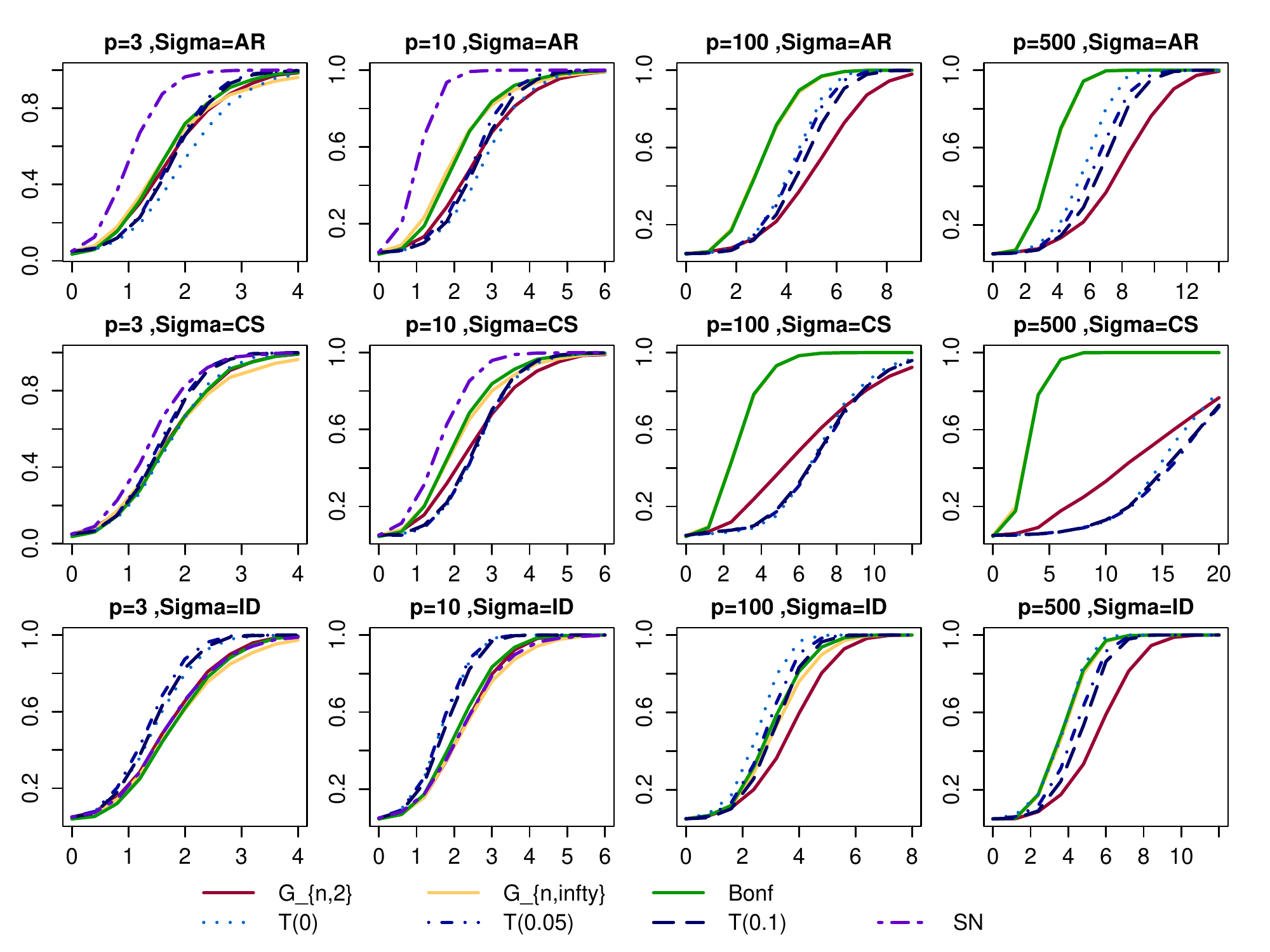}
    \caption{Power curves (size-adjusted) against a single sparse change point when $n=200$}
    \label{Fig:SparsePowerCurve}
\end{figure}

According to Figure \ref{Fig:SparsePowerCurve}, the ranking of all the methods in terms of power depends on the dependence structure $\Sigma$ and the dimension. Specifically, \texttt{SN} in \cite{shao2010testing} achieves the best power under the existence of cross-sectional dependence (AR or CS) when $p$ is small ($p=3,10$), which is contrary to our observation in the dense alternative. We observe that $G_{n,\infty}$ has a comparable empirical rejection rate with $\texttt{Bonf}$, and both outperform $G_{n,2}$ as expected. Although our methods do not show much advantage in the ID case, it is interesting to see that the power of $G_{n,\infty}$ and $\texttt{Bonf}$ significantly outperform all others in both AR and CS cases when $p=100, 500$. The performance of $T(\eta_0)$ with multiple $\eta_0$ seems similar across all the settings.

\subsection{Power Analysis for Multiple Change Points Testing}\label{Appendix:AddNumResults-power-multiCP}

Lastly, we examine the power behavior against the multiple change points alternative. We still generate a sample of iid vectors $\{\widetilde{X}_t\}_{t=1}^{n} \in \br^p$ from $\cn_p(0,\Sigma)$ with three types of $\Sigma$. For the AR(1) structure, we set $\rho=0.5$. Under the scenario where multiple change points, dense or sparse, are present, we consider various possible combinations as follows.
\begin{enumerate}[label=(\roman*)]
    \item {\bf Two dense change points}, denoted by \texttt{D-D}.
    \BEqn
    & & \mu_1 = \cdots = \mu_{k_1} = (0,\cdots,0)^{\top} \quad\mbox{with } k_1=\lrfl{0.3n},\\
    & & \mu_{k_1+1} = \cdots = \mu_{k_2} = c (1,\cdots,1)^{\top} / \sqrt{p} \quad\mbox{with } k_2=\lrfl{0.7n},\\
    & & \mu_{k_2+1} = \cdots = \mu_{n} = 2c (1,\cdots,1)^{\top} / \sqrt{p},
    \EEqn

    \item {\bf Two sparse change points}, denoted by \texttt{S-S}.
    \BEqn
    & & \mu_1 = \cdots = \mu_{k_1} = (0,\cdots,0)^{\top} \quad\mbox{with } k_1=\lrfl{0.3n},\\
    & & \mu_{k_1+1} = \cdots = \mu_{k_2} = c (0,0,1,0,\cdots,0)^{\top} \quad\mbox{with } k_2=\lrfl{0.7n},\\
    & & \mu_{k_2+1} = \cdots = \mu_{n} = 2c (0,0,1,0,\cdots,0)^{\top},
    \EEqn
    
    \item {\bf One dense change point and one sparse change point}, denoted by \texttt{D-S}.
    \BEqn
    & & \mu_1 = \cdots = \mu_{k_1} = (0,\cdots,0)^{\top} \quad\mbox{with } k_1=\lrfl{0.3n},\\
    & & \mu_{k_1+1} = \cdots = \mu_{k_2} = c (1,\cdots,1)^{\top} / \sqrt{p} \quad\mbox{with } k_2=\lrfl{0.7n},\\
    & & \mu_{k_2+1} = \cdots = \mu_{n} = 2c (1,1,1,1,1,0,\cdots,0)/\sqrt{5}.
    \EEqn
    
    \item {\bf Three dense change points}, denoted by \texttt{D-D-D}.
    \BEqn
    & & \mu_1 = \cdots = \mu_{k_1} = (0,\cdots,0)^{\top} \quad\mbox{with } k_1=\lrfl{0.2n},\\
    & & \mu_{k_1+1} = \cdots = \mu_{k_2} = c (1,\cdots,1)^{\top} / \sqrt{p} \quad\mbox{with } k_2=\lrfl{0.4n},\\
    & & \mu_{k_2+1} = \cdots = \mu_{k_3} = 2c (1,\cdots,1)^{\top} / \sqrt{p} \quad\mbox{with } k_3=\lrfl{0.8n},\\
    & & \mu_{k_3+1} = \cdots = \mu_{n} = 3c (1,\cdots,1)^{\top} / \sqrt{p}.
    \EEqn
\end{enumerate}

For all these scenarios, we fix $n=100$ and consider $p\in\{10,50,100,500\}$. For each case, a respective grid of signal-noise-ratio $c$'s is selected. 

\begin{figure}[h!]
    \centering
    \includegraphics[width=0.9\textwidth]{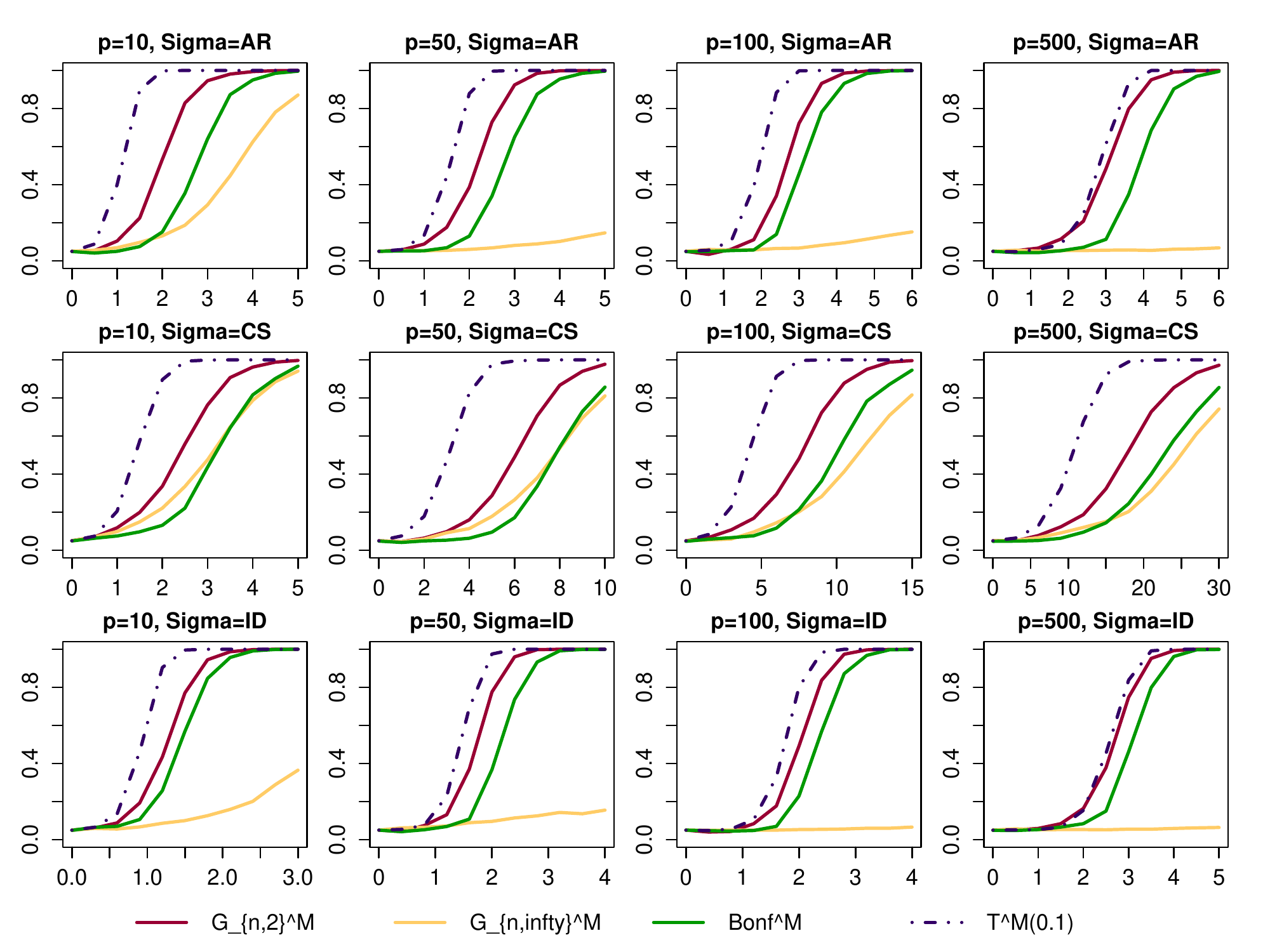}
    \caption{Power curves (size-adjusted) against multiple change points of type \texttt{D-D} when $n=100$}
    \label{Fig:PowerCurve-M-DD}
\end{figure}

\begin{figure}[h!]
    \centering
    \includegraphics[width=0.9\textwidth]{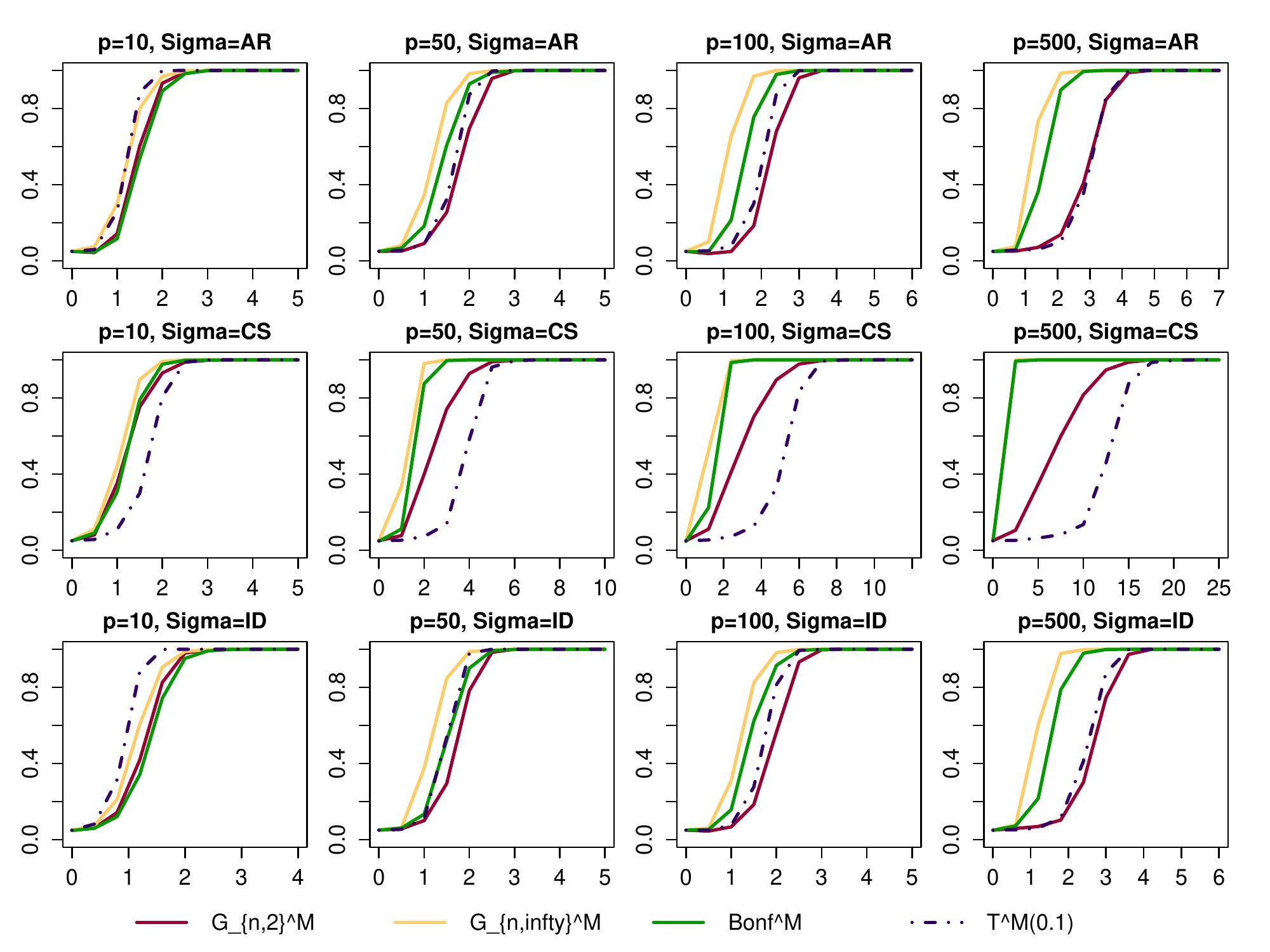}
    \caption{Power curves (size-adjusted) against multiple change points of type \texttt{S-S} when $n=100$}
    \label{Fig:PowerCurve-M-SS}
\end{figure}

\begin{figure}[!h]
    \centering
    \includegraphics[width=0.9\textwidth]{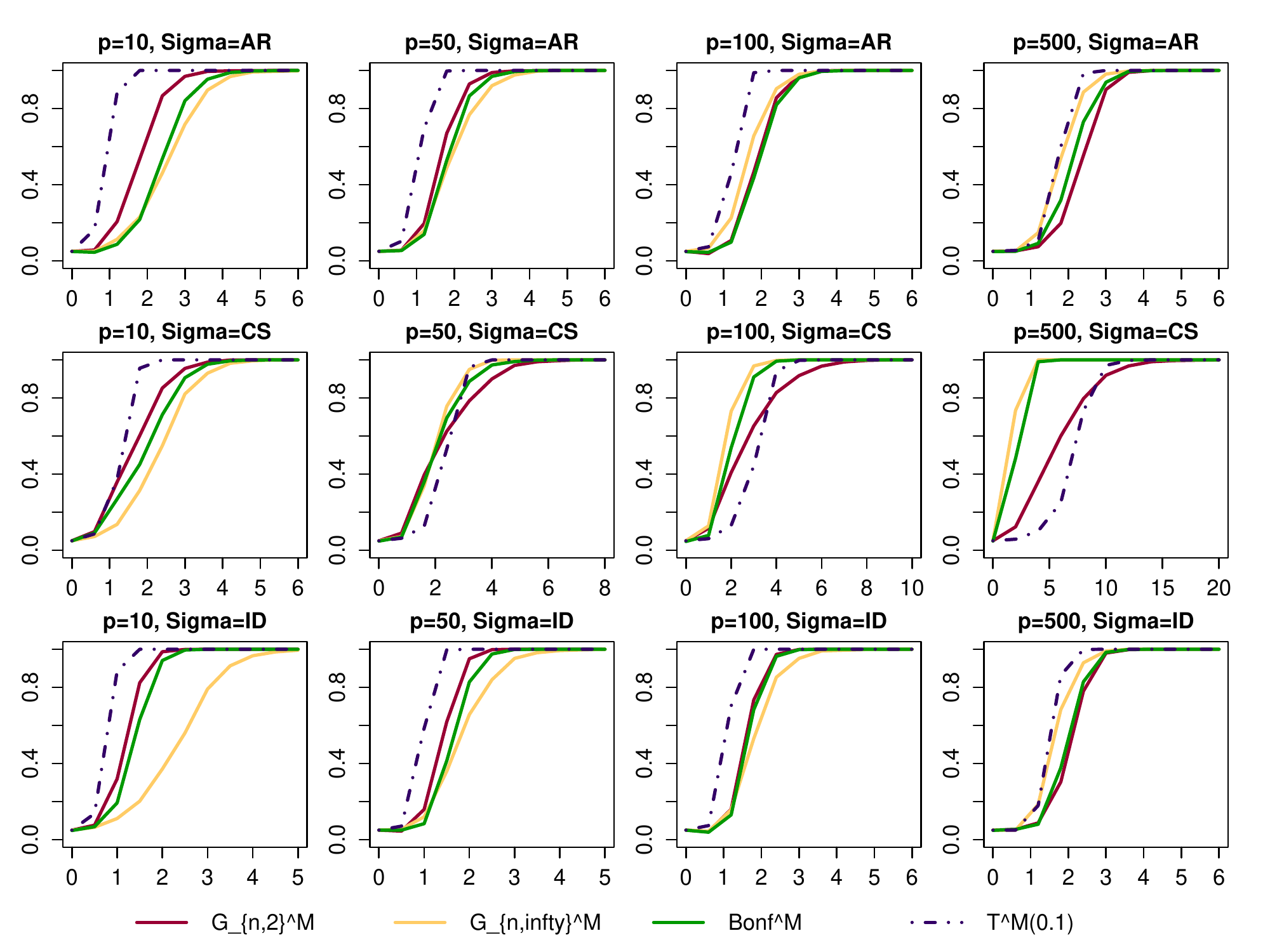}
    \caption{Power curves (size-adjusted) against multiple change points of type \texttt{D-S} when $n=100$}
    \label{Fig:PowerCurve-M-DS}
\end{figure}

Figure \ref{Fig:PowerCurve-M-DD}-\ref{Fig:PowerCurve-M-DS} present the size-adjusted power curves corresponding to all the four scenarios of interest, and all the simulation results are based on $T=1000$ MC replicates. In each plot, the four columns in each plot correspond to the cases when $p=10,50,100$ and $500$ whereas the three rows refer to the three types of $\Sigma$. 

The patterns shown from Figure \ref{Fig:PowerCurve-M-DD} and Figure \ref{Fig:PowerCurve-M-SS} generally match those in Section \ref{Sec:NumResults-simul-power-singleCP}. In the \texttt{D-D} case, the sparse test $G_{n,\infty}^{M}$ does not show much power against multiple dense change points, and  its power loss is relatively much less in the CS case, which is also observed in the case of single dense alternative. Also, we see a slight advantage of $G_{n,2}^M$ over $\texttt{Bonf}^{M}$, mostly under the non-ID case, which is expected and both tests are consistent and their powers go to $1$ when the signal-to-noise ratio is sufficiently large. The most powerful test is $T^M(0.1)$ and the tradeoff between size distortion and power loss is apparent from these plots.

In the \texttt{S-S} case, the proposed $G_{n,\infty}^{M}$ and $\texttt{Bonf}^{M}$ achieve significant power gain over others, and their performance dominates in most cases, especially when $p$ is large or the data has strong cross-sectional dependence. In contrast, $G_{n,2}^{M}$ has comparable performance with $G_{n,\infty}^{M}$ and $\texttt{Bonf}^{M}$ under the low-dimensional setting, but has some significantly more power loss as $p$ grows. As for $T^{M}(0.1)$, though it has competitive power behavior in the ID case or the low-dimensional case, we observe that it is outperformed by $G_{n,\infty}^{M}$ and $\texttt{Bonf}^{M}$ as $p$ increases under all cases and is outperformed by $G_{n,2}^{M}$ under the existence of cross-sectional dependence, which is generally consistent with what we observe in the single sparse alternative; see Section \ref{Sec:NumResults-simul-power-singleCP}.

\begin{figure}[!h]
    \centering
    \includegraphics[width=0.9\textwidth]{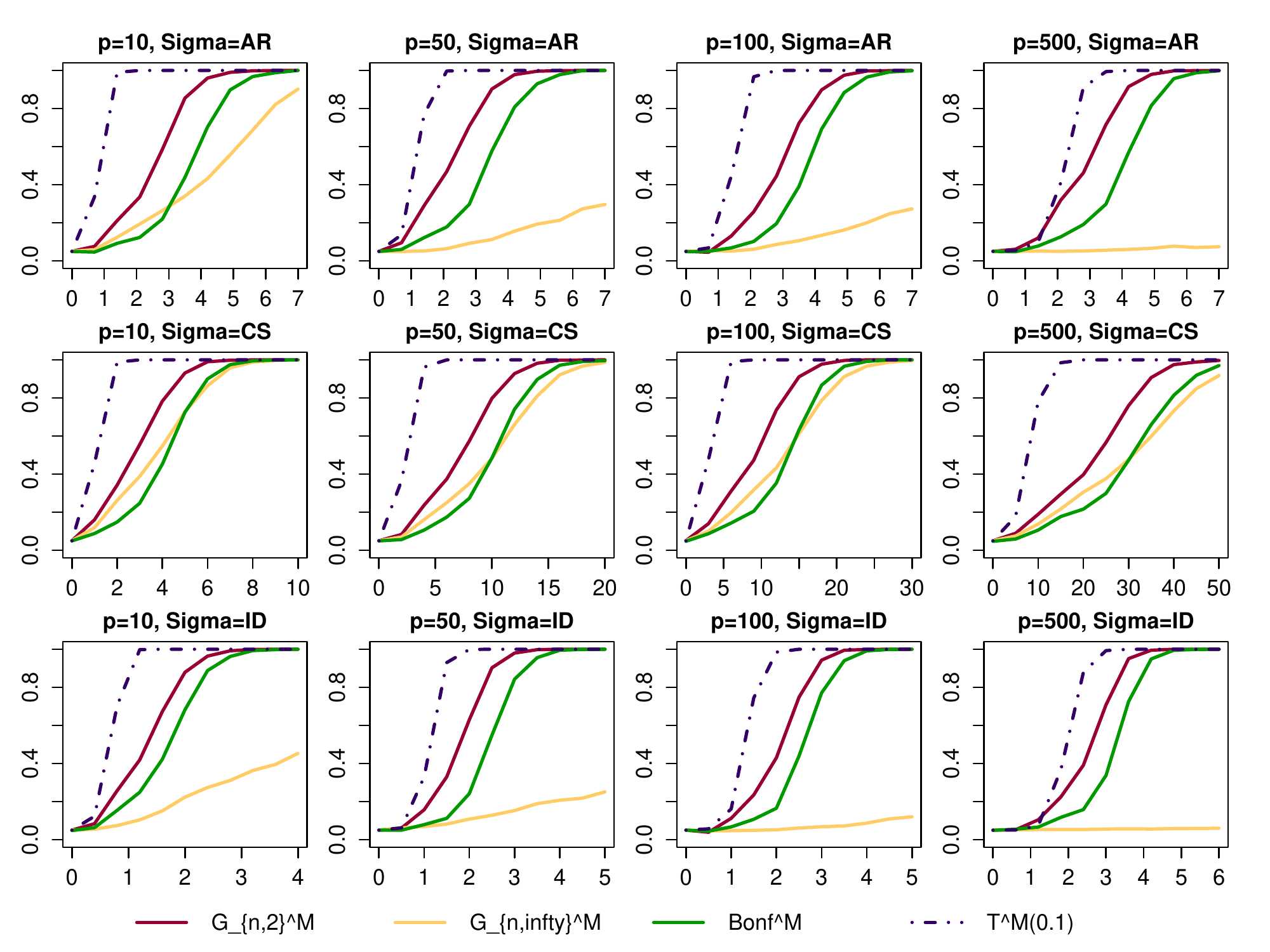}
    \caption{Power curves (size-adjusted) against multiple change points of type \texttt{D-D-D} when $n=100$}
    \label{Fig:PowerCurve-M-DDD}
\end{figure}

It is not surprising that $G_{n,\infty}^{M}$ still has severe power loss in the case of \texttt{D-D-D}, especially when $\Sigma$ takes an identity matrix or an $AR(1)$ matrix. As expected, the power of $G_{n,\infty}^{M}$ improves under the existence of sparse mean shift, i.e., in the $\texttt{D-S}$ case. The performance of $G_{n,2}^{M}$ and \texttt{Bonf} are similar in most scenarios, though both of them still have a noticeable power loss when compared to $T^M(0.1)$ in \cite{wang2022inference}, except when $p$ is large and the cross-sectional dependence is strong.

\section{Proofs of Main Theorems}\label{Appdix:MainThm}

\subsection{Proof of Theorem \ref{Thm:NullDist_Gn_Fixedp}}


It follows from Definition \ref{Def:Model_Fixedp} that $\frac{1}{\sqrt{n}} \sum\limits_{t=1}^{\lfloor{nr}\rfloor} X_t \leadsto (\Omega^{(1)})^{1/2} B_p(r)$ in $D^p[0,1]$ as $n\rightarrow\infty$, where $\{B_p(r)\}_{r\in[0,1]}$ denotes the standard $p$-dimensional Brownian motion. For any $r\in[0,1]$, we have that
\BEqn
    S_{1,\lrfl{Nr}} 
&=& \sum\limits_{j=1}^{\lrfl{Nr}} \lrag{\hat\mu_1-\hat\mu_n, X_{j+m}} \\
&=& \frac{1}{m_1} \lrp{\sum\limits_{i=1}^{m_1} X_i}^{\top} \lrp{\sum\limits_{j=1}^{\lrfl{Nr}} X_{j+m}}
 -  \frac{1}{m_1} \lrp{\sum\limits_{i=1}^{m_1} X_{n+1-i}}^{\top} \lrp{\sum\limits_{j=1}^{\lrfl{Nr}} X_{j+m}} \\
&\leadsto& \frac{1}{\varepsilon-\eta} \lrp{B_p(\varepsilon-\eta) - B_p(1) + B_p(1-\varepsilon+\eta)}^{\top} \Omega^{(1)} \lrp{B_p(\varepsilon+(1-2\varepsilon)r) - B_p(\varepsilon)}.
\EEqn

Define $\cb_p = \{B_p(s)\}_{\varepsilon \le s \le 1-\varepsilon}$ and $b(\varepsilon,\eta) := \frac{1}{\varepsilon-\eta}\lrp{B_p(\varepsilon-\eta) - B_p(1) + B_p(1-\varepsilon+\eta)}$, then it holds for $k = \lrfl{Nr}$ that
\BEqn
  S_{1,k} 
&\leadsto& b^{\top}(\varepsilon,\eta) \Omega^{(1)} \lrp{B_p(\varepsilon+(1-2\varepsilon)r) - B_p(\varepsilon)}, \\
  \frac{k}{N}S_{1,N} 
&\leadsto& r b^{\top}(\varepsilon,\eta) \Omega^{(1)} \lrp{B_p(1-\varepsilon) - B_p(\varepsilon)},
\EEqn
both of which imply that
\BEqn
    \sqrt{N} T_n(k)
&=& S_{1,k} - \frac{k}{N} S_{1,N} \\
&\leadsto& b^{\top}(\varepsilon,\eta) \Omega^{(1)} \lrp{B_p(\varepsilon+(1-2\varepsilon)r) - B_p(\varepsilon) - r\lrp{B_p(1-\varepsilon) - B_p(\varepsilon)}} \\
&=:& T(r,b(\varepsilon,\eta),\cb_p).
\EEqn

Similarly, we have that
\BEqn
& & \frac{1}{N} \sum\limits_{t=1}^{k} \lrp{S_{1,t} - \frac{t}{k}S_{1,k}}^2 \\
&\leadsto& \int_{0}^{r} 
           \lrp{b^{\top}(\varepsilon,\eta) \Omega^{(1)} \bigp{B_p(\varepsilon+(1-2\varepsilon)s) - B_p(\varepsilon) - \frac{s}{r}\lrp{B_p(\varepsilon+(1-2\varepsilon)r) - B_p(\varepsilon)}}}^2 ds, \\[4mm]
& & \frac{1}{N} \sum\limits_{t=k+1}^{N} \lrp{S_{t,N} - \frac{N-t+1}{N-k}S_{k+1,N}}^2 \\
&=& \frac{1}{N} \sum\limits_{t=k+1}^{N} \lrp{\lrp{S_{1,N}-S_{1,t-1}} - \frac{N-t+1}{N-k}\lrp{S_{1,N}-S_{1,k}}}^2 \\
&\leadsto& \int_{r}^{1} 
           \left(
             b^{\top} (\varepsilon,\eta) \Omega^{(1)} 
             \big(B_p(1-\varepsilon) - B_p(\varepsilon+(1-2\varepsilon)s)
           \right. \\
& & \hspace{8em}           
           \left.
             - \frac{1-s}{1-r} \lrp{B_p(1-\varepsilon) - B_p(\varepsilon+(1-2\varepsilon)r)} \big)
           \right)^2 ds,
\EEqn
and it follows that
\BEqn
& & N V_n(k) \\
&=& \frac{1}{N} \sum\limits_{t=1}^{k} \lrp{S_{1,t} - \frac{t}{k}S_{1,k}}^2 
 +  \frac{1}{N} \sum\limits_{t=k+1}^{N} \lrp{S_{t,N} - \frac{N-t+1}{N-k}S_{k+1,N}}^2 \\
&\leadsto& \int_{0}^{r} \lrp{b^{\top}(\varepsilon,\eta) \Omega^{(1)} \bigp{B_p(\varepsilon+(1-2\varepsilon)s) - B_p(\varepsilon) - \frac{s}{r}\lrp{B_p(\varepsilon+(1-2\varepsilon)r) - B_p(\varepsilon)}}}^2 ds \\
& & + \int_{r}^{1} \lrp{b^{\top}(\varepsilon,\eta) \Omega^{(1)} \bigp{B_p(1-\varepsilon) - B_p(\varepsilon+(1-2\varepsilon)s) - \frac{1-s}{1-r}\lrp{B_p(1-\varepsilon) - B_p(\varepsilon+(1-2\varepsilon)r)}}}^2 ds \\
&=:& V(r,b(\varepsilon,\eta),\cb_p).
\EEqn

By applying the continuous mapping theorem, we obtain that
\begin{equation*}
    G_n
 = \sup\limits_{k=1,\cdots,N-1} T_n(k)V_n^{-1/2}(k)
 \stra{d} M(b(\varepsilon,\eta),\cb_p)
 := \sup\limits_{r\in[0,1]} T(r,b(\varepsilon,\eta),\cb_p) V_n^{-1/2}(r,b(\varepsilon,\eta),\cb_p).
\end{equation*}

Recall that $b(\varepsilon,\eta)$ is independent of $\cb_p$ since Brownian motion has independent increments, hence if conditioning on $b(\varepsilon,\eta)=b_0$, it holds that the processes $\{b_0^{\top} \Omega^{(1)} B_p(s)\}_{\varepsilon \le s \le 1-\varepsilon}$ and $\{\omega B(s)\}_{\varepsilon \le s \le 1-\varepsilon}$ are equal in distribution, where $\omega = \sqrt{b_0^{\top} (\Omega^{(1)})^2 b_0}$. Additionally, the process $\{B(\varepsilon+(1-2\varepsilon)r)-B(\varepsilon)\}_{0\le r\le 1}$ is equal in distribution with the process $\{\sqrt{1-2\varepsilon} B(r)\}_{0\le r\le 1}$. Consequently, we have that
\BEqn
& & M(b(\varepsilon,\eta),\cb) |_{b(\varepsilon,\eta)=b_0} \\
&=& \sup_{r\in[0,1]}
    \frac{b_0^{\top} \Omega^{(1)} \bigp{B_p(\varepsilon+(1-2\varepsilon)r) - B_p(\varepsilon) - r\lrp{B_p(1-\varepsilon) - B_p(\varepsilon)}}}
    {\lrp{
     \begin{array}{l}
     ~ ~ \int_{0}^{r} \lrp{b_0^{\top} \Omega^{(1)} \bigp{B_p(\varepsilon+(1-2\varepsilon)s) - B_p(\varepsilon) - \frac{s}{r}\lrp{B_p(\varepsilon+(1-2\varepsilon)r) - B_p(\varepsilon)}}}^2 ds \\[1mm]
     + \int_{r}^{1} \lrp{b_0^{\top} \Omega^{(1)} \bigp{B_p(1-\varepsilon) - B_p(\varepsilon+(1-2\varepsilon)s) - \frac{1-s}{1-r}\lrp{B_p(1-\varepsilon) - B_p(\varepsilon+(1-2\varepsilon)r)}}}^2 ds
     \end{array}
    }^{1/2}} \\[2mm]
&=^d& \sup\limits_{r\in[0,1]} \frac{\omega \lrp{B(\varepsilon+(1-2\varepsilon)r) - B(\varepsilon) - r\lrp{B(1-2\varepsilon) - B(\varepsilon)}}}
{\lrp{
     \begin{array}{l}
     ~~ \omega^2 \int_{0}^{r} \lrp{B(\varepsilon+(1-2\varepsilon)s) - B(\varepsilon) - \frac{s}{r} \lrp{B(\varepsilon+(1-2\varepsilon)r)-B(\varepsilon)}}^2 ds \\[1mm]
     + \omega^2 \int_{r}^{1} \lrp{B(1-\varepsilon)-B(\varepsilon+(1-2\varepsilon)s) - \frac{1-s}{1-r} \lrp{B(1-\varepsilon)-B(\varepsilon+(1-2\varepsilon)r)}}^2 ds
     \end{array}
}^{1/2}} \\[2mm]
&=^d& \sup\limits_{r\in[0,1]} \frac{B(r) - rB(1)}
{\lrp{\int_{0}^{r} \lrp{B(s) - \frac{s}{r}B(r)}^2ds + \int_{r}^{1} \lrp{B(1-s) - \frac{1-s}{1-r} B(1-r)}^2 ds}^{1/2}} \\
&=^d& G,
\EEqn
which is independent of $b(\varepsilon,\eta)=b_0$. This  implies that $M(b(\varepsilon,\eta),\cb) =^d G$ and consequently, $G_n \stra{d} G$ as $n\rightarrow\infty$, which completes the proof.

\subsection{Proof of Proposition \ref{Prop:NullDist_Wn}}

Proposition \ref{Prop:NullDist_Wn} is a direct consequence of Lemma \ref{Lemma:NullDist_Wn_1} and Lemma \ref{Lemma:NullDist_Wn_2}.

\subsection{Proof of Proposition \ref{Prop:NullDist_Wn_lp}}

It follows from the definition of $\{W_n(r)\}_{\varepsilon\le r\le 1-\varepsilon}$ and the Beveridge-Nelson decomposition that
\BEqn
& & W_n(r) \\
&=& \sum\limits_{i=1}^{m_1} \lrp{X_i - X_{n+1-i}}^{\top} \lrp{\sum\limits_{j=1}^{\lfloor{nr}\rfloor - \lfloor{n\varepsilon}\rfloor} X_{j+m}} \\
&=& \sum\limits_{i=1}^{m_1} \lrp{D_i - D_{n+1-i}}^{\top} \lrp{\sum\limits_{j=1}^{\lfloor{nr}\rfloor - \lfloor{n\varepsilon}\rfloor} D_{j+m}}
   - \sum\limits_{i=1}^{m_1} \lrp{R_i - R_{n+1-i}}^{\top} \lrp{\sum\limits_{j=1}^{\lfloor{nr}\rfloor - \lfloor{n\varepsilon}\rfloor} D_{j+m}} \\
& & - \sum\limits_{i=1}^{m_1} \lrp{D_i - D_{n+1-i}}^{\top} \lrp{\sum\limits_{j=1}^{\lfloor{nr}\rfloor - \lfloor{n\varepsilon}\rfloor} R_{j+m}} 
    + \sum\limits_{i=1}^{m_1} \lrp{R_i - R_{n+1-i}}^{\top} \lrp{\sum\limits_{j=1}^{\lfloor{nr}\rfloor - \lfloor{n\varepsilon}\rfloor} R_{j+m}}
\EEqn

Note that $\{D_t\}_{t=1}^{n}$ is an iid sequence with mean zero and covariance matrix $A^{(0)} \Gamma^{(2)} (A^{(0)})^{\top}$, then the limiting null distribution of the process $\lrcp{\frac{1}{\widetilde{N}_n} \sum\limits_{i=1}^{m_1} \lrp{D_i - D_{n+1-i}}^{\top} \lrp{\sum\limits_{j=1}^{\lfloor{nr}\rfloor - \lfloor{n\varepsilon}\rfloor} D_{j+m}}}_{\varepsilon\le r\le 1-\varepsilon}$ directly follows from Proposition \ref{Prop:NullDist_Wn}. Also note that it follows from Lemma \ref{Lemma:NullDist_DR-1_lp}, Lemma \ref{Lemma:NullDist_DR-2_lp}, Lemma \ref{Lemma:NullDist_RD_lp} and Lemma \ref{Lemma:NullDist_RR_lp} that
\BEqn
& & \sup\limits_{r\in[\varepsilon,1-\varepsilon]} \lrabs{\frac{1}{\widetilde{N}_n} \sum\limits_{i=1}^{m_1} \lrp{D_i + D_{n+1-i}}^{\top} \lrp{\sum\limits_{j=1}^{\lrfl{nr} - \lrfl{n\varepsilon}} R_{j+m}}} = o_p(1), \\
& & \sup\limits_{r\in[\varepsilon,1-\varepsilon]} \lrabs{\frac{1}{\widetilde{N}_n} \sum\limits_{i=1}^{m_1} \lrp{R_i + R_{n+1-i}}^{\top} \lrp{\sum\limits_{j=1}^{\lrfl{nr} - \lrfl{n\varepsilon}} D_{j+m}}} = o_p(1), \\
& & \sup\limits_{r\in[\varepsilon,1-\varepsilon]} \lrabs{\frac{1}{\widetilde{N}_n} \sum\limits_{i=1}^{m_1} \lrp{R_i + R_{n+1-i}}^{\top} \lrp{\sum\limits_{j=1}^{\lrfl{nr} - \lrfl{n\varepsilon}} R_{j+m}}} = o_p(1)
\EEqn
as $n\rightarrow\infty$, then we obtain the desired result.

\subsection{Proof of Theorem \ref{Thm:NullDist_Gn_Growingp_lp}}

For any $\varepsilon \le u \le 1-\varepsilon$, it holds that $m_1 S_{1, \lrfl{nu}-\lrfl{n\varepsilon}} = W_n(u)$. From Proposition \ref{Prop:NullDist_Wn_lp}, we have that
\begin{equation*}
    \frac{m_1}{\widetilde{N}_n} S_{1, \lrfl{nu}-\lrfl{n\varepsilon}}
\leadsto B(u) - B(\varepsilon)~\mbox{in}~D[\varepsilon,1-\varepsilon].
\end{equation*}
Recall that
\BEqn
    G_n
&=& \sup\limits_{k=1,\cdots,N-1}
    \frac{\frac{m_1}{\widetilde{N}_n} \lrp{S_{1,k} - \frac{k}{N}S_{1,N}}}{\sqrt{\frac{m_1^2}{\widetilde{N}_n^2}\frac{1}{N}\lrp{ \sum\limits_{t=1}^{k}\lrp{S_{1,t} - \frac{t}{k}S_{1,k}}^2 + \sum\limits_{t=k+1}^{N}\lrp{(S_{1,N}-S_{1,t-1}) - \frac{N-t+1}{N-k}(S_{1,N}-S_{1,k})}^2}}},
\EEqn
then with $k = \lrfl{nu}-\lrfl{n\varepsilon}$, it follows that
\begin{equation*}
    \frac{m_1}{\widetilde{N}_n} \lrp{S_{1,k} - \frac{k}{N}S_{1,N}}
\leadsto B(u) - B(\varepsilon) - \frac{u-\varepsilon}{1-2\varepsilon}\lrp{B(1-\varepsilon) - B(\varepsilon)}.
\end{equation*}
Similarly, we can also show that
\begin{equation*}
    \frac{m_1^2}{\widetilde{N}_n^2} \cdot \frac{1}{N} \lrp{\sum\limits_{t=1}^{k} \lrp{S_{1,t} - \frac{t}{k}S_{1,k}}^2} \\
\leadsto
  \frac{1}{1-2\varepsilon}
  \int_{\varepsilon}^{u} 
  \lrp{B(v) - B(\varepsilon) 
  - \frac{v-\varepsilon}{u-\varepsilon}\lrp{B(u) - B(\varepsilon)}}^2 dv,
\end{equation*}
and
\BEqn
& & \frac{m_1^2}{\widetilde{N}_n^2} \cdot \frac{1}{N} \lrp{\sum\limits_{t=1}^{k} \lrp{(S_{1,N}-S_{1,t-1}) - \frac{N-t+1}{N-k}(S_{1,N}-S_{1,k})}^2} \\
&\leadsto&
  \frac{1}{1-2\varepsilon}
  \int_{u}^{1-\varepsilon} 
  \lrp{B(1-\varepsilon) - B(v) 
  - \frac{1-\varepsilon-v}{1-\varepsilon-u} \lrp{B(1-\varepsilon) - B(u)}}^2 dv
\EEqn

Therefore, it follows from the continuous mapping theorem that
\BEqn
& & G_n \\
&\stra{d}& \sup\limits_{u\in[\varepsilon,1-\varepsilon]} \frac{B(u)-B(\varepsilon) - \frac{u-\varepsilon}{1-2\varepsilon} (B(1-\varepsilon)-B(\varepsilon))}
{\lrp{
     \begin{array}{l}
     ~ ~ \frac{1}{1-2\varepsilon} \int_{\varepsilon}^{u} \lrp{B(v)-B(\varepsilon) - \frac{v-\varepsilon}{u-\varepsilon} (B(u) - B(\varepsilon))}^2 dv \\[2mm]
     + \frac{1}{1-2\varepsilon} \int_{u}^{1-\varepsilon} \lrp{B(1-\varepsilon)-B(v) - \frac{1-\varepsilon-v}{1-\varepsilon-u} \lrp{B(1-\varepsilon) - B(u)}}^2 dv
     \end{array}
    }^{1/2}} \\
&=^d& \sup\limits_{r\in[0,1]} \frac{B(\varepsilon+(1-2\varepsilon)r)-B(\varepsilon) - r (B(1-\varepsilon)-B(\varepsilon))}
{\lrp{
     \begin{array}{l}
     ~ ~ \int_{0}^{r} \lrp{B(\varepsilon+(1-2\varepsilon)s)-B(\varepsilon) - \frac{s}{r} (B(\varepsilon+(1-2\varepsilon)r) - B(\varepsilon))}^2 ds \\[2mm]
     + \int_{r}^{1} \lrp{B(1-\varepsilon)-B(\varepsilon+(1-2\varepsilon)s) - \frac{1-s}{1-r} \lrp{B(1-\varepsilon) - B(\varepsilon+(1-2\varepsilon)r)}}^2 ds
     \end{array}
    }^{1/2}} \\
&=^d& 
  \sup\limits_{r\in[0,1]} \frac{B(r) - rB(1)}{\sqrt{\int_{0}^{r} 
  \lrp{B(s) - \frac{s}{r}B(r)}^2 ds + \int_{r}^{1} 
  \lrp{B(1) - B(s) - \frac{1-s}{1-r}\lrp{B(1) - B(r)}}^2 ds}} \\
&=^d& G,
\EEqn
where in the second step, we change the variable by $r = (u-\varepsilon)/(1-2\varepsilon)$, $s=(v-\varepsilon)/(1-2\varepsilon)$ and in the second from the last step, we use the fact that the process $\{B(\varepsilon + (1-2\varepsilon)r) - B(\varepsilon)\}_{0\le r\le 1}$ is equal in distribution with $\{\sqrt{1-2\varepsilon}B(r)\}_{0\le r\le 1}$. This completes the proof.

\subsection{Proof of Theorem \ref{Thm:NullDist_Gn_factor}}

Recall the data structure $\{X_t\}_{t=1}^{n}$ defined as Definition \ref{Def:Model_factor}, we can decompose the process $W_n(r)$ as below
\BEqn
& & W_n(r) \\
&=& \sum\limits_{i=1}^{m_1} \lrp{X_i - X_{n+1-i}}^{\top} \lrp{\sum\limits_{j=1}^{\lfloor{nr}\rfloor - \lfloor{n\varepsilon}\rfloor} X_{j+m}} \\
&=& \sum\limits_{i=1}^{m_1} \sum\limits_{j=1}^{\lrfl{nr} - \lrfl{n\varepsilon}} \lrp{\Lambda(F_{i} - F_{n+1-i}) + (Z_{i} - Z_{n+1-i})}^{\top} \lrp{\Lambda F_{j+m} + Z_{j+m}} \\
&=& \sum\limits_{i=1}^{m_1} \sum\limits_{j=1}^{\lrfl{nr} - \lrfl{n\varepsilon}} (F_{i} - F_{n+1-i})^{\top} \Lambda^{\top} \Lambda F_{j+m}  
    + \sum\limits_{i=1}^{m_1} \sum\limits_{j=1}^{\lrfl{nr} - \lrfl{n\varepsilon}}  (Z_{i} - Z_{n+1-i})^{\top} Z_{j+m} \\
& & + \sum\limits_{i=1}^{m_1} \sum\limits_{j=1}^{\lrfl{nr} - \lrfl{n\varepsilon}} Z_{j+m}^{\top} \Lambda(F_{i} - F_{n+1-i})
    + \sum\limits_{i=1}^{m_1} \sum\limits_{j=1}^{\lrfl{nr} - \lrfl{n\varepsilon}} (Z_{i} - Z_{n+1-i})^{\top} \Lambda F_{j+m} \\
&=:& W_{n,1}(r) + W_{n,2}(r) + W_{n,3}(r) + W_{n,4}(r).
\EEqn
With properly selected normalizers, we have derived in Lemma \ref{Lemma:factor-aux-1-1}, Lemma \ref{Lemma:factor-aux-1-2} and Lemma \ref{Lemma:factor-aux-1-3} the asymptotic distribution of each process $\{W_{n,i}(r)\}_{\varepsilon\le r\le 1-\varepsilon}$. Based on the relationship of $\|\Lambda\|$ and $\|\Gamma^{(3)}\|_F^{1/2}$, we divide into the following three cases to prove the proposed theorem.

\begin{enumerate}[label=(\roman*)]
    \item If $\|\Lambda\| = o\lrp{\|\Gamma^{(3)}\|_F^{1/2}}$, with Assumption \ref{Assumpt:NullDist_Wn_lp}\ref{Assumpt:NullDist_Wn_Lp-3}, we have that  $\|\Lambda^{\top} \Lambda\| \le \|\Lambda\|^2 = o\lrp{\|A^{(0)} \Gamma^{(3)} (A^{(0)})^{\top}\|_F}$ since $\|\Gamma^{(3)}\|_F = O_s\lrp{\|A^{(0)} \Gamma^{(3)} (A^{(0)})^{\top}\|_F}$. It follows from Lemma \ref{Lemma:factor-aux-1-1} that,
    \begin{equation*}
        \frac{W_{n,1}(r)}{\sqrt{2n\lrfl{n(\varepsilon-\eta)}} \|A^{(0)} \Gamma^{(3)} (A^{(0)})^{\top}\|_F}
      = \frac{\|\Lambda^{\top}\Lambda\|}{\sqrt{2\lrfl{n(\varepsilon-\eta)}/n} \|A^{(0)} \Gamma^{(3)} (A^{(0)})^{\top}\|_F} \cdot \frac{W_{n,1}(r)}{n\|\Lambda^{\top}\Lambda\|} \\
      \leadsto 0
    \end{equation*}
    in $D[\varepsilon,1-\varepsilon]$, where we use the fact that $\Omega^{(3)},L_0\in\br^{s\times s}$ are both independent of $p$ and can be viewed as constant matrices.
    
    Additionally, it follows from Lemma \ref{Lemma:factor-aux-1-2} and Lemma \ref{Lemma:factor-aux-1-3} that
    \BEqn
    & & \frac{W_{n,2}(r)}{\sqrt{2n\lrfl{n(\varepsilon-\eta)}} \|A^{(0)} \Gamma^{(3)} (A^{(0)})^{\top}\|_F} 
    \leadsto \tilde{B}(r) - \tilde{B}(\varepsilon)
    \quad\mbox{ in } D[\varepsilon,1-\varepsilon], \\
    & & \frac{W_{n,3}(r) + W_{n,4}(r)}{\sqrt{2n\lrfl{n(\varepsilon-\eta)}} \|A^{(0)} \Gamma^{(3)} (A^{(0)})^{\top}\|_F} 
    \leadsto 0
    \quad\mbox{ in } D[\varepsilon,1-\varepsilon].
    \EEqn
    
    Therefore, it holds that when $\|\Lambda\| = o\lrp{\|\Gamma^{(3)}\|_F^{1/2}}$,
    \begin{equation*}
        \frac{W_n(r)}{\sqrt{2n\lrfl{n(\varepsilon-\eta)}} \|A^{(0)} \Gamma^{(3)} (A^{(0)})^{\top}\|_F} 
    \leadsto \tilde{B}(r) - \tilde{B}(\varepsilon)
    \quad \mbox{in } D[\varepsilon,1-\varepsilon].
    \end{equation*}
    Consequently, the asymptotic null distribution of $G_n$ can be shown to be $G$ using the similar argument as that for Theorem \ref{Thm:NullDist_Gn_Growingp_lp}.
    
    \item If $\|\Gamma^{(3)}\|_F^{1/2} = o\lrp{\|\Lambda\|}$, we have that $\|A^{(0)} \Gamma^{(3)} (A^{(0)})^{\top}\|_F = \|\Lambda\|^2$ under Assumption \ref{Assumpt:NullDist_Wn_lp}\ref{Assumpt:NullDist_Wn_Lp-3}. It is shown in Lemma \ref{Lemma:factor-aux-2} that $\|\Lambda^{\top}\Lambda\| = O_s\lrp{\|\Lambda\|_F^2} = O_s\lrp{\|\Lambda\|^2}$. Also, it follows from Lemma \ref{Lemma:factor-aux-1-1}, Lemma \ref{Lemma:factor-aux-1-2} and Lemma \ref{Lemma:factor-aux-1-3} that in $D[\varepsilon,1-\varepsilon]$,
    \BEqn
    & & \frac{W_{n,1}(r)}{n\|\Lambda^{\top}\Lambda\|} 
     \leadsto b^{\top}(\varepsilon,\eta) ((\Omega^{(3)})^{1/2})^{\top} L_0 (\Omega^{(3)})^{1/2} \lrp{B_s(r)-B_s(\varepsilon)}, \\
    & & \frac{W_{n,2}(r)}{n\|\Lambda^{\top}\Lambda\|} 
     =  \frac{\sqrt{2n\lrfl{n(\varepsilon-\eta)}} \|A^{(0)} \Gamma^{(3)} (A^{(0)})^{\top}\|_F}{n\|\Lambda^{\top}\Lambda\|} \cdot \frac{W_{n,2}(r)}{\sqrt{2n\lrfl{n(\varepsilon-\eta)}} \|A^{(0)} \Gamma^{(3)} (A^{(0)})^{\top}\|_F} 
     \leadsto 0, \\
    & & \frac{W_{n,3}(r)+W_{n,4}(r)}{n\|\Lambda^{\top}\Lambda\|} 
     \leadsto 0.
    \EEqn
    This implies that 
    \begin{equation*}
        \frac{W_n(r)}{n\|\Lambda^{\top}\Lambda\|} 
    \leadsto b^{\top}(\varepsilon,\eta) ((\Omega^{(3)})^{1/2})^{\top} L_0 (\Omega^{(3)})^{1/2} \lrp{B_s(r)-B_s(\varepsilon)}
    \mbox{ in } D[\varepsilon,1-\varepsilon].
    \end{equation*}
    
    For any $u\in[\varepsilon,1-\varepsilon]$, let $k=\lrfl{nu}-\lrfl{n\varepsilon}$, then it holds that $m_1 S_{1,k} = W_n(u)$. Using the  results established in the proof of Theorem \ref{Thm:NullDist_Gn_Growingp_lp}, we have that
    \BEqn
    & & G_n \\
    &=& \sup\limits_{k=1,\cdots,N-1}
        \frac{\frac{m_1}{n\|\Lambda^{\top}\Lambda\|} \lrp{S_{1,k} - \frac{k}{N}S_{1,N}}}{\sqrt{\frac{m_1^2}{n^2\|\Lambda^{\top}\Lambda\|^2}\frac{1}{N}\lrp{ \sum\limits_{t=1}^{k}\lrp{S_{1,t} - \frac{t}{k}S_{1,k}}^2 + \sum\limits_{t=k+1}^{N}\lrp{(S_{1,N}-S_{1,t-1}) - \frac{N-t+1}{N-k}(S_{1,N}-S_{1,k})}^2}}} \\
    &\stra{d}& 
        \sup\limits_{u\in[\varepsilon,1-\varepsilon]}
        \frac{b^{\top}(\varepsilon,\eta) ((\Omega^{(3)})^{1/2})^{\top} L_0 (\Omega^{(3)})^{1/2} \lrp{B_s(u)-B_s(\varepsilon) - \frac{u-\varepsilon}{1-2\varepsilon} \lrp{B_s(1-\varepsilon)-B_s(\varepsilon)}}}
        {\lrp{
          \begin{array}{l}
            ~~ \frac{1}{1-2\varepsilon} \int_{\varepsilon}^{u} \Big{(} b^{\top}(\varepsilon,\eta) ((\Omega^{(3)})^{1/2})^{\top} L_0 (\Omega^{(3)})^{1/2} \\
            \hspace{6em} \times (B_s(v)-B_s(\varepsilon)-\frac{v-\varepsilon}{u-\varepsilon}(B_s(u)-B_s(\varepsilon))) \Big{)}^2 dv \\[2mm]
            + \frac{1}{1-2\varepsilon} \int_{u}^{1-\varepsilon} \Big{(} b^{\top}(\varepsilon,\eta) ((\Omega^{(3)})^{1/2})^{\top} L_0 (\Omega^{(3)})^{1/2} \\
            \hspace{6em} \times (B_s(1-\varepsilon)-B_s(v)-\frac{1-\varepsilon-v}{1-\varepsilon-u}(B_s(1-\varepsilon)-B_s(u))) \Big{)}^2 dv
          \end{array}
        }^{1/2}} \\
    &=:& M(b(\varepsilon,\eta),\cb_s),
    \EEqn
    where $\cb_s := \{B_s(u)\}_{\varepsilon\le u\le 1-\varepsilon}$ and is independent of $b(\varepsilon,\eta)$. If conditioning on $b(\varepsilon,\eta)=b_0$, the processes $\{b^{\top}(\varepsilon,\eta) ((\Omega^{(3)})^{1/2})^{\top} L_0 (\Omega^{(3)})^{1/2} B_s(u)\}_{\varepsilon\le u\le 1-\varepsilon}$ and $\{\omega B(u)\}_{\varepsilon\le u\le 1-\varepsilon}$ are equal in distribution, where $\omega = \sqrt{b_0^{\top} ((\Omega^{(3)})^{1/2})^{\top} L_0 \Omega^{(3)} L_0^{\top} (\Omega^{(3)})^{1/2} b_0}$. 
    
    By applying the same techniques in the proof of Theorem \ref{Thm:NullDist_Gn_Fixedp}, it follows from the same conditioning arguments that
    \begin{equation*}
        M(b(\varepsilon,\eta),\cb_s)|_{b(\varepsilon,\eta)=b_0} = G,
    \end{equation*}
    which is independent of $b_0$. Therefore, we conclude that $G_n \stra{d} G$ as $\min\{n,p\}\rightarrow\infty$ in this case.

    \item If $\|\Gamma^{(3)}\|_F^{1/2} = O_s\lrp{\|\Lambda\|}$. By Lemma \ref{Lemma:factor-aux-1-1}, Lemma \ref{Lemma:factor-aux-1-2} and Lemma \ref{Lemma:factor-aux-1-3}, it holds in $D[\varepsilon,1-\varepsilon]$ that
    \BEqn
    & & \frac{W_{n,1}(r)}{n\|\Lambda^{\top}\Lambda\|} 
     \leadsto b^{\top}(\varepsilon,\eta) ((\Omega^{(3)})^{1/2})^{\top} L_0 (\Omega^{(3)})^{1/2} \lrp{B_s(r)-B_s(\varepsilon)}
     =: M_1(r), \\
    & & \frac{W_{n,2}(r)}{\sqrt{2n\lrfl{n(\varepsilon-\eta)}} \|A^{(0)} \Gamma^{(3)} (A^{(0)})^{\top}\|_F} 
    \leadsto \tilde{B}(r)-\tilde{B}(\varepsilon)
     =: M_2(r),
    \EEqn
    and $\displaystyle \frac{W_{n,3}(r)+W_{n,4}(r)}{n\max\{\|\Lambda\|^2,\|\Gamma^{(3)}\|_F\}} \leadsto 0$, where $\{B_s(r)\}_{0\le r\le 1}$ is a $s$-dimensional Brownian motion whereas $\{\tilde{B}(r)\}_{0\le r\le 1}$ is a $1$-dimensional Brownian motion. From Definition \ref{Def:Model_factor}, $\{F_t\}_{t=1}^{n}$ is independent of $\{Z_t\}_{t=1}^{n}$, therefore it can be easily shown that
    \begin{equation*}
        \lrp{\frac{W_{n,1}(r)}{n\|\Lambda^{\top}\Lambda\|}, \frac{W_{n,2}(r)}{\sqrt{2n\lrfl{n(\varepsilon-\eta)}} \|A^{(0)} \Gamma^{(3)} (A^{(0)})^{\top}\|_F} }
    \leadsto \lrp{M_1(r), M_2(r)}
    \quad\mbox{ jointly in }D^2[\varepsilon,1-\varepsilon]
    \end{equation*}
    where $M_1(r)$ and $M_2(r)$ are also independent.
    
    Assume that $\displaystyle \frac{n\|\Lambda^{\top} \Lambda\|}{\sqrt{2n\lrfl{n(\varepsilon-\eta)}\|A^{(0)}\Gamma^{(3)}(A^{(0)})^{\top}\|_F}} \rightarrow c$ as $\min\{n,p\}\rightarrow\infty$, then it holds in $D[\varepsilon,1-\varepsilon]$ that
    \BEqn
    & & \frac{1}{\sqrt{2n\lrfl{n(\varepsilon-\eta)}\|A^{(0)} \Gamma^{(3)} (A^{(0)})^{\top}\|_F}} W_n(r) \\
    &=& \frac{n\|\Lambda^{\top}\Lambda\|}{\sqrt{2n\lrfl{n(\varepsilon-\eta)}} \|A^{(0)} \Gamma^{(3)} (A^{(0)})^{\top}\|_F} \cdot \frac{W_{n,1}(r)}{n\|\Lambda^{\top}\Lambda\|} 
    + \frac{W_{n,2}(r)}{\sqrt{2n\lrfl{n(\varepsilon-\eta)}} \|A^{(0)} \Gamma^{(3)} (A^{(0)})^{\top}\|_F} \\
    & & + \frac{n \max\{\|\Lambda\|^2,\|\Gamma^{(3)}\|_F\}}{\sqrt{2n\lrfl{n(\varepsilon-\eta)}}\|A^{(0)} \Gamma^{(3)} (A^{(0)})^{\top}\|_F} \cdot \frac{W_{n,3}(r)+W_{n,4}(r)}{n \max\{\|\Lambda\|^2,\|\Gamma^{(3)}\|_F\}} \\
    &\leadsto& c b^{\top}(\varepsilon,\eta) ((\Omega^{(3)})^{1/2})^{\top} L_0 (\Omega^{(3)})^{1/2} \lrp{B_s(r)-B_s(\varepsilon)} + \tilde{B}(r)-\tilde{B}(\varepsilon).
    \EEqn
    
Therefore, with $k=\lrfl{nu}-\lrfl{n\varepsilon}$ for $u\in[\varepsilon,1-\varepsilon]$, we have that
    \begin{equation*}
        G_n
    \stra{d} M(b(\varepsilon,\eta),\cb_s,\tilde{\cb})
    := \sup\limits_{u\in[\varepsilon,1-\varepsilon]} T(u,b(\varepsilon,\eta),\cb_s,\tilde{\cb}) V^{-1/2}(u,b(\varepsilon,\eta),\cb_s,\tilde{\cb}),
    \end{equation*}
    where $\cb_s=\{B_s(u)\}_{\varepsilon\le u\le 1-\varepsilon}$ and $\tilde{\cb} = \{\tilde{B}(u)\}_{\varepsilon\le u\le 1-\varepsilon}$, 
    \BEqn
    & & T(u,b(\varepsilon,\eta),\cb_s,\tilde{\cb}) \\
    &=& c b^{\top}(\varepsilon,\eta) ((\Omega^{(3)})^{1/2})^{\top} L_0 (\Omega^{(3)})^{1/2} \lrp{B_s(u) - B_s(\varepsilon) + \frac{u-\varepsilon}{1-2\varepsilon} (B_s(1-\varepsilon) - B_s(\varepsilon)) } \\
    & & + \tilde{B}(u) - \tilde{B}(\varepsilon) - \frac{u-\varepsilon}{1-2\varepsilon} (\tilde{B}(1-\varepsilon) - \tilde{B}(\varepsilon)),
    \EEqn
    and
    \BEqn
    & & V(u,b(\varepsilon,\eta),\cb_s,\tilde{\cb}) \\
    &=& \frac{1}{1-2\varepsilon} 
        \int_{\varepsilon}^{u} 
        \left(
          c b^{\top}(\varepsilon,\eta) ((\Omega^{(3)})^{1/2})^{\top} L_0 (\Omega^{(3)})^{1/2} \lrp{B_s(v) - B_s(\varepsilon) + \frac{v-\varepsilon}{u-\varepsilon} (B_s(u) - B_s(\varepsilon)) } 
        \right. \\
    & & \hspace{6em}
        \left.
        + \tilde{B}(v) - \tilde{B}(\varepsilon) - \frac{v-\varepsilon}{u-\varepsilon} (\tilde{B}(u) - \tilde{B}(\varepsilon))
        \right)^2 ds \\
    & & + \frac{1}{1-2\varepsilon} 
        \int_{u}^{1-\varepsilon}
        \left(
          c b^{\top}(\varepsilon,\eta) ((\Omega^{(3)})^{1/2})^{\top} L_0 (\Omega^{(3)})^{1/2} 
        \right. \\
    & & \hspace{8em} \times
          \bigp{B_s(1-\varepsilon) - B_s(v) + \frac{1-\varepsilon-v}{1-\varepsilon-u} (B_s(1-\varepsilon) - B_s(u)) } \\
    & & \hspace{7em}
        \left.
        + \tilde{B}(1-\varepsilon) - \tilde{B}(v) - \frac{1-\varepsilon-v}{1-\varepsilon-u} (\tilde{B}(1-\varepsilon) - \tilde{B}(u))
        \right)^2 ds.
    \EEqn
    
    Note that $b(\varepsilon,\eta),\cb_s$ and $\tilde{\cb}$ are mutually independent, thus if conditioning on $b(\varepsilon,\eta_0)$, the processes $\{c b_0^{\top} ((\Omega^{(3)})^{1/2})^{\top} L_0 (\Omega^{(3)})^{1/2} B_s(u) + \tilde{B}(u)\}_{\varepsilon\le u\le 1-\varepsilon}$ and $\{\omega B(u)\}_{\varepsilon\le u\le 1-\varepsilon}$ are equal in distribution, where $\omega = \sqrt{c^2b_0^{\top} ((\Omega^{(3)})^{1/2})^{\top} L_0 \Omega^{(3)} L_0^{\top} (\Omega^{(3)})^{1/2} b_0 + 1}$. By using the conditioning argument in Theorem \ref{Thm:NullDist_Gn_Fixedp} again, we obtain that
    \BEqn
    & & M(b(\varepsilon,\eta),\cb_s,\tilde{\cb})|_{b(\varepsilon,\eta)=b_0} \\
    &=^d& \sup\limits_{u\in[\varepsilon,1-\varepsilon]} 
    \frac{\omega \lrp{B(u)-B(\varepsilon)-\frac{u-\varepsilon}{1-2\varepsilon}(B(1-\varepsilon)-B(\varepsilon)}}
    {\lrp{
          \begin{array}{l}
            ~~ \frac{1}{1-2\varepsilon} \int_{\varepsilon}^{u} \lrp{\omega \lrp{B(v)-B(\varepsilon)-\frac{v-\varepsilon}{u-\varepsilon}(B(u)-B(\varepsilon))}}^2 dv   \\[2mm]
            + \frac{1}{1-2\varepsilon} \int_{u}^{1-\varepsilon} \lrp{\omega \lrp{B(1-\varepsilon)-B(v)-\frac{1-\varepsilon-v}{1-\varepsilon-u}(B(1-\varepsilon)-B(u))}}^2 dv
          \end{array}
        }^{1/2}} \\
    &=^d& \sup\limits_{r\in[0,1]}
        \frac{B(r)-rB(1)}
        {\sqrt{\int_{0}^{r} \lrp{B(s)-\frac{s}{r}B(r)}^2 dv + \int_{r}^{1} \lrp{B(1-s) - \frac{1-s}{1-r}B(1-r)}^2 dv}} \\
    &=^d& G,
    \EEqn
    which does not depend on $b_0$. Hence, we may conclude that $G_n \stra{d} M(b(\varepsilon,\eta),\cb_s,\tilde{\cb}) =^d G$ as $\min\{n,p\}\rightarrow\infty$.
\end{enumerate}
In summary, by combining the results from all the three cases, we complete the proof of the proposed statement.

\subsection{Proof of Proposition \ref{Prop:PowerExpression}}

It follows from the definition of $\tilde{X}_t$ that $\bbe{\tilde{X}_t} = 0$ for $1\le t\le n$, and $\hat\nu_1$ and $\hat\nu_n$ denote the sample mean estimate of $\{\tilde{X}_t\}_{t=1}^{n}$ over the blocks $\cx_{11}$ and $\cx_{32}$ respectively. Define 
\begin{equation*}
    \tilde{Y}_j = \lrag{\hat\nu_1-\hat\nu_n, \tilde{X}_{j+m}}, \quad j=1,\cdots,N,
\end{equation*}
and $\bar{\tilde{Y}}_N = \frac{1}{N} \sum\limits_{j=1}^{N} \tilde{Y}_j$. Additionally, for $k=1,\cdots,N$, we define
\begin{equation*}
    \tilde{T}_n(k)
  = N^{-1/2} \sum\limits_{t=1}^{k} \lrp{\tilde{Y}_t - \bar{\tilde{Y}}_N}, 
\end{equation*}
and
\begin{equation*}
    \tilde{V}_n(k)
  = N^{-2} 
    \lrp{ \sum\limits_{t=1}^{k} (\tilde{S}_{1,t} - \frac{t}{k}\tilde{S}_{1,k})^2
         -\sum\limits_{t=k+1}^{N} (\tilde{S}_{t,N} - \frac{N-t+1}{N-k}\tilde{S}_{k+1,N})^2},
\end{equation*}
where $\tilde{S}_{a,b} = \sum\limits_{j=a}^{b} \tilde{Y}_j$ for any $1 \le a \le b \le N$.

Now we can rewrite $\hat\mu_1$ and $\hat\nu_n$ in terms of the newly introduced notations, that is,
\BEqn
    \hat\mu_1
&=& \frac{1}{m_1} \sum\limits_{i=1}^{m_1} (\tilde{X}_i+\mu) 
 =  \hat\nu_1 + \mu, \\
    \hat\mu_n
&=& \frac{1}{m_1} \sum\limits_{i=1}^{m_1} (\tilde{X}_{n+1-i}+\mu+\delta)
 =  \hat\nu_n + \mu + \delta.
\EEqn
Similarly, we have that
\BEqn
    Y_j
&=& \lrag{\hat\mu_1-\hat\mu_n,X_{j+m}}
 =  \lrag{\hat\nu_1-\hat\nu_n-\delta, \tilde{X}_{j+m}+\mu+\delta\bone\{j+m>k_0\}} \\
&=& \left\{
    \begin{array}{ll}
        \lrag{\hat\nu_1-\hat\nu_n-\delta, \tilde{X}_{j+m}+\mu}, & 1 \le j \le k_0-m, \\
        \lrag{\hat\nu_1-\hat\nu_n-\delta, \tilde{X}_{j+m}+\mu+\delta}, & k_0-m < j \le N,
    \end{array}
    \right.
\EEqn
and it follows that $\bar{Y}_N = \lrag{\hat\nu_1-\hat\nu_n-\delta, \bar{\tilde{X}}_N + \mu + \frac{N-k_0+m}{N}\delta}$, where $\bar{\tilde{X}}_N = \frac{1}{N}\sum\limits_{j=1}^{N} \tilde{X}_{j+m}$.

Next we express $T_n(k)$ in terms of the new notations. Recall that $T_n(k) = N^{-1/2} \sum\limits_{t=1}^{k}(Y_t-\bar{Y}_N)$ for $1 \le k \le N$. If $k \le k_0-m$, it holds that
\BEqn
    N^{1/2} T_n(k)
&=& \sum\limits_{t=1}^{k} (Y_t-\bar{Y}_N)
 =  \sum\limits_{t=1}^{k}
    \lrag{\hat\nu_1-\hat\nu_n-\delta, \tilde{X}_{j+m} - \bar{\tilde{X}}_N - \frac{N-k_0+m}{N} \delta} \\
&=& \lrag{\hat\nu_1-\hat\nu_n-\delta, 
    \sum\limits_{t=1}^{k}\tilde{X}_{j+m} - \frac{k}{N} \sum\limits_{t=1}^{N}\tilde{X}_{j+m}}
 -  \frac{k(N-k_0+m)}{N} \lrag{\hat\nu_1-\hat\nu_n-\delta, \delta}.
\EEqn
If $k > k_0-m$, then we have that
\BEqn
    N^{1/2} T_n(k)
&=& \sum\limits_{t=1}^{k_0-m} (Y_t-\bar{Y}_N) + \sum\limits_{t=k_0-m+1}^{k} (Y_t-\bar{Y}_N) \\
&=& \sum\limits_{t=1}^{k_0-m}
    \lrag{\hat\nu_1-\hat\nu_n-\delta, \tilde{X}_{j+m} - \bar{\tilde{X}}_N - \frac{N-k_0+m}{N}\delta} \\
& & + \sum\limits_{t=k_0-m+1}^{k}
    \lrag{\hat\nu_1-\hat\nu_n-\delta, \tilde{X}_{j+m} - \bar{\tilde{X}}_N + \frac{k_0-m}{N}\delta} \\
&=& \lrag{\hat\nu_1-\hat\nu_n-\delta, 
    \sum\limits_{t=1}^{k}\tilde{X}_{j+m} - \frac{k}{N} \sum\limits_{t=1}^{N}\tilde{X}_{j+m}}
 -  \frac{(N-k)(k_0-m)}{N}\lrag{\hat\nu_1-\hat\nu_n-\delta,\delta}.
\EEqn
By unifying these two expressions, we obtain the desired uniform expression of $N^{1/2} T_n(k)$ against the alternative $\delta$.

Similarly, we can derive the expression of $N^2 V_n(k) = \sum\limits_{t=1}^{k}\lrp{S_{1,t}-\frac{t}{k}S_{1,k}}^2 + \sum\limits_{t=k+1}^{N}\lrp{S_{t,N}-\frac{N-t+1}{N-k}S_{k+1,N}}^2$. If $t \le k_0-m$, we have that
\BEqn
    S_{1,t}
&=& \sum\limits_{j=1}^{t} \lrag{\hat\nu_1-\hat\nu_n-\delta, \tilde{X}_{j+m}+\mu}, \\
    S_{t,N}
&=& \sum\limits_{j=t}^{N} \lrag{\hat\nu_1-\hat\nu_n-\delta, \tilde{X}_{j+m}+\mu} 
 +  (N-k_0+m)\lrag{\hat\nu_1-\hat\nu_n-\delta, \delta}.
\EEqn
If $t > k_0-m$, we have that
\BEqn
    S_{1,t}
&=& \sum\limits_{j=1}^{t} \lrag{\hat\nu_1-\hat\nu_n-\delta, \tilde{X}_{j+m}+\mu}
 +  (t-k_0+m) \lrag{\hat\nu_1-\hat\nu_n-\delta, \delta}, \\
    S_{t,N}
&=& \sum\limits_{j=t}^{N} \lrag{\hat\nu_1-\hat\nu_n-\delta, \tilde{X}_{j+m}+\mu} 
 +  (N-t+1)\lrag{\hat\nu_1-\hat\nu_n-\delta, \delta}.
\EEqn
It follows that, when $k < k_0-m$, we have that
{\small
\BEqn
& & N^2 V_n(k) \\
&=& \sum\limits_{t=1}^{k} (S_{1,t} - \frac{t}{k}S_{1,k})^2 
 +  \sum\limits_{t=k+1}^{k_0-m} (S_{t,N} - \frac{N-t+1}{N-k}S_{k+1,N})^2
 +  \sum\limits_{t=k_0-m+1}^{N} (S_{t,N} - \frac{N-t+1}{N-k}S_{k+1,N})^2 \\
&=& \sum\limits_{t=1}^{k} 
    \lrp{ \lrag{\hat\nu_1-\hat\nu_n-\delta, \sum\limits_{j=1}^{t}\tilde{X}_{j+m} - \frac{t}{k}\sum\limits_{j=1}^{k}\tilde{X}_{j+m}} }^2 \\
& & + \sum\limits_{t=k+1}^{k_0-m}
   \left(
   \lrag{\hat\nu_1-\hat\nu_n-\delta, \sum\limits_{j=t}^{N}\tilde{X}_{j+m} - \frac{N-t+1}{N-k} \sum\limits_{j=k+1}^{N}\tilde{X}_{j+m}}   + \frac{(N-k_0+m)(t-k+1)}{N-k}\lrag{\hat\nu_1-\hat\nu_n-\delta,\delta} 
   \right)^2 \\
& & + \sum\limits_{t=k_0-m+1}^{N}
   \left(
   \lrag{\hat\nu_1-\hat\nu_n-\delta, \sum\limits_{j=t}^{N}\tilde{X}_{j+m} - \frac{N-t+1}{N-k}\sum\limits_{j=k+1}^{N}\tilde{X}_{j+m}}
   + \frac{(N-t+1)(k_0-m-k)}{N-k}\lrag{\hat\nu_1-\hat\nu_n-\delta,\delta} 
   \right)^2.
\EEqn}
When $k = k_0-m$, we have that
\BEqn
& & N^2 V_n(k) \\
&=& \sum\limits_{t=1}^{k_0-m} (S_{1,t} - \frac{t}{k_0-m}S_{1,k_0-m})^2 
 +  \sum\limits_{t=k_0-m+1}^{N} (S_{t,N} - \frac{N-t+1}{N-k_0+m}S_{k_0-m+1,N})^2 \\
&=& \sum\limits_{t=1}^{k_0-m}
    \lrp{ \lrag{\hat\nu_1-\hat\nu_n-\delta, \sum\limits_{j=1}^{t}\tilde{X}_{j+m} - \frac{t}{k_0-m}\sum\limits_{j=1}^{k_0-m}\tilde{X}_{j+m}} }^2 \\
& & + \sum\limits_{t=k_0-m+1}^{N} 
    \lrp{ \lrag{\hat\nu_1-\hat\nu_n-\delta, \sum\limits_{j=t}^{N}\tilde{X}_{j+m} - \frac{N-t+1}{N-k_0+m}\sum\limits_{j=k_0-m+1}^{N}\tilde{X}_{j+m}} }^2.
\EEqn
Finally, when $k > k_0-m$, we have that
{\small
\BEqn
& & N^2 V_n(k) \\
&=& \sum\limits_{t=1}^{k_0-m} (S_{1,t} - \frac{t}{k}S_{1,k})^2 
 +  \sum\limits_{t=k_0-m+1}^{k} (S_{1,t} - \frac{t}{k}S_{1,k})^2
 +  \sum\limits_{t=k+1}^{N} (S_{t,N} - \frac{N-t+1}{N-k}S_{k+1,N})^2 \\
&=& \sum\limits_{t=1}^{k_0-m} 
    \lrp{
    \lrag{\hat\nu_1-\hat\nu_n-\delta, \sum\limits_{j=1}^{t}\tilde{X}_{j+m} - \frac{t}{k}\sum\limits_{j=1}^{k}\tilde{X}_{j+m}} - \frac{t(k-k_0+m)}{k}\lrag{\hat\nu_1-\hat\nu_n-\delta, \delta}}^2 \\
& & + \sum\limits_{t=k_0-m+1}^{k} 
    \lrp{ \lrag{\hat\nu_1-\hat\nu_n-\delta, \sum\limits_{j=1}^{t}\tilde{X}_{j+m} - \frac{t}{k} \sum\limits_{j=1}^{k}\tilde{X}_{j+m}} - \frac{(k_0-m)(k-t)}{k}\lrag{\hat\nu_1-\hat\nu_n-\delta,\delta} }^2 \\
& & + \sum\limits_{t=k+1}^{N} 
    \lrp{ \lrag{\hat\nu_1-\hat\nu_n-\delta, \sum\limits_{j=t}^{N}\tilde{X}_{j+m} - \frac{N-t+1}{N-k}\sum\limits_{j=k+1}^{N}\tilde{X}_{j+m}} }^2.
\EEqn}
Again, by unifying the expressions of all the three cases, we obtain the uniform expression of $N^2 V_n(k)$ against the alternative $\delta$.

\subsection{Proof of Theorem \ref{Thm:DensePower_Fixedp}}

By the definitions of $\hat\nu_1$ and $\hat\nu_n$ and with $\delta = \Delta/\sqrt{n}$, we have that 
\begin{equation*}
    \sqrt{n} \lrp{\hat\nu_1-\hat\nu_n-\delta}
  = \frac{1}{\sqrt{n}(\varepsilon-\eta)} \sum\limits_{i=1}^{m_1} \lrp{\tilde{X}_{i}-\tilde{X}_{n+1-i}}
  - \Delta,
\end{equation*}
and it follows that for any $0\le r\le 1$,
\BEqn
& & \lrag{\hat\nu_1-\hat\nu_n-\delta, \sum\limits_{j=1}^{\lrfl{Nr}} \tilde{X}_{j+m}} \\
&=& \frac{1}{\varepsilon-\eta}
    \lrp{\frac{1}{\sqrt{n}} \sum\limits_{i=1}^{m_1} \lrp{\tilde{X}_{i}-\tilde{X}_{n+1-i}}}^{\top}
    \lrp{\frac{1}{\sqrt{n}} \sum\limits_{j=1}^{\lrfl{Nr}} \tilde{X}_{j+m}}
  - \lrp{\frac{1}{\sqrt{n}} \sum\limits_{j=1}^{\lrfl{Nr}} \tilde{X}_{j+m}}^{\top}
    \Delta
\EEqn
and
\begin{equation*}
    n \lrag{\hat\nu_1-\hat\nu_n-\delta, \delta}
  = \frac{1}{\varepsilon-\eta}
    \lrp{\frac{1}{\sqrt{n}} \sum\limits_{i=1}^{m_1} \lrp{\tilde{X}_{i}-\tilde{X}_{n+1-i}}}^{\top}
    \Delta
  - \|\Delta\|_2^2.
\end{equation*}

To investigate the limiting distribution of $G_n$ against the alternative $\delta$, we need to divide into the following three cases depending on the limit of $\|\Delta\|_2$.

\begin{enumerate}[label=(\roman*)]
    \item If $\|\Delta\|_2 = o(1)$ as $n\rightarrow\infty$, we equivalently have that $\Delta\rightarrow0 \in \br^{p}$ as $n\rightarrow\infty$. From Definition \ref{Def:Model_Fixedp}, we have that $n \lrag{\hat\nu_1-\hat\nu_n-\delta, \delta} \leadsto 0$ and
    \begin{equation*}
         \lrag{\hat\nu_1-\hat\nu_n-\delta, \sum\limits_{j=1}^{\lrfl{Nr}} \tilde{X}_{j+m}} \\
    \leadsto b^{\top}(\varepsilon,\eta) \Omega^{(1)}
               \lrp{B_p(\varepsilon+(1-2\varepsilon)r) - B_p(\varepsilon)}
    \quad \mbox{in } D[0,1],
    \end{equation*}
    where $b(\varepsilon,\eta) = \frac{1}{\varepsilon-\eta} \lrp{B_p(\varepsilon-\eta)-B_p(1)+B_p(1-\varepsilon+\eta)}$.
    
    For any $r\in[0,1]$, let $k = \lrfl{Nr}$, then it follows from Proposition \ref{Prop:PowerExpression} that
    \BEqn
    & & N^{1/2} T_n(k) \\
    &\leadsto& b^{\top}(\varepsilon,\eta) \Omega^{(1)}
               \lrp{B_p((1-2\varepsilon)r+\varepsilon) - B_p(\varepsilon) - r\bigp{B_p(1-\varepsilon)-B_p(\varepsilon)}}, \\[5mm]
    & & N V_n(k) \\
    &\leadsto& \int_{0}^{r} 
               \lrp{ b^{\top}(\varepsilon,\eta) \Omega^{(1)}
                     \bigp{B_p((1-2\varepsilon)s+\varepsilon) - B_p(\varepsilon) -
                         \frac{s}{r}\lrp{B_p((1-2\varepsilon)r+\varepsilon)-B_p(\varepsilon)}}}^2 ds \\
    & & + \int_{r}^{1}
          \lrp{ b^{\top}(\varepsilon,\eta) \Omega^{(1)}
                \bigp{B_p(1-\varepsilon) - B_p((1-2\varepsilon)s+\varepsilon) -
                     \frac{1-s}{1-r}\lrp{B_p(1-\varepsilon) - B_p((1-2\varepsilon)r+\varepsilon)}}}^2 ds.
    \EEqn
    
    We observe that the limiting distributions of $N^{1/2} T_n(k)$ and $N V_n(k)$ in this case are exactly the same as those under the null. By using the same conditional arguments as used for Theorem \ref{Thm:NullDist_Gn_Fixedp}, we obtain that $G_n \stra{d} G$ as $n\rightarrow\infty$, which further implies that $\blrp{G_n > G_{1-\alpha}} \rightarrow \alpha$ as $n\rightarrow\infty$.

    \item If $\|\Delta\|_2 \rightarrow \infty$, then in this case we have that 
    \begin{equation*}
        \frac{1}{\|\Delta\|_2^2} 
        \lrag{\hat\nu_1-\hat\nu_n-\delta, \sum\limits_{j=1}^{\lrfl{Nr}}\tilde{X}_{j+m}}
    \leadsto 0
    \quad \mbox{in }D[0,1]
    \end{equation*}
    and 
    \begin{equation*}
        \frac{n}{\|\Delta\|_2^2}
        \lrag{\hat\nu_1-\hat\nu_n-\delta, \delta}
    \stra{p} -1.
    \end{equation*}
    Recall that $r_0 = \lim_{n\rightarrow\infty}(k_0-m)/N$, then it follows from Proposition \ref{Prop:PowerExpression} that
    \begin{equation*}
        \frac{N^{1/2}}{\|\Delta\|_2^2} T_n(k)
    \leadsto (1-2\varepsilon) \lrp{r_0 \wedge r} \lrp{(1-r_0) \wedge (1-r)},
    \end{equation*}
    and similarly,
    \BEqn
        \frac{N}{\|\Delta\|_2^4} V_n(k)
    &\leadsto& \int_{0}^{r} 
                  \lrp{(1-2\varepsilon)\frac{\lrp{s \wedge r_0} \lrp{(r-s)\wedge(r-r_0} \vee 0}{r}}^2 ds \\
    & & + \int_{r}^{1}
          \lrp{(1-2\varepsilon) \frac{\lrp{(1-s)\wedge(1-r_0)} \lrp{(s-r)\wedge(r_0-r)} \vee 0}{1-r}}^2 ds.
    \EEqn
    Therefore, 
    \begin{equation*}
        G_n
      = \sup\limits_{k=1,\cdots,N-1} T_n(k) V_n^{-1/2}(k)
      \geq T_n(\lrfl{Nr_0}) V_n^{-1/2}(\lrfl{Nr_0})
      \stra{d} \infty,
    \end{equation*}
    which yields $\blrp{G_n>G_{1-\alpha}} \rightarrow 1$ as $n\rightarrow\infty$.

    \item If $\|\Delta\|_2 \rightarrow c \in (0,\infty)$ as $n\rightarrow\infty$, then under the assumption that $\frac{\Delta}{\|\Delta\|_2} \rightarrow \Delta_0 \in \br^{p}$, we have that
    \begin{equation*}
        \lrag{\hat\nu_1-\hat\nu_n-\delta, \sum\limits_{j=1}^{\lrfl{Nr}} \tilde{X}_{j+m}} 
    \leadsto \lrp{(\Omega^{(1)})^{1/2}b(\varepsilon,\eta) - c\Delta_0}^{\top} (\Omega^{(1)})^{1/2}
               \lrp{B_p(\varepsilon+(1-2\varepsilon)r) - B_p(\varepsilon)}
    \end{equation*}
    in $D[0,1]$ and 
    \begin{equation*}
        n \lrag{\hat\nu_1-\hat\nu_n-\delta, \delta}
    \stra{d} \lrp{(\Omega^{(1)})^{1/2}b(\varepsilon,\eta) - c\Delta_0}^{\top} (c\Delta_0),
    \end{equation*}
    where $b(\varepsilon,\eta) = \frac{1}{\varepsilon-\eta} \lrp{B_p(\varepsilon-\eta)-B_p(1)+B_p(1-\varepsilon+\eta)}$.
    
    Define $\cb_p=\{B_p(s)\}_{\varepsilon\le s\le 1-\varepsilon}$, then with $k=\lrfl{Nr}$, we obtain from Proposition \ref{Prop:PowerExpression} that
    \BEqn
    & & N^{1/2} T_n(k) \\
    &\leadsto& \lrp{(\Omega^{(1)})^{1/2} b(\varepsilon,\eta) - c\Delta_0}^{\top}
                (\Omega^{(1)})^{1/2}               
                \lrp{\bigp{B_p((1-2\varepsilon)r+\varepsilon) - B_p(\varepsilon)} - r\bigp{B_p(1-\varepsilon) - B_p(\varepsilon)}} \\
    & & - (1-2\varepsilon) \lrp{r_0\wedge r} \lrp{(1-r_0)\wedge(1-r)}
       \lrp{(\Omega^{(1)})^{1/2} b(\varepsilon,\eta) - c\Delta_0}^{\top} (c\Delta_0) \\
    &=:& \tilde{T}(r,\Delta,b(\varepsilon,\eta),\cb_p),
    \EEqn
    and 
    \BEqn
    & & N V_n(k) \\
    &\leadsto& \int_{0}^{r}
        \Large\left(
          \lrp{(\Omega^{(1)})^{1/2} b(\varepsilon,\eta) - c\Delta_0}^{\top}
          (\Omega^{(1)})^{1/2} 
        \right. \\
    & & \hspace{4em} \times
        \left.
          \lrp{\bigp{B_p((1-2\varepsilon)s+\varepsilon) - B_p(\varepsilon)} - \frac{s}{r}\bigp{B_p((1-2\varepsilon)r+\varepsilon) - B_p(\varepsilon)}}
        \right. \\
    & & \hspace{3em}
        \left.
        - (1-2\varepsilon) \frac{\lrp{s \wedge r_0} \lrp{(r-s) \wedge (r-r_0)} \vee 0}{r}
        \lrp{(\Omega^{(1)})^{1/2} b(\varepsilon,\eta) - c\Delta_0}^{\top} (c\Delta_0)
        \Large\right)^2 ds \\
    & & + \int_{r}^{1}
        \Large\left(
          \lrp{(\Omega^{(1)})^{1/2} b(\varepsilon,\eta) - c\Delta_0}^{\top}
          (\Omega^{(1)})^{1/2} 
        \right. \\
    & & \hspace{5em} \times
          \lrp{\bigp{B_p(1-\varepsilon) - B_p((1-2\varepsilon)s+\varepsilon)} - \frac{1-s}{1-r}\bigp{B_p(1-\varepsilon) - B_p((1-2\varepsilon)r+\varepsilon)}} \\
    & & \hspace{4em}
        \left.
        + (1-2\varepsilon) \frac{\lrp{(1-s) \wedge (1-r_0)} \lrp{(s-r) \wedge (r_0-r)} \vee 0}{1-r}
        \lrp{(\Omega^{(1)})^{1/2} b(\varepsilon,\eta) - c\Delta_0}^{\top} (c\Delta_0)
        \Large\right)^2 ds \\
    &=:& \tilde{V}(r,\Delta,b(\varepsilon,\eta),\cb_p),
    \EEqn
    
    By applying the continuous mapping theorem (CMT), we obtain that
    \begin{equation*}
        G_n
    \stra{d} \tilde{M}(\Delta,b(\varepsilon,\eta),\cb_p)
    := \sup\limits_{r\in[0,1]} \tilde{T}(r,\Delta,b(\varepsilon,\eta),\cb_p) \tilde{V}^{-1/2}(r,\Delta,b(\varepsilon,\eta),\cb_p).
    \end{equation*}
    
    We can again apply the conditional distribution arguments developed in the proof of Theorem \ref{Thm:NullDist_Gn_Fixedp}. If conditioning on $b(\varepsilon,\eta)=b_0$, it holds that the processes $\{((\Omega^{(1)})^{1/2}b_0 - c\Delta_0)^{\top} (\Omega^{(1)})^{1/2} B_p(s)\}_{\varepsilon\le s\le 1-\varepsilon}$ and $\{\sqrt{u^{\top}\Omega^{(1)} u} B(s)\}_{\varepsilon\le s\le 1-\varepsilon}$ are equal in distribution, where $u=u(b_0,\Delta) = (\Omega^{(1)})^{1/2}b_0-c\Delta_0$. Consequently, we have that
    \BEqn
    & & \tilde{M}(\Delta,b(\varepsilon,\eta),\cb_p)|_{b(\varepsilon,\eta)=b_0} \\
    &=^d& {\footnotesize
          \sup\limits_{r\in[0,1]} 
          \frac{\sqrt{u^{\top}\Omega^{(1)} u} \lrp{B(r)-rB(1)} - \sqrt{1-2\varepsilon} \lrp{r_0\wedge r} \lrp{(1-r_0)\wedge(1-r)} u^{\top} (c\Delta_0)}
          {\lrp{\begin{array}{l}
            ~~ \int_{0}^{r} \lrp{\sqrt{u^{\top}\Omega^{(1)} u}\lrp{B(s)-\frac{s}{r}B(r)} - \sqrt{1-2\varepsilon} \frac{\lrp{s \wedge r_0} \lrp{(r-s) \wedge (r-r_0)} \vee 0}{r} u^{\top} (c\Delta_0) }^2 ds \\
            + \int_{r}^{1} \lrp{\sqrt{u^{\top}\Omega^{(1)} u}\lrp{B(1-s)-\frac{1-s}{1-r}B(1-r)} + \sqrt{1-2\varepsilon} \frac{\lrp{(1-s) \wedge (1-r_0)} \lrp{(s-r) \wedge (r_0-r)} \vee 0}{1-r} u^{\top} (c\Delta_0) }^2 ds
          \end{array}}^{1/2}}} \\
    &=^d& {\footnotesize
          \sup\limits_{r\in[0,1]} 
          \frac{B(r)-rB(1) - c \lrp{r_0\wedge r} \lrp{(1-r_0)\wedge(1-r)} \sqrt{\frac{1-2\varepsilon}{u^{\top}\Omega^{(1)} u}} u^{\top} \Delta_0}
          {\lrp{\begin{array}{l}
            ~~ \int_{0}^{r} \lrp{ B(s)-\frac{s}{r}B(r) - c \frac{\lrp{s \wedge r_0} \lrp{(r-s) \wedge (r-r_0)} \vee 0}{r} \sqrt{\frac{1-2\varepsilon}{u^{\top}\Omega^{(1)} u}} u^{\top} \Delta_0 }^2 ds \\
            + \int_{r}^{1} \lrp{ B(1-s)-\frac{1-s}{1-r}B(1-r) + c \frac{\lrp{(1-s) \wedge (1-r_0)} \lrp{(s-r) \wedge (r_0-r)} \vee 0}{1-r} \sqrt{\frac{1-2\varepsilon}{u^{\top}\Omega^{(1)} u}} u^{\top} \Delta_0 }^2 ds
          \end{array}}^{1/2}}} \\
    &=^d& \sup\limits_{r\in[0,1]} T(r,\Delta,b_0) V^{-1/2}(r,\Delta,b_0) \\
    &=:& M(\Delta,b_0),
    \EEqn
    where $T(r,\Delta,b_0)$, $V(r,\Delta,b_0)$ are defined as Theorem \ref{Thm:DensePower_Fixedp}.

    By noting that 
    \begin{equation*}
        b(\varepsilon,\eta) 
    =^d \frac{1}{\varepsilon-\eta} \cn_p(0,2(\varepsilon-\eta)I_p)
    =^d \cn_p\lrp{0, \frac{2}{\varepsilon-\eta}I_p},
    \end{equation*}
    we have that
    \BEqn
    & & \blrp{G_n > G_{1-\alpha}} \\
    &\rightarrow& \blrp{\tilde{M}(\Delta,b(\varepsilon,\eta),\cb_p) > G_{1-\alpha}} \\
    &=& \int_{\br^{p}} \blrp{\tilde{M}(\Delta,b(\varepsilon,\eta),\cb_p)|_{b(\varepsilon,\eta)=b_0}  > G_{1-\alpha}} dF_{b(\varepsilon,\eta)}(b_0) \\
    &=& \int_{b_0\in\br^{p}} \lrp{\frac{4\pi}{\varepsilon-\eta}}^{-p/2} \exp\lrp{-\frac{1}{4}(\varepsilon-\eta) b_0^{\top}b_0} \blrp{M(\Delta,b_0) > G_{1-\alpha}} db_0,
    \EEqn
    which completes the proof.
    
\end{enumerate}

\subsection{Proof of Theorem \ref{Thm:DensePower_lp}}

It holds for any $\varepsilon\le u\le 1-\varepsilon$ that
\BEqn
& & \lrag{\hat\nu_1-\hat\nu_n-\delta, \sum\limits_{j=1}^{\lrfl{nu}-\lrfl{n\varepsilon}} \tilde{X}_{j+m}} \\
&=& \frac{1}{\varepsilon-\eta}
    \lrp{\frac{1}{\sqrt{n}} \sum\limits_{i=1}^{m_1} \lrp{\tilde{X}_{i}-\tilde{X}_{n+1-i}}}^{\top}
    \lrp{\frac{1}{\sqrt{n}} \sum\limits_{j=1}^{\lrfl{nu}-\lrfl{n\varepsilon}} \tilde{X}_{j+m}}
  - \lrp{\frac{1}{\sqrt{n}} \sum\limits_{j=1}^{\lrfl{nu}-\lrfl{n\varepsilon}} \tilde{X}_{j+m}}^{\top}
    \Delta \\
&=& \frac{1}{n(\varepsilon-\eta)} W_n(u) 
  - \lrp{\frac{1}{\sqrt{n}} \sum\limits_{j=1}^{\lrfl{nu}-\lrfl{n\varepsilon}} \tilde{X}_{j+m}}^{\top}
    \Delta,
\EEqn
and
\begin{equation*}
    n \lrag{\hat\nu_1-\hat\nu_n-\delta, \delta}
  = \frac{1}{\varepsilon-\eta}
    \lrp{\frac{1}{\sqrt{n}} \sum\limits_{i=1}^{m_1} \lrp{\tilde{X}_{i}-\tilde{X}_{n+1-i}}}^{\top}
    \Delta
  - \|\Delta\|_2^2.
\end{equation*}

From Proposition \ref{Prop:NullDist_Wn_lp}, if $\rho^{m_2/4} \|\Gamma\|_F = o\lrp{\frac{n}{\log(n)}}$, we have that
\begin{equation*}
    \frac{1}{\sqrt{2n\lrfl{(\varepsilon-\eta)n}} \|A^{(0)} \Gamma^{(2)} (A^{(0)})^{\top}\|_F} W_n(u)
\leadsto B\lrp{u}-B\lrp{\varepsilon}
\quad \mbox{in }D[\varepsilon,1-\varepsilon].
\end{equation*}
By Lemma S9.8 in \cite{wang2022inference}, we additionally have that
\begin{equation*}
    \sup\limits_{u\in[\varepsilon,1-\varepsilon]}
    \lrabs{\frac{1}{\|A^{(0)} \Gamma^{(2)} (A^{(0)})^{\top}\|_F} 
    \lrp{\frac{1}{\sqrt{n}} \sum\limits_{j=1}^{\lrfl{nu}-\lrfl{n\varepsilon}} \tilde{X}_{j+m}}^{\top} \Delta} 
  = o_p\lrp{\frac{\|\Delta\|_2}{\|A^{(0)} \Gamma^{(2)} (A^{(0)})^{\top}\|_F^{1/2}}}.
\end{equation*}

Based on the limit of $\frac{\|\Delta\|_2}{\|A^{(0)} \Gamma^{(2)} (A^{(0)})^{\top}\|_F^{1/2}}$, we divide into the following three cases to derive the asymptotic distribution of $G_n$.

\begin{enumerate}[label=(\roman*)]
    \item
    If $\displaystyle\frac{\|\Delta\|_2^2}{\|A^{(0)} \Gamma^{(2)} (A^{(0)})^{\top}\|_F} \rightarrow 0$ as $\min\{n,p\}\rightarrow\infty$, then it holds that 
    \begin{equation*}
        \frac{1}{\|A^{(0)} \Gamma^{(2)} (A^{(0)})^{\top}\|_F} 
    \lrp{\frac{1}{\sqrt{n}} \sum\limits_{j=1}^{\lrfl{nu}-\lrfl{n\varepsilon}} \tilde{X}_{j+m}}^{\top} \Delta
    \leadsto 0
    \quad \mbox{in }D[\varepsilon,1-\varepsilon],
    \end{equation*}
    which directly implies that
    \begin{equation*}
        \frac{n}{\|A^{(0)} \Gamma^{(2)} (A^{(0)})^{\top}\|_F} \lrag{\hat\nu_1-\hat\nu_n-\delta, \sum\limits_{j=1}^{\lrfl{nu}-\lrfl{n\varepsilon}} \tilde{X}_{j+m}}
    \leadsto \sqrt{\frac{2}{\varepsilon-\eta}} \lrp{B(u)-B(\varepsilon)}
    \quad \mbox{in }D[\varepsilon,1-\varepsilon],
    \end{equation*}
    and 
    \begin{equation*}
        \frac{n}{\|A^{(0)} \Gamma^{(2)} (A^{(0)})^{\top}\|_F} 
        \lrag{\hat\nu_1-\hat\nu_n-\delta, \delta} 
    \stra{p} 0.
    \end{equation*}
    
    Let $k=\lrfl{nu}-\lrfl{n\varepsilon}$, thus it follows from Proposition \ref{Prop:PowerExpression} that
    \begin{equation*}
        \frac{N^{1/2}}{\|A^{(0)} \Gamma^{(2)} (A^{(0)})^{\top}\|_F} T_n(k)
    \leadsto \sqrt{\frac{2}{\varepsilon-\eta}} \lrp{B(u)-B(\varepsilon) - \frac{u-\varepsilon}{1-2\varepsilon}(B(1-\varepsilon)-B(\varepsilon))}
    \end{equation*}
    and
    \BEqn
    & & \frac{N}{\|A^{(0)} \Gamma^{(2)} (A^{(0)})^{\top}\|_F^2} V_n(k) \\
    &\leadsto& \frac{2}{\varepsilon-\eta} \cdot \frac{1}{1-2\varepsilon} \int_{\varepsilon}^{u} \lrp{B(v)-B(\varepsilon) - \frac{v-\varepsilon}{u-\varepsilon} (B(u) - B(\varepsilon))}^2 dv \\
    & & + \frac{2}{\varepsilon-\eta} \cdot \frac{1}{1-2\varepsilon} \lrp{B(1-\varepsilon)-B(v) - \frac{1-\varepsilon-v}{1-\varepsilon-u} \lrp{B(1-\varepsilon) - B(u)}}^2 dv
    \EEqn
    Note that the limiting distributions of $\frac{N^{1/2}}{\|A^{(0)} \Gamma^{(2)} (A^{(0)})^{\top}\|_F} T_n(k)$ and $\frac{N}{\|A^{(0)} \Gamma^{(2)} (A^{(0)})^{\top}\|_F^2} V_n(k)$ match their counterparts under the null up to a constant, then following the same steps as those under the null, we obtain that $G_n \stra{d} G$ as $\min\{n,p\}\rightarrow\infty$ which directly implies that $\blrp{G_n>G_{1-\alpha}} \rightarrow \alpha$.

    \item 
    If $\displaystyle\frac{\|\Delta\|_2^2}{\|A^{(0)} \Gamma^{(2)} (A^{(0)})^{\top}\|_F} \rightarrow \infty$, we have that
    \BEqn
    & & \frac{1}{\|\Delta\|_2^2}
        \lrag{\hat\nu_1-\hat\nu_n-\delta, \sum\limits_{j=1}^{\lrfl{nu}-\lrfl{n\varepsilon}} \tilde{X}_{j+m}} \\
    &=& \frac{\|A^{(0)} \Gamma^{(2)} (A^{(0)})^{\top}\|_F}{\|\Delta\|_2^2} 
        \cdot \frac{1}{\|A^{(0)} \Gamma^{(2)} (A^{(0)})^{\top}\|_F}
        \cdot \lrag{\hat\nu_1-\hat\nu_n-\delta, \sum\limits_{j=1}^{\lrfl{nu}-\lrfl{n\varepsilon}} \tilde{X}_{j+m}} \\
    &\leadsto& 0,
    \EEqn
    and
    \begin{equation*}
        \frac{n}{\|\Delta\|_2^2}
        \lrag{\hat\nu_1-\hat\nu_n-\delta, \delta}
     =  \frac{\|A^{(0)} \Gamma^{(2)} (A^{(0)})^{\top}\|_F}{\|\Delta\|_2^2} 
        \cdot \frac{1}{\|A^{(0)} \Gamma^{(2)} (A^{(0)})^{\top}\|_F}
        \cdot \lrp{n\lrag{\hat\nu_1-\hat\nu_n-\delta, \delta}}
     \stra{p} -1.
    \end{equation*}
    Then by Proposition \ref{Prop:PowerExpression}, we obtain that
    \BEqn
        \frac{N^{1/2}}{\|\Delta\|_2^2} T_n(k)
    &\leadsto& (1-2\varepsilon)\lrp{r_0 \wedge \lrp{\frac{u-\varepsilon}{1-2\varepsilon}}} \lrp{(1-r_0)\wedge\lrp{\frac{1-\varepsilon-u}{1-2\varepsilon}}} \\
    &=& (1-2\varepsilon) \lrp{r_0 \wedge r} \lrp{(1-r_0) \wedge (1-r)}
    \EEqn
    with variable change $r = (u-\varepsilon)/(1-2\varepsilon)$. Similarly, we can show that
    and 
    \BEqn
        \frac{N}{\|\Delta\|_2^4} V_n(k)
    &\leadsto& \int_{0}^{r} 
        \lrp{ (1-2\varepsilon) \frac{(s \wedge r_0) \lrp{(r-s)\wedge(r-r_0)}\vee0}{r} }^2 ds \\
    & & + \int_{r}^{1} 
        \lrp{ (1-2\varepsilon) \frac{\lrp{(1-s)\wedge(1-r_0)} \lrp{(s-r)\wedge(r_0-r)}\vee0}{1-r} }^2 ds.
    \EEqn
    Consequently, again by CMT, 
    \begin{equation*}
        G_n 
      = \sup\limits_{k=1,\cdots,N-1} T_n(k) V_n^{-1/2}(k)
      \geq T_n(\lrfl{Nr_0}) V_n^{-1/2}(\lrfl{Nr_0})
      \stra{p} \infty
    \end{equation*}
    as $\min\{n,p\}\rightarrow\infty$, which implies that $\blrp{G_n > G_{1-\alpha}} \rightarrow 1$.

    \item 
    If $\displaystyle\frac{\|\Delta\|_2^2}{\|A^{(0)} \Gamma^{(2)} (A^{(0)})^{\top}\|_F} \rightarrow c \in (0,\infty)$, then we have that
    \begin{equation*}
        \frac{1}{\|A^{(0)} \Gamma^{(2)} (A^{(0)})^{\top}\|_F} 
        \lrag{\hat\nu_1-\hat\nu_n-\delta, \sum\limits_{j=1}^{\lrfl{nu}-\lrfl{n\varepsilon}} \tilde{X}_{j+m}}
    \leadsto \sqrt{\frac{2}{\varepsilon-\eta}} \lrp{B\lrp{u} - B\lrp{\varepsilon}}
    \quad \mbox{in }D[\varepsilon,1-\varepsilon].
    \end{equation*}
    
    Also, it holds that
    \begin{equation*}
        \frac{n}{\|A^{(0)} \Gamma^{(2)} (A^{(0)})^{\top}\|_F} 
        \lrag{\hat\nu_1-\hat\nu_n-\delta, \delta}
    \stra{p} -c.
    \end{equation*}
    
    Again, with $k=\lrfl{nu}-\lrfl{n\varepsilon}$, it follows from Proposition \ref{Prop:PowerExpression} that
    \BEqn
       \frac{N^{1/2}}{\|A^{(0)} \Gamma^{(2)} (A^{(0)})^{\top}\|_F} T_n(k)
    &\leadsto& \sqrt{\frac{2}{\varepsilon-\eta}} \lrp{B(u) - B(\varepsilon) - \frac{u-\varepsilon}{1-2\varepsilon} (B(1-\varepsilon) - B(\varepsilon))} \\
    & & + c(1-2\varepsilon)\lrp{r_0 \wedge \lrp{\frac{u-\varepsilon}{1-2\varepsilon}}} \lrp{(1-r_0)\wedge\lrp{\frac{1-\varepsilon-u}{1-2\varepsilon}}} \\
    &=:& \tilde{T}(u,\Delta),
    \EEqn
    and
    {\small
    \BEqn
    & & \frac{N}{\|A^{(0)} \Gamma^{(2)} (A^{(0)})^{\top}\|_F^2} V_n(k) \\
    &\leadsto& \frac{1}{1-2\varepsilon}
        \int_{\varepsilon}^{u} 
        \left( \sqrt{\frac{2}{\varepsilon-\eta}} \lrp{B(v) - B(\varepsilon) - \frac{v-\varepsilon}{u-\varepsilon}\lrp{B(u) - B(\varepsilon)}}
        \right. \\
    & & \hspace{6em}        
        \left.
             + c(1-2\varepsilon) \frac{\lrp{ \lrp{r_0 \wedge \lrp{\frac{v-\varepsilon}{1-2\varepsilon}}} \lrp{ \lrp{\frac{u-\varepsilon}{1-2\varepsilon}-r_0} \wedge \lrp{\frac{u-v}{1-2\varepsilon}}}} \vee 0}{\frac{u-\varepsilon}{1-2\varepsilon}} 
        \right)^2 ds \\
    & & + \frac{1}{1-2\varepsilon}
        \int_{u}^{1-\varepsilon} 
        \left( \sqrt{\frac{2}{\varepsilon-\eta}} \lrp{B(1-\varepsilon) - B(v) - \frac{1-\varepsilon-v}{1-\varepsilon-u}\lrp{B(1-\varepsilon) - B(u)}} 
        \right. \\
    & & \hspace{8em}        
        \left.    
             - c(1-2\varepsilon) \frac{\lrp{\lrp{\lrp{r_0-\frac{u-\varepsilon}{1-2\varepsilon}}\wedge\lrp{\frac{v-u}{1-2\varepsilon}}} \lrp{\lrp{1-r_0}\wedge\lrp{1-\frac{v-\varepsilon}{1-2\varepsilon}}}} \vee 0}{1-\frac{u-\varepsilon}{1-2\varepsilon}} 
        \right)^2 ds \\
    &=:& \tilde{V}(u,\Delta).
    \EEqn}
    By applying CMT, we obtain that $G_n\stra{d} M(\Delta):= \sup\limits_{u\in[\varepsilon,1-\varepsilon]} \tilde{T}(u,\Delta) \tilde{V}^{-1/2}(u,\Delta)$. Next, we change the variables by $u = \varepsilon+(1-2\varepsilon)r$ and $v = \varepsilon+(1-2\varepsilon)s$, and obtain that
    {\small
    \BEqn
    & & M(c) \\
    &=^d& \sup\limits_{r\in[0,1]} 
    \frac{\sqrt{\frac{2(1-2\varepsilon)}{\varepsilon-\eta}} \lrp{B(r) - rB(1)} + c(1-2\varepsilon)\lrp{r_0\wedge r}\lrp{(1-r_0)\wedge(1-r)}}
    {\lrp{
    \begin{array}{l}
      ~~ \int_{0}^{r} \lrp{ \sqrt{\frac{2(1-2\varepsilon)}{\varepsilon-\eta}} \lrp{B(s)-\frac{s}{r}B(r)} + c(1-2\varepsilon)\frac{\lrp{\lrp{r_0\wedge s}\lrp{(r-r_0)\wedge(r-s)}}\vee0}{r}  }^2 ds  \\
      + \int_{r}^{1} \lrp{ \sqrt{\frac{2(1-2\varepsilon)}{\varepsilon-\eta}} \lrp{B(1-s)} - \frac{1-s}{1-r}B(1-r) - c(1-2\varepsilon) \frac{\lrp{\lrp{(1-s) \wedge (1-r_0)} \lrp{(r_0-r) \wedge (s-r)}} \vee 0}{1-r} }^2 ds
    \end{array}
    }^{1/2}} \\
    &=^d& \sup\limits_{r\in[0,1]} 
    \frac{B(r) - rB(1) + c \sqrt{\frac{(1-2\varepsilon)(\varepsilon-\eta)}{2}} \lrp{r_0\wedge r}\lrp{(1-r_0)\wedge(1-r)}}
    {\lrp{
    \begin{array}{l}
      ~~ \int_{0}^{r} 
         \lrp{ B(s) - \frac{s}{r}B(r)
              + c \sqrt{\frac{(1-2\varepsilon)(\varepsilon-\eta)}{2}} \frac{(s \wedge r_0) \lrp{(r-s)\wedge(r-r_0)}\vee0}{r} }^2 ds \\
      + \int_{r}^{1} 
        \lrp{  B(1-s) - \frac{1-s}{1-r}B(1-r)
             - c \sqrt{\frac{(1-2\varepsilon)(\varepsilon-\eta)}{2}} \frac{\lrp{(1-s)\wedge(1-r_0)} \lrp{(s-r)\wedge(r_0-r)}\vee0}{1-r} }^2 ds
    \end{array}
    }^{1/2}} \\
    &=:& \sup\limits_{r\in[0,1]} T(r,c) V^{-1/2}(r,c),
    \EEqn}
    where $T(r,c)$ and $V(r,c)$ are defined in Theorem \ref{Thm:DensePower_lp}. Therefore, it is trivial that $\blrp{G_n>G_{1-\alpha}} \rightarrow \blrp{M(c)>G_{1-\alpha}}$, which completes the proof.
\end{enumerate}

\subsection{Proof of Theorem \ref{Thm:DensePower_factor}}

We have shown in the proof of Theorem \ref{Thm:DensePower_lp} that for any $\varepsilon \le u \le 1-\varepsilon$, 
\begin{equation*}
    \lrag{\hat\nu_1-\hat\nu_n-\delta, \sum\limits_{j=1}^{\lrfl{nu}-\lrfl{n\varepsilon}} \tilde{X}_{j+m}} \\
  = \frac{1}{n(\varepsilon-\eta)} W_n(u) 
  - \lrp{\frac{1}{\sqrt{n}} \sum\limits_{j=1}^{\lrfl{nu}-\lrfl{n\varepsilon}} \tilde{X}_{j+m}}^{\top}
    \Delta,
\end{equation*}
and
\begin{equation*}
    n \lrag{\hat\nu_1-\hat\nu_n-\delta, \delta}
  = \frac{1}{\varepsilon-\eta}
    \lrp{\frac{1}{\sqrt{n}} \sum\limits_{i=1}^{m_1} \lrp{\tilde{X}_{i}-\tilde{X}_{n+1-i}}}^{\top}
    \Delta
  - \|\Delta\|_2^2.
\end{equation*}

Recall that $\tilde{X}_t = \Lambda F_t + Z_t$, and it is shown in the proof of Theorem \ref{Thm:NullDist_Gn_factor} that
\BEqn
    W_n(r)
&=& \sum\limits_{i=1}^{m_1} \sum\limits_{j=1}^{\lrfl{nr} - \lrfl{n\varepsilon}} (F_{i} - F_{n+1-i})^{\top} \Lambda^{\top} \Lambda F_{j+m}  
    + \sum\limits_{i=1}^{m_1} \sum\limits_{j=1}^{\lrfl{nr} - \lrfl{n\varepsilon}}  (Z_{i} - Z_{n+1-i})^{\top} Z_{j+m} \\
& & + \sum\limits_{i=1}^{m_1} \sum\limits_{j=1}^{\lrfl{nr} - \lrfl{n\varepsilon}} Z_{j+m}^{\top} \Lambda(F_{i} - F_{n+1-i})
    + \sum\limits_{i=1}^{m_1} \sum\limits_{j=1}^{\lrfl{nr} - \lrfl{n\varepsilon}} (Z_{i} - Z_{n+1-i})^{\top} \Lambda F_{j+m} \\
&=:& W_{n,1}(r) + W_{n,2}(r) + W_{n,3}(r) + W_{n,4}(r).
\EEqn
Furthermore, it follows from Lemma \ref{Lemma:factor-aux-1-1}, Lemma \ref{Lemma:factor-aux-1-2} and Lemma \ref{Lemma:factor-aux-1-3} that
\BEqn
& & \frac{1}{n\|\Lambda^{\top} \Lambda\|_F} W_{n,1}(r)
 \leadsto (\varepsilon-\eta) b^{\top}(\varepsilon,\eta)
          ((\Omega^{(3)})^{1/2})^{\top} L_0 (\Omega^{(3)})^{1/2}
          \lrp{B_s(r) - B_s(\varepsilon)}, \\
& & \frac{1}{\sqrt{2n\lrfl{n(\varepsilon-\eta)}} \|A^{(0)} \Gamma^{(3)} (A^{(0)})^{\top}\|_F} W_{n,2}(r)
 \leadsto \tilde{B}(r) - \tilde{B}(\varepsilon), \\
& & \frac{1}{n \max\{\|\Lambda\|^2, \|\Gamma^{(3)}\|_F\}} \lrp{W_{n,3}(r) + W_{n,4}(r)}
 \leadsto 0,
\EEqn
where $\{B_s(r)\}_{0\le r\le 1}$ and $\{\tilde{B}(r)\}_{r\in[0,1]}$ are two independent Brownian processes in $\br^s$ and $\br$ respectively, and $b(\varepsilon,\eta) = \frac{1}{\varepsilon-\eta} \lrp{B_s(\varepsilon-\eta) - B_s(1) + B_s(1-\varepsilon+\eta)}$.

Additionally, with $k=\lrfl{nu}-\lrfl{n\varepsilon}$, it holds that
\begin{equation*}
    \lrp{\frac{1}{\sqrt{n}} \sum\limits_{j=1}^{k} \tilde{X}_{j+m}}^{\top} \Delta
  = \frac{1}{\sqrt{n}} \sum\limits_{j=1}^{k} F_{j+m}^{\top} \Lambda^{\top} \Delta
  + \frac{1}{\sqrt{n}} \sum\limits_{j+1}^{k} Z_{j+m}^{\top} \Delta.
\end{equation*}
From Assumption \ref{Assumpt:NullDist_Wn_factor}\ref{Assumpt:NullDist_Wn_factor-1}, we have that 
\begin{equation*}
    \lrp{\frac{1}{\sqrt{n}} \sum\limits_{j=1}^{k} F_{j+m}}^{\top} \Lambda^{\top} \Delta
\leadsto \lrp{B_s(u-\varepsilon) - B_s(\varepsilon)}^{\top} ((\Omega^{(3)})^{1/2})^{\top}
            \Lambda^{\top} \Delta,
\end{equation*}
On the other hand, by Lemma S9.8 in \cite{wang2022inference}, we have that
\begin{equation*}
    \sup\limits_{u\in[\varepsilon,1-\varepsilon]}
    \lrabs{\frac{1}{\|A^{(0)} \Gamma^{(3)} (A^{(0)})^{\top}\|_F} \lrp{\frac{1}{\sqrt{n}} \sum\limits_{j=1}^{k} Z_{j+m}}^{\top}\Delta} 
  = o_p\lrp{\frac{\|\Delta\|_2}{\|A^{(0)} \Gamma^{(3)} (A^{(0)})^{\top}\|_F^{1/2}}}.
\end{equation*}
Then by discussing the relationship between $\|\Delta\|_2$ and $\max\{\|\Lambda\|,\|\Gamma^{(3)}\|_F^{1/2}\}$, we are ready to prove the proposed results.

\begin{enumerate}[label=(\roman*)]
    \item If $\|\Delta\|_2 = o\lrp{\max\{\|\Lambda\|,\|\Gamma^{(3)}\|_F^{1/2}\}}$ as $\min\{n,p\}\rightarrow\infty$, that is $\frac{\|\Delta\|_2}{\max\{\|\Lambda\|,\|\Gamma^{(3)}\|_F^{1/2}\}} \rightarrow 0$, then we have that
    \begin{equation*}
        \frac{1}{\max\{\|\Lambda\|^2,\|\Gamma^{(3)}\|_F\}} \cdot \lrp{\frac{1}{\sqrt{n}} \sum\limits_{j=1}^{\lrfl{nu}-\lrfl{n\varepsilon}} \tilde{X}_{j+m}}^{\top} \Delta
    \leadsto 0
    \quad \mbox{in }D[\varepsilon,1-\varepsilon]
    \end{equation*}
    which further implies that
    \begin{equation*}
        \frac{n}{\max\{\|\Lambda\|^2,\|\Gamma^{(3)}\|_F\}} \lrag{\hat\nu_1-\hat\nu_n-\delta, \delta}
    \leadsto 0.
    \end{equation*}
    
    Note that the asymptotically dominant term of $\lrag{\hat\nu_1-\hat\nu_n-\delta, \sum\limits_{j=1}^{\lrfl{nu}-\lrfl{n\varepsilon}} \tilde{X}_{j+m}}$ is $W_n(u)-W_n(\varepsilon)$, and in the proof of Theorem \ref{Thm:NullDist_Gn_factor}, we have shown that there exists some functions $c_1(\Lambda,\Gamma^{(3)})$ and $c_2(\Lambda,\Gamma^{(3)})$, such that
    \BEqn
    & & \frac{1}{n(\varepsilon-\eta)\max\{\|\Lambda\|^2,\|\Gamma^{(3)}\|_F\}} W_n(r) \\
    &\leadsto& c_1(\Lambda,\Gamma^{(3)}) b^{\top}(\varepsilon,\eta) ((\Omega^{(3)})^{1/2})^{\top} L_0 (\Omega^{(3)})^{1/2} \lrp{B_s(r)-B_s(\varepsilon)} 
    + c_2(\Lambda,\Gamma^{(3)}) \lrp{\tilde{B}(r)-\tilde{B}(\varepsilon)}.
    \EEqn
    Therefore, we have that
    \BEqn
    & & \frac{1}{\max\{\|\Lambda\|^2,\|\Gamma^{(3)}\|_F\}} \lrag{\hat\nu_1-\hat\nu_n-\delta, \sum\limits_{j=1}^{\lrfl{nu}-\lrfl{n\varepsilon}} \tilde{X}_{j+m}} \\
    &=& \frac{1}{n(\varepsilon-\eta)\max\{\|\Lambda\|^2,\|\Gamma^{(3)}\|_F\}} W_n(u) \\
    & & - \frac{1}{\max\{\|\Lambda\|^2,\|\Gamma^{(3)}\|_F\}} \lrp{\frac{1}{\sqrt{n}} \sum\limits_{j=1}^{\lrfl{nu}-\lrfl{n\varepsilon}} F_{j+m}}^{\top} \Lambda^{\top} \Delta \\
    & & - \frac{1}{\max\{\|\Lambda\|^2,\|\Gamma^{(3)}\|_F\}} \lrp{\frac{1}{\sqrt{n}} \sum\limits_{j=1}^{\lrfl{nu}-\lrfl{n\varepsilon}} Z_{j+m}}^{\top} \Delta \\
    &\leadsto& c_1(\Lambda,\Gamma^{(3)}) b^{\top}(\varepsilon,\eta) ((\Omega^{(3)})^{1/2})^{\top} L_0 (\Omega^{(3)})^{1/2} \lrp{B_s(u)-B_s(\varepsilon)} + c_2(\Lambda,\Gamma^{(3)}) \lrp{\tilde{B}(u)-\tilde{B}(\varepsilon)}.
    \EEqn
    
    Again, we let $k=\lrfl{nu}-\lrfl{n\varepsilon}$ for $u\in[\varepsilon,1-\varepsilon]$. From Proposition \ref{Prop:PowerExpression}, we further obtain that
    \BEqn
    & & \frac{N^{1/2}}{\max\{\|\Lambda\|^2,\|\Gamma^{(3)}\|_F\}} T_n(k) \\
    &\leadsto& c_1(\Lambda,\Gamma^{(3)}) b^{\top}(\varepsilon,\eta) ((\Omega^{(3)})^{1/2})^{\top} L_0 (\Omega^{(3)})^{1/2} \lrp{B_s(u)-B_s(\varepsilon)-\frac{u-\varepsilon}{1-2\varepsilon}(B_s(1-\varepsilon)-B_s(\varepsilon))} \\
    & & + c_2(\Lambda,\Gamma^{(3)}) \lrp{\tilde{B}(u)-\tilde{B}(\varepsilon)-\frac{u-\varepsilon}{1-2\varepsilon}(\tilde{B}(1-\varepsilon)-\tilde{B}(\varepsilon))} \\
    &=:& T(u,b(\varepsilon,\eta),\cb_s,\tilde{\cb}),
    \EEqn
    and
    \BEqn
    & & \frac{N}{\max\{\|\Lambda\|^2,\|\Gamma^{(3)}\|_F\}^2} V_n(k) \\
    &=& \frac{1}{1-2\varepsilon}
        \int_{\varepsilon}^{u}
        \left(
          c_1(\Lambda,\Gamma^{(3)}) b^{\top}(\varepsilon,\eta) ((\Omega^{(3)})^{1/2})^{\top} L_0 (\Omega^{(3)})^{1/2} \lrp{B_s(v)-B_s(\varepsilon)-\frac{v-\varepsilon}{u-\varepsilon}(B_s(u)-B_s(\varepsilon))}
        \right. \\
    & & \hspace{5em}
        \left.
          + c_2(\Lambda,\Gamma^{(3)}) \lrp{\tilde{B}(v)-\tilde{B}(\varepsilon)-\frac{v-\varepsilon}{u-\varepsilon}(\tilde{B}(u)-\tilde{B}(\varepsilon))}
        \right)^2 dv \\
    & & + \frac{1}{1-2\varepsilon} 
        \int_{u}^{1-\varepsilon}
        \left(
          c_1(\Lambda,\Gamma^{(3)}) b^{\top}(\varepsilon,\eta) ((\Omega^{(3)})^{1/2})^{\top} L_0 (\Omega^{(3)})^{1/2} 
        \right. \\
    & & \hspace{8em}  
        \times 
        \lrp{B_s(1-\varepsilon)-B_s(v)-\frac{1-\varepsilon-v}{1-\varepsilon-u}(B_s(1-\varepsilon)-B_s(u))} \\
    & & \hspace{7em}
        \left.
          + c_2(\Lambda,\Gamma^{(3)}) \lrp{\tilde{B}(1-\varepsilon)-\tilde{B}(v)-\frac{1-\varepsilon-v}{1-\varepsilon-u}(\tilde{B}(1-\varepsilon)-\tilde{B}(u))}
        \right)^2 dv \\
    &=:& V(u,b(\varepsilon,\eta),\cb_s,\tilde{\cb}),
    \EEqn
    where $\cb_s = \{B_s(u)\}_{\varepsilon\le u\le 1-\varepsilon}$ and $\tilde{\cb} = \{\tilde{B}(u)\}_{\varepsilon\le u\le 1-\varepsilon}$. Using the same conditioning arguments in the proof of Theorem \ref{Thm:NullDist_Gn_factor}, we can show that $G_n \stra{d} G$ as $\min\{n,p\}\rightarrow\infty$, which implies that $\blrp{G_n>G_{1-\alpha}} \rightarrow \alpha$.

    \item If $\max\{\|\Lambda\|,\|\Gamma^{(3)}\|_F^{1/2}\} = o\lrp{\|\Delta\|_2}$, then we have that
    \begin{equation*}
        \frac{1}{n \|\Delta\|_2^2} W_n(u) \leadsto 0
    \quad \mbox{in }D[\varepsilon,1-\varepsilon],
    \end{equation*}
    and
    \begin{equation*}
        \frac{1}{\|\Delta\|_2^2} \lrp{\frac{1}{\sqrt{n}} \sum\limits_{j=1}^{\lrfl{nu}-\lrfl{n\varepsilon}} \tilde{X}_{j+m}}^{\top} \Delta
    \leadsto 0
    \quad \mbox{in }D[\varepsilon,1-\varepsilon]
    \end{equation*}
    This implies that
    \begin{equation*}
        \frac{1}{\|\Delta\|_2^2} \lrag{\hat\nu_1-\hat\nu_n-\delta, \sum\limits_{j=1}^{\lrfl{nu}-\lrfl{n\varepsilon}} \tilde{X}_{j+m}}
    \leadsto 0,
    \end{equation*}
    and 
    \begin{equation*} 
        \frac{n}{\|\Delta\|_2^2} \lrag{\hat\nu_1-\hat\nu_n-\delta, \delta}
    \stra{p} -1.
    \end{equation*}
    
    Therefore, with $k=\lrfl{nu}-\lrfl{n\varepsilon}$ and $r=(u-\varepsilon)/(1-2\varepsilon)$, we have that
    \begin{equation*}
        \frac{N^{1/2}}{\|\Delta\|_2^2} T_n(k)
    \leadsto (1-2\varepsilon) (r_0 \wedge r) \lrp{(1-r_0) \wedge (1-r)},
    \end{equation*}
    and
    \BEqn
        \frac{N}{\|\Delta\|_2^4} V_n(k)
    &\leadsto& \int_{0}^{r} 
        \lrp{ (1-2\varepsilon) \frac{(s \wedge r_0) \lrp{(r-s)\wedge(r-r_0)}\vee0}{r} }^2 ds \\
    & & + \int_{r}^{1} 
        \lrp{ (1-2\varepsilon) \frac{\lrp{(1-s)\wedge(1-r_0)} \lrp{(s-r)\wedge(r_0-r)}\vee0}{1-r} }^2 ds.
    \EEqn
    Following similar analysis used in the proof of Theorem \ref{Thm:DensePower_Fixedp}\ref{Thm:DensePower_Fixedp-2} and Theorem \ref{Thm:DensePower_lp}\ref{Thm:DensePower_lp-2}, we obtain that $\blrp{G_n>G_{1-\alpha}}\rightarrow1$ as $\min\{n,p\}\rightarrow\infty$.

    \item If $\|\Delta\|_2 \sim \max\{\|\Lambda\|,\|\Gamma^{(3)}\|_F^{1/2}\}$, then we further divide the discussion into two sub-cases based on the leading term between $\|\Lambda\|$ and $\|\Gamma^{(3)}\|_F^{1/2}$.
    
    \begin{enumerate}[label=(\arabic*)]
        \item when $\|\Lambda\| = o\lrp{\|\Gamma^{(3)}\|_F^{1/2}}$, then we have $\|\Delta\|_2 \sim \|\Gamma^{(3)}\|_F^{1/2} \sim \|A^{(0)} \Gamma^{(3)} (A^{(0)})^{\top}\|_F^{1/2}$. We assume that
        \begin{equation*}
            \frac{\|\Delta\|_2^2}{\sqrt{2(\varepsilon-\eta)} \|A^{(0)} \Gamma^{(3)} (A^{(0)})^{\top}\|_F}
        \rightarrow c:= c(\Delta)
        \end{equation*}
        as $\min\{n,p\}\rightarrow\infty$. It follows that
        \BEqn
        & & \frac{1}{\sqrt{2n\lrfl{n(\varepsilon-\eta)}} \|A^{(0)} \Gamma^{(3)} (A^{(0)})^{\top}\|_F} W_{n,1}(r) \\
        &=& \frac{n\|\Lambda^{\top}\Lambda\|}{\sqrt{2n\lrfl{n(\varepsilon-\eta)}} \|A^{(0)} \Gamma^{(3)} (A^{(0)})^{\top}\|_F} \cdot \frac{1}{n\|\Lambda^{\top}\Lambda\|} W_{n,1}(r) \\
        &\leadsto& 0,
        \EEqn
        and 
        \BEqn
        & & \frac{1}{\sqrt{2n\lrfl{n(\varepsilon-\eta)}} \|A^{(0)} \Gamma^{(3)} (A^{(0)})^{\top}\|_F} \lrp{W_{n,3}(r) + W_{n,4}(r)} \\
        &=& \frac{n \max\{\Lambda\|^2, \|\Gamma^{(3)}\|_F\}}{\sqrt{2n\lrfl{n(\varepsilon-\eta)}} \|A^{(0)} \Gamma^{(3)} (A^{(0)})^{\top}\|_F} \cdot \frac{1}{n \max\{\Lambda\|^2, \|\Gamma^{(3)}\|_F\}} \lrp{W_{n,3}(r) + W_{n,4}(r)} \\
        &\leadsto& 0,
        \EEqn
        both of which together imply that 
        \begin{equation*}
            \frac{1}{\sqrt{2n\lrfl{n(\varepsilon-\eta)}} \|A^{(0)} \Gamma^{(3)} (A^{(0)})^{\top}\|_F} W_n(r)
        \leadsto \tilde{B}(r) - \tilde{B}(\varepsilon)
        \quad \mbox{in } D[\varepsilon,1-\varepsilon].
        \end{equation*}
        
        Additionally, from previous analysis we obtain that
        \begin{equation*}
            \frac{1}{\sqrt{2(\varepsilon-\eta)} \|A^{(0)} \Gamma^{(3)} (A^{(0)})^{\top}\|_F} 
            \lrp{\frac{1}{\sqrt{n}} \sum\limits_{j=1}^{k} \tilde{X}_{j+m}}^{\top} \Delta
        \leadsto 0.
        \end{equation*}
        Consequently, it holds in $D[\varepsilon,1-\varepsilon]$ that
        \begin{equation*}
            \frac{1}{\sqrt{2(\varepsilon-\eta)} \|A^{(0)} \Gamma^{(3)} (A^{(0)})^{\top}\|_F}
            \lrag{\hat\nu_1-\hat\nu_n-\delta, \sum\limits_{j=1}^{\lrfl{nu}-\lrfl{n\varepsilon}} \tilde{X}_{j+m}}
        \leadsto \sqrt{\frac{2}{\varepsilon-\eta}} \lrp{\tilde{B}(u) - \tilde{B}(\varepsilon)},
        \end{equation*}
        and $\displaystyle \frac{n}{\sqrt{2(\varepsilon-\eta)} \|A^{(0)} \Gamma^{(3)} (A^{(0)})^{\top}\|_F} \lrag{\hat\nu_1-\hat\nu_n-\delta,\delta} \stra{p} -c$. 
        
        It follows that with $k=\lrfl{nu}-\lrfl{n\varepsilon}$, we have that
        \BEqn
            \frac{N^{1/2}}{\|A^{(0)} \Gamma^{(3)} (A^{(0)})^{\top}\|_F} T_n(k)
        &\leadsto& \sqrt{\frac{2}{\varepsilon-\eta}} \lrp{B(u) - B(\varepsilon) - \frac{u-\varepsilon}{1-2\varepsilon} (B(1-\varepsilon) - B(\varepsilon))} \\
        & & + c(1-2\varepsilon)\lrp{r_0 \wedge \lrp{\frac{u-\varepsilon}{1-2\varepsilon}}} \lrp{(1-r_0)\wedge\lrp{\frac{1-\varepsilon-u}{1-2\varepsilon}}} \\
        &=:& \tilde{T}(u,\Delta),
        \EEqn
        and
        {\small
        \BEqn
        & & \frac{N}{\|A^{(0)} \Gamma^{(3)} (A^{(0)})^{\top}\|_F^2} V_n(k) \\
        &\leadsto& \frac{1}{1-2\varepsilon}
            \int_{\varepsilon}^{u} 
            \left( \sqrt{\frac{2}{\varepsilon-\eta}} \lrp{B(v) - B(\varepsilon) -   \frac{v-\varepsilon}{u-\varepsilon}\lrp{B(u) - B(\varepsilon)}}
            \right. \\
        & & \hspace{6em}        
            \left.
            + c(1-2\varepsilon) \frac{\lrp{ \lrp{r_0 \wedge \lrp{\frac{v-\varepsilon}{1-2\varepsilon}}} \lrp{ \lrp{\frac{u-\varepsilon}{1-2\varepsilon}-r_0} \wedge \lrp{\frac{u-v}{1-2\varepsilon}}}} \vee 0}{\frac{u-\varepsilon}{1-2\varepsilon}} 
            \right)^2 ds \\
        & & + \frac{1}{1-2\varepsilon}
            \int_{u}^{1-\varepsilon} 
            \left( \sqrt{\frac{2}{\varepsilon-\eta}} \lrp{B(1-\varepsilon) - B(v) - \frac{1-\varepsilon-v}{1-\varepsilon-u}\lrp{B(1-\varepsilon) - B(u)}} 
            \right. \\
        & & \hspace{7em}        
            \left.    
            - c(1-2\varepsilon) \frac{\lrp{\lrp{\lrp{r_0-\frac{u-\varepsilon}{1-2\varepsilon}}\wedge\lrp{\frac{v-u}{1-2\varepsilon}}} \lrp{\lrp{1-r_0}\wedge\lrp{1-\frac{v-\varepsilon}{1-2\varepsilon}}}} \vee 0}{1-\frac{u-\varepsilon}{1-2\varepsilon}} 
            \right)^2 ds \\
        &=:& \tilde{V}(u,\Delta).
        \EEqn}
        both of which exactly match the analysis of Theorem \ref{Thm:DensePower_lp}\ref{Thm:DensePower_lp-3}, and we spare the remaining details.
        
        \item when $\|\Gamma^{(3)}\|_F^{1/2} = O\lrp{\|\Lambda\|}$, i.e. we have either $\|\Gamma^{(3)}\|_F^{1/2} \sim \|\Lambda\|$ or $\|\Gamma^{(3)}\|_F^{1/2} = o\lrp{\|\Lambda\|}$. In this case, it holds that $\|\Delta\|_2 \sim \|\Lambda\| \sim \|\Lambda^{\top}\Lambda\|^{1/2}$. We assume that
        \begin{equation*}
            \frac{\sqrt{2(\varepsilon-\eta)} \|A^{(0)} \Gamma^{(3)} (A^{(0)})^{\top}\|_F}{\|\Lambda^{\top} \Lambda\|} 
        \rightarrow c_1:=c_1(\Delta), 
        \end{equation*}
        and $\frac{\|\Lambda^{\top} \Delta\|_2}{\|\Lambda^{\top}\Lambda\|} \rightarrow c_2:=c_2(\Delta)$, $\frac{\|\Delta\|_2^2}{\|\Lambda^{\top}\Lambda\|} \rightarrow c_3:=c_3(\Delta)$. Then it holds in $D[\varepsilon,1-\varepsilon]$ that
        \begin{equation*}
            \frac{1}{n\|\Lambda^{\top}\Lambda\|} W_n^{(1)}(u) \\
        \leadsto (\varepsilon-\eta) b(\varepsilon,\eta)^{\top} 
            ((\Omega^{(3)})^{1/2})^{\top} L_0 (\Omega^{(3)})^{1/2} \lrp{B_s(u)-B_s(\varepsilon)},
        \end{equation*}
        and
        \BEqn
        & & \frac{1}{n\|\Lambda^{\top}\Lambda\|} W_n^{(2)}(u) \\
        &=& \frac{\sqrt{2n\lrfl{n(\varepsilon-\eta)}} \|A^{(0)} \Gamma^{(3)} (A^{(0)})^{\top}\|_F}{n \|\Lambda^{\top} \Lambda\|} \cdot \frac{1}{\sqrt{2n\lrfl{n(\varepsilon-\eta)}} \|A^{(0)} \Gamma^{(3)} (A^{(0)})^{\top}\|_F} W_n^{(2)}(u) \\
        &\leadsto& c_1 \lrp{\tilde{B}(u) - \tilde{B}(\varepsilon)},
        \EEqn
        as well as
        \BEqn
        & & \frac{1}{n\|\Lambda^{\top}\Lambda\|} \lrp{W_{n,3}(r)+W_{n,4}(r)} \\
        &=& \frac{\max\{\|\Lambda\|^2,\|\Gamma^{(3)}\|_F\}}{\|\Lambda^{\top} \Lambda\|} \cdot \frac{1}{n\max\{\|\Lambda\|^2,\|\Gamma^{(3)}\|_F\}} \lrp{W_{n,3}(r)+W_{n,4}(r)} \\
        &\leadsto& 0.
        \EEqn
        Note that $\{W_{n,1}(r)\}_{\varepsilon\le r\le 1-\varepsilon}$ is independent of $\{W_{n,2}(r)\}_{\varepsilon\le r\le 1-\varepsilon}$ due to the independence between $\{F_t\}_{t=1}^{n}$ and $\{Z_t\}_{t=1}^{n}$, then we have that
        \BEqn
        & & \frac{1}{n\|\Lambda^{\top}\Lambda\|} W_n(r) \\
        &\leadsto& (\varepsilon-\eta) b(\varepsilon,\eta)^{\top} ((\Omega^{(3)})^{1/2})^{\top} L_0 (\Omega^{(3)})^{1/2} 
                 \lrp{B_s(r)-B_s(\varepsilon)}
            + c_1 \lrp{\tilde{B}(r) - \tilde{B}(\varepsilon)},
        \EEqn
        where $\cb_s := \{B_s(r)\}_{\varepsilon\le r\le 1-\varepsilon}$ and $\tilde\cb := \{\tilde{B}(r)\}_{\varepsilon\le r\le 1-\varepsilon}$ are two independent Brownian processes in $\br^s$ and $\br$ respectively.
        
        Additionally, under Assumption \ref{Assumpt:DensePower_factor}, we have that
        \BEqn
            \frac{1}{\|\Lambda^{\top}\Lambda\|} 
            \lrp{\frac{1}{\sqrt{n}}\sum\limits_{j=1}^{\lrfl{nu}-\lrfl{n\varepsilon}} \tilde{X}_{j+m}}^{\top} \Delta
        &\leadsto& c_2 \lambda_0^{\top}  (\Omega^{(3)})^{1/2} 
                      \lrp{B_s(u) - B(\varepsilon)},
        \EEqn
        which further implies that
        \BEqn
        & & \frac{1}{\|\Lambda^{\top}\Lambda\|} 
            \lrag{\hat\nu_1-\hat\nu_n-\delta, \sum\limits_{j=1}^{\lrfl{nu}-\lrfl{n\varepsilon}} \tilde{X}_{j+m}} \\
        &=& \frac{1}{n(\varepsilon-\eta)\|\Lambda^{\top}\Lambda\|} W_n(u) - \frac{1}{\|\Lambda^{\top}\Lambda\|} 
            \lrp{\frac{1}{\sqrt{n}}\sum\limits_{j=1}^{\lrfl{nu}-\lrfl{n\varepsilon}} \tilde{X}_{j+m}}^{\top} \Delta \\
        &\leadsto& \lrp{b(\varepsilon,\eta)^{\top} ((\Omega^{(3)})^{1/2})^{\top} L_0 - c_2 \lambda_0^{\top}} (\Omega^{(3)})^{1/2} \lrp{B_s(u)-B_s(\varepsilon)}
        + \frac{c_1}{\varepsilon-\eta} \lrp{\tilde{B}(u) - \tilde{B}(\varepsilon)} 
        \EEqn
        and
        \begin{equation*}
            \frac{n}{\|\Lambda^{\top}\Lambda\|}
            \lrag{\hat\nu_1-\hat\nu_n-\delta, \delta}
        \stra{d} c_2 \lambda_0^{\top} (\Omega^{(3)})^{1/2} b(\varepsilon,\eta) - c_3.
        \end{equation*}
            
        Consequently, we have that
        \BEqn
        & & \frac{N^{1/2}}{\|\Lambda^{\top}\Lambda\|} T_n(\lrfl{nu}-\lrfl{n\varepsilon}) \\
        &\leadsto& \lrp{b(\varepsilon,\eta)^{\top} ((\Omega^{(3)})^{1/2})^{\top} L_0 - c_2 \lambda_0^{\top}} (\Omega^{(3)})^{1/2} \lrp{B_s(u)-B_s(\varepsilon)-\frac{u-\varepsilon}{1-2\varepsilon}(B_s(1-\varepsilon)-B_s(\varepsilon))} \\
        & & + \frac{c_1}{\varepsilon-\eta} \lrp{\tilde{B}(u) - \tilde{B}(\varepsilon) - \frac{u-\varepsilon}{1-2\varepsilon}(\tilde{B}(1-\varepsilon)-\tilde{B}(\varepsilon))} \\
        & & - (1-2\varepsilon)\lrp{r_0 \wedge \lrp{\frac{u-\varepsilon}{1-2\varepsilon}}} \lrp{(1-r_0)\wedge\lrp{\frac{1-\varepsilon-u}{1-2\varepsilon}}} \lrp{c_2 \lambda_0^{\top} (\Omega^{(3)})^{1/2} b(\varepsilon,\eta) - c_3} \\
        &=:& \tilde{T}(u,\Delta,b(\varepsilon,\eta),\cb_s,\tilde{\cb}),
        \EEqn
        and
        \BEqn
        & & \frac{N}{\|\Lambda^{\top}\Lambda\|^2} V_n(k) \\
        &\leadsto& \frac{1}{1-2\varepsilon}
            \int_{\varepsilon}^{u} 
            \left( 
              \lrp{b(\varepsilon,\eta)^{\top} ((\Omega^{(3)})^{1/2})^{\top} L_0 - c_2 \lambda_0^{\top}} (\Omega^{(3)})^{1/2}
            \right. \\  
        & & \hspace{8em}   
            \times \lrp{B_s(v)-B_s(\varepsilon)-\frac{v-\varepsilon}{u-\varepsilon}(B_s(u)-B_s(\varepsilon))} \\
        & & \hspace{6em} 
              + \frac{c_1}{\varepsilon-\eta} \lrp{\tilde{B}(v) - \tilde{B}(\varepsilon) - \frac{v-\varepsilon}{u-\varepsilon}(\tilde{B}(u)-\tilde{B}(\varepsilon))} \\
        & & \hspace{6em}       
            - (1-2\varepsilon) \frac{\lrp{ \lrp{r_0 \wedge \lrp{\frac{v-\varepsilon}{1-2\varepsilon}}} \lrp{ \lrp{\frac{u-\varepsilon}{1-2\varepsilon}-r_0} \wedge \lrp{\frac{u-v}{1-2\varepsilon}}}} \vee 0}{\frac{u-\varepsilon}{1-2\varepsilon}} \\ 
        & & \hspace{8em}    
            \left.
            \times \lrp{c_2 \lambda_0^{\top} (\Omega^{(3)})^{1/2} b(\varepsilon,\eta) - c_3}  
            \right)^2 ds \\
        & & + \frac{1}{1-2\varepsilon}
            \int_{u}^{1-\varepsilon} 
            \left( 
              \lrp{b(\varepsilon,\eta)^{\top} ((\Omega^{(3)})^{1/2})^{\top} L_0 - c_2 \lambda_0^{\top}} (\Omega^{(3)})^{1/2} 
            \right. \\  
        & & \hspace{9em}   
            \times
            \lrp{B_s(1-\varepsilon)-B_s(v)-\frac{1-\varepsilon-v}{1-\varepsilon-u}(B_s(1-\varepsilon)-B_s(u))} \\
        & & \hspace{8em} 
              + \frac{c_1}{\varepsilon-\eta} \lrp{\tilde{B}(1-\varepsilon) - \tilde{B}(v) - \frac{1-\varepsilon-v}{1-\varepsilon-u}(\tilde{B}(1-\varepsilon)-\tilde{B}(u))} \\
        & & \hspace{8em}     
            + (1-2\varepsilon) \frac{\lrp{\lrp{\lrp{r_0-\frac{u-\varepsilon}{1-2\varepsilon}}\wedge\lrp{\frac{v-u}{1-2\varepsilon}}} \lrp{\lrp{1-r_0}\wedge\lrp{1-\frac{v-\varepsilon}{1-2\varepsilon}}}} \vee 0}{1-\frac{u-\varepsilon}{1-2\varepsilon}} \\
        & & \hspace{9em}
            \left.    
            \times
            \lrp{c_2 \lambda_0^{\top} (\Omega^{(3)})^{1/2} b(\varepsilon,\eta) - c_3}  
            \right)^2 ds \\
        &=:& \tilde{V}(u,\Delta,b(\varepsilon,\eta),\cb_s,\tilde{\cb}).
        \EEqn
        
        Then by applying CMT, we have that 
        \begin{equation*}
            G_n 
        \stra{d} \tilde{M}(\Delta,b(\varepsilon,\eta),\cb_s,\tilde{\cb})
        := \sup\limits_{u\in[\varepsilon,1-\varepsilon]} \tilde{T}(u,b(\varepsilon,\eta),\Delta,\cb_s,\tilde{\cb}) \tilde{V}^{-1/2}(u,\Delta,b(\varepsilon,\eta),\cb_s,\tilde{\cb}).
        \end{equation*}
        
        By using the argument used to prove the last case in Theorem \ref{Thm:NullDist_Gn_factor}, if conditioning on $b(\varepsilon,\eta)=b_0$, with 
        \begin{equation*}
            \omega_1 
          = \sqrt{\lrp{b_0^{\top} ((\Omega^{(3)})^{1/2})^{\top} L_0 - c_2 \lambda_0^{\top}} \Omega \lrp{b_0^{\top} ((\Omega^{(3)})^{1/2})^{\top} L_0 - c_2 \lambda_0^{\top}}^{\top}},
        \end{equation*}
        and $\omega_2 = \frac{c_1}{\varepsilon-\eta}$, $\omega_3 = c_2 \lambda_0^{\top} (\Omega^{(3)})^{1/2} b(\varepsilon,\eta) - c_3$, we have that
        {\small
        \BEqn
        & & \tilde{M}(\Delta,b(\varepsilon,\eta),\cb_s,\tilde{\cb})|_{b(\varepsilon,\eta)=b_0} \\
        &=^d& \sup\limits_{u\in[\varepsilon,1-\varepsilon]}
        \frac{\begin{array}{l}
              ~~ \sqrt{ \omega_1^2+\omega_2^2} \lrp{B(u)-B(\varepsilon)-\frac{u-\varepsilon}{1-2\varepsilon}(B(1-\varepsilon)-B(\varepsilon))}  \\
              - \omega_3(1-2\varepsilon)\lrp{r_0 \wedge \lrp{\frac{u-\varepsilon}{1-2\varepsilon}}} \lrp{(1-r_0)\wedge\lrp{\frac{1-\varepsilon-u}{1-2\varepsilon}}}
              \end{array}}
        {\lrp{
          \begin{array}{l}
          ~~\frac{1}{1-2\varepsilon} 
            \int_{\varepsilon}^{u} 
            \left( 
              \sqrt{\omega_1^2+\omega_2^2} \lrp{B_s(v)-B_s(\varepsilon)-\frac{v-\varepsilon}{u-\varepsilon}(B_s(u)-B_s(\varepsilon))} 
            \right.   \\
          \hspace{6em} 
          \left.
            - \omega_3 (1-2\varepsilon) \frac{\lrp{ \lrp{r_0 \wedge \lrp{\frac{v-\varepsilon}{1-2\varepsilon}}} \lrp{ \lrp{\frac{u-\varepsilon}{1-2\varepsilon}-r_0} \wedge \lrp{\frac{u-v}{1-2\varepsilon}}}} \vee 0}{\frac{u-\varepsilon}{1-2\varepsilon}} 
          \right)^2 \\
          + \frac{1}{1-2\varepsilon}
            \int_{u}^{1-\varepsilon} 
            \left(   
              \sqrt{\omega_1^2+\omega_2^2}
              \lrp{B_s(1-\varepsilon)-B_s(v)-\frac{1-\varepsilon-v}{1-\varepsilon-u}(B_s(1-\varepsilon)-B_s(u))}
            \right) \\
          \hspace{6em} 
          \left.
            + \omega_3 (1-2\varepsilon) \frac{\lrp{\lrp{\lrp{r_0-\frac{u-\varepsilon}{1-2\varepsilon}}\wedge\lrp{\frac{v-u}{1-2\varepsilon}}} \lrp{\lrp{1-r_0}\wedge\lrp{1-\frac{v-\varepsilon}{1-2\varepsilon}}}} \vee 0}{1-\frac{u-\varepsilon}{1-2\varepsilon}} 
          \right)^2
          \end{array}
        }^{1/2}} \\
        &=^d& \sup\limits_{r\in[0,1]} 
        \frac{B(r)-rB(1) - \omega_3\sqrt{\frac{1-2\varepsilon}{\omega_1^2+\omega_2^2}} \lrp{r_0\wedge r} \lrp{(1-r_0)\wedge(1-r)}}
        {\lrp{
        \begin{array}{l}
          ~~\int_{0}^{r} \lrp{B(s)-\frac{s}{r}B(r) -  \omega_3\sqrt{\frac{1-2\varepsilon}{\omega_1^2+\omega_2^2}} \frac{(s \wedge r_0) \lrp{(r-s)\wedge(r-r_0)}\vee0}{r}}^2 ds  \\
          + \int_{r}^{1} \lrp{B(1-s) - \frac{1-s}{1-r}B(1-r) + \omega_3\sqrt{\frac{1-2\varepsilon}{\omega_1^2+\omega_2^2}} \frac{\lrp{(1-s)\wedge(1-r_0)} \lrp{(s-r)\wedge(r_0-r)}\vee0}{1-r} }^2 ds 
        \end{array}}^{1/2}} \\
        &=^d& \sup\limits_{r\in[0,1]} T(r,\Delta,b_0) V^{-1/2}(r,\Delta,b_0) \\
        &=:& M(\Delta,b_0)
        \EEqn}
        where $T(r,\Delta,b_0)$ and $V(r,\Delta,b_0)$ are defined as Theorem \ref{Thm:DensePower_factor}\ref{Thm:DensePower_factor-3-2}. 
        
        Finally, noting that $b(\varepsilon,\eta) = \frac{1}{\varepsilon-\eta} \lrp{B_s(\varepsilon-\eta)-B_s(1)+B_s(1-\varepsilon+\eta)} =^d \cn_{s}\lrp{0, \frac{2}{\varepsilon-\eta}}$, the desired result follows the similar steps used in Theorem \ref{Thm:DensePower_Fixedp}\ref{Thm:DensePower_Fixedp-3}.
    \end{enumerate}
\end{enumerate}

\subsection{Proof of Theorem \ref{Thm:NullDist_multiple}}

Recall that
\begin{equation*}
    W_n(r)
  = m_1 \sum\limits_{j=1}^{\lrfl{nr}-\lrfl{n\varepsilon}} Y_j
  = \sum\limits_{i=1}^{m_1} \lrp{X_i-X_{n+1-i}}^{\top} \lrp{\sum\limits_{j=1}^{\lrfl{nr}-\lrfl{n\varepsilon}} X_{j+m}},
  \quad r\in[\varepsilon,1-\varepsilon],
\end{equation*}
then with $j_1 = \lrfl{nr_1} - \lrfl{n\varepsilon}$, $j_2 = \lrfl{nr_2} - \lrfl{n\varepsilon}$, $j_3 = \lrfl{nr_3} - \lrfl{n\varepsilon}$ for some $\varepsilon \le r_1 \le r_2 \le r_3 \le 1-\varepsilon$, we can express $T_n^{f}(j_1,j_2,j_3), T_n^{b}(j_1,j_2,j_3), R_n^{f}(j_1,j_2,j_3)$ and $R_n^{b}(j_1,j_2,j_3)$ in terms of $\{W_n(r)\}_{\varepsilon\le r\le 1-\varepsilon}$. 

Assume that $\frac{1}{N_n}W_n(r) \leadsto W(r)$ in $D[\varepsilon,1-\varepsilon]$ for some normalizer $N_n$ and limiting process $\{W(r)\}_{\varepsilon\le r\le 1-\varepsilon}$, then it directly follows from the CMT that
\BEqn
& & \max\limits_{(k_1,k_2)\in\Xi_n(\varepsilon)} \lrabs{\frac{T_n^{f}(1,k_1,k_2)}{\lrp{V_n^{f}(1,k_1,k_2)}^{1/2}}} \\
&=& \sup\limits_{(u_1,u_2)\in\Xi(\varepsilon)} \lrabs{\frac{T_n^{f}(1, \lrfl{nu_1}-\lrfl{n\varepsilon}, \lrfl{nu_2}-\lrfl{n\varepsilon})}{\lrp{V_n^{f}(1, \lrfl{nu_1}-\lrfl{n\varepsilon}, \lrfl{nu_2}-\lrfl{n\varepsilon})}^{1/2}}} \\
&\stra{d}& \sup^{\ast}\limits_{(u_1,u_2)\in\Xi(\varepsilon)}
\frac{\sqrt{u_2-\varepsilon} \lrp{W(u_1)-W(\varepsilon) - \frac{u_1-\varepsilon}{u_2-\varepsilon} \lrp{W(u_2)-W(\varepsilon)}}}
{\lrp{
\begin{array}{l}
  ~~ \int_{\varepsilon}^{u_1} \lrp{W(t)-W(\varepsilon) - \frac{t-\varepsilon}{u_1-\varepsilon}(W(u_1)-W(\varepsilon))}^2 dt \\
   + \int_{u_1}^{u_2} \lrp{W(u_2)-W(t) - \frac{u_2-t}{u_2-u_1}(W(u_2)-W(u_1))}^2 dt
\end{array}
}^{1/2}} \\
&=:& M^{(f)}(\{W(r)\}_{\varepsilon\le r\le 1-\varepsilon}),
\EEqn
where $\Xi_n(\varepsilon) = \{(\ell_1,\ell_2): 1<\ell_1<\ell_2<N=\lrfl{(1-2\varepsilon)n}, \ell_2-\ell_1>1\}$ and $\Xi(\varepsilon) = \{(r_1,r_2): \varepsilon \le r_1 \le r_2 \le 1-\varepsilon\}$ as defined in the appendix and we use $\sup^{\ast}\limits_{x\in \ca} g(x) = \sup\limits_{x\in\ca} |g(x)|$ to denote the supremum of the absolute values. Similarly,
\BEqn
& & \max\limits_{(\ell_1,\ell_2)\in\Xi_n(\varepsilon)} \lrabs{\frac{T_n^{b}(\ell_1,\ell_2,N)}{\lrp{V_n^{b}(\ell_1,\ell_2,N)}^{1/2}}} \\
&\stra{d}& \sup^{\ast}\limits_{(v_1,v_2)\in\Xi(\varepsilon)}
\frac{\sqrt{1-\varepsilon-v_1} \lrp{W(1-\varepsilon)-W(v_2) - \frac{1-\varepsilon-v_2}{1-\varepsilon-v_1} \lrp{W(1-\varepsilon)-W(v_1)}}}
{\lrp{
\begin{array}{l}
  ~~ \int_{v_1}^{v_2} \lrp{W(t)-W(v_1) - \frac{t-v_1}{v_2-v_1}(W(v_2)-W(v_1))}^2 dt \\
   + \int_{v_2}^{1-\varepsilon} \lrp{W(1-\varepsilon)-W(t) - \frac{1-\varepsilon-t}{1-\varepsilon-v_2}(W(1-\varepsilon)-W(v_2))}^2 dt
\end{array}
}^{1/2}} \\
&=:& M^{(b)}(\{W(r)\}_{\varepsilon\le r\le 1-\varepsilon}),
\EEqn
and consequently, $G_n^{M} \stra{d} M^{(f)}(\{W(r)\}_{\varepsilon\le r\le 1-\varepsilon}) + M^{(b)}(\{W(r)\}_{\varepsilon\le r\le 1-\varepsilon})$.

For the three data generating processes discussed in this article, we have derived the respective normalizer $N_n$ and the limiting null process $\{W(r)\}_{\varepsilon\le r\le 1-\varepsilon}$, then it remains to analyze $M^{(f)}(\{W(r)\}_{\varepsilon\le r\le 1-\varepsilon})$ and $M^{(b)}(\{W(r)\}_{\varepsilon\le r\le 1-\varepsilon})$ case by case.

\begin{enumerate}[label=(\roman*)]
    \item 
    We first consider the case that the observed data is a stationary time series with fixed $p$. Under the conditions in Theorem \ref{Thm:NullDist_Gn_Fixedp}, we have that with $N_n = n$, it holds under the null that
    \begin{equation*}
        \frac{1}{N_n} W_n(r) 
    \leadsto b^{\top}(\varepsilon,\eta) \Omega^{(1)} \lrp{B_p(r)-B_p(\varepsilon)}
    \quad \mbox{in } D[\varepsilon,1-\varepsilon],
    \end{equation*}
    where $\{B_p(r)\}_{0\le r\le 1}$ is a standard Brownian motion in $\br^p$ and $b(\varepsilon,\eta) = B_p(\varepsilon-\eta)-B_p(1)+B_p(1-\varepsilon+\eta)$. Therefore, in this case, we have that $W(r) = b^{\top}(\varepsilon,\eta) \Omega^{(1)} \lrp{B_p(r)-B_p(\varepsilon)}$ and it follows that
    \BEqn
    & & M^{(f)}(\{W(r)\}_{\varepsilon\le r\le 1-\varepsilon}) \\
    &=& \sup^{\ast}\limits_{(u_1,u_2)\in\Xi(\varepsilon)}
    \frac{\sqrt{u_2-\varepsilon} b^{\top}(\varepsilon,\eta) \Omega^{(1)} \lrp{B_p(u_1)-B_p(\varepsilon) - \frac{u_1-\varepsilon}{u_2-\varepsilon} \lrp{B_p(u_2)-B_p(\varepsilon)}}}
    {\lrp{
    \begin{array}{l}
      ~~ \int_{\varepsilon}^{u_1} \lrp{ b^{\top}(\varepsilon,\eta) \Omega^{(1)} \lrp{B_p(t)-B_p(\varepsilon) -
      \frac{t-\varepsilon}{u_1-\varepsilon}(B_p(u_1)-B_p(\varepsilon))}}^2 dt \\
      + \int_{u_1}^{u_2} \lrp{ b^{\top}(\varepsilon,\eta) \Omega^{(1)} \lrp{B_p(u_2)-B_p(t) - \frac{u_2-t}{u_2-u_1}(B_p(u_2)-B_p(u_1))}}^2 dt
    \end{array}
    }^{1/2}}.
    \EEqn
    
    Define $\cb_p = \{B_p(r)\}_{\varepsilon\le r\le 1-\varepsilon}$, it follows from the property of Brownian motion that $b(\varepsilon,\eta)$ is independent of $\cb_p$. Hence by using the same conditional arguments as in Theorem \ref{Thm:NullDist_Gn_Fixedp}, we have that if conditioning on $b(\varepsilon,\eta)=b_0$, the process $\{b_0^{\top}\Omega^{(1)} B_p(r)\}_{\varepsilon\le r\le 1-\varepsilon}$ equals in distribution to the process $\{\omega B(r)\}_{\varepsilon\le r\le 1-\varepsilon}$ with $\omega = \sqrt{b_0^{\top}(\Omega^{(1)})^2 b_0}$, where $\{B(r)\}_{0\le r\le 1}$ is a standard Brownian motion in $\br$. 
    
    Consequently, if conditioning on $b(\varepsilon,\eta)=b_0$, we have that
    \BEqn
    & & M^{(f)}(\{W(r)\}_{\varepsilon\le r\le 1-\varepsilon})|_{b(\varepsilon,\eta)=b_0} \\
    &=^d& \sup^{\ast}\limits_{(u_1,u_2)\in\Xi(\varepsilon)}
    \frac{\sqrt{u_2-\varepsilon} \lrp{B(u_1)-B(\varepsilon) - \frac{u_1-\varepsilon}{u_2-\varepsilon} \lrp{B(u_2)-B(\varepsilon)}}}
    {\lrp{
    \begin{array}{l}
      ~~ \int_{\varepsilon}^{u_1} \lrp{ B(t)-B(\varepsilon) -
      \frac{t-\varepsilon}{u_1-\varepsilon}(B(u_1)-B(\varepsilon))}^2 dt \\
      + \int_{u_1}^{u_2} \lrp{ B(u_2)-B(t) - \frac{u_2-t}{u_2-u_1}(B(u_2)-B(u_1))}^2 dt
    \end{array}
    }^{1/2}} \\
    &=^d& \sup^{\ast}\limits_{(r_1,r_2)\in\Xi}
    \frac{\sqrt{r_2} \lrp{B((1-2\varepsilon)r_1+\varepsilon)-B(\varepsilon) - \frac{r_1}{r_2} \lrp{B((1-2\varepsilon)r_2+\varepsilon)-B(\varepsilon)}}}
    {\lrp{
    \begin{array}{l}
      ~~ \int_{0}^{r_1} \lrp{ B((1-2\varepsilon)s+\varepsilon)-B(\varepsilon) -
      \frac{s}{r_1}(B((1-2\varepsilon)r_1+\varepsilon)-B(\varepsilon))}^2 ds \\
      + \int_{r_1}^{r_2} \Big{(} B((1-2\varepsilon)r_2+\varepsilon)-B((1-2\varepsilon)s+\varepsilon) \\
      \hspace{4em} - \frac{r_2-s}{r_2-r_1}(B((1-2\varepsilon)r_2+\varepsilon)-B((1-2\varepsilon)r_1+\varepsilon))\Big{)}^2 ds
    \end{array}
    }^{1/2}} \\
    &=^d& \sup^{\ast}\limits_{(r_1,r_2)\in\Xi}
    \frac{\sqrt{r_2} \lrp{B(r_1) - \frac{r_1}{r_2} B(r_2)}}
    {\lrp{
    \begin{array}{l}
      ~~ \int_{0}^{r_1} \lrp{ B(s) -
      \frac{s}{r_1}(B(r_1) }^2 ds \\
      + \int_{r_1}^{r_2} \lrp{ B(r_2) - B(s) - \frac{r_2-s}{r_2-r_1}(B(r_2)-B(r_1)) }^2 ds
    \end{array}
    }^{1/2}} \\
    &=^d& \sup^{\ast}\limits_{(r_1,r_2)\in\Xi} \sqrt{r_2} T^{f}(r_1,r_2) \lrp{V^{f}(r_1,r_2)}^{-1/2} \\
    &=^d& \sup\limits_{(r_1,r_2)\in\Xi} \lrabs{\sqrt{r_2} T^{f}(r_1,r_2) \lrp{V^{f}(r_1,r_2)}^{-1/2}},
    \EEqn
    where in the second step we change the variables by $r_1 = (u_1-\varepsilon)/(1-2\varepsilon)$, $r_2 = (u_2-\varepsilon)/(1-2\varepsilon)$, and $s = (t-\varepsilon)/(1-2\varepsilon)$, and the last step follows from the fact that $\{B((1-2\varepsilon)s+\varepsilon)\}_{\varepsilon\le s\le 1-\varepsilon}$ is equal in distribution with $\{B(r)\}_{0\le r\le 1}$.
    
    Note that the distribution of $M^{(f)}(\{W(r)\}_{\varepsilon\le r\le 1-\varepsilon})|_{b(\varepsilon,\eta)=b_0}$ is independent of $b_0$, then we may conclude that $M^{(f)}(\{W(r)\}_{\varepsilon\le r\le 1-\varepsilon}) =^d \sup\limits_{(r_1,r_2)\in\Xi} \lrabs{\sqrt{r_2} T^{f}(r_1,r_2) \lrp{V^{f}(r_1,r_2)}^{-1/2}}$. Using the similar arguments, we also show that 
    \begin{equation*}
        M^{(b)}(\{W(r)\}_{\varepsilon\le r\le 1-\varepsilon}) 
    =^d \sup\limits_{(r_1,r_2)\in\Xi} \lrabs{\sqrt{1-r_1} T^{b}(r_1,r_2) \lrp{V^{b}(r_1,r_2)}^{-1/2}},
    \end{equation*}
    which completes the proof.

    \item 
    As for the linear process, it is shown in Proposition \ref{Prop:NullDist_Wn_lp} that under the conditions of Theorem \ref{Thm:NullDist_Gn_Growingp_lp}, we have
    \begin{equation*}
        \frac{1}{N_n} W_n(r)
    \leadsto W(r) 
    := B(r) - B(\varepsilon)
    \quad \mbox{in } D[\varepsilon,1-\varepsilon],
    \end{equation*}
    where the normalizer is $N_n = \sqrt{2nm_1} \|A^{(0)} \Gamma^{(2)} (A^{(0)})^{\top}\|_F$. Thus in this case, we have that
    \BEqn
    & & M^{(f)}(\{W(r)\}_{\varepsilon\le r\le 1-\varepsilon}) \\
    &=^d& \sup^{\ast}\limits_{(u_1,u_2)\in\Xi(\varepsilon)}
    \frac{\sqrt{u_2-\varepsilon} \lrp{B(u_1)-B(\varepsilon) - \frac{u_1-\varepsilon}{u_2-\varepsilon} \lrp{B(u_2)-B(\varepsilon)}}}
    {\lrp{
      \begin{array}{l}
        ~~ \int_{\varepsilon}^{u_1} \lrp{B(t)-B(\varepsilon) -\frac{t-\varepsilon}{u_1-\varepsilon}(B(u_1)-B(\varepsilon))}^2 dt \\
        + \int_{u_1}^{u_2} \lrp{B(u_2)-B(t) - \frac{u_2-t}{u_2-u_1}(B(u_2)-B(u_1))}^2 dt
      \end{array}
    }^{1/2}} \\
    &=^d& \sup\limits_{(r_1,r_2)\in\Xi} \lrabs{\sqrt{r_2} T^{f}(r_1,r_2) \lrp{V^{f}(r_1,r_2)}^{-1/2}},
    \EEqn
    where the last step is obtained following the exactly same steps used for $M^{(f)}(\{W(r)\}_{\varepsilon\le r\le 1-\varepsilon})|_{b(\varepsilon,\eta)=b_0}$ in the previous case.
    
    Similarly, we can also show that $M^{(b)}(\{W(r)\}_{\varepsilon\le r\le 1-\varepsilon}) =^d \sup\limits_{(r_1,r_2)\in\Xi} \lrabs{\sqrt{1-r_1} T^{b}(r_1,r_2) \lrp{V^{b}(r_1,r_2)}^{-1/2}}$, which leads to the desired result.

    \item 
    It remains to consider the case when $\{X_t\}_{t=1}^{n}$ admits a factor model defined as Definition \ref{Def:Model_factor}. Under the conditions of Theorem \ref{Thm:NullDist_Gn_factor}, we have shown in the proof of Theorem \ref{Thm:NullDist_Gn_factor} that the normalizer $N_n$ and limiting process $\{W(r)\}_{\varepsilon\le r\le 1-\varepsilon}$ are determined jointly by $\Lambda$ and $\Gamma^{(3)}$. 
    
    If $\|\Lambda\| = o\lrp{\|\Gamma^{(3)}\|_F^{1/2}}$, with $N_n = \sqrt{2nm_1} \|A^{(0)} \Gamma^{(3)} (A^{(0)})^{\top}\|_F$, it holds that $\frac{1}{N_n}W_n(r) \leadsto W(r) = B(r)-B(\varepsilon)$ in $D[\varepsilon,1-\varepsilon]$. If $\|\Gamma^{(3)}\|_F^{1/2} = o\lrp{\|\Lambda\|}$, then 
    \begin{equation*}
        \frac{1}{N_n}W_n(r) 
    \leadsto W(r)
    = b^{\top}(\varepsilon,\eta) (\Omega^{(3)})^{1/2})^{\top} L_0 \Omega^{(3)})^{1/2} (B_s(r) - B_s(\varepsilon))
    \quad \mbox{in } D[\varepsilon,1-\varepsilon]
    \end{equation*}
    with $N_n = n \|\Lambda^{\top} \Lambda\|$, where $L_0$ is defined as Assumption \ref{Assumpt:DensePower_factor}, $\{B_s(r)\}_{0\le r\le 1}$ is a standard Brownian motion in $\br^s$ and $b(\varepsilon,\eta) = B_s(\varepsilon-\eta)-B_s(1)+B_s(1-\varepsilon+\eta)$. If $\|\Gamma^{(3)}\|_F^{1/2} = O_s(\|\Lambda\|)$, then there exists a deterministic constant $c$, s.t. it holds in $D[\varepsilon,1-\varepsilon]$ that
    \begin{equation*}
        \frac{1}{N_n}W_n(r) 
    \leadsto W(r) 
    = c b^{\top}(\varepsilon,\eta) ((\Omega^{(3)})^{1/2})^{\top} L_0 (\Omega^{(3)})^{1/2} (B_s(r) - B_s(\varepsilon)) + \tilde{B}(r) - \tilde{B}(\varepsilon),
    \end{equation*}
    where $\{B_s(r)\}_{0\le r\le 1}$ and $\{\tilde{B}(r)\}_{0\le r\le 1}$ are two independent Brownian motions in $\br^s$ and $\br$ respectively, and $b(\varepsilon,\eta) = B_s(\varepsilon-\eta) - B_s(1) + B_s(1-\varepsilon+\eta)$.
    
    In summary, there always exists two deterministic constants $c_1,c_2$, such that
    \begin{equation*}
        \frac{1}{N_n} W_n(r)
    \leadsto W(r)
    = c_1 b^{\top}(\varepsilon,\eta) ((\Omega^{(3)})^{1/2})^{\top} L_0 (\Omega^{(3)})^{1/2} (B_s(r) - B_s(\varepsilon)) 
    + c_2 \lrp{\tilde{B}(r) - \tilde{B}(\varepsilon)}
    \end{equation*}
    in $D[\varepsilon,1-\varepsilon]$, where $N_n$ is a deterministic normalizer depending only on $n,\Lambda,\Gamma^{(3)}$, and $\{B_s(r)\}_{0\le r\le 1}$, $\{\tilde{B}(r)\}_{0\le r\le 1}$ are two independent Brownian motions in $\br^s$ and $\br$ respectively and $b(\varepsilon,\eta) = B_s(\varepsilon-\eta) - B_s(1) + B_s(1-\varepsilon+\eta)$.
    
    Therefore, we have that
    \BEqn
    & & M^{(f)}(\{W(r)\}_{\varepsilon\le r\le 1-\varepsilon}) \\
    &=^d& \sup^{\ast}\limits_{(u_1,u_2)\in\Xi(\varepsilon)}
    \frac{\sqrt{u_2-\varepsilon}
          \lrp{\begin{array}{l}
            ~~ c_1 b^{\top}(\varepsilon,\eta) ((\Omega^{(3)})^{1/2})^{\top} L_0 (\Omega^{(3)})^{1/2} \\
            \hspace{2em} \times \lrp{B_s(u_1)-B_s(\varepsilon) - \frac{u_1-\varepsilon}{u_2-\varepsilon} \lrp{B_s(u_2)-B_s(\varepsilon)}}  \\
             + c_2 \lrp{\tilde{B}(u_1)-\tilde{B}(\varepsilon) - \frac{u_1-\varepsilon}{u_2-\varepsilon} \lrp{\tilde{B}(u_2)-\tilde{B}(\varepsilon)}}
          \end{array}}}
    {\lrp{
      \begin{array}{l}
       ~~ \int_{\varepsilon}^{u_1} \Big{(} c_1 b^{\top}(\varepsilon,\eta) ((\Omega^{(3)})^{1/2})^{\top} L_0 (\Omega^{(3)})^{1/2} \\
       \hspace{5em} \times \lrp{B_s(t)-B_s(\varepsilon) -\frac{t-\varepsilon}{u_1-\varepsilon}(B_s(u_1)-B_s(\varepsilon))} \\
       \hspace{4em} + c_2 \lrp{\tilde{B}(t)-\tilde{B}(\varepsilon) -\frac{t-\varepsilon}{u_1-\varepsilon}(\tilde{B}(u_1)-\tilde{B}(\varepsilon))} \Big{)}^2 dt \\
        + \int_{u_1}^{u_2} \Big{(} c_1 b^{\top}(\varepsilon,\eta) ((\Omega^{(3)})^{1/2})^{\top} L_0 (\Omega^{(3)})^{1/2} \\
       \hspace{5em} \times \lrp{B_s(u_2)-B_s(t) - \frac{u_2-t}{u_2-u_1}(B_s(u_2)-B_s(u_1))} \\
        \hspace{4em} + c_2 \lrp{\tilde{B}(u_2)-\tilde{B}(t) - \frac{u_2-t}{u_2-u_1}(\tilde{B}(u_2)-\tilde{B}(u_1))} \Big{)}^2 dt \\
      \end{array}
    }^{1/2}}.
    \EEqn
    
    Note that $b(\varepsilon,\eta)$, $\cb_s=\{B_s(r)\}_{\varepsilon\le r\le 1-\varepsilon}$ and $\tilde{\cb} = \{\tilde{B}(r)\}_{\varepsilon\le r\le 1-\varepsilon}$ are mutually independent, if conditioning on $b(\varepsilon,\eta) = b_0$, the process $\{b^{\top}(\varepsilon,\eta) ((\Omega^{(3)})^{1/2})^{\top} L_0 (\Omega^{(3)})^{1/2} B_s(r)\}_{\varepsilon\le r\le 1-\varepsilon}$ is equal in distribution with $\{\omega B(r)\}_{\varepsilon\le r\le 1-\varepsilon}$, where 
    \begin{equation*}
        \omega 
      = \sqrt{b^{\top}(\varepsilon,\eta) ((\Omega^{(3)})^{1/2})^{\top} L_0 \Omega^{(3)} L_0^{\top} ((\Omega^{(3)})^{1/2}) b(\varepsilon,\eta)},
    \end{equation*}
    and $\{B(r)\}_{0\le r\le 1}$ is another standard Brownian motion in $\br$ that is independent of $\{\tilde{B}(r)\}_{0\le r\le 1}$. Furthermore, we have that $\{c_1 \omega B(r) + c_2 \tilde{B}(r)\}_{\varepsilon\le r\le 1-\varepsilon} =^d \{c B(r)\}_{\varepsilon\le r\le 1-\varepsilon}$, where $c = \sqrt{c_1^2 \omega^2 + c_2^2}$. 
    
    Using these observations, if conditioning on $b(\varepsilon,\eta) = b_0$, we have that
    \BEqn
    & &  M^{(f)}\lrp{\{W(r)\}_{\varepsilon\le r\le 1-\varepsilon}} |_{b(\varepsilon,\eta)=b_0} \\
    &=^d& \sup^{\ast}\limits_{(u_1,u_2)\in\Xi(\varepsilon)}
    \frac{\sqrt{u_2-\varepsilon}
          \lrp{\begin{array}{l}
            ~~ c_1 \omega \lrp{B(u_1)-B(\varepsilon) - \frac{u_1-\varepsilon}{u_2-\varepsilon} \lrp{B(u_2)-B(\varepsilon)}}  \\
             + c_2 \lrp{\tilde{B}(u_1)-\tilde{B}(\varepsilon) - \frac{u_1-\varepsilon}{u_2-\varepsilon} \lrp{\tilde{B}(u_2)-\tilde{B}(\varepsilon)}}
          \end{array}}}
    {\lrp{
      \begin{array}{l}
       ~~ \int_{\varepsilon}^{u_1} \Big{(} c_1 \omega \lrp{B(t)-B(\varepsilon) -\frac{t-\varepsilon}{u_1-\varepsilon}(B(u_1)-B(\varepsilon))} \\
       \hspace{4em} + c_2 \lrp{\tilde{B}(t)-\tilde{B}(\varepsilon) -\frac{t-\varepsilon}{u_1-\varepsilon}(\tilde{B}(u_1)-\tilde{B}(\varepsilon))} \Big{)}^2 dt \\
        + \int_{u_1}^{u_2} \Big{(} c_1 \omega \lrp{B(u_2)-B(t) - \frac{u_2-t}{u_2-u_1}(B(u_2)-B(u_1))} \\
        \hspace{4em} + c_2 \lrp{\tilde{B}(u_2)-\tilde{B}(t) - \frac{u_2-t}{u_2-u_1}(\tilde{B}(u_2)-\tilde{B}(u_1))} \Big{)}^2 dt \\
      \end{array}
    }^{1/2}} \\
    &=^d& \sup^{\ast}\limits_{(u_1,u_2)\in\Xi(\varepsilon)}
    \frac{\sqrt{u_2-\varepsilon} \lrp{B(u_1)-B(\varepsilon) - \frac{u_1-\varepsilon}{u_2-\varepsilon} \lrp{B(u_2)-B(\varepsilon)}}}
    {\lrp{
      \begin{array}{l}
       ~~ \int_{\varepsilon}^{u_1} \lrp{B(t)-B(\varepsilon) - \frac{t-\varepsilon}{u_1-\varepsilon} (B(u_1)-B(\varepsilon))}^2 dt \\
        + \int_{u_1}^{u_2} \lrp{B(u_2)-B(t) - \frac{u_2-t}{u_2-u_1} (B(u_2)-B(u_1))}^2 dt
      \end{array}
    }^{1/2}} \\
    &=^d& \sup^{\ast}\limits_{(r_1,r_2)\in\Xi} \sqrt{r_2} T^{f}(r_1,r_2) \lrp{V^{f}(r_1,r_2)}^{-1/2},
    \EEqn
    where the last step uses the same techniques as previous cases. This implies that the distribution of $M^{(f)}\lrp{\{W(r)\}_{\varepsilon\le r\le 1-\varepsilon}} |_{b(\varepsilon,\eta)=b_0}$ is independent of $b_0$, which further implies that
    \begin{equation*}
        M^{(f)}\lrp{\{W(r)\}_{\varepsilon\le r\le 1-\varepsilon}}
    =^d \sup\limits_{(r_1,r_2)\in\Xi} \lrabs{\sqrt{r_2} T^{f}(r_1,r_2) \lrp{V^{f}(r_1,r_2)}^{-1/2}},
    \end{equation*}
    and similarly, 
    \begin{equation*}
        M^{(b)}\lrp{\{W(r)\}_{\varepsilon\le r\le 1-\varepsilon}}
    =^d \sup\limits_{(r_1,r_2)\in\Xi} \lrabs{\sqrt{1-r_1} T^{b}(r_1,r_2) \lrp{V^{b}(r_1,r_2)}^{-1/2}},
    \end{equation*}    
    both of which jointly lead to the desired result and complete the proof.
\end{enumerate}

\subsection{Proof of Theorem \ref{Thm:MultiplePower}}

The proofs of Theorem \ref{Thm:MultiplePower}\ref{Thm:MultiplePower-1}-\ref{Thm:MultiplePower-3} are similar, and we only provide the details of Theorem \ref{Thm:MultiplePower}\ref{Thm:MultiplePower-1}.

If $\{\tilde{X}_t\}_{t=1}^{n} \in \br^{p}$ is a stationary sequence defined as Definition \ref{Def:Model_Fixedp}. Under Assumption \ref{Assumpt:MultiplePower}, we have either Assumption \ref{Assumpt:MultiplePower-1} or Assumption \ref{Assumpt:MultiplePower-2}, and consequently, it follows from Lemma \ref{Lemma:G-DGP1} that we have either
\begin{equation*}
    \lrabs{T_n^f(1,\lrfl{n\xi_i}-m,\lrfl{n\xi_{i+1}}-m) \lrp{V_n^f(1,\lrfl{n\xi_i}-m,\lrfl{n\xi_{i+1}}-m)}^{-1/2}}
\stra{p} \infty.
\end{equation*}
or
\begin{equation*}
    \lrabs{T_n^b(\lrfl{n\xi_{i-1}}-m,\lrfl{n\xi_i}-m,N) \lrp{V_n^b(\lrfl{n\xi_{i-1}}-m,\lrfl{n\xi_i}-m,N)}^{-1/2}}
\stra{p} \infty.
\end{equation*}
Thus it follows from the definition of $G_n^{M}$ that
\BEqn
& & G_n^{M} \\
&=& \max\limits_{(\ell_1,\ell_2)\in\Xi_n(\varepsilon)} \lrabs{\frac{T_n^{f}(1,\ell_1,\ell_2)}{(V_n^{f}(1,\ell_1,\ell_2))^{1/2}}}
  + \max\limits_{(\ell_1,\ell_2)\in\Xi_n(\varepsilon)} \lrabs{\frac{T_n^{b}(\ell_1,\ell_2,N)}{(V_n^{b}(\ell_1,\ell_2,N))^{1/2}}} \\
&\geq& \lrabs{\frac{T_n^f(1,\lrfl{n\xi_i}-m,\lrfl{n\xi_{i+1}}-m)}{\lrp{V_n^f(1,\lrfl{n\xi_i}-m,\lrfl{n\xi_{i+1}}-m)}^{1/2}}} 
  + \lrabs{\frac{T_n^b(\lrfl{n\xi_{i-1}}-m,\lrfl{n\xi_i}-m,N)}{\lrp{V_n^b(\lrfl{n\xi_{i-1}}-m,\lrfl{n\xi_i}-m,N)}^{1/2}}} \\
&\stra{p}& \infty,
\EEqn
which implies that $\blrp{G_n^{M} > G_{1-\alpha}^{M}} \rightarrow 1$ as $n\rightarrow\infty$.
    
If $\{\tilde{X}_t\}_{t=1}^{n} \in \br^{p}$ is a linear process as defined in Definition \ref{Def:Model_lp} or is generated from a factor model as defined in Definition \ref{Def:Model_factor}, the same conclusion can be obtained using Lemma \ref{Lemma:G-DGP2} and Lemma \ref{Lemma:G-DGP3} respectively, and we spare the details.

\section{Auxiliary Lemmas I}\label{Appdix:AuxLemma-1}

\subsection{Lemmas for Proposition \ref{Prop:NullDist_Wn}}\label{Appdix:lemma_null_iid_Growingp}

To facilitate the analysis of the linear process, we first investigate the case of an iid data $\{X_t\}_{t=1}^{n}$ with mean zero and covariance matrix $\Sigma^{(2)}$.

\begin{lemma}\label{Lemma:Y_martingale}
Let $\cf_j = \sigma(X_1,\cdots,X_{j+m})$ for $j=1,\cdots,N$, then it holds under the null that $\{Y_j\}_{j=1}^{N}$ is a martingale difference sequence w.r.t. $\cf_{j}$ with mean $\be[Y_j] = 0$ and variance $\Var(Y_j) = \frac{2}{m_1}\|\Sigma^{(2)}\|_F^2$.
\end{lemma}
\begin{Proof}
It is trivial that $Y_j$ is $\cf_j$-measurable for each $j=1,\cdots,N$. By the definition of $Y_j$ and the independence of sequence $\{X_t\}_{t=1}^{n}$, we have
\begin{eqnarray*}
    \blre{Y_j | \cf_{j-1}}
&=& \frac{1}{m_1} \sum\limits_{i=1}^{m_1} \blre{\lrp{X_i - X_{n+1-i}}^{\top} X_{j+m} | \cf_{j-1}} \\
&=& \frac{1}{m_1} \sum\limits_{i=1}^{m_1} X_i^{\top} \blre{X_{j+m}} - \frac{1}{m_1} \sum\limits_{i=1}^{n} \blre{X_{n+1-t}}^{\top} \blre{X_{j+m}} \\
&=& 0,
\end{eqnarray*}
which implies that $\{Y_j\}_{j=1}^{N}$ is a martingale difference sequence w.r.t. $\cf_{j}$ and $\be[Y_j] = 0$.

Again, by noting that $\{X_t\}_{t=1}^{n}$ is an iid sequence, we can compute the variance of $Y_j$.
\begin{eqnarray*}
    \Var(Y_j)
&=& \frac{1}{m_1^2} \Var\lrp{\sum\limits_{i=1}^{m_1} \lrp{X_i - X_{n+1-i}}^{\top} X_{j+m}} \\
&=& \frac{1}{m_1} \Var\lrp{\lrp{X_1-X_n}^{\top} X_{j+m}} \\
&=& \frac{1}{m_1} \blre{\lrp{X_1-X_n}^{\top} X_{j+m} X_{j+m}^{\top} \lrp{X_1-X_n}} \\
&=& \frac{2}{m_1} \blre{X_1^{\top} X_{j+m} X_{j+m}^{\top} X_1} \\
&=& \frac{2}{m_1} \tr\lrp{\blre{X_{j+m} X_{j+m}^{\top}} \blre{X_1 X_1^{\top}}} \\
&=& \frac{2}{m_1} \|\Sigma^{(2)}\|_F^2.
\end{eqnarray*}
\end{Proof}

\begin{lemma}\label{Lemma:Wn_martingale}
Let
\begin{equation*}
    \xi_j = \left\{
    \begin{array}{ll}
      \displaystyle{\frac{(\alpha_1 + \alpha_2)m_1}{N_n} Y_j} & 1 \le j \le \lfloor{nr_1}\rfloor - \lfloor{n\varepsilon}\rfloor \\
      \displaystyle{\frac{\alpha_2 m_1}{N_n} Y_j} & \lfloor{nr_1}\rfloor - \lfloor{n\varepsilon}\rfloor + 1 \le j \le \lfloor{nr_2}\rfloor - \lfloor{n\varepsilon}\rfloor
    \end{array}
    \right.
\end{equation*}
Then for any $\alpha_1,\alpha_2>0$ and $\varepsilon \le r_1<r_2 \le 1-\varepsilon$, it holds that $\frac{1}{N_n}\lrp{\alpha_1 W_n(r_1) + \alpha_2 W_n(r_2)} = \sum\limits_{j=1}^{\lrfl{nr_2}-\lrfl{n\varepsilon}} \xi_j$ and $\lrcp{\frac{1}{N_n}\lrp{\alpha_1 W_n(r_1) + \alpha_2 W_n(r_2)}:\ r_1<r_2\le1-\varepsilon}$ forms a martingale w.r.t. $\{F_j\}_{j=1}^{N}$.
\end{lemma}
\begin{Proof}
For any fixed $\varepsilon \le r_1 < r_2 \le 1-\varepsilon$ and any fixed $\alpha_1,\alpha_2 > 0$, we have
\BEqn
& & \frac{1}{N_n} \lrp{\alpha_1 W_n(r_1) + \alpha_2 W_n(r_2)} \\
&=& \sum\limits_{i=1}^{m_1} \lrp{X_i - X_{n+1-i}}^{\top} \lrp{\frac{\alpha_1}{N_n} \sum\limits_{j=1}^{\lfloor{nr_1}\rfloor - \lfloor{n\varepsilon}\rfloor} X_{j+m} + \frac{\alpha_2}{N_n} \sum\limits_{j=1}^{\lfloor{nr_2}\rfloor - \lfloor{n\varepsilon}\rfloor} X_{j+m}} \\
&=& \sum\limits_{i=1}^{m_1} \lrp{X_i - X_{n+1-i}}^{\top} \lrp{\frac{\alpha_1 + \alpha_2}{N_n} \sum\limits_{j=1}^{\lfloor{nr_1}\rfloor - \lfloor{n\varepsilon}\rfloor} X_{j+m} + \frac{\alpha_2}{N_n} \sum\limits_{j=\lfloor{nr_1}\rfloor - \lfloor{n\varepsilon}\rfloor+1}^{\lfloor{nr_2}\rfloor - \lfloor{n\varepsilon}\rfloor} X_{j+m}} \\
&=& \sum\limits_{j=1}^{\lfloor{nr_2}\rfloor - \lfloor{n\varepsilon}\rfloor} \xi_j.
\EEqn

Note that we have shown in Lemma \ref{Lemma:Y_martingale} that $\{Y_j\}_{j=1}^{N}$ is a martingale difference sequence, then it follows that the sequence $\{\alpha_1 W_n(r_1) + \alpha_2 W_n(r_2)\}_{r\in[\varepsilon, 1-\varepsilon]}$ forms a martingale w.r.t. $\{\cf_j\}_{j=1}^{N}$, which completes the proof.
\end{Proof}

\begin{lemma}\label{Lemma:NullDist_Wn_1_clt1}
Under Assumption \ref{Assumpt:NullDist_Wn_iid_Growingp},it holds under the null that for any $\delta>0$, we have that
\begin{equation*}
    \sum\limits_{j=1}^{\lfloor{nr_2}\rfloor - \lfloor{n\varepsilon}\rfloor} \blre{\xi_j^2 \bone\{|\xi_j| > \delta\} | \cf_{j-1}} 
\stra{p} 0,
\end{equation*}
where $\xi_j$ is defined as Lemma \ref{Lemma:Wn_martingale}.
\end{lemma}
\begin{Proof}
It suffices to show that $\sum\limits_{j=1}^{\lrfl{nr_2} - \lrfl{n\varepsilon}} \blre{\xi_j^4} \rightarrow 0$, where
\begin{equation*}
    \xi_j^4
  = \left\{
    \begin{array}{ll}
      \displaystyle{\frac{(\alpha_1 + \alpha_2)^4 m_1^4}{N_n^4} Y_j^4} & 1 \le j \le \lrfl{nr_1} - \lrfl{n\varepsilon} \\
      \displaystyle{\frac{\alpha_2^4 m_1^4}{N_n^4} Y_j^4} & \lrfl{nr_1} - \lrfl{n\varepsilon} + 1 \le j \le \lrfl{nr_2} - \lrfl{n\varepsilon}
    \end{array}
    \right. \\
\end{equation*}
and $Y_j^4 = m_1^{-4} \lrp{\sum\limits_{i=1}^{m_1} \lrp{X_i - X_{n+1-i}}^{\top} X_{j+m}}^4$.

By using the iid property and applying the $c_r$ inequality, it holds for any $j=1,\cdots,N$ that
\BEqn
& & \blre{Y_j^4} \\
&\lesssim& m_1^{-4} \blre{\lrp{\sum\limits_{i=1}^{m_1} X_i^{\top} X_{j+m}}^4} + m_1^{-4} \blre{\lrp{\sum\limits_{i=1}^{m_1} X_{n+1-i}^{\top} X_{j+m}}^4} \\
&=& 2m_1^{-4} \blre{\lrp{\sum\limits_{i=1}^{m_1} X_i^{\top} X_{j+m}}^4} \\
&=& 2m_1^{-4} \sum\limits_{i_1,i_2,i_3,i_4}^{m_1} \sum\limits_{\ell_1,\ell_2,\ell_3,\ell_4=1}^{p} \blre{X_{i_1,\ell_1}, X_{i_2,\ell_2}, X_{i_3,\ell_3}, X_{i_4,\ell_4}} \blre{X_{j+m,\ell_1}, X_{j+m,\ell_2}, X_{j+m,\ell_3}, X_{j+m,\ell_4}}
\EEqn
If follows from the cumulant formula that
\BEqn
    \blre{X_{i_1,\ell_1}, X_{i_2,\ell_2}, X_{i_3,\ell_3}, X_{i_4,\ell_4}} 
&=& \bone\{i_1=i_2=i_3=i_4\} \cum\lrp{X_{1,\ell_1}, X_{1,\ell_2}, X_{1,\ell_3}, X_{1,\ell_4}} \\
& & + \bone\{i_1=i_2,i_3=i_4\} \Sigma^{(2)}_{\ell_1,\ell_2} \Sigma^{(2)}_{\ell_3,\ell_4} \\
& & + \bone\{i_1=i_3,i_2=i_4\} \Sigma^{(2)}_{\ell_1,\ell_3}
\Sigma^{(2)}_{\ell_2,\ell_4} \\
& & + \bone\{i_1=i_4,i_2=i_3\} \Sigma^{(2)}_{\ell_1,\ell_4} \Sigma^{(2)}_{\ell_2,\ell_3}
\EEqn
and
\begin{equation*}
    \blre{X_{j+m,\ell_1}, X_{j+m,\ell_2}, X_{j+m,\ell_3}, X_{j+m,\ell_4}}
  = \cum\lrp{X_{1,\ell_1}, X_{1,\ell_2}, X_{1,\ell_3}, X_{1,\ell_4}} + \Sigma^{(2)}_{\ell_1,\ell_2} + \Sigma^{(2)}_{\ell_1,\ell_3} + \Sigma^{(2)}_{\ell_1,\ell_4}.
\end{equation*}

If $\|\Sigma^{(2)}\| = o(\|\Sigma^{(2)}\|_F)$, we have
\begin{equation*}
    \sum\limits_{\ell_1,\ell_2,\ell_3,\ell_4=1}^{p} \Sigma^{(2)}_{\ell_1,\ell_2} \Sigma^{(2)}_{\ell_1,\ell_3} \Sigma^{(2)}_{\ell_2,\ell_4} \Sigma^{(2)}_{\ell_3,\ell_4}
  = \|(\Sigma^{(2)})^2\|_F^2
  \le \|\Sigma^{(2)}\|^2 \|\Sigma^{(2)}\|_F^2  
  = o(\|\Sigma^{(2)}\|_F^4).
\end{equation*}
Consequently, we obtain that for $j=1,\cdots,N$,
\BEqn
    \blre{Y_j^4}
&\lesssim& m_1^{-3} \sum\limits_{\ell_1,\ell_2,\ell_3,\ell_4=1}^{p} \cum^2\lrp{X_{1,\ell_1}, X_{1,\ell_2}, X_{1,\ell_3}, X_{1,\ell_4}} \\
& & + m_1^{-2} \sum\limits_{\ell_1,\ell_2,\ell_3,\ell_4=1}^{p} \Sigma^{(2)}_{\ell_1,\ell_2} \Sigma^{(2)}_{\ell_3,\ell_4} \cum\lrp{X_{1,\ell_1}, X_{1,\ell_2}, X_{1,\ell_3}, X_{1,\ell_4}} \\
& & + m_1^{-2} \sum\limits_{\ell_1,\ell_2,\ell_3,\ell_4=1}^{p} \Sigma^{(2)}_{\ell_1,\ell_2} \Sigma^{(2)}_{\ell_1,\ell_3} \Sigma^{(2)}_{\ell_2,\ell_4} \Sigma^{(2)}_{\ell_3,\ell_4} \\
& & + m_1^{-2} \sum\limits_{\ell_1,\ell_2,\ell_3,\ell_4=1}^{p} \lrp{\Sigma^{(2)}_{\ell_1,\ell_2}}^2 \lrp{\Sigma^{(2)}_{\ell_3,\ell_4}}^2 \\
&\lesssim& m_1^{-2} \|\Sigma^{(2)}\|_F^4
\EEqn
as long as the condition $\sum\limits_{\ell_1,\ell_2,\ell_3,\ell_4=1}^{p} \cum^2\lrp{X_{1,\ell_1}, X_{1,\ell_2}, X_{1,\ell_3}, X_{1,\ell_4}} = O\lrp{\|\Sigma^{(2)}\|_F^4}$ is satisfied.

Finally, we have that
\BEqn
    \sum\limits_{j=1}^{\lrfl{nr_2} - \lrfl{n\varepsilon}} \blre{\xi_j^4}
&=& \frac{(\alpha_1 + \alpha_2)^4 m_1^4}{N_n^4} \sum\limits_{j=1}^{\lrfl{nr_1} - \lrfl{n\varepsilon}} \blre{\xi_j^4} + \frac{\alpha_2^4 m_1^4}{N_n^4} \sum\limits_{j=\lrfl{nr_1} - \lrfl{n\varepsilon}+1}^{\lrfl{nr_2} - \lrfl{n\varepsilon}} \blre{\xi_j^4} \\
&\lesssim& \frac{(\alpha_1 + \alpha_2)^4 m_1^4}{N_n^4} \sum\limits_{j=1}^{\lrfl{nr_2} - \lrfl{n\varepsilon}} \blre{\xi_j^4} \\
&\lesssim& \frac{m_1^4\lrp{\lrfl{nr_2} - \lrfl{n\varepsilon}}}{n^2 m_1^2 \|\Sigma^{(2)}\|_F^4} \cdot \frac{\|\Sigma^{(2)}\|_F^4}{m_1^2} \\
&\lesssim& \frac{1}{n} \rightarrow 0,
\EEqn
which completes the proof of the lemma.
\end{Proof}

\begin{lemma}\label{Lemma:NullDist_Wn_1_clt2}
For any fixed $\alpha_1,\alpha_2>0$ and $\varepsilon \le r_1<r_2 \le 1-\varepsilon$, define $\sigma^2 = (\alpha_1^2 + 2\alpha_1\alpha_2)(r_1-\varepsilon) + \alpha_2^2(r_2-\varepsilon)$. Under Assumption \ref{Assumpt:NullDist_Wn_iid_Growingp}, it holds under the null that
\begin{equation*}
    V_n
  = \sum\limits_{j=1}^{\lfloor{nr_2}\rfloor - \lfloor{n\varepsilon}\rfloor} \blre{\xi_j^2 | \cf_{j-1}} 
  \stra{p} \sigma^2,
\end{equation*}
where $\xi_j$ is defined as Lemma \ref{Lemma:Wn_martingale}.
\end{lemma}
\begin{Proof}
By the definition of $Y_j$, we have
\BEqn
    \blre{Y_j | \cf_{j-1}}
&=& \frac{1}{m_1^2} \sum\limits_{i_1,i_2=1}^{m_1} \blre{ \lrp{X_{i_1} - X_{n+1-i_1}}^{\top} X_{j+m} X_{j+m}^{\top} \lrp{X_{i_2} - X_{n+1-i_2}} | \cf_{j-1} } \\
&=& \frac{1}{m_1^2} \sum\limits_{i_1,i_2=1}^{m_1} X_{i_1}^{\top} \blre{X_{j+m} X_{j+m}^{\top}} X_{i_2} + \frac{1}{m_1^2} \sum\limits_{i=1}^{m_1} \blre{X_{n+1-i}^{\top} X_{j+m} X_{j+m}^{\top} X_{n+1-i}} \\
&=& \frac{1}{m_1^2} \sum\limits_{i_1,i_2=1}^{m_1} X_{i_1}^{\top} \Sigma^{(2)} X_{i_2} + \frac{1}{m_1} \|\Sigma^{(2)}\|_F^2,
\EEqn
which is independent of $j$, then for each $j=1,\cdots,N$, we have
\begin{equation*}
    \blre{Y_j | \cf_{j-1}}
  = \frac{1}{m_1^2} \sum\limits_{i_1,i_2=1}^{m_1} X_{i_1}^{\top} \Sigma^{(2)} X_{i_2} + \frac{1}{m_1} \|\Sigma^{(2)}\|_F^2,
\end{equation*}
and it follows that
\begin{equation*}
    V_n^2
  = \lrp{\alpha_1^2+2\alpha_1\alpha_2} \frac{m_1^2 \lrp{\lrfl{nr_1} - \lrfl{n\varepsilon}}}{N_n^2} \blre{Y_2^2 | \cf_1} 
  + \alpha_2^2 \frac{m_1^2 \lrp{\lrfl{nr_2} - \lrfl{n\varepsilon}}}{N_n^2} \blre{Y_2^2 | \cf_1}. 
\end{equation*}

Define
\BEqn
    L_n^{(1)} 
&=& \frac{m_1^2 \lrp{\lrfl{nr_1} - \lrfl{n\varepsilon}}}{N_n^2} \lrp{\frac{1}{m_1^2} \sum\limits_{i_1,i_2=1}^{m_1} X_{i_1}^{\top} \Sigma^{(2)} X_{i_2} + \frac{1}{m_1} \|\Sigma^{(2)}\|_F^2}, \\
    L_n^{(2)} 
&=& \frac{m_1^2 \lrp{\lrfl{nr_2} - \lrfl{n\varepsilon}}}{N_n^2} \lrp{\frac{1}{m_1^2} \sum\limits_{i_1,i_2=1}^{m_1} X_{i_1}^{\top} \Sigma^{(2)} X_{i_2} + \frac{1}{m_1} \|\Sigma^{(2)}\|_F^2}.
\EEqn
To show $V_n \stra{p} \sigma^2$ is equivalent to show that $L_n^{(1)} \stra{p} r_1-\varepsilon$ and $L_n^{(2)} \stra{p} r_2-\varepsilon$.

Note that 
\begin{equation*}
    \blre{\frac{1}{m_1^2} \sum\limits_{i_1,i_2=1}^{m_1} X_{i_1}^{\top} \Sigma^{(2)} X_{i_2} + \frac{1}{m_1} \|\Sigma^{(2)}\|_F^2}
  = \frac{2}{m_1} \|\Sigma^{(2)}\|_F^2,
\end{equation*}
and we can compute that
\BEqn
& & \blre{\lrp{\frac{1}{m_1^2} \sum\limits_{i_1,i_2=1}^{m_1} X_{i_1}^{\top} \Sigma^{(2)} X_{i_2} + \frac{1}{m_1} \|\Sigma^{(2)}\|_F^2}^2} \\
&=& \blre{\lrp{\frac{1}{m_1^2} \sum\limits_{i_1,i_2=1}^{m_1} X_{i_1}^{\top} \Sigma^{(2)} X_{i_2}}^2} + \frac{2}{m_1}\|\Sigma^{(2)}\|_F^2 \blre{\frac{1}{m_1^2} \sum\limits_{i_1,i_2=1}^{m_1} X_{i_1}^{\top} \Sigma^{(2)} X_{i_2}} + \frac{1}{m_1^2} \|\Sigma^{(2)}\|_F^4 \\
&=& \frac{1}{m_1^4} \sum\limits_{i_1,i_2,i_3,i_4=1}^{m_1} \sum\limits_{\ell_1,\ell_2,\ell_3,\ell_4=1}^{p} \Sigma^{(2)}_{\ell_1,\ell_2} \Sigma^{(2)}_{\ell_3,\ell_4} \blre{X_{i_1,\ell_1}, X_{i_2,\ell_2}, X_{i_3,\ell_3}, X_{i_4,\ell_4}} + \frac{3}{m_1^2} \|\Sigma^{(2)}\|_F^4 \\
&=& \frac{4}{m_1^2}(1+o(1)) \|\Sigma^{(2)}\|_F^4,
\EEqn
as long as $\sum\limits_{\ell_1,\ell_2,\ell_3,\ell_4=1}^{p} \cum^2\lrp{X_{1,\ell_1}, X_{1,\ell_2}, X_{1,\ell_3}, X_{1,\ell_4}} = O\lrp{\|\Sigma^{(2)}\|_F^4}$ and $\|\Sigma^{(2)}\| = o(\|\Sigma^{(2)}\|_F)$. Note that  the last step is obtained from the calculation as follows
\BEqn
& & \sum\limits_{i_1,i_2,i_3,i_4=1}^{m_1} \sum\limits_{\ell_1,\ell_2,\ell_3,\ell_4=1}^{p} \Sigma^{(2)}_{\ell_1,\ell_2} \Sigma^{(2)}_{\ell_3,\ell_4} \blre{X_{i_1,\ell_1}, X_{i_2,\ell_2}, X_{i_3,\ell_3}, X_{i_4,\ell_4}} \\
&=& m_1 \sum\limits_{\ell_1,\ell_2,\ell_3,\ell_4=1}^{p} \Sigma^{(2)}_{\ell_1,\ell_2} \Sigma^{(2)}_{\ell_3,\ell_4} \cum\lrp{X_{1,\ell_1}, X_{1,\ell_2}, X_{1,\ell_3}, X_{1,\ell_4}} \\
& & + 2m_1^2 \sum\limits_{\ell_1,\ell_2,\ell_3,\ell_4=1}^{p} \Sigma^{(2)}_{\ell_1,\ell_2} \Sigma^{(2)}_{\ell_1,\ell_3} \Sigma^{(2)}_{\ell_2,\ell_4} \Sigma^{(2)}_{\ell_3,\ell_4} \\
& & + m_1^2 \sum\limits_{\ell_1,\ell_2,\ell_3,\ell_4=1}^{p} \lrp{\Sigma^{(2)}_{\ell_1,\ell_2}}^2 \lrp{\Sigma^{(2)}_{\ell_3,\ell_4}}^2 \\
&=& m_1^2 (1+o(1)) \|\Sigma^{(2)}\|_F^4.
\EEqn

Next we compute $\blre{\lrp{L_n^{(1)} - (r_1-\varepsilon)}^2}$.
\BEqn
& & \blre{\lrp{L_n^{(1)} - (r_1-\varepsilon)}^2} \\
&=& \blre{\lrp{L_n^{(1)}}^2} - 2(r_1-\varepsilon)\blre{L_n^{(1)}} + (r_1-\varepsilon)^2 \\
&=& \frac{m_1^4 \lrp{\lrfl{nr_1} - \lrfl{n\varepsilon}}^2}{N_n^4} \blre{\lrp{\frac{1}{m_1^2} \sum\limits_{i_1,i_2=1}^{m_1} X_{i_1}^{\top} \Sigma^{(2)} X_{i_2} + \frac{1}{m_1} \|\Sigma^{(2)}\|_F^2}^2} \\
& & - \frac{2m_1^2 (r_1-\varepsilon)\lrp{\lrfl{nr_1} - \lrfl{n\varepsilon}}}{N_n^2} \blre{\frac{1}{m_1^2} \sum\limits_{i_1,i_2=1}^{m_1} X_{i_1}^{\top} \Sigma^{(2)} X_{i_2} + \frac{1}{m_1} \|\Sigma^{(2)}\|_F^2} \\
& & + (r_1-\varepsilon)^2 \\
&=& \frac{m_1^4 \lrp{\lrfl{nr_1} - \lrfl{n\varepsilon}}^2}{4n^2 m_1^2 \|\Sigma^{(2)}\|_F^4} \cdot \frac{4}{m_1^2}(1+o(1)) \|\Sigma^{(2)}\|_F^4 
 - \frac{2m_1^2 (r_1-\varepsilon)\lrp{\lrfl{nr_1} - \lrfl{n\varepsilon}}}{2n m_1 \|\Sigma^{(2)}\|_F^2} \cdot \frac{2}{m_1} \|\Sigma^{(2)}\|_F^2
 + (r_1-\varepsilon)^2 \\
 &=& (r_1-\varepsilon)^2 - 2(r_1-\varepsilon)^2 + (r_1-\varepsilon)^2 + o(1) \\
 &\rightarrow& 0,
\EEqn
which implies that $L_n^{(1)} \stra{p} r_1-\varepsilon$. Similarly, we can show that $L_n^{(2)} \stra{p} r_2-\varepsilon$, jointly with which we finish the proof.
\end{Proof}

\begin{lemma}\label{Lemma:NullDist_Wn_1}
Under Assumption \ref{Assumpt:NullDist_Wn_iid_Growingp}, it holds under the null for any $\alpha_1,\alpha_2 > 0$ and $\varepsilon \le r_1 < r_2 \le 1-\varepsilon$ that
\begin{equation*}
    \frac{1}{N_n} \lrp{\alpha_1 W_n(r_1) + \alpha_2 W_n(r_2)} 
\stra{d} \alpha_1 W(r_1) + \alpha_2 W(r_2),
\end{equation*}
where the process $\{W(r)\}_{r\in[\varepsilon,1-\varepsilon]}$ is defined as $W(r) = B(r)-B(\varepsilon)$.
\end{lemma}
\begin{Proof}
It's shown in Lemma \ref{Lemma:Wn_martingale} that, for any fixed $\varepsilon \le r_1 < r_2 \le 1-\varepsilon$ and $\alpha_1,\alpha_2>0$, the sequence $\{\frac{1}{N_n}\lrp{\alpha_1 W_n(r_1) + \alpha_2 W_n(r_2)}\}$ forms a martingale w.r.t. $\{\cf_j\}_{j=1}^{\lrfl{nr_2} - \lrfl{n\varepsilon}}$. Consequently, the desired result directly follows from the martingale CLT (Theorem 35.12 of \cite{billingsley2008probability}), Lemma \ref{Lemma:NullDist_Wn_1_clt1} and Lemma \ref{Lemma:NullDist_Wn_1_clt2}.
\end{Proof}

\begin{lemma}\label{Lemma:NullDist_Wn_2}
Under Assumption \ref{Assumpt:NullDist_Wn_iid_Growingp}, it holds under the null that the process $\lrcp{\frac{W_n(r)}{N_n}}_{r\in[\varepsilon,1-\varepsilon]}$ is tight, where the process $\{W(r)\}_{r\in[\varepsilon,1-\varepsilon]}$ is defined as $W(r) = B(r)-B(\varepsilon)$.
\end{lemma}
\begin{Proof}
By Lemma 9.8 of \cite{wang2020hypothesis} with $\gamma=4$ and $\alpha=2$, it suffices to show that for any $\varepsilon \le a < b \le 1-\varepsilon$, it holds that
\begin{equation*}
    \blre{\lrabs{\frac{W_n(b)}{N_n} - \frac{W_n(a)}{N_n}}^4}
\lesssim \lrp{\frac{\lrfl{nb} - \lrfl{na}}{n}}^2.
\end{equation*}

Note that
\begin{equation*}
    W_n(b)-W_n(a)
  = m_1 \sum\limits_{j=\lrfl{na} - \lrfl{n\varepsilon} + 1}^{\lrfl{nb} - \lrfl{n\varepsilon}} Y_j
\end{equation*}
and we have shown that $\{Y_j\}_{j=1}^{N}$ is a martingale difference sequence w.r.t. $\{\cf_j\}_{j=1}^{N}$, then by applying Burkholder's inequality (Theorem 2.10 of \cite{hall2014martingale}), we obtain that
\begin{equation*}
    \blre{\lrp{\sum\limits_{j=\lrfl{na} - \lrfl{n\varepsilon} + 1}^{\lrfl{nb} - \lrfl{n\varepsilon}} Y_j}^4}
\lesssim \blre{\lrp{\sum\limits_{j=\lrfl{na} - \lrfl{n\varepsilon} + 1}^{\lrfl{nb} - \lrfl{n\varepsilon}} Y_j^2}^2},
\end{equation*}
thus it remains to compute the RHS.

It follows from the definition of $Y_j$ and the $c_r$ inequality that
\BEqn
& & \blre{\lrp{\sum\limits_{j=\lrfl{na} - \lrfl{n\varepsilon} + 1}^{\lrfl{nb} - \lrfl{n\varepsilon}} Y_j^2}^2} \\
&=& m_1^{-4} \blre{\lrp{\sum\limits_{j=\lrfl{na} - \lrfl{n\varepsilon} + 1}^{\lrfl{nb} - \lrfl{n\varepsilon}} \sum\limits_{i_1,i_2=1}^{m_1} \lrp{X_{i_1} - X_{n+1-i_1}}^{\top} X_{j+m} (X_{i_2} - X_{n+1-i_2})^{\top} X_{j+m}}^2} \\
&\lesssim& m_1^{-4} \blre{\lrp{\sum\limits_{j=\lrfl{na} - \lrfl{n\varepsilon} + 1}^{\lrfl{nb} - \lrfl{n\varepsilon}} \sum\limits_{i_1,i_2=1}^{m_1} X_{i_1}^{\top} X_{j+m} X_{i_2}^{\top} X_{j+m}}^2} \\
& & + m_1^{-4} \blre{\lrp{\sum\limits_{j=\lrfl{na} - \lrfl{n\varepsilon} + 1}^{\lrfl{nb} - \lrfl{n\varepsilon}} \sum\limits_{i_1,i_2=1}^{m_1} X_{i_1}^{\top} X_{j+m} X_{n+1-i_2}^{\top} X_{j+m}}^2} \\
& & + m_1^{-4} \blre{\lrp{\sum\limits_{j=\lrfl{na} - \lrfl{n\varepsilon} + 1}^{\lrfl{nb} - \lrfl{n\varepsilon}} \sum\limits_{i_1,i_2=1}^{m_1} X_{n+1-i_1}^{\top} X_{j+m} X_{i_2}^{\top} X_{j+m}}^2} \\
& & + m_1^{-4} \blre{\lrp{\sum\limits_{j=\lrfl{na} - \lrfl{n\varepsilon} + 1}^{\lrfl{nb} - \lrfl{n\varepsilon}} \sum\limits_{i_1,i_2=1}^{m_1} X_{n+1-i_1}^{\top} X_{j+m} X_{n+1-i_2}^{\top} X_{j+m}}^2} \\
&\lesssim& m_1^{-4} \blre{\lrp{\sum\limits_{j=\lrfl{na} - \lrfl{n\varepsilon} + 1}^{\lrfl{nb} - \lrfl{n\varepsilon}} \sum\limits_{i_1,i_2=1}^{m_1} X_{i_1}^{\top} X_{j+m} X_{i_2}^{\top} X_{j+m}}^2} \\
& & + m_1^{-4} \blre{\lrp{\sum\limits_{j=\lrfl{na} - \lrfl{n\varepsilon} + 1}^{\lrfl{nb} - \lrfl{n\varepsilon}} \sum\limits_{i_1,i_2=1}^{m_1} X_{i_1}^{\top} X_{j+m} X_{n+1-i_2}^{\top} X_{j+m}}^2}
\EEqn
where the last step uses the iid property.

Furthermore, we apply the cumulant formula and obtain that
\BEqn
& & \blre{\lrp{\sum\limits_{j=\lrfl{na} - \lrfl{n\varepsilon} + 1}^{\lrfl{nb} - \lrfl{n\varepsilon}} \sum\limits_{i_1,i_2=1}^{m_1} X_{i_1}^{\top} X_{j+m} X_{i_2}^{\top} X_{j+m}}^2} \\
&=& \sum\limits_{\ell_1,\ell_2,\ell_3,\ell_4=1}^{p} 
\lrp{\sum\limits_{i_1,i_2,i_3,i_4=1}^{m_1} \blre{X_{i_1,\ell_1} X_{i_2,\ell_2} X_{i_3,\ell_3} X_{i_4,\ell_4}}} \\
& & \hspace{5em}\lrp{\sum\limits_{j_1,j_2=\lrfl{na} - \lrfl{n\varepsilon} + 1}^{\lrfl{nb} - \lrfl{n\varepsilon}} \blre{X_{j_1+m,\ell_1} X_{j_1+m,\ell_2} X_{j_2+m,\ell_3} X_{j_2+m,\ell_4}}} \\
&\lesssim& m_1 \lrp{\lrfl{nb}-\lrfl{na}}
\sum\limits_{\ell_1,\ell_2,\ell_3,\ell_4=1}^{p} 
 \cum^2(X_{1,\ell_1},X_{1,\ell_2},X_{1,\ell_3},X_{1,\ell_4}) \\
& & + m_1^2 \lrp{\lrfl{nb}-\lrfl{na}}
\sum\limits_{\ell_1,\ell_2,\ell_3,\ell_4=1}^{p}  
     \Sigma^{(2)}_{\ell_1,\ell_2} \Sigma^{(2)}_{\ell_3,\ell_4} \cum(X_{1,\ell_1},X_{1,\ell_2},X_{1,\ell_3},X_{1,\ell_4}) \\
& & + m_1 \lrp{\lrfl{nb}-\lrfl{na}}^2 
\sum\limits_{\ell_1,\ell_2,\ell_3,\ell_4=1}^{p}  
      \Sigma^{(2)}_{\ell_1,\ell_2} \Sigma^{(2)}_{\ell_3,\ell_4} \cum(X_{1,\ell_1},X_{1,\ell_2},X_{1,\ell_3},X_{1,\ell_4}) \\
& & + m_1^2 \lrp{\lrfl{nb}-\lrfl{na}}^2 
\sum\limits_{\ell_1,\ell_2,\ell_3,\ell_4=1}^{p}  
      \lrp{\Sigma^{(2)}_{\ell_1,\ell_2}}^2 \lrp{\Sigma^{(2)}_{\ell_3,\ell_4}}^2 \\
& & + m_1^2 \lrp{\lrfl{nb}-\lrfl{na}}^2 
\sum\limits_{\ell_1,\ell_2,\ell_3,\ell_4=1}^{p}  
      \Sigma^{(2)}_{\ell_1,\ell_2} \Sigma^{(2)}_{\ell_1,\ell_3}  \Sigma^{(2)}_{\ell_2,\ell_4}  \Sigma^{(2)}_{\ell_3,\ell_4} \\
&\lesssim& m_1^2 \lrp{\lrfl{nb}-\lrfl{na}}^2 \|\Sigma^{(2)}\|_F^4
\EEqn
as long as $\sum\limits_{\ell_1,\ell_2,\ell_3,\ell_4=1}^{p} \cum^2\lrp{X_{1,\ell_1}, X_{1,\ell_2}, X_{1,\ell_3}, X_{1,\ell_4}} = O\lrp{\|\Sigma^{(2)}\|_F^4}$ and $\|\Sigma^{(2)}\| = o(\|\Sigma^{(2)}\|_F)$, where the process $\{W(r)\}_{r\in[\varepsilon,1-\varepsilon]}$ is defined as $W(r) = B(r)-B(\varepsilon)$.

Under the same conditions, we can show that
\BEqn
& & \blre{\lrp{\sum\limits_{j=\lrfl{na} - \lrfl{n\varepsilon} + 1}^{\lrfl{nb} - \lrfl{n\varepsilon}} \sum\limits_{i_1,i_2=1}^{m_1} X_{i_1}^{\top} X_{j+m} X_{n+1-i_2}^{\top} X_{j+m}}^2} \\
&=& \sum\limits_{\ell_1,\ell_2,\ell_3,\ell_4=1}^{p} 
\lrp{\sum\limits_{i_1,i_2=1}^{m_1} \blre{X_{i_1,\ell_1} X_{i_2,\ell_2}}}
\lrp{\sum\limits_{i_3,i_4=1}^{m_1} \blre{X_{n+1-i_3,\ell_3} X_{n+1-i_4,\ell_4}}} \\
& & \hspace{5em}\lrp{\sum\limits_{j_1,j_2=\lrfl{na} - \lrfl{n\varepsilon} + 1}^{\lrfl{nb} - \lrfl{n\varepsilon}} \blre{X_{j_1+m,\ell_1} X_{j_1+m,\ell_2} X_{j_2+m,\ell_3} X_{j_2+m,\ell_4}}} \\
&\lesssim& m_1^2 \lrp{\lrfl{nb}-\lrfl{na}} 
\sum\limits_{\ell_1,\ell_2,\ell_3,\ell_4=1}^{p}  
    \Sigma^{(2)}_{\ell_1,\ell_2} \Sigma^{(2)}_{\ell_3,\ell_4} \cum(X_{1,\ell_1}, X_{1,\ell_2}, X_{1,\ell_3}, X_{1,\ell_4}) \\
& & + m_1^2 \lrp{\lrfl{nb}-\lrfl{na}}^2
\sum\limits_{\ell_1,\ell_2,\ell_3,\ell_4=1}^{p}  
    \lrp{\Sigma^{(2)}_{\ell_1,\ell_2}}^2 \lrp{\Sigma^{(2)}_{\ell_3,\ell_4}}^2 \\
& & + m_1^2 \lrp{\lrfl{nb}-\lrfl{na}}^2
\sum\limits_{\ell_1,\ell_2,\ell_3,\ell_4=1}^{p}  
    \Sigma^{(2)}_{\ell_1,\ell_2} \Sigma^{(2)}_{\ell_1,\ell_3} \Sigma^{(2)}_{\ell_2,\ell_4} \Sigma^{(2)}_{\ell_3,\ell_4} \\
&\lesssim& m_1^2 \lrp{\lrfl{nb}-\lrfl{na}}^2 \|\Sigma^{(2)}\|_F^4.
\EEqn

Consequently, we obtain that
\BEqn
    \blre{\lrabs{\frac{W_n(b)}{N_n} - \frac{W_n(a)}{N_n}}^4}
&\lesssim& \frac{m_1^4}{N_n^4} \blre{\lrp{\sum\limits_{j=\lrfl{na} - \lrfl{n\varepsilon} + 1}^{\lrfl{nb} - \lrfl{n\varepsilon}} Y_j^2}^2} \\
&\lesssim& \frac{m_1^4}{n^2 m_1^2 \|\Sigma^{(2)}\|_F^4} \cdot \frac{\lrp{\lrfl{nb}-\lrfl{na}}^2 \|\Sigma^{(2)}\|_F^4}{m_1^2} \\
&\lesssim& \lrp{\frac{\lrfl{nb} - \lrfl{na}}{n}}^2,
\EEqn
which thus completes the proof of tightness.
\end{Proof}

\subsection{Lemmas for Proposition \ref{Prop:NullDist_Wn_lp}}\label{Appdix:lemma_null_lp}

\begin{lemma}\label{Lemma:NullDist_DR-1_lp}
Under Assumption \ref{Assumpt:NullDist_Wn_lp}, if $\rho^{m_2/4} \|\Gamma^{(2)}\|_F = o\lrp{\frac{n}{\log(n)}}$, then it holds that
\begin{equation*}
    \left\|\sup\limits_{r\in[\varepsilon,1-\varepsilon]} \lrabs{\frac{1}{\widetilde{N}_n} \sum\limits_{i=1}^{m_1} D_i^{\top} \lrp{\sum\limits_{j=1}^{\lrfl{nr} - \lrfl{n\varepsilon}} R_{j+m}}}\right\|_4 = o(1)
\end{equation*}
as $n\rightarrow\infty$.
\end{lemma}
\begin{Proof}
Let $d$ denote the integer such that $2^d \le N < 2^{d+1}$ and $\tilde{N} = N-2^d$, then it follows from the fact $\sum\limits_{j=a}^{b} R_{j} = \widetilde{D}_{b} - \widetilde{D}_{a-1}$ and Proposition 1 of \cite{wu2007strong} that
\BEqn
& & \left\|\sup\limits_{r\in[\varepsilon,1-\varepsilon]} \lrabs{\sum\limits_{i=1}^{m_1} D_{n+1-i}^{\top} \lrp{\sum\limits_{j=1}^{\lrfl{nr} - \lrfl{n\varepsilon}} R_{j+m}}}\right\|_4 \\
&\le& \left\|\max\limits_{1\le h\le N} \lrabs{\sum\limits_{i=1}^{m_1} D_i^{\top} \lrp{\sum\limits_{j=1}^{h} R_{j+m}}}\right\|_4 \\
&\le& \left\| \max\limits_{1\le h\le 2^d} \lrabs{\sum\limits_{i=1}^{m_1} D_i^{\top} \lrp{\sum\limits_{j=1}^{h} R_{j+m}}} \right\|_4 
 +    \left\| \max\limits_{N-2^d+1\le h\le N} \lrabs{\sum\limits_{i=1}^{m_1} D_i^{\top} \lrp{\sum\limits_{j=1}^{h} R_{j+m}}} \right\|_4 \\
&\le& \left\| \max\limits_{1\le h\le 2^d} \lrabs{\sum\limits_{i=1}^{m_1} D_i^{\top} \lrp{\sum\limits_{j=1}^{h} R_{j+m}}} \right\|_4 
 +    \left\| \max\limits_{1\le h\le 2^d} \lrabs{\sum\limits_{i=1}^{m_1} D_i^{\top} \lrp{\sum\limits_{j=1}^{h+N-2^d} R_{j+m}}} \right\|_4 \\
&\le& \sum\limits_{h=0}^{d} \lrp{ \sum\limits_{u=1}^{2^{d-h}} \left\| \sum\limits_{i=1}^{m_1} D_i^{\top} \lrp{\sum\limits_{j=2^h(u-1)+1}^{2^h u} R_{j+m}} \right\|_4^4}^{1/4} \\
& & + \sum\limits_{h=0}^{d} \lrp{ \sum\limits_{u=1}^{2^{d-h}} \left\| \sum\limits_{i=1}^{m_1} D_i^{\top} \lrp{\sum\limits_{j=N-2^d+2^h(u-1)+1}^{N-2^d+2^h u} R_{j+m}} \right\|_4^4}^{1/4} \\
&\lesssim& \sum\limits_{h=0}^{d} \lrp{ \sum\limits_{u=1}^{2^{d-h}} S_1}^{1/4} 
 + \sum\limits_{h=0}^{d} \lrp{ \sum\limits_{u=1}^{2^{d-h}} S_2}^{1/4}
 + \sum\limits_{h=0}^{d} \lrp{ \sum\limits_{u=1}^{2^{d-h}} S_3}^{1/4}
 + \sum\limits_{h=0}^{d} \lrp{ \sum\limits_{u=1}^{2^{d-h}} S_4}^{1/4},
\EEqn
where
\BEqn
    S_1 
&=& \sum\limits_{i_1,i_2,i_3,i_4=1}^{m_1} \blre{D_{i_1}^{\top} \widetilde{D}_{2^h u+m} D_{i_2}^{\top} \widetilde{D}_{2^h u+m} D_{i_3}^{\top} \widetilde{D}_{2^h u+m} D_{i_4}^{\top} \widetilde{D}_{2^h u+m}}, \\
    S_2
&=& \sum\limits_{i_1,i_2,i_3,i_4=1}^{m_1} \blre{D_{i_1}^{\top} \widetilde{D}_{2^h(u-1)+m} D_{i_2}^{\top} \widetilde{D}_{2^h(u-1)+m} D_{i_3}^{\top} \widetilde{D}_{2^h(u-1)+m} D_{i_4}^{\top} \widetilde{D}_{2^h(u-1)+m}} , \\
    S_3
&=& \sum\limits_{i_1,i_2,i_3,i_4=1}^{m_1} \blre{D_{i_1}^{\top} \widetilde{D}_{\tilde{N}+2^h u+m} D_{i_2}^{\top} \widetilde{D}_{\tilde{N}+2^h u+m} D_{i_3}^{\top} \widetilde{D}_{\tilde{N}+2^h u+m} D_{i_4}^{\top} \widetilde{D}_{\tilde{N}+2^h u+m}} ,
\EEqn
and
\begin{equation*}
    S_4
  = \sum\limits_{i_1,i_2,i_3,i_4=1}^{m_1} \blre{D_{i_1}^{\top} \widetilde{D}_{\tilde{N}+2^h(u-1)+m} D_{i_2}^{\top} \widetilde{D}_{\tilde{N}+2^h(u-1)+m} D_{i_3}^{\top} \widetilde{D}_{\tilde{N}+2^h(u-1)+m} D_{i_4}^{\top} \widetilde{D}_{\tilde{N}+2^h(u-1)+m}}.
\end{equation*}

We only show the details regarding $S_1$, and all the analysis of $S_2,S_3$ and $S_4$ can be performed in a similar way. It follows from direct calculations and the decomposition proposed in Lemma \ref{Lemma:LP-aux-3} that 
\BEqn
    S_1
&=& \sum\limits_{i_1,i_2,i_3,i_4=1}^{m_1} 
    \sum\limits_{s_1,s_2,s_3,s_4=0}^{\infty} 
    \sum\limits_{\ell_1,\ell_2,\ell_3,\ell_4=1}^{p}
    \be\left[ (A^{(0)} \varepsilon_{i_1})_{\ell_1}
              (A^{(s_1+1)} \varepsilon_{2^h u+m-s_1})_{\ell_1}
              (A^{(0)} \varepsilon_{i_2})_{\ell_2}
              (A^{(s_2+1)} \varepsilon_{2^h u+m-s_2})_{\ell_2}
       \right. \\
& & \hspace{12em} \times
       \left. (A^{(0)} \varepsilon_{i_3})_{\ell_3} 
              (A^{(s_3+1)} \varepsilon_{2^h u+m-s_3})_{\ell_3}
              (A^{(0)} \varepsilon_{i_4})_{\ell_4} 
              (A^{(s_4+1)} \varepsilon_{2^h u+m-s_4})_{\ell_4}
       \right] \\
&=& \sum\limits_{i_1,i_2,i_3,i_4=1}^{m_1}
    \sum\limits_{s_1,s_2,s_3,s_4=0}^{\infty} 
    \sum\limits_{\ell_1,\ell_2,\ell_3,\ell_4=1}^{p}
    \lrp{S_{1,1} + S_{1,2} + S_{1,3} + S_{1,4} + S_{1,5} + S_{1,6} + S_{1,7}},
\EEqn
where $S_{1,1},\cdots,S_{1,7}$ correspond to the individual terms in the decomposition of Lemma \ref{Lemma:LP-aux-3}. To find the upper bound of $S_1$, it suffices to apply Lemma \ref{Lemma:LP-aux-3-1}-Lemma \ref{Lemma:LP-aux-3-7} to derive the upper bound for each $S_{1,i}$.

For $S_{1,1}$, it follows from Lemma \ref{Lemma:LP-aux-3-1} that
\BEqn
& & \lrabs{\sum\limits_{i_1,i_2,i_3,i_4=1}^{m_1}
           \sum\limits_{s_1,s_2,s_3,s_4=0}^{\infty} 
           \sum\limits_{\ell_1,\ell_2,\ell_3,\ell_4=1}^{p} S_{1,1}} \\
&\lesssim& \sum\limits_{i_1,i_2,i_3,i_4=1}^{m_1}
           \sum\limits_{s_1,s_2,s_3,s_4=0}^{\infty} 
           \bone\{i_1=\cdots=i_4=2^h u+m-s_1=\cdots=2^h u+m-s_4\}
           \|A^{(0)}\|^4 \bigp{ \prod\limits_{i=1}^{4} \|A^{(s_i+1)}\| } \|\Gamma^{(2)}\|_F^8 \\
&=& \|A^{(0)}\|^4 \|\Gamma^{(2)}\|_F^8 \sum\limits_{i_1=1}^{m_1} \|A^{(2^h u+m-i_1+1)}\|^4 \\
&\lesssim& \rho^{4(2^h u+m_2)} \|\Gamma^{(2)}\|_F^8,
\EEqn
where the last step follows from the fact that $m=m_1+m_2$ and Assumption \ref{Assumpt:NullDist_Wn_lp}\ref{Assumpt:NullDist_Wn_Lp-7}.

Next we consider $S_{1,2}$, and it follows from Lemma \ref{Lemma:LP-aux-3-2} that
\BEqn
& & \lrabs{\sum\limits_{i_1,i_2,i_3,i_4=1}^{m_1}
           \sum\limits_{s_1,s_2,s_3,s_4=0}^{\infty} 
           \sum\limits_{\ell_1,\ell_2,\ell_3,\ell_4=1}^{p} S_{1,2}} \\
&\lesssim& \sum\limits_{i_1,i_2,i_3,i_4=1}^{m_1}
           \sum\limits_{s_1,s_2,s_3,s_4=0}^{\infty}
           \|A^{(0)}\|^4 \bigp{ \prod\limits_{i=1}^{4} \|A^{(s_i+1)}\| } \|\Gamma^{(2)}\|_F^7 \\
& & \hspace{8em} \times
         \Big{(}
           \bone\{i_1=2^h u+m-s_1,\ i_2=\cdots=i_4=2^h u+m-s_2=\cdots=2^h u+m-s_4\} \sqrt{p} \\
& & \hspace{9em}
         + \bone\{i_1=i_2,\ i_3=i_4=2^h u+m-s_1=\cdots=2^h u+m-s_4\} \\
& & \hspace{9em}
         + \bone\{2^h u+m-s_1=2^h u+m-s_2,\ i_1=\cdots=i_4=2^h u+m-s_3=2^h u+m-s_4\} \Big{)} \\
&\lesssim& \|A^{(0)}\|^4 \|\Gamma^{(2)}\|_F^7 
           \Big{(}
              \bigp{\sum\limits_{i_1,i_2=1}^{m_1} \|A^{(2^h u+m-i_1+1)}\| \cdot \|A^{(2^h u+m-i_2+1)}\|^3}\sqrt{p} \\
& & \hspace{6em}      
            + \sum\limits_{i_1,i_3=1}^{m_1} \|A^{(2^h u+m-i_3+1)}\|^4 \\
& & \hspace{6em}            
            + \sum\limits_{i_1=1}^{m_1} \sum\limits_{s_1=0}^{\infty} \|A^{(s_1+1)}\|^2 \|A^{(2^h u+m-i_1+1)}\|^2
           \Big{)} \\
&\lesssim& \lrp{\rho^{4(2^h u+m_2)}(\sqrt{p} + m_1) + \rho^{2(2^h u+m_2)}} \|\Gamma^{(2)}\|_F^7.
\EEqn

As for $S_{1,3}$, we use the result derived in Lemma \ref{Lemma:LP-aux-3-3} and obtain that
\BEqn
& & \lrabs{\sum\limits_{i_1,i_2,i_3,i_4=1}^{m_1}
           \sum\limits_{s_1,s_2,s_3,s_4=0}^{\infty} 
           \sum\limits_{\ell_1,\ell_2,\ell_3,\ell_4=1}^{p} S_{1,3}} \\
&\lesssim& \sum\limits_{i_1,i_2,i_3,i_4=1}^{m_1}
           \sum\limits_{s_1,s_2,s_3,s_4=0}^{\infty}
           \|A^{(0)}\|^4 \bigp{ \prod\limits_{i=1}^{4} \|A^{(s_i+1)}\| } \|\Gamma^{(2)}\|_F^8 \\
& & \hspace{7em} \times
         \Big{(}
           \bone\{i_1=i_2=i_3,\ i_4=2^h u+m-s_1=\cdots=2^h u+m-s_4\} \\
& & \hspace{8em}
         + \bone\{i_1=i_2=2^h u+m-s_1,\ i_3=i_4=2^h u+m-s_2=\cdots=2^h u+m-s_4\} \\
& & \hspace{8em}
         + \bone\{i_1=2^h u+m-s_1=2^h u+m-s_2,\ i_2=\cdots=i_4=2^h u+m-s_3=2^h u+m-s_4\} \\
& & \hspace{8em}
         + \bone\{2^h u+m-s_1=\cdots=2^h u+m-s_3,\ i_1=\cdots=i_4=2^h u+m-s_4\} \Big{)} \\
&\lesssim& \|A^{(0)}\|^4 \|\Gamma^{(2)}\|_F^8
           \Big{(}
              \bigp{\sum\limits_{i_1,i_4=1}^{m_1} \|A^{(2^h u+m-i_4+1)}\|^4} \\
& & \hspace{6em}      
            + \sum\limits_{i_1,i_3=1}^{m_1} \|A^{(2^h u+m-i_1+1)}\| \cdot \|A^{(2^h u+m-i_3+1)}\|^3 \\
& & \hspace{6em}            
            + \sum\limits_{i_1,i_2=1}^{m_1} \|A^{(2^h u+m-i_1+1)}\|^2 \cdot\|A^{(2^h u+m-i_2+1)}\|^2 \\
& & \hspace{6em}           
            + \sum\limits_{i_1=1}^{m_1} \sum\limits_{s_1=0}^{\infty} \|A^{(s_1+1)}\|^3 \|A^{(2^h u+m-i_1+1)}\|
           \Big{)} \\
&\lesssim& \lrp{\rho^{4(2^h u+m_2)}m_1 + \rho^{2^h u+m_2}} \|\Gamma^{(2)}\|_F^8.
\EEqn

Similarly, it follows from Lemma \ref{Lemma:LP-aux-3-4} that
\BEqn
& & \lrabs{\sum\limits_{i_1,i_2,i_3,i_4=1}^{m_1}
           \sum\limits_{s_1,s_2,s_3,s_4=0}^{\infty} 
           \sum\limits_{\ell_1,\ell_2,\ell_3,\ell_4=1}^{p} S_{1,4}} \\
&\lesssim& \sum\limits_{i_1,i_2,i_3,i_4=1}^{m_1}
           \sum\limits_{s_1,s_2,s_3,s_4=0}^{\infty}
           \|A^{(0)}\|^4 \bigp{ \prod\limits_{i=1}^{4} \|A^{(s_i+1)}\| } \|\Gamma^{(2)}\|_F^4 \\
& & \hspace{2em} \times
         \Big{(}
           \bone\{i_1=\cdots=i_4,\ 2^h u+m-s_1=\cdots=2^h u+m-s_4\} \\
& & \hspace{3em}
         + \bone\{i_1=i_2=i_3=2^h u+m-s_4,\ i_4=2^h u+m-s_1=\cdots=2^h u+m-s_3\}  \\
& & \hspace{3em}
         + \bone\{i_1=i_2=i_3=2^h u+m-s_1,\ i_4=2^h u+m-s_2=\cdots=2^h u+m-s_4\} \|\Gamma^{(2)}\|_F^4 \\
& & \hspace{3em}
         + \bone\{i_1=i_2=2^h u+m-s_3=2^h u+m-s_4,\ i_3=i_4=2^h u+m-s_1=2^h u+m-s_2\} \\
& & \hspace{3em}
         + \bone\{i_1=i_2=2^h u+m-s_1=2^h u+m-s_2,\ i_3=i_4=2^h u+m-s_3=2^h u+m-s_4\} \|\Gamma^{(2)}\|_F^4
         \Big{)} \\
&\lesssim& \|A^{(0)}\|^4 \|\Gamma^{(2)}\|_F^4
           \Big{(}
              \bigp{\sum\limits_{i_1=1}^{m_1} \sum\limits_{s_1=0}^{\infty} \|A^{(s_2+1)}\|^4} \\
& & \hspace{6em}      
            + \sum\limits_{i_1,i_4=1}^{m_1} \|A^{(2^h u+m-i_1+1)}\| \cdot \|A^{(2^h u+m-i_4+1)}\|^3 \\
& & \hspace{6em}      
            + \sum\limits_{i_1,i_4=1}^{m_1} \|A^{(2^h u+m-i_1+1)}\| \cdot \|A^{(2^h u+m-i_4+1)}\|^3 \|\Gamma^{(2)}\|_F^4 \\
& & \hspace{6em}            
            + \sum\limits_{i_1,i_3=1}^{m_1} \|A^{(2^h u+m-i_1+1)}\|^2 \cdot\|A^{(2^h u+m-i_3+1)}\|^2 \\
& & \hspace{6em}            
            + \sum\limits_{i_1,i_3=1}^{m_1} \|A^{(2^h u+m-i_1+1)}\|^2 \cdot\|A^{(2^h u+m-i_3+1)}\|^2 \|\Gamma^{(2)}\|_F^4
           \Big{)} \\
&\lesssim& \rho^{4(2^h u+m_2)} \|\Gamma^{(2)}\|_F^8 + m_1 \|\Gamma^{(2)}\|_F^4.
\EEqn

To study $S_{1,5}$, we apply Lemma \ref{Lemma:LP-aux-3-5} and it holds that
\BEqn
& & \lrabs{\sum\limits_{i_1,i_2,i_3,i_4=1}^{m_1}
           \sum\limits_{s_1,s_2,s_3,s_4=0}^{\infty} 
           \sum\limits_{\ell_1,\ell_2,\ell_3,\ell_4=1}^{p} S_{1,5}} \\
&\lesssim& \sum\limits_{i_1,i_2,i_3,i_4=1}^{m_1}
           \sum\limits_{s_1,s_2,s_3,s_4=0}^{\infty}
           \|A^{(0)}\|^4 \bigp{ \prod\limits_{i=1}^{4} \|A^{(s_i+1)}\| } \|\Gamma^{(2)}\|_F^4 \\
& & \hspace{2em} \times
         \Big{(}
           \bone\{i_1=2^h u+m-s_1,\ i_2=i_3=i_4,\ 2^h u+m-s_2=2^h u+m-s_3=2^h u+m-s_4\} \sqrt{p} \\
& & \hspace{3em}
         + \bone\{i_1=2^h u+m-s_1,\ i_2=i_3=2^h u+m-s_4,\ i_4=2^h u+m-s_2=2^h u+m-s_3\} \sqrt{p} \\
& & \hspace{3em}
         + \bone\{i_1=2^h u+m-s_1,\ i_2=i_3=2^h u+m-s_2,\ i_4=2^h u+m-s_3=2^h u+m-s_4\} \|\Gamma^{(2)}\|_F^3 \sqrt{p} \\
& & \hspace{3em}
         + \bone\{i_1=i_2,\ i_3=i_4=2^h u+m-s_1,\ 2^h u+m-s_2=2^h u+m-s_3=2^h u+m-s_4\} \\
& & \hspace{3em}
         + \bone\{i_1=i_2,\ i_3=2^h u+m-s_1=2^h u+m-s_2,\ i_4=2^h u+m-s_3=2^h u+m-s_4\} \\    
& & \hspace{3em}
         + \bone\{i_1=i_2,\ i_3=2^h u+m-s_1=2^h u+m-s_3,\ i_4=2^h u+m-s_2=2^h u+m-s_4\} \|\Gamma^{(2)}\|_F^3 \\
& & \hspace{3em}
         + \bone\{i_1=2^h u+m-s_2,\ i_2=i_3=i_4,\ 2^h u+m-s_1=2^h u+m-s_3=2^h u+m-s_4\} \\
& & \hspace{3em}
         + \bone\{i_1=2^h u+m-s_2,\ i_2=i_3=2^h u+m-s_1,\ i_4=2^h u+m-s_3=2^h u+m-s_4\} \\
& & \hspace{3em}
         + \bone\{i_1=2^h u+m-s_2,\ i_2=i_3=2^h u+m-s_3,\ i_4=2^h u+m-s_1=2^h u+m-s_4\} \|\Gamma^{(2)}\|_F^3 \\
& & \hspace{3em}
         + \bone\{2^h u+m-s_1=2^h u+m-s_2,\ i_1=i_2=i_3,\ i_4=2^h u+m-s_3=2^h u+m-s_4\} \\
& & \hspace{3em}
         + \bone\{2^h u+m-s_1=2^h u+m-s_2,\ i_1=i_2=2^h u+m-s_3,\ i_3=i_4=2^h u+m-s_4\} \\
& & \hspace{3em}
         + \bone\{2^h u+m-s_1=2^h u+m-s_2,\ i_1=i_3=2^h u+m-s_3,\ i_2=i_4=2^h u+m-s_4\} \|\Gamma^{(2)}\|_F^3
         \Big{)} \\
&\lesssim& \|A^{(0)}\|^4 \|\Gamma^{(2)}\|_F^4
           \Big{(}
              \bigp{\sum\limits_{i_1,i_2=1}^{m_1} \sum\limits_{s_2=0}^{\infty} \|A^{(2^h u+m-i_1+1)}\| \|A^{(s_2+1)}\|^3 \sqrt{p}} \\
& & \hspace{6em}      
            + \sum\limits_{i_1,i_2,i_4=1}^{m_1} \|A^{(2^h u+m-i_1+1)}\| \cdot \|A^{(2^h u+m-i_2+1)}\| \cdot \|A^{(2^h u+m-i_4+1)}\|^2 \sqrt{p} \\
& & \hspace{6em}      
            + \sum\limits_{i_1,i_2,i_4=1}^{m_1} \|A^{(2^h u+m-i_1+1)}\| \cdot \|A^{(2^h u+m-i_2+1)}\| \cdot \|A^{(2^h u+m-i_4+1)}\|^2 \|\Gamma^{(2)}\|_F^3 \sqrt{p} \\
& & \hspace{6em}            
            + \sum\limits_{i_1,i_3=1}^{m_1} \sum\limits_{s_2=0}^{\infty} \|A^{(2^h u+m-i_3+1)}\| \cdot \|A^{(s_2+1)}\|^3 \\
& & \hspace{6em}            
            + \sum\limits_{i_1,i_3,i_4=1}^{m_1} \|A^{(2^h u+m-i_3+1)}\|^2 \cdot \|A^{(2^h u+m-i_4+1)}\|^2 \\    
& & \hspace{6em}            
            + \sum\limits_{i_1,i_3,i_4=1}^{m_1} \|A^{(2^h u+m-i_3+1)}\|^2 \cdot \|A^{(2^h u+m-i_4+1)}\|^2 \|\Gamma^{(2)}\|_F^3 \\
& & \hspace{6em}            
            + \sum\limits_{i_1,i_2=1}^{m_1} \sum\limits_{s_1=0}^{\infty} \|A^{(2^h u+m-i_1+1)}\| \cdot \|A^{(s_1+1)}\|^3 \\            
& & \hspace{6em}            
            + \sum\limits_{i_1,i_2,i_4=1}^{m_1} \|A^{(2^h u+m-i_1+1)}\| \cdot \|A^{(2^h u+m-i_2+1)}\| \cdot \|A^{(2^h u+m-i_4+1)}\|^2 \\ 
& & \hspace{6em}            
            + \sum\limits_{i_1,i_2,i_4=1}^{m_1} \|A^{(2^h u+m-i_1+1)}\| \cdot \|A^{(2^h u+m-i_2+1)}\| \cdot \|A^{(2^h u+m-i_4+1)}\|^2 \|\Gamma^{(2)}\|_F^3 \\             
& & \hspace{6em}            
            + \sum\limits_{i_1,i_4=1}^{m_1} \sum\limits_{s_1=0}^{\infty} \|A^{(s_1+1)}\|^2 \|A^{(2^h u+m-i_4+1)}\|^2 \\  
& & \hspace{6em}            
            + \sum\limits_{i_1,i_3=1}^{m_1} \sum\limits_{s_1=0}^{\infty} \|A^{(s_1+1)}\|^2 \|A^{(2^h u+m-i_1+1)}\|^2 \cdot \|A^{(2^h u+m-i_3+1)}\| \\ 
& & \hspace{6em}            
            + \sum\limits_{i_1,i_2=1}^{m_1} \sum\limits_{s_1=0}^{\infty} \|A^{(s_1+1)}\|^2 \|A^{(2^h u+m-i_1+1)}\| \cdot \|A^{(2^h u+m-i_2+1)}\| \|\Gamma^{(2)}\|_F^3
           \Big{)} \\
&\lesssim& \rho^{2^h u+m_2} m_1 \sqrt{p} \|\Gamma^{(2)}\|_F^4 + \rho^{4(2^h u+m_2)} \sqrt{p} \|\Gamma^{(2)}\|_F^7.
\EEqn

As for $S_{1,6}$, it follows from Lemma \ref{Lemma:LP-aux-3-6} that
\BEqn
& & \lrabs{\sum\limits_{i_1,i_2,i_3,i_4=1}^{m_1}
           \sum\limits_{s_1,s_2,s_3,s_4=0}^{\infty} 
           \sum\limits_{\ell_1,\ell_2,\ell_3,\ell_4=1}^{p} S_{1,6}} \\
&\lesssim& \sum\limits_{i_1,i_2,i_3,i_4=1}^{m_1}
           \sum\limits_{s_1,s_2,s_3,s_4=0}^{\infty}
           \|A^{(0)}\|^4 \bigp{ \prod\limits_{i=1}^{4} \|A^{(s_i+1)}\| } \|\Gamma^{(2)}\|_F^4 \\
& & \hspace{2em} \times
         \Big{(}
           \bone\{i_1=2^h u+m-s_1,\ i_2=2^h u+m-s_2,\ i_3=i_4=2^h u+m-s_3=2^h u+m-s_4\} \|\Gamma^{(2)}\|_F^2 p \\
& & \hspace{3em}
         + \bone\{i_1=2^h u+m-s_1,\ i_2=i_3,\ i_4=2^h u+m-s_2=2^h u+m-s_3=2^h u+m-s_4\} \|\Gamma^{(2)}\|_F^2 \sqrt{p} \\
& & \hspace{3em}
         + \bone\{i_1=2^h u+m-s_1,\ i_2=2^h u+m-s_3,\ i_3=i_4=2^h u+m-s_2=2^h u+m-s_4\} \|\Gamma^{(2)}\|_F^2 \sqrt{p} \\         
& & \hspace{3em}
         + \bone\{i_1=2^h u+m-s_1,\ 2^h u+m-s_2=2^h u+m-s_3,\ i_2=i_3=i_4=2^h u+m-s_4\} \|\Gamma^{(2)}\|_F^2 \sqrt{p} \\
& & \hspace{3em}
         + \bone\{i_1=i_2,\ i_3=2^h u+m-s_1,\ i_4=2^h u+m-s_2=2^h u+m-s_3=2^h u+m-s_4\} \|\Gamma^{(2)}\|_F^2 \\ 
& & \hspace{3em}
         + \bone\{i_1=i_2,\ 2^h u+m-s_1=2^h u+m-s_2,\ i_3=i_4=2^h u+m-s_3=2^h u+m-s_4\} \|\Gamma^{(2)}\|_F^2 \\
& & \hspace{3em}
         + \bone\{i_1=2^h u+m-s_2,\ i_2=2^h u+m-s_1,\ i_3=i_4=2^h u+m-s_3=2^h u+m-s_4\} \|\Gamma^{(2)}\|_F^2 \\
& & \hspace{3em}
         + \bone\{i_1=2^h u+m-s_2,\ 2^h u+m-s_1=2^h u+m-s_3,\ i_2=i_3=i_4=2^h u+m-s_4\} \|\Gamma^{(2)}\|_F^2 \\
& & \hspace{3em}
         + \bone\{i_1=i_2,\ i_3=i_4,\ 2^h u+m-s_1=2^h u+m-s_2=2^h u+m-s_3=2^h u+m-s_4\} \\
& & \hspace{3em}
         + \bone\{i_1=i_2,\ i_3=2^h u+m-s_4,\ i_4=2^h u+m-s_1=2^h u+m-s_2=2^h u+m-s_3\} \\
& & \hspace{3em}
         + \bone\{i_1=i_2,\ 2^h u+m-s_3=2^h u+m-s_4,\ i_3=i_4=2^h u+m-s_1=2^h u+m-s_2\} \\         
& & \hspace{3em}
         + \bone\{i_1=2^h u+m-s_2,\ i_3=2^h u+m-s_4,\ i_2=i_4=2^h u+m-s_1=2^h u+m-s_3\} \\
& & \hspace{3em}
         + \bone\{i_1=2^h u+m-s_2,\ 2^h u+m-s_3=2^h u+m-s_4,\ i_2=i_3=i_4=2^h u+m-s_1\} \\         
& & \hspace{3em}
         + \bone\{2^h u+m-s_1=2^h u+m-s_2,\ 2^h u+m-s_3=2^h u+m-s_4,\ i_1=i_2=i_3=i_4\} 
         \Big{)} \\
&\lesssim& \|A^{(0)}\|^4 \|\Gamma^{(2)}\|_F^4    
         \Big{(}
           \sum\limits_{i_1,i_2,i_3=1}^{m_1} \|A^{(2^h u+m-i_1+1)}\| \cdot \|A^{(2^h u+m-i_2+1)}\| \cdot \|A^{(2^h u+m-i_3+1)}\|^2 \|\Gamma^{(2)}\|_F^2 p \\
& & \hspace{6em}
         + \sum\limits_{i_1,i_2,i_4=1}^{m_1} \|A^{(2^h u+m-i_1+1)}\| \cdot \|A^{(2^h u+m-i_4+1)}\|^3 \|\Gamma^{(2)}\|_F^2 \sqrt{p} \\
& & \hspace{6em}
         + \sum\limits_{i_1,i_2,i_3=1}^{m_1} \|A^{(2^h u+m-i_1+1)}\| \cdot \|A^{(2^h u+m-i_2+1)}\| \cdot \|A^{(2^h u+m-i_3+1)}\|^2 \|\Gamma^{(2)}\|_F^2 \sqrt{p} \\         
& & \hspace{6em}
         + \sum\limits_{i_1,i_2=1}^{m_1} \sum\limits_{s_2=0}^{\infty} \|A^{(2^h u+m-i_1+1)}\| \cdot \|A^{(s_2+1)}\|^2 \cdot \|A^{(2^h u+m-i_2+1)}\| \|\Gamma^{(2)}\|_F^2 \sqrt{p} \\
& & \hspace{6em}
         + \sum\limits_{i_1,i_3,i_4=1}^{m_1} \|A^{(2^h u+m-i_3+1)}\| \cdot \|A^{(2^h u+m-i_4+1)}\|^3 \|\Gamma^{(2)}\|_F^2 \\ 
& & \hspace{6em}
         + \sum\limits_{i_1,i_3=1}^{m_1} \sum\limits_{s_1=0}^{\infty} \|A^{(s_1+1)}\|^2 \|A^{(2^h u+m-i_3+1)}\|^2 \|\Gamma^{(2)}\|_F^2 \\
& & \hspace{6em}
         + \sum\limits_{i_1,i_2,i_3=1}^{m_1} \|A^{(2^h u+m-i_1+1)}\| \cdot \|A^{(2^h u+m-i_2+1)}\| \cdot \|A^{(2^h u+m-i_3+1)}\|^2 \|\Gamma^{(2)}\|_F^2 \\
& & \hspace{6em}
         + \sum\limits_{i_1,i_2=1}^{m_1} \sum\limits_{s_1=0}^{\infty} \|A^{(2^h u+m-i_1+1)}\| \cdot  \|A^{(s_1+1)}\|^2 \|A^{(2^h u+m-i_2+1)}\| \|\Gamma^{(2)}\|_F^2 \\
& & \hspace{6em}
         + \sum\limits_{i_1,i_3=1}^{m_1} \sum\limits_{s_1=0}^{\infty} \|A^{(s_1+1)}\|^4 \\
& & \hspace{6em}
         + \sum\limits_{i_1,i_3,i_4=1}^{m_1} \|A^{(2^h u+m-i_3+1)}\| \cdot \|A^{(2^h u+m-i_4+1)}\|^3 \\
& & \hspace{6em}
         + \sum\limits_{i_1,i_3=1}^{m_1} \sum\limits_{s_3=0}^{\infty} \|A^{(2^h u+m-i_3+1)}\|^2 \cdot  \|A^{(s_3+1)}\|^2 \\         
& & \hspace{6em}
         + \sum\limits_{i_1,i_2,i_3=1}^{m_1} \|A^{(2^h u+m-i_1+1)}\| \cdot \|A^{(2^h u+m-i_2+1)}\|^2 \cdot \|A^{(2^h u+m-i_3+1)}\| \\
& & \hspace{6em}
         + \sum\limits_{i_1,i_2=1}^{m_1} \sum\limits_{s_3=0}^{\infty} \|A^{(2^h u+m-i_1+1)}\| \cdot \|A^{(2^h u+m-i_2+1)}\| \cdot \|A^{(s_3+1)}\|^2 \\         
& & \hspace{6em}
         + \sum\limits_{i_1=1}^{m_1} \sum\limits_{s_1,s_3=0}^{\infty} \|A^{(s_1+1)}\|^2 \cdot \|A^{(s_3+1)}\|^2
         \Big{)} \\
&\lesssim& \rho^{4(2^h u+m_2)} (p + m_1 \sqrt{p}) \|\Gamma^{(2)}\|_F^6 
+ \rho^{2(2^h u+m_2)} (\sqrt{p} + m_1) \|\Gamma^{(2)}\|_F^6 
+ m_1^2 \|\Gamma^{(2)}\|_F^4.
\EEqn

It remains to consider $S_{1,7}$. By Lemma \ref{Lemma:LP-aux-3-7}, we have that
\BEqn
& & \lrabs{\sum\limits_{i_1,i_2,i_3,i_4=1}^{m_1}
           \sum\limits_{s_1,s_2,s_3,s_4=0}^{\infty} 
           \sum\limits_{\ell_1,\ell_2,\ell_3,\ell_4=1}^{p} S_{1,7}} \\
&\lesssim& \sum\limits_{i_1,i_2,i_3,i_4=1}^{m_1}
           \sum\limits_{s_1,s_2,s_3,s_4=0}^{\infty}
           \|A^{(0)}\|^4 \bigp{ \prod\limits_{i=1}^{4} \|A^{(s_i+1)}\| } \|\Gamma^{(2)}\|_F^4 \\
& & \hspace{4em} \times
         \Big{(}
           \bone\{i_1=2^h u+m-s_1,\ i_2=2^h u+m-s_2,\ i_3=2^h u+m-s_3,\ i_4=2^h u+m-s_4\} p^2 \\
& & \hspace{5em}
         + \bone\{i_1=2^h u+m-s_1,\ i_2=2^h u+m-s_2,\ i_3=i_4,\ 2^h u+m-s_3=2^h u+m-s_4\} p \\
& & \hspace{5em}
         + \bone\{i_1=2^h u+m-s_1,\ i_2=2^h u+m-s_2,\ i_3=2^h u+m-s_4,\ i_4=2^h u+m-s_3\} p \\  
& & \hspace{5em}
         + \bone\{i_1=2^h u+m-s_1,\ i_2=i_3,\ i_4=2^h u+m-s_2,\ 2^h u+m-s_3=2^h u+m-s_4\} \sqrt{p} \\
& & \hspace{5em}
         + \bone\{i_1=2^h u+m-s_1,\ i_2=2^h u+m-s_3,\ i_3=2^h u+m-s_4,\ i_4=2^h u+m-s_2\} \sqrt{p} \\
& & \hspace{5em}
         + \bone\{i_1=i_2,\ i_3=i_4,\ 2^h u+m-s_1=2^h u+m-s_2,\ 2^h u+m-s_3=2^h u+m-s_4\} \\
& & \hspace{5em}
         + \bone\{i_1=i_2,\ i_3=2^h u+m-s_1,\ i_4=2^h u+m-s_2,\ 2^h u+m-s_3=2^h u+m-s_4\} \\
& & \hspace{5em}
         + \bone\{i_1=2^h u+m-s_2,\ i_2=2^h u+m-s_3,\ i_3=2^h u+m-s_4,\ i_4=2^h u+m-s_1\}
         \Big{)} \\
&\lesssim& \|A^{(0)}\|^4 \|\Gamma^{(2)}\|_F^4
         \Big{(}
           \sum\limits_{i_1,i_2,i_3,i_4=1}^{m_1} \|A^{(2^h u+m_1-i_1+1)}\| \cdot \|A^{(2^h u+m_1-i_2+1)}\| \cdot \|A^{(2^h u+m_1-i_3+1)}\| \cdot \|A^{(2^h u+m_1-i_4+1)}\| p^2 \\
& & \hspace{4em}
         + \sum\limits_{i_1,i_2,i_3=1}^{m_1} \sum\limits_{s_3=0}^{\infty} \|A^{(2^h u+m_1-i_1+1)}\| \cdot \|A^{(2^h u+m_1-i_2+1)}\| \cdot \|A^{(s_3+1)}\|^2 p \\
& & \hspace{4em}
         + \sum\limits_{i_1,i_2,i_3,i_4=1}^{m_1} \|A^{(2^h u+m_1-i_1+1)}\| \cdot \|A^{(2^h u+m_1-i_2+1)}\| \cdot \|A^{(2^h u+m_1-i_3+1)}\| \cdot \|A^{(2^h u+m_1-i_4+1)}\| p \\  
& & \hspace{4em}
         + \sum\limits_{i_1,i_2,i_4=1}^{m_1} \sum\limits_{s_3=0}^{\infty} \|A^{(2^h u+m_1-i_1+1)}\| \cdot \|A^{(2^h u+m_1-i_4+1)}\| \cdot \|A^{(s_3+1)}\|^2 \sqrt{p} \\
& & \hspace{4em}
         + \sum\limits_{i_1,i_2,i_3,i_4=1}^{m_1} \|A^{(2^h u+m_1-i_1+1)}\| \cdot \|A^{(2^h u+m_1-i_2+1)}\| \cdot \|A^{(2^h u+m_1-i_3+1)}\| \cdot \|A^{(2^h u+m_1-i_4+1)}\| \sqrt{p} \\
& & \hspace{4em}
         + \sum\limits_{i_1,i_3=1}^{m_1} \sum\limits_{s_1,s_3=0}^{\infty} \|A^{(s_1+1)}\|^2 \cdot \|A^{(s_3+1)}\|^2 \\
& & \hspace{4em}
         + \sum\limits_{i_1,i_3,i_4=1}^{m_1} \sum\limits_{s_3=0}^{\infty} \|A^{(2^h u+m_1-i_3+1)}\| \cdot \|A^{(2^h u+m_1-i_4+1)}\| \cdot \|A^{(s_3+1)}\|^2 \\
& & \hspace{4em}
         + \sum\limits_{i_1,i_2,i_3,i_4=1}^{m_1} \|A^{(2^h u+m_1-i_1+1)}\| \cdot \|A^{(2^h u+m_1-i_2+1)}\| \cdot \|A^{(2^h u+m_1-i_3+1)}\| \cdot \|A^{(2^h u+m_1-i_4+1)}\|   
         \Big{)} \\
&\lesssim& \lrp{ \rho^{4(2^h u+m_2)} p^2 + \rho^{2(2^h u+m_2)} m_1 p + m_1^2 } \|\Gamma^{(2)}\|_F^4.
\EEqn

In summary, we have that
\BEqn
    \lrabs{S_1}
&\lesssim& \lrp{\rho^{4(2^h u+m_2)}m_1 + \rho^{2^h u+m_2}} \|\Gamma^{(2)}\|_F^8 \\
& & + \rho^{4(2^h u+m_2)}(\sqrt{p} + m_1) \|\Gamma^{(2)}\|_F^7 \\
& & + \lrp{\rho^{4(2^h u+m_2)} (p + m_1 \sqrt{p}) + \rho^{2(2^h u+m_2)} (\sqrt{p} + m_1)} \|\Gamma^{(2)}\|_F^6 \\
& & + \lrp{ \rho^{4(2^h u+m_2)} p^2 + \rho^{2(2^h u+m_2)} m_1 p + \rho^{2^h u+m_2} m_1 \sqrt{p} + m_1^2 } \|\Gamma^{(2)}\|_F^4.
\EEqn
Note that it holds naturally that $\sqrt{p} \lesssim \|\Gamma^{(2)}\|_F \lesssim p$ under Assumption \ref{Assumpt:NullDist_Wn_lp}\ref{Assumpt:NullDist_Wn_Lp-4},\ref{Assumpt:NullDist_Wn_Lp-6}, then we further obtain that
\begin{equation*}
    \lrabs{S_1}
\lesssim \lrp{\rho^{4(2^h u+m_2)}m_1 + \rho^{2^h u+m_2}} \|\Gamma^{(2)}\|_F^8
       + \rho^{2(2^h u+m_2)} m_1 \|\Gamma^{(2)}\|_F^6 
       + \lrp{\rho^{2^h u+m_2} m_1 \sqrt{p} + m_1^2 } \|\Gamma^{(2)}\|_F^4,
\end{equation*}
and similarly, we obtain that
\BEqn
  \lrabs{S_2}
&\lesssim& \lrp{\rho^{4(2^h(u-1)+m_2)}m_1 + \rho^{2^h(u-1)+m_2}} \|\Gamma^{(2)}\|_F^8
       + \rho^{2(2^h(u-1)+m_2)} m_1 \|\Gamma^{(2)}\|_F^6 \\
& &    + \lrp{\rho^{2^h(u-1)+m_2} m_1 \sqrt{p} + m_1^2 } \|\Gamma^{(2)}\|_F^4, \\
  \lrabs{S_3}
&\lesssim& \lrp{\rho^{4(\tilde{N}+2^h u+m_2)}m_1 + \rho^{\tilde{N}+2^h u+m_2}} \|\Gamma^{(2)}\|_F^8
       + \rho^{2(\tilde{N}+2^h u+m_2)} m_1 \|\Gamma^{(2)}\|_F^6 \\
& &    + \lrp{\rho^{\tilde{N}+2^h u+m_2} m_1 \sqrt{p} + m_1^2 } \|\Gamma^{(2)}\|_F^4, 
\EEqn      
and
\BEqn
    \lrabs{S_4}
&\lesssim& \lrp{\rho^{4(\tilde{N}+2^h(u-1)+m_2)}m_1 + \rho^{\tilde{N}+2^h(u-1)+m_2}} \|\Gamma^{(2)}\|_F^8
       + \rho^{2(\tilde{N}+2^h(u-1)+m_2)} m_1 \|\Gamma^{(2)}\|_F^6 \\
& &    + \lrp{\rho^{\tilde{N}+2^h(u-1)+m_2} m_1 \sqrt{p} + m_1^2 } \|\Gamma^{(2)}\|_F^4.
\EEqn

It follows that
\BEqn
& & \left\|\sup\limits_{r\in[\varepsilon,1-\varepsilon]} \lrabs{\sum\limits_{i=1}^{m_1} D_{n+1-i}^{\top} \lrp{\sum\limits_{j=1}^{\lrfl{nr} - \lrfl{n\varepsilon}} R_{j+m}}}\right\|_4 \\
&\lesssim& \sum\limits_{h=0}^{d} \lrp{ \sum\limits_{u=1}^{2^{d-h}} S_1}^{1/4} 
 + \sum\limits_{h=0}^{d} \lrp{ \sum\limits_{u=1}^{2^{d-h}} S_2}^{1/4}
 + \sum\limits_{h=0}^{d} \lrp{ \sum\limits_{u=1}^{2^{d-h}} S_3}^{1/4}
 + \sum\limits_{h=0}^{d} \lrp{ \sum\limits_{u=1}^{2^{d-h}} S_4}^{1/4} \\
&\lesssim& \sum\limits_{h=0}^{d} \lrp{\rho^{m_2} m_1^{1/4} + \rho^{(m_2)/4}} \|\Gamma^{(2)}\|_F^2
         + \sum\limits_{h=0}^{d} \lrp{\rho^{m_2/2} m_1^{1/4}} \|\Gamma^{(2)}\|_F^{3/2} \\
& &      + \sum\limits_{h=0}^{d} \lrp{\rho^{m_2/4} m_1^{1/4} p^{1/8} + 2^{(d-h)/4} m_1^{1/2}} \|\Gamma^{(2)}\|_F \\
&\lesssim& \lrp{ \rho^{m_2} m_1^{1/4} + \rho^{m_2/4} } d \|\Gamma^{(2)}\|_F^2
         + \rho^{m_2/2} m_1^{1/4} d \|\Gamma^{(2)}\|_F^{3/2} 
         + \lrp{ \rho^{m_2/4} m_1^{1/4} p^{1/8} d + 2^{d/4} m_1^{1/2}} \|\Gamma^{(2)}\|_F \\
&\lesssim& \lrp{ \rho^{m_2} n^{1/4} + \rho^{m_2/4} } \log(n) \|\Gamma^{(2)}\|_F^2
         + \rho^{m_2/2} n^{1/4} \log(n) \|\Gamma^{(2)}\|_F^{3/2} \\
& &      + \lrp{ \rho^{m_2/4} n^{1/4} \log(n) p^{1/8}  + n^{3/4} } \|\Gamma^{(2)}\|_F,
\EEqn
where the last step uses the fact that $d\lesssim\log(n)$ and $2^d \lesssim n$.

Recall that under Assumption \ref{Assumpt:NullDist_Wn_lp}\ref{Assumpt:NullDist_Wn_Lp-3}, we have that $\widetilde{N} \sim n \|\Gamma^{(2)}\|_F$, then 
\begin{equation*}
    \left\|\sup\limits_{r\in[\varepsilon,1-\varepsilon]} \lrabs{\frac{1}{\widetilde{N}_n} \sum\limits_{i=1}^{m_1} D_i^{\top} \lrp{\sum\limits_{j=1}^{\lrfl{nr} - \lrfl{n\varepsilon}} R_{j+m}}}\right\|_4 = o(1)
\end{equation*}
as long as
\begin{equation*}
\begin{array}{ll}
    \displaystyle \frac{\rho^{m_2} \log(n)}{n^{3/4}} \|\Gamma^{(2)}\|_F = o(1), \quad & \displaystyle \frac{\rho^{m_2/4} \log(n)}{n} \|\Gamma^{(2)}\|_F = o(1), \\[3mm]
    \displaystyle \frac{\rho^{m_2/2} \log(n)}{n^{3/4}} \|\Gamma^{(2)}\|_F^{1/2} = o(1), \quad & \displaystyle \frac{\rho^{m_2/4} \log(n) p^{1/8} }{n^{3/4}} = o(1).
\end{array}
\end{equation*}
Note that $\rho^{\alpha n} n^{\beta} = o(1)$ holds for any $\alpha,\beta>0$, then the desired bound is achieved as long as $\rho^{m_2/4} \|\Gamma^{(2)}\|_F = o\lrp{\frac{n}{\log(n)}}$.
\end{Proof}

\begin{lemma}\label{Lemma:NullDist_DR-2_lp}
Under Assumption \ref{Assumpt:NullDist_Wn_lp}, it holds that
\begin{equation*}
    \left\|\sup\limits_{r\in[\varepsilon,1-\varepsilon]} \lrabs{\frac{1}{\widetilde{N}_n} \sum\limits_{i=1}^{m_1} D_{n+1-i}^{\top} \lrp{\sum\limits_{j=1}^{\lrfl{nr} - \lrfl{n\varepsilon}} R_{j+m}}}\right\|_4 = o(1)
\end{equation*}
as $n\rightarrow\infty$.
\end{lemma}
\begin{Proof}
Let $d$ denote the integer such that $2^d \le N < 2^{d+1}$ and $\tilde{N} = N-2^d$, then it follows from the fact $\sum\limits_{j=a}^{b} R_{j} = \widetilde{D}_{b} - \widetilde{D}_{a-1}$ and Proposition 1 of \cite{wu2007strong} that
\BEqn
& & \left\|\sup\limits_{r\in[\varepsilon,1-\varepsilon]} \lrabs{\sum\limits_{i=1}^{m_1} D_{n+1-i}^{\top} \lrp{\sum\limits_{j=1}^{\lrfl{nr} - \lrfl{n\varepsilon}} R_{j+m}}}\right\|_4 \\
&\le& \left\|\max\limits_{1\le h\le N} \lrabs{\sum\limits_{i=1}^{m_1} D_{n+1-i}^{\top} \lrp{\sum\limits_{j=1}^{h} R_{j+m}}}\right\|_4 \\
&\le& \left\| \max\limits_{1\le h\le 2^d} \lrabs{\sum\limits_{i=1}^{m_1} D_{n+1-i}^{\top} \lrp{\sum\limits_{j=1}^{h} R_{j+m}}} \right\|_4 
 +    \left\| \max\limits_{N-2^d+1\le h\le N} \lrabs{\sum\limits_{i=1}^{m_1} D_{n+1-i}^{\top} \lrp{\sum\limits_{j=1}^{h} R_{j+m}}} \right\|_4 \\
&\le& \left\| \max\limits_{1\le h\le 2^d} \lrabs{\sum\limits_{i=1}^{m_1} D_{n+1-i}^{\top} \lrp{\sum\limits_{j=1}^{h} R_{j+m}}} \right\|_4 
 +    \left\| \max\limits_{1\le h\le 2^d} \lrabs{\sum\limits_{i=1}^{m_1} D_{n+1-i}^{\top} \lrp{\sum\limits_{j=1}^{h+N-2^d} R_{j+m}}} \right\|_4 \\
&\le& \sum\limits_{h=0}^{d} \lrp{ \sum\limits_{u=1}^{2^{d-h}} \left\| \sum\limits_{i=1}^{m_1} D_{n+1-i}^{\top} \lrp{\sum\limits_{j=2^h(u-1)+1}^{2^h u} R_{j+m}} \right\|_4^4}^{1/4} \\
& & + \sum\limits_{h=0}^{d} \lrp{ \sum\limits_{u=1}^{2^{d-h}} \left\| \sum\limits_{i=1}^{m_1} D_{n+1-i}^{\top} \lrp{\sum\limits_{j=N-2^d+2^h(u-1)+1}^{N-2^d+2^h u} R_{j+m}} \right\|_4^4}^{1/4} \\
&\lesssim& \sum\limits_{h=0}^{d} \lrp{ \sum\limits_{u=1}^{2^{d-h}} S_1}^{1/4} 
 + \sum\limits_{h=0}^{d} \lrp{ \sum\limits_{u=1}^{2^{d-h}} S_2}^{1/4}
 + \sum\limits_{h=0}^{d} \lrp{ \sum\limits_{u=1}^{2^{d-h}} S_3}^{1/4}
 + \sum\limits_{h=0}^{d} \lrp{ \sum\limits_{u=1}^{2^{d-h}} S_4}^{1/4},
\EEqn
where
\BEqn
    S_1 
&=& \sum\limits_{i_1,i_2,i_3,i_4=1}^{m_1} \blre{D_{n+1-i1}^{\top} \widetilde{D}_{2^h u+m} D_{n+1-i2}^{\top} \widetilde{D}_{2^h u+m} D_{n+1-i3}^{\top} \widetilde{D}_{2^h u+m} D_{n+1-i4}^{\top} \widetilde{D}_{2^h u+m}}, \\
    S_2
&=& \sum\limits_{i_1,i_2,i_3,i_4=1}^{m_1} \blre{D_{n+1-i1}^{\top} \widetilde{D}_{2^h(u-1)+m} D_{n+1-i2}^{\top} \widetilde{D}_{2^h(u-1)+m} D_{n+1-i3}^{\top} \widetilde{D}_{2^h(u-1)+m} D_{n+1-i4}^{\top} \widetilde{D}_{2^h(u-1)+m}}, \\
    S_3
&=& \sum\limits_{i_1,i_2,i_3,i_4=1}^{m_1} \blre{D_{n+1-i1}^{\top} \widetilde{D}_{\tilde{N}+2^h u+m} D_{n+1-i2}^{\top} \widetilde{D}_{\tilde{N}+2^h u+m} D_{n+1-i3}^{\top} \widetilde{D}_{\tilde{N}+2^h u+m} D_{n+1-i4}^{\top} \widetilde{D}_{\tilde{N}+2^h u+m}},
\EEqn
and
\BEqn
    S_4
&=& \sum\limits_{i_1,i_2,i_3,i_4=1}^{m_1} 
    \be\left[
      D_{n+1-i1}^{\top} \widetilde{D}_{\tilde{N}+2^h(u-1)+m} D_{n+1-i2}^{\top} \widetilde{D}_{\tilde{N}+2^h(u-1)+m} 
    \right. \\
& & \hspace{6em} \times
    \left.
      D_{n+1-i3}^{\top} \widetilde{D}_{\tilde{N}+2^h(u-1)+m} D_{n+1-i4}^{\top} \widetilde{D}_{\tilde{N}+2^h(u-1)+m}
    \right].
\EEqn
Note that for $1\le u\le 2^{d-h}$ and $1\le i_1,i_2,i_3,i_4\le m_1$, it holds that $\{D_{n+1-i_1}, D_{n+1-i_2}, D_{n+1-i_3}, D_{n+1-i_4}\}$ and $\{\widetilde{D}_{2^h u+m}, \widetilde{D}_{2^h(u-1)+m}, \widetilde{D}_{\tilde{N}+2^h u+m}, \widetilde{D}_{\tilde{N}+2^h(u-1)+m}\}$ are independent, then it follows that
\BEqn
& & S_1 \\
&=& \sum\limits_{i_1,i_2,i_3,i_4=1}^{m_1} \sum\limits_{s_1,s_2,s_3,s_4=0}^{\infty}
    \be 
    \left[ \bigp{A^{(0)} \varepsilon_{n+1-i_1}}^{\top} \bigp{A^{(s_1+1)} \varepsilon_{2^h u+m-s_1}} 
           \bigp{A^{(0)} \varepsilon_{n+1-i_2}}^{\top} \bigp{A^{(s_2+1)} \varepsilon_{2^h u+m-s_2}}
    \right. \\
& & \hspace{11em} 
    \left. \bigp{A^{(0)} \varepsilon_{n+1-i_3}}^{\top} \bigp{A^{(s_3+1)} \varepsilon_{2^h u+m-s_3}}
           \bigp{A^{(0)} \varepsilon_{n+1-i_4}}^{\top} \bigp{A^{(s_4+1)} \varepsilon_{2^h u+m-s_4}}
    \right] \\
&=&  \sum\limits_{i_1,i_2,i_3,i_4=1}^{m_1} \sum\limits_{s_1,s_2,s_3,s_4=0}^{\infty} \sum\limits_{\ell_1,\ell_2,\ell_3,\ell_4=1}^{p} \sum\limits_{k_1,\cdots,k_8=1}^{p} 
     A_{\ell_1,k_1}^{(0)} A_{\ell_1,k_2}^{(s_1+1)} A_{\ell_2,k_3}^{(0)} A_{\ell_2,k_4}^{(s_2+1)} A_{\ell_3,k_5}^{(0)} A_{\ell_3,k_6}^{(s_3+1)} A_{\ell_4,k_7}^{(0)} A_{\ell_4,k_8}^{(s_4+1)} \\
& & \hspace{4em} \times     
    \blre{\varepsilon_{n+1-i_1,k_1} \varepsilon_{n+1-i_2,k_3} \varepsilon_{n+1-i_3,k_5} \varepsilon_{n+1-i_4,k_7}} \\
& & \hspace{4em} \times  
    \blre{\varepsilon_{2^h u+m-s_1,k_2} \varepsilon_{2^h u+m-s_2,k_4} \varepsilon_{2^h u+m-s_3,k_6} \varepsilon_{2^h u+m-s_4,k_8}} \\
&=&  \sum\limits_{i_1,i_2,i_3,i_4=1}^{m_1} \sum\limits_{s_1,s_2,s_3,s_4=0}^{\infty} \sum\limits_{\ell_1,\ell_2,\ell_3,\ell_4=1}^{p} \sum\limits_{k_1,\cdots,k_8=1}^{p} 
     A_{\ell_1,k_1}^{(0)} A_{\ell_1,k_2}^{(s_1+1)} A_{\ell_2,k_3}^{(0)} A_{\ell_2,k_4}^{(s_2+1)} A_{\ell_3,k_5}^{(0)} A_{\ell_3,k_6}^{(s_3+1)} A_{\ell_4,k_7}^{(0)} A_{\ell_4,k_8}^{(s_4+1)} \\
& & \hspace{4em} \times     
    \big( \bone\{i_1=i_2=i_3=i_4\} \cum(\varepsilon_{0,k_1}, \varepsilon_{0,k_3}, \varepsilon_{0,k_5}, \varepsilon_{0,k_7}) 
        + \bone\{i_1=i_2,i_3=i_4\} \Gamma^{(2)}_{k_1,k_3} \Gamma^{(2)}_{k_5,k_7} \\
& & \hspace{5em} 
        + \bone\{i_1=i_3,i_2=i_4\} \Gamma^{(2)}_{k_1,k_5} \Gamma^{(2)}_{k_3,k_7} 
        + \bone\{i_1=i_4,i_2=i_3\} \Gamma^{(2)}_{k_1,k_7} \Gamma^{(2)}_{k_3,k_5} \big) \\
& & \hspace{4em} \times     
    \big( \bone\{s_1=s_2=s_3=s_4\} \cum(\varepsilon_{0,k_2}, \varepsilon_{0,k_4}, \varepsilon_{0,k_6}, \varepsilon_{0,k_8}) 
        + \bone\{s_1=s_2,s_3=s_4\} \Gamma^{(2)}_{k_2,k_4} \Gamma^{(2)}_{k_6,k_8} \\
& & \hspace{5em} 
        + \bone\{s_1=s_3,s_2=s_4\} \Gamma^{(2)}_{k_2,k_6} \Gamma^{(2)}_{k_4,k_8} 
        + \bone\{s_1=s_4,s_2=s_3\} \Gamma^{(2)}_{k_2,k_8} \Gamma^{(2)}_{k_4,k_6} \big) \\
&\lesssim& \sum\limits_{i_1,i_2,i_3,i_4=1}^{m_1} \sum\limits_{s_1,s_2,s_3,s_4=0}^{\infty} \bone\{i_1=i_2=i_3=i_4\} \bone\{s_1=s_2=s_3=s_4\} \\
& & \hspace{2em} \times
    \sum\limits_{\ell_1,\ell_2,\ell_3,\ell_4=1}^{p} \sum\limits_{k_1,\cdots,k_8=1}^{p} 
    A_{\ell_1,k_1}^{(0)} A_{\ell_1,k_2}^{(s_1+1)} A_{\ell_2,k_3}^{(0)} A_{\ell_2,k_4}^{(s_2+1)} A_{\ell_3,k_5}^{(0)} A_{\ell_3,k_6}^{(s_3+1)} A_{\ell_4,k_7}^{(0)} A_{\ell_4,k_8}^{(s_4+1)} \\
& & \hspace{12em} \times 
    \cum(\varepsilon_{0,k_1}, \varepsilon_{0,k_3}, \varepsilon_{0,k_5}, \varepsilon_{0,k_7}) \cum(\varepsilon_{0,k_2}, \varepsilon_{0,k_4}, \varepsilon_{0,k_6}, \varepsilon_{0,k_8}) \\
& & + \sum\limits_{i_1,i_2,i_3,i_4=1}^{m_1} \sum\limits_{s_1,s_2,s_3,s_4=0}^{\infty} \bone\{i_1=i_2=i_3=i_4\} \bone\{s_1=s_2,s_3=s_4\} \\
& & \hspace{2em} \times
    \sum\limits_{\ell_1,\ell_2,\ell_3,\ell_4=1}^{p} \sum\limits_{k_1,\cdots,k_8=1}^{p} 
    A_{\ell_1,k_1}^{(0)} A_{\ell_1,k_2}^{(s_1+1)} A_{\ell_2,k_3}^{(0)} A_{\ell_2,k_4}^{(s_2+1)} A_{\ell_3,k_5}^{(0)} A_{\ell_3,k_6}^{(s_3+1)} A_{\ell_4,k_7}^{(0)} A_{\ell_4,k_8}^{(s_4+1)} \\
& & \hspace{12em} \times 
    \cum(\varepsilon_{0,k_1}, \varepsilon_{0,k_3}, \varepsilon_{0,k_5}, \varepsilon_{0,k_7}) \Gamma^{(2)}_{k_2,k_4} \Gamma^{(2)}_{k_6,k_8} \\
& & + \sum\limits_{i_1,i_2,i_3,i_4=1}^{m_1} \sum\limits_{s_1,s_2,s_3,s_4=0}^{\infty} \bone\{i_1=i_2,i_3=i_4\} \bone\{s_1=s_2=s_3=s_4\} \\
& & \hspace{2em} \times
    \sum\limits_{\ell_1,\ell_2,\ell_3,\ell_4=1}^{p} \sum\limits_{k_1,\cdots,k_8=1}^{p} 
    A_{\ell_1,k_1}^{(0)} A_{\ell_1,k_2}^{(s_1+1)} A_{\ell_2,k_3}^{(0)} A_{\ell_2,k_4}^{(s_2+1)} A_{\ell_3,k_5}^{(0)} A_{\ell_3,k_6}^{(s_3+1)} A_{\ell_4,k_7}^{(0)} A_{\ell_4,k_8}^{(s_4+1)} \\
& & \hspace{12em} \times 
    \cum(\varepsilon_{0,k_2}, \varepsilon_{0,k_4}, \varepsilon_{0,k_6}, \varepsilon_{0,k_8}) \Gamma^{(2)}_{k_1,k_3} \Gamma^{(2)}_{k_5,k_7} \\
& & + \sum\limits_{i_1,i_2,i_3,i_4=1}^{m_1} \sum\limits_{s_1,s_2,s_3,s_4=0}^{\infty} \bone\{i_1=i_2,i_3=i_4\} \bone\{s_1=s_2,s_3=s_4\} \\
& & \hspace{2em} \times
    \sum\limits_{\ell_1,\ell_2,\ell_3,\ell_4=1}^{p} \sum\limits_{k_1,\cdots,k_8=1}^{p} 
    A_{\ell_1,k_1}^{(0)} A_{\ell_1,k_2}^{(s_1+1)} A_{\ell_2,k_3}^{(0)} A_{\ell_2,k_4}^{(s_2+1)} A_{\ell_3,k_5}^{(0)} A_{\ell_3,k_6}^{(s_3+1)} A_{\ell_4,k_7}^{(0)} A_{\ell_4,k_8}^{(s_4+1)} \Gamma^{(2)}_{k_1,k_3} \Gamma^{(2)}_{k_5,k_7} \Gamma^{(2)}_{k_2,k_4} \Gamma^{(2)}_{k_6,k_8} \\
&\lesssim&  \sum\limits_{i_1=1}^{m_1} \sum\limits_{s_1=0}^{\infty} 
    \lrp{\sum\limits_{\ell_1,\ell_2,\ell_3,\ell_4=1}^{p} \lrp{ \sum\limits_{k_1,k_3,k_5,k_7=1}^{p} 
    A_{\ell_1,k_1}^{(0)} A_{\ell_2,k_3}^{(0)} A_{\ell_3,k_5}^{(0)} A_{\ell_4,k_7}^{(0)}
    \cum(\varepsilon_{0,k_1}, \varepsilon_{0,k_3}, \varepsilon_{0,k_5}, \varepsilon_{0,k_7}) }^2}^{1/2} \\
& & \hspace{3em} \times
    \lrp{\sum\limits_{\ell_1,\ell_2,\ell_3,\ell_4=1}^{p} \lrp{ \sum\limits_{k_2,k_4,k_6,k_8=1}^{p} 
    A_{\ell_1,k_2}^{(s_1+1)} A_{\ell_2,k_4}^{(s_2+1)} A_{\ell_3,k_6}^{(s_3+1)} A_{\ell_4,k_8}^{(s_4+1)} \cum(\varepsilon_{0,k_2}, \varepsilon_{0,k_4}, \varepsilon_{0,k_6}, \varepsilon_{0,k_8}) }^2}^{1/2} \\
& & + \sum\limits_{i_1=1}^{m_1} \sum\limits_{s_1,s_3=0}^{\infty}    
    \lrp{\sum\limits_{\ell_1,\ell_2,\ell_3,\ell_4=1}^{p} \lrp{ \sum\limits_{k_1,k_3,k_5,k_7=1}^{p} 
    A_{\ell_1,k_1}^{(0)} A_{\ell_2,k_3}^{(0)} A_{\ell_3,k_5}^{(0)} A_{\ell_4,k_7}^{(0)}
    \cum(\varepsilon_{0,k_1}, \varepsilon_{0,k_3}, \varepsilon_{0,k_5}, \varepsilon_{0,k_7}) }^2}^{1/2} \\
& & \hspace{5em} \times
    \lrp{\sum\limits_{\ell_1,\ell_2,\ell_3,\ell_4=1}^{p} \lrp{A^{(s_1+1)} \Gamma^{(2)} (A^{(s_1+1)})^{\top}}_{\ell_1,\ell_2}^2 \lrp{A^{(s_3+1)} \Gamma^{(2)} (A^{(s_3+1)})^{\top}}_{\ell_3,\ell_4}^2 }^{1/2} \\
& & + \sum\limits_{i_1,i_3=1}^{m_1} \sum\limits_{s_1=0}^{\infty}    
    \lrp{\sum\limits_{\ell_1,\ell_2,\ell_3,\ell_4=1}^{p} \lrp{ \sum\limits_{k_2,k_4,k_6,k_8=1}^{p} 
    A_{\ell_1,k_2}^{(s_1+1)} A_{\ell_2,k_4}^{(s_1+1)} A_{\ell_3,k_6}^{(s_1+1)} A_{\ell_4,k_8}^{(s_1+1)} \cum(\varepsilon_{0,k_2}, \varepsilon_{0,k_4}, \varepsilon_{0,k_6}, \varepsilon_{0,k_8}) }^2}^{1/2} \\
& & \hspace{5em} \times
    \lrp{\sum\limits_{\ell_1,\ell_2,\ell_3,\ell_4=1}^{p} \lrp{A^{(0)} \Gamma^{(2)} (A^{(0)})^{\top}}_{\ell_1,\ell_2}^2 \lrp{A^{(0)} \Gamma^{(2)} (A^{(0)})^{\top}}_{\ell_3,\ell_4}^2 }^{1/2} \\
& & + \sum\limits_{i_1,i_3=1}^{m_1} \sum\limits_{s_1,s_3=0}^{\infty}
    \tr\lrp{A^{(0)} \Gamma^{(2)} (A^{(0)})^{\top} A^{(s_1+1)} (\Gamma^{(2)})^{\top} (A^{(s_1+1)})^{\top}} \\
& & \hspace{6em} \times
    \tr\lrp{A^{(0)} \Gamma^{(2)} (A^{(0)})^{\top} A^{(s_3+1)} (\Gamma^{(2)})^{\top} (A^{(s_3+1)})^{\top}} \\
&\lesssim& m_1 \sum\limits_{s_1=0}^{\infty} \|A^{(0)}\|^4 \|A^{(s_1+1)}\|^4 \|\Gamma^{(2)}\|_F^4 \\
& & + m_1 \sum\limits_{s_1,s_3=0}^{\infty} \|A^{(0)}\|^4 \|A^{(s_1+1)}\|^2 \|A^{(s_3+1)}\|^2 \|\Gamma^{(2)}\|_F^4 \\
& & + m_1^2 \sum\limits_{s_1=0}^{\infty} \|A^{(0)}\|^4 \|A^{(s_1+1)}\|^4 \|\Gamma^{(2)}\|_F^4 \\
& & + m_1^2 \sum\limits_{s_1,s_3=0}^{\infty} \|A^{(0)}\|^4 \|A^{(s_1+1)}\|^2 \|A^{(s_3+1)}\|^2 \|\Gamma^{(2)}\|_F^4 \\
&\lesssim& m_1^2 \|\Gamma^{(2)}\|_F^4,
\EEqn
where the third to the last step follows from the Cauchy-Schwarz inequality and the second to the last step  is a direct result of Lemma \ref{Lemma:LP-aux-2} and the last step follows from Assumption \ref{Assumpt:NullDist_Wn_lp}\ref{Assumpt:NullDist_Wn_Lp-7}.

Similarly, we can also show that $\max\{S_2,S_3,S_4\} \lesssim m_1^2 \|\Gamma^{(2)}\|_F^4$, and it follows that
\BEqn
& & \left\|\sup\limits_{r\in[\varepsilon,1-\varepsilon]} \lrabs{\sum\limits_{i=1}^{m_1} D_{n+1-i}^{\top} \lrp{\sum\limits_{j=1}^{\lrfl{nr} - \lrfl{n\varepsilon}} R_{j+m}}}\right\|_4 \\
&\lesssim& \sum\limits_{h=0}^{d} \lrp{\sum\limits_{u=1}^{2^{d-h}}  m_1^2 \|\Gamma^{(2)}\|_F^4}^{1/4} 
\lesssim 2^{d/4} m_1^{1/2} \|\Gamma^{(2)}\|_F
= O(n^{3/4} \|\Gamma^{(2)}\|_F),
\EEqn
which further implies the proposed result under Assumption \ref{Assumpt:NullDist_Wn_lp}\ref{Assumpt:NullDist_Wn_Lp-3}.
\end{Proof}
\begin{lemma}\label{Lemma:NullDist_RD_lp}
Under Assumption \ref{Assumpt:NullDist_Wn_lp}, if $\rho^{m_2}\|\Gamma^{(2)}\|_F = o\lrp{\frac{n}{\log(n)}}$, then it holds that
\begin{equation*}
    \left\|\sup\limits_{r\in[\varepsilon,1-\varepsilon]}
    \lrabs{\frac{1}{\widetilde{N}_n} \sum\limits_{i=1}^{m_1} R_i^{\top} \lrp{\sum\limits_{j=1}^{\lrfl{nr} - \lrfl{n\varepsilon}} D_{j+m}}}\right\|_2 = o(1)
\end{equation*}
and 
\begin{equation*}
    \left\|\sup\limits_{r\in[\varepsilon,1-\varepsilon]} \lrabs{\frac{1}{\widetilde{N}_n} \sum\limits_{i=1}^{m_1} R_{n+1-i}^{\top} \lrp{\sum\limits_{j=1}^{\lrfl{nr} - \lrfl{n\varepsilon}} D_{j+m}}}\right\|_2 = o(1)
\end{equation*}
as $n\rightarrow\infty$.
\end{lemma}
\begin{Proof}
It is trivial that there exists $d$ s.t. $2^d \le N < 2^{d+1}$, then applying Proposition 1 of \cite{wu2007strong}, we have
\BEqn
& & \lrnorm{\sup\limits_{r\in[\varepsilon,1-\varepsilon]}
    \lrabs{\sum\limits_{i=1}^{m_1} R_i^{\top} \lrp{\sum\limits_{j=1}^{\lrfl{nr} - \lrfl{n\varepsilon}} D_{j+m}}}}_2 \\
&\le& \lrnorm{\max\limits_{1\le h\le 2^d} \lrabs{\sum\limits_{j=1}^{h} \lrp{\sum\limits_{i=1}^{m_1} R_i}^{\top} D_{j+m}}}_2 + \lrnorm{\max\limits_{N-2^d+1\le h\le N} \lrabs{\sum\limits_{j=1}^{h} \lrp{\sum\limits_{i=1}^{m_1} R_i}^{\top} D_{j+m}}}_2 \\
&\le& \sum\limits_{h=0}^{d} \lrp{\sum\limits_{u=1}^{2^{d-h}} \lrnorm{\sum\limits_{j=2^h(u-1)+1}^{2^h u} \lrp{\widetilde{D}_{m_1} - \widetilde{D}_0}^{\top} D_{j+m}}_2^2}^{1/2} \\
& & + \sum\limits_{h=0}^{d} \lrp{\sum\limits_{u=1}^{2^{d-h}} \lrnorm{\sum\limits_{j=N-2^d+2^h(u-1)+1}^{N-2^d+2^hu} \lrp{\widetilde{D}_{m_1} - \widetilde{D}_0}^{\top} D_{j+m}}_2^2}^{1/2}.
\EEqn
For each $u$, by noting the iid property of $\{D_t\}_{t=0}^{\infty}$, we have 
\BEqn
& & \lrnorm{\sum\limits_{j=2^h(u-1)+1}^{2^h u} \lrp{\widetilde{D}_{m_1} - \widetilde{D}_0}^{\top} D_{j+m}}_2^2 \\
&\lesssim& \sum\limits_{j_1,j_2=2^h(u-1)+1}^{2^h u} \lrp{\blre{\widetilde{D}_{m_1}^{\top} D_{j_1+m} \widetilde{D}_{m_1}^{\top} D_{j_2+m}} + \blre{\widetilde{D}_{0}^{\top} D_{j_1+m} \widetilde{D}_{0}^{\top} D_{j_2+m}}} \\
&=& \sum\limits_{j=2^h(u-1)+1}^{2^h u} \lrp{\tr\lrp{\bbe{\widetilde{D}_{m_1} \widetilde{D}_{m_1}^{\top}} \bbe{D_{j+m} D_{j+m}^{\top}}} + \tr\lrp{\bbe{\widetilde{D}_0 \widetilde{D}_0^{\top}} \bbe{D_{j+m} D_{j+m}^{\top}}}} \\
&=& 2\sum\limits_{j=2^h(u-1)+1}^{2^h u} \tr\lrp{\sum\limits_{s=0}^{\infty} A^{(s+1)} \Gamma^{(2)} \bigp{A^{(s+1)}}^{\top} A^{(0)} \Gamma^{(2)} \bigp{A^{(0)}}^{\top}} \\
&\lesssim& \sum\limits_{j=2^h(u-1)+1}^{2^h u} \sum\limits_{s=0}^{\infty} \|A^{(s+1)}\|^2 \|A^{(0)}\|^2 \|\Gamma^{(2)}\|_F^2 \\
&\lesssim& 2^h \|\Gamma^{(2)}\|_F^2,
\EEqn
where the second to the last step follows from Lemma 9.1 of \cite{wang2020hypothesis} and the last step follows from Assumption \ref{Assumpt:NullDist_Wn_lp}. Similarly, it also holds that
\begin{equation*}
    \lrnorm{\sum\limits_{j=N-2^d+2^h(u-1)+1}^{N-2^d+2^hu} \lrp{\widetilde{D}_{m_1} - \widetilde{D}_0}^{\top} D_{j+m}}_2^2
\lesssim 2^h \|\Gamma^{(2)}\|_F^2.    
\end{equation*}
Therefore, we obtain that
\begin{equation*}
    \lrnorm{\sup\limits_{r\in[\varepsilon,1-\varepsilon]}
    \lrabs{\sum\limits_{i=1}^{m_1} R_i^{\top} \lrp{\sum\limits_{j=1}^{\lrfl{nr} - \lrfl{n\varepsilon}} D_{j+m}}}}_2 
\lesssim d 2^{d/2} \|\Gamma^{(2)}\|_F
\lesssim \sqrt{n} \log(n) \|\Gamma^{(2)}\|_F,
\end{equation*}
then under Assumption \ref{Assumpt:NullDist_Wn_lp}\ref{Assumpt:NullDist_Wn_Lp-3}, we have that
\begin{equation*}
    \left\|\sup\limits_{r\in[\varepsilon,1-\varepsilon]}
    \lrabs{\frac{1}{\widetilde{N}_n} \sum\limits_{i=1}^{m_1} R_i^{\top} \lrp{\sum\limits_{j=1}^{\lrfl{nr} - \lrfl{n\varepsilon}} D_{j+m}}}\right\|_2 
\lesssim \frac{\log(n)}{\sqrt{n}}
= o(1).
\end{equation*}

Now it remains to consider $\left\|\sup\limits_{r\in[\varepsilon,1-\varepsilon]} \lrabs{\frac{1}{\widetilde{N}_n} \sum\limits_{i=1}^{m_1} R_{n+1-i}^{\top} \lrp{\sum\limits_{j=1}^{\lrfl{nr} - \lrfl{n\varepsilon}} D_{j+m}}}\right\|_2$. It follows from Proposition 1 of \cite{wu2007strong} that
\BEqn
& & \lrnorm{\sup\limits_{r\in[\varepsilon,1-\varepsilon]} \lrabs{\sum\limits_{i=1}^{m_1} R_{n+1-i}^{\top} \lrp{\sum\limits_{j=1}^{\lrfl{nr} - \lrfl{n\varepsilon}} D_{j+m}}}}_2 \\
&\le& \lrnorm{\max\limits_{1\le h\le 2^d} \lrabs{\sum\limits_{j=1}^{h} \lrp{\widetilde{D}_{n} - \widetilde{D}_{n-m_1}}^{\top} D_{j+m}}}_2 + \lrnorm{\max\limits_{N-2^d+1\le h\le N} \lrabs{\sum\limits_{j=1}^{h} \lrp{\widetilde{D}_{n} - \widetilde{D}_{n-m_1}}^{\top} D_{j+m}}}_2 \\
&\le& \sum\limits_{h=0}^{d} \lrp{\sum\limits_{u=1}^{2^{d-h}} \lrnorm{\sum\limits_{j=2^h(u-1)+1}^{2^hu} \lrp{\widetilde{D}_{n} - \widetilde{D}_{n-m_1}}^{\top} D_{j+m}}_2^2}^{1/2} \\ 
& & + \sum\limits_{h=0}^{d} \lrp{\sum\limits_{u=1}^{2^{d-h}} \lrnorm{\sum\limits_{j=N-2^d+2^h(u-1)+1}^{N-2^d+2^hu} \lrp{\widetilde{D}_{n} - \widetilde{D}_{n-m_1}}^{\top} D_{j+m}}_2^2}^{1/2}.
\EEqn
Note that by $c_r$ inequality, we have
\BEqn
& & \lrnorm{\sum\limits_{j=2^h(u-1)+1}^{2^hu} \lrp{\widetilde{D}_{n} - \widetilde{D}_{n-m_1}}^{\top} D_{j+m}}_2^2 \\
&\lesssim& \sum\limits_{j_1,j_2=2^h(u-1)+1}^{2^hu} \lrp{\blre{\widetilde{D}_{n}^{\top} D_{j_1+m} \widetilde{D}_{n}^{\top} D_{j_2+m}} + \blre{\widetilde{D}_{n-m_1}^{\top} D_{j_1+m} \widetilde{D}_{n-m_1}^{\top} D_{j_2+m}}}.
\EEqn
Consequently, to investigate the order of the entire term, it suffices to look into $\blre{\widetilde{D}_{n}^{\top} D_{j_1+m} \widetilde{D}_{n}^{\top} D_{j_2+m}}$ and $\blre{\widetilde{D}_{n-m_1}^{\top} D_{j_1+m} \widetilde{D}_{n-m_1}^{\top} D_{j_2+m}}$. 

After some calculations, we have
\BEqn
& & \blre{\widetilde{D}_{n}^{\top} D_{j_1+m} \widetilde{D}_{n}^{\top} D_{j_2+m}} \\
&=& \sum\limits_{\ell_1,\ell_2=1}^{p} \blre{\widetilde{D}_{n,\ell_1} D_{j_1+m,\ell_1} \widetilde{D}_{n,\ell_2} D_{j_2+m,\ell_2}} \\
&=& \sum\limits_{\ell_1,\ell_2=1}^{p} \sum\limits_{s_1,s_2=0}^{\infty} \sum\limits_{k_1,k_2,k_3,k_4=1}^{p} A_{\ell_1,k_1}^{(s_1+1)} A_{\ell_1,k_2}^{(0)} A_{\ell_2,k_3}^{(s_2+1)} A_{\ell_2,k_4}^{(0)} \blre{\varepsilon_{n-s_1,k_1} \varepsilon_{j_1+m,k_2} \varepsilon_{n-s_2,k_3} \varepsilon_{j_2+m,k_4}} \\
&=& \sum\limits_{\ell_1,\ell_2=1}^{p} \sum\limits_{s_1,s_2=0}^{\infty} \sum\limits_{k_1,k_2,k_3,k_4=1}^{p} A_{\ell_1,k_1}^{(s_1+1)} A_{\ell_1,k_2}^{(0)} A_{\ell_2,k_3}^{(s_2+1)} A_{\ell_2,k_4}^{(0)} \\
& & \hspace{10em}
    \times\Big{(}
      \bone\{n-s_1=j_1+m=n-s_2=j_2+m\} \cum(\varepsilon_{0,k_1}, \varepsilon_{0,k_2}, \varepsilon_{0,k_3}, \varepsilon_{0,k_4}) \\
& & \hspace{10em}      
    + \bone\{n-s_1=j_1+m,\ n-s_2=j_2+m\} \Gamma^{(2)}_{k_1,k_2} \Gamma^{(2)}_{k_3,k_4} \\
& & \hspace{10em}    
    + \bone\{n-s_1=n-s_2,\ j_1+m=j_2+m\} \Gamma^{(2)}_{k_1,k_3} \Gamma^{(2)}_{k_2,k_4} \\
& & \hspace{10em}    
    + \bone\{n-s_1=j_2+m,\ n-s_2=j_1+m\} \Gamma^{(2)}_{k_1,k_4} \Gamma^{(2)}_{k_2,k_3}
    \Big{)}
\EEqn
and it follows that
\BEqn
& & \sum\limits_{j_1,j_2=2^h(u-1)+1}^{2^hu} \blre{\widetilde{D}_{n}^{\top} D_{j_1+m} \widetilde{D}_{n}^{\top} D_{j_2+m}} \\
&=& \sum\limits_{s=n-m-2^hu}^{n-m-2^h(u-1)-1} \sum\limits_{k_1,\cdots,k_4=1}^{p} \lrp{\lrp{A^{(s+1)} A^{(0)}}^{\top}}_{k_1,k_2} \lrp{\lrp{A^{(s+1)} A^{(0)}}^{\top}}_{k_3,k_4} \cum(\varepsilon_{0,k_1}, \varepsilon_{0,k_2}, \varepsilon_{0,k_3}, \varepsilon_{0,k_4}) \\
& & + \lrp{\sum\limits_{s=n-m-2^hu}^{n-m-2^h(u-1)-1} \tr\lrp{A^{(s+1)} \Gamma^{(2)} \lrp{A^{(0)}}^{\top}}}^2 \\
& & + \sum\limits_{j=2^h(u-1)+1}^{2^hu} \sum\limits_{s=0}^{\infty} \tr\lrp{A^{(s+1)} \Gamma^{(2)} \lrp{A^{(s+1)}}^{\top} A^{(0)} \Gamma^{(2)} \lrp{A^{(0)}}^{\top}} \\
& & + \sum\limits_{s_1,s_2=n-m-2^hu}^{n-m-2^h(u-1)-1} \tr\lrp{A^{(s_1+1)} \Gamma^{(2)} \lrp{A{(0)}}^{\top} A^{(s_2+1)} \Gamma^{(2)} \lrp{A^{(0)}}^{\top}} \\
&\lesssim& \sum\limits_{s=n-m-2^h u}^{n-m-2^h(u-1)-1} \|\bigp{A^{(s+1)}}^{\top} A^{(0)}\|^2 \sum\limits_{k_1,\cdots,k_4=1}^{p} \lrabs{\cum\lrp{\varepsilon_{0,k_1}, \varepsilon_{0,k_2}, \varepsilon_{0,k_3}, \varepsilon_{0,k_4}}} \\
& & + \tr^2\lrp{\sum\limits_{s=n-m-2^h u}^{n-m-2^h(u-1)-1} A^{(s+1)} \Gamma^{(2)} \bigp{A^{(0)}}^{\top}} \\
& & + 2^h \sum\limits_{s=0}^{\infty} \tr\lrp{\bigp{A^{(0)}}^{\top} A^{(s+1)} \Gamma^{(2)} \bigp{A^{(s+1)}}^{\top} A^{(0)} \Gamma^{(2)}} \\
& & + \tr\lrp{\bigp{A^{(0)}}^{\top} \lrp{\sum\limits_{s=n-m-2^h u}^{n-m-2^h(u-1)-1} A^{(s_1+1)}} \Gamma^{(2)} \bigp{A^{(0)}}^{\top} \lrp{\sum\limits_{s=n-m-2^h u}^{n-m-2^h(u-1)-1} A^{(s_2+1)}} \Gamma^{(2)}} \\
&\lesssim& \rho^{2(n-m-2^h u)}\|\Gamma^{(2)}\|_F^4 + \rho^{2(n-m-2^h u)}\|\Gamma^{(2)}\|_F^2 p + 2^h\|\Gamma^{(2)}\|_F^2  \\
&\lesssim& \rho^{2(n-m-2^h u)}\|\Gamma^{(2)}\|_F^4 + 2^h\|\Gamma^{(2)}\|_F^2,
\EEqn
where in the second to the last inequality, we apply Lemma \ref{Lemma:LP-aux-1-2} to bound the second term in the previous step, and in the last inequality we use the assumption that $\|\Gamma^{(2)}\|_F\gtrsim\sqrt{p}$.

Similarly, we obtain that
\begin{equation*}
    \sum\limits_{j_1,j_2=2^h(u-1)+1}^{2^hu} \blre{\widetilde{D}_{n-m_1}^{\top} D_{j_1+m} \widetilde{D}_{n-m_1}^{\top} D_{j_2+m}} \\
\lesssim \rho^{2(n-m-m_1-2^h u)}\|\Gamma^{(2)}\|_F^4 + 2^h\|\Gamma^{(2)}\|_F^2,
\end{equation*}
which implies that
\begin{equation*}
    \lrnorm{\sum\limits_{j=2^h(u-1)+1}^{2^hu} \lrp{\widetilde{D}_{n} - \widetilde{D}_{n-m_1}}^{\top} D_{j+m}}_2^2
\lesssim \rho^{2(N+m_2-2^h u)}\|\Gamma^{(2)}\|_F^4 + 2^h\|\Gamma^{(2)}\|_F^2.
\end{equation*} 
Using a similar technique, we further obtain that
\begin{equation*}
    \lrnorm{\sum\limits_{j=N-2^d+2^h(u-1)+1}^{N-2^d+2^hu} \lrp{\widetilde{D}_{n} - \widetilde{D}_{n-m_1}}^{\top} D_{j+m}}_2^2
\lesssim \rho^{2(m_2+2^d-2^h u)}\|\Gamma^{(2)}\|_F^4 + 2^h\|\Gamma^{(2)}\|_F^2.
\end{equation*} 

Consequently, we have that
\BEqn
& & \lrnorm{\sup\limits_{r\in[\varepsilon,1-\varepsilon]}
    \lrabs{\sum\limits_{i=1}^{m_1} R_{n+1-i}^{\top} \lrp{\sum\limits_{j=1}^{\lrfl{nr} - \lrfl{n\varepsilon}} D_{j+m}}}}_2 \\
&\lesssim& \sum\limits_{h=0}^{d} \lrp{\sum\limits_{u=1}^{2^{d-h}} \lrnorm{\sum\limits_{j=2^h(u-1)+1}^{2^hu} \lrp{\widetilde{D}_{n} - \widetilde{D}_{n-m_1}}^{\top} D_{j+m}}_2^2}^{1/2} \\ 
& & + \sum\limits_{h=0}^{d} \lrp{\sum\limits_{u=1}^{2^{d-h}} \lrnorm{\sum\limits_{j=N-2^d+2^h(u-1)+1}^{N-2^d+2^hu} \lrp{\widetilde{D}_{n} - \widetilde{D}_{n-m_1}}^{\top} D_{j+m}}_2^2}^{1/2}  \\
&\lesssim& \sum\limits_{h=0}^{d} \lrp{\sum\limits_{u=1}^{2^{d-h}} \lrp{\rho^{2(N+m_2-2^h u)}\|\Gamma^{(2)}\|_F^4 + 2^h\|\Gamma^{(2)}\|_F^2 }}^{1/2} \\
& & + \sum\limits_{h=0}^{d} \lrp{\sum\limits_{u=1}^{2^{d-h}} \lrp{ \rho^{2(m_2+2^d-2^h u)}\|\Gamma^{(2)}\|_F^4 + 2^h\|\Gamma^{(2)}\|_F^2 }}^{1/2} \\
&\lesssim& \log(n)\rho^{m_2} \|\Gamma^{(2)}\|_F^2 + n^{1/2}\log(n)\|\Gamma^{(2)}\|_F.
\EEqn

Finally, under the assumption that $\|A^{(0)} \Gamma^{(2)} \bigp{A^{(0)}}^{\top}\|_F = O_s\lrp{\|\Gamma^{(2)}\|_F}$, we have $\tilde{N}_n = O_s\lrp{n\|\Gamma^{(2)}\|_F}$, which implies that
\begin{equation*}
    \lrnorm{\sup\limits_{r\in[\varepsilon,1-\varepsilon]}
    \lrabs{\frac{1}{\tilde{N}_n} \sum\limits_{i=1}^{m_1} R_{n+1-i}^{\top} \lrp{\sum\limits_{j=1}^{\lrfl{nr} - \lrfl{n\varepsilon}} D_{j+m}}}}_2 
\lesssim \frac{\log(n)\rho^{m_2}\|\Gamma^{(2)}\|_F}{n} + \frac{\log(n)}{n^{1/2}} 
\rightarrow 0
\end{equation*}
as long as $\rho^{m_2}\|\Gamma^{(2)}\|_F = o\lrp{\frac{n}{\log(n)}}$.
\end{Proof}

\begin{lemma}\label{Lemma:NullDist_RR_lp}
Under Assumption \ref{Assumpt:NullDist_Wn_lp}, if $\rho^{m_2}\|\Gamma^{(2)}\|_F = o\lrp{\frac{\log(n)}{n}}$, then it holds that
\begin{equation*}
    \left\|\sup\limits_{r\in[\varepsilon,1-\varepsilon]}
    \lrabs{\frac{1}{\widetilde{N}_n} \sum\limits_{i=1}^{m_1} R_i^{\top} \lrp{\sum\limits_{j=1}^{\lrfl{nr} - \lrfl{n\varepsilon}} R_{j+m}}}\right\|_2 = o(1)
\end{equation*}
and 
\begin{equation*}
    \left\|\sup\limits_{r\in[\varepsilon,1-\varepsilon]}
    \lrabs{\frac{1}{\widetilde{N}_n} \sum\limits_{i=1}^{m_1} R_{n+1-i}^{\top} \lrp{\sum\limits_{j=1}^{\lrfl{nr} - \lrfl{n\varepsilon}} R_{j+m}}}\right\|_2 = o(1)
\end{equation*}
as $n\rightarrow\infty$.
\end{lemma}
\begin{Proof}
Let $d$ denote the integer such that $2^d \le N < 2^{d+1}$, then by applying Proposition 1 of \cite{wu2007strong} again, we have that
\BEqn
& & \left\|\sup\limits_{r\in[\varepsilon,1-\varepsilon]}
    \lrabs{\sum\limits_{i=1}^{m_1} R_i^{\top} \lrp{\sum\limits_{j=1}^{\lrfl{nr} - \lrfl{n\varepsilon}} R_{j+m}}}\right\|_2 \\
&\le& \lrnorm{\max\limits_{1\le h\le 2^d} \lrabs{\sum\limits_{j=1}^{h} \lrp{\sum\limits_{i=1}^{m_1} R_i}^{\top} R_{j+m}}}_2 + \lrnorm{\max\limits_{N-2^d+1\le h\le N} \lrabs{\sum\limits_{j=1}^{h} \lrp{\sum\limits_{i=1}^{m_1} R_i}^{\top} R_{j+m}}}_2 \\
&\le& \sum\limits_{h=0}^{d} \lrp{\sum\limits_{u=1}^{2^{d-h}} \lrnorm{\sum\limits_{j=2^h(u-1)+1}^{2^h u} \lrp{\sum\limits_{i=1}^{m_1} R_i}^{\top} R_{j+m}}_2^2}^{1/2} \\
& & + \sum\limits_{h=0}^{d} \lrp{\sum\limits_{u=1}^{2^{d-h}} \lrnorm{\sum\limits_{j=N-2^d+2^h(u-1)+1}^{N-2^d+2^hu} \lrp{\sum\limits_{i=1}^{m_1} R_i}^{\top} R_{j+m}}_2^2}^{1/2},
\EEqn
where
\BEqn
& & \lrnorm{\sum\limits_{j=2^h(u-1)+1}^{2^h u} \lrp{\sum\limits_{i=1}^{m_1} R_i}^{\top} R_{j+m}}_2^2 \\
&\lesssim& \blre{\tilde{D}_{m_1}^{\top} \tilde{D}_{2^h u+m} \tilde{D}_{m_1}^{\top} \tilde{D}_{2^h u+m}} 
+ \blre{\tilde{D}_{0}^{\top} \tilde{D}_{2^h u+m} \tilde{D}_{0}^{\top} \tilde{D}_{2^h u+m}} \\
& & + \blre{\tilde{D}_{m_1}^{\top} \tilde{D}_{2^h(u-1)+m} \tilde{D}_{m_1}^{\top} \tilde{D}_{2^h(u-1)+m}}
+ \blre{\tilde{D}_{0}^{\top} \tilde{D}_{2^h(u-1)+m} \tilde{D}_{0}^{\top} \tilde{D}_{2^h(u-1)+m}},
\EEqn
and
\BEqn
& & \lrnorm{\sum\limits_{j=N-2^d+2^h(u-1)+1}^{N-2^d+2^hu} \lrp{\sum\limits_{i=1}^{m_1} R_i}^{\top} R_{j+m}}_2^2 \\
&\lesssim& \blre{\tilde{D}_{m_1}^{\top} \tilde{D}_{N-2^d+2^h u+m} \tilde{D}_{m_1}^{\top} \tilde{D}_{N-2^d+2^h u+m}} 
+ \blre{\tilde{D}_{0}^{\top} \tilde{D}_{N-2^d+2^h u+m} \tilde{D}_{0}^{\top} \tilde{D}_{N-2^d+2^h u+m}} \\
& & + \blre{\tilde{D}_{m_1}^{\top} \tilde{D}_{N-2^d+2^h(u-1)+m} \tilde{D}_{m_1}^{\top} \tilde{D}_{N-2^d+2^h(u-1)+m}}
+ \blre{\tilde{D}_{0}^{\top} \tilde{D}_{N-2^d+2^h(u-1)+m} \tilde{D}_{0}^{\top} \tilde{D}_{N-2^d+2^h(u-1)+m}}.
\EEqn

After some tedious calculations, we obtain that
\BEqn
& & \blre{\tilde{D}_{m_1}^{\top} \tilde{D}_{2^h u+m} \tilde{D}_{m_1}^{\top} \tilde{D}_{2^h u+m}} \\
&=& \sum\limits_{s_1,\cdots,s_4=0}^{\infty} \sum\limits_{k_1,\cdots,k_4=1}^{p} \bone\lrcp{m_1-s_1 = 2^h{u}+m-s_2 = m_1-s_3 = 2^h{u}+m-s_4} \\
& & \hspace{8em}
\lrp{\bigp{A^{(s_1+1)}}^{\top} A^{(s_2+1)}}^{k_1,k_2} \lrp{\bigp{A^{(s_3+1)}}^{\top} A^{(s_4+1)}}^{k_3,k_4} \cum\lrp{\varepsilon_{0,k_1}, \varepsilon_{0,k_2}, \varepsilon_{0,k_3}, \varepsilon_{0,k_4}} \\
& & + \sum\limits_{s_1,\cdots,s_4=0}^{\infty} \bone\lrcp{m_1-s_1 = 2^h{u}+m-s_2,\ m_1-s_3 = 2^h{u}+m-s_4} \\
& & \hspace{8em}
\tr\lrp{A^{(s_1+1)} \Gamma^{(2)} \bigp{A^{(s_2+1)}}^{\top}} \tr\lrp{A^{(s_3+1)} \Gamma^{(2)} \bigp{A^{(s_4+1)}}^{\top}} \\
& & + \sum\limits_{s_1,\cdots,s_4=0}^{\infty} \bone\lrcp{m_1-s_1 = m_1-s_3,\ 2^h{u}+m-s_2 = 2^h{u}+m-s_4} \\
& & \hspace{8em}
\tr\lrp{A^{(s_1+1)} \Gamma^{(2)} \bigp{A^{(s_3+1)}}^{\top} A^{(s_4+1)} \Gamma^{(2)} \bigp{A^{(s_2+1)}}^{\top}} \\
& & + \sum\limits_{s_1,\cdots,s_4=0}^{\infty} \bone\lrcp{m_1-s_1 = 2^h{u}+m-s_4,\ m_1-s_3 = 2^h{u}+m-s_2} \\
& & \hspace{8em}
\tr\lrp{A^{(s_1+1)} \Gamma^{(2)} \bigp{A^{(s_4+1)}}^{\top} A^{(s_3+1)} \Gamma^{(2)} \bigp{A^{(s_2+1)}}^{\top}} \\
&\lesssim& \sum\limits_{s=0}^{\infty} \|A^{(s+1)}\|^2 \|A^{(2^h{u}+m-m_1+s+1)}\|^2 \|\Gamma^{(2)}\|_F^4 
+ \tr^2\lrp{\sum\limits_{s=0}^{\infty} A^{(s+1)} \Gamma^{(2)} \bigp{A^{(2^h{u}+m-m_1+s+1)}}^{\top}} \\
& & + \lrp{\sum\limits_{s=0}^{\infty} \|A^{(s+1)}\|^2}^2 \|\Gamma^{(2)}\|_F^2 + \lrp{\sum\limits_{s=0}^{\infty} \|A^{(s+1)}\| \cdot \|A^{(2^h{u}+m-m_1+s+1)}\|}^2 \|\Gamma^{(2)}\|_F^2 \\
&\lesssim& \rho^{2(2^h{u}+m_2)} \|\Gamma^{(2)}\|_F^4 + \rho^{2(2^h{u}+m_2)} \|\Gamma^{(2)}\|_F^2 p + \|\Gamma^{(2)}\|_F^2 \\
&\lesssim& rho^{2(2^h{u}+m_2)} \|\Gamma^{(2)}\|_F^4 + \|\Gamma^{(2)}\|_F^2
\EEqn
since $\|\Gamma^{(2)}\|_G\gtrsim\sqrt{p}$ by Assumption \ref{Assumpt:NullDist_Wn_lp}\ref{Assumpt:NullDist_Wn_Lp-5}. Following the similar steps, we also obtain that
\BEqn
& & \blre{\tilde{D}_{0}^{\top} \tilde{D}_{2^h u+m} \tilde{D}_{0}^{\top} \tilde{D}_{2^h u+m}} 
\lesssim \rho^{2(2^h{u}+m)} \|\Gamma^{(2)}\|_F^4 + \|\Gamma^{(2)}\|_F^2, \\
& & \blre{\tilde{D}_{m_1}^{\top} \tilde{D}_{2^h(u-1)+m} \tilde{D}_{m_1}^{\top} \tilde{D}_{2^h(u-1)+m}}
\lesssim \rho^{2(2^h(u-1)+m_2)} \|\Gamma^{(2)}\|_F^4 + \|\Gamma^{(2)}\|_F^2, \\
& & \blre{\tilde{D}_{0}^{\top} \tilde{D}_{2^h(u-1)+m} \tilde{D}_{0}^{\top} \tilde{D}_{2^h(u-1)+m}}
\lesssim \rho^{2(2^h(u-1)+m)} \|\Gamma^{(2)}\|_F^4 + \|\Gamma^{(2)}\|_F^2, \\
& & \blre{\tilde{D}_{m_1}^{\top} \tilde{D}_{N-2^d+2^h u+m} \tilde{D}_{m_1}^{\top} \tilde{D}_{N-2^d+2^h u+m}
\lesssim \rho^{2(N-2^d+2^h{u}+m_2)} \|\Gamma^{(2)}\|_F^4 + \|\Gamma^{(2)}\|_F^2, } \\
& & \blre{\tilde{D}_{0}^{\top} \tilde{D}_{N-2^d+2^h u+m} \tilde{D}_{0}^{\top} \tilde{D}_{N-2^d+2^h u+m}}
\lesssim \rho^{2(N-2^d+2^h{u}+m)} \|\Gamma^{(2)}\|_F^4 + \|\Gamma^{(2)}\|_F^2, \\
& & \blre{\tilde{D}_{m_1}^{\top} \tilde{D}_{N-2^d+2^h(u-1)+m} \tilde{D}_{m_1}^{\top} \tilde{D}_{N-2^d+2^h(u-1)+m}}
\lesssim \rho^{2(N-2^d+2^h(u-1)+m_2)} \|\Gamma^{(2)}\|_F^4 + \|\Gamma^{(2)}\|_F^2, \\
& & \blre{\tilde{D}_{0}^{\top} \tilde{D}_{N-2^d+2^h(u-1)+m} \tilde{D}_{0}^{\top} \tilde{D}_{N-2^d+2^h(u-1)+m}}
\lesssim \rho^{2(N-2^d+2^h(u-1)+m)} \|\Gamma^{(2)}\|_F^4 + \|\Gamma^{(2)}\|_F^2.
\EEqn

It follows that 
\BEqn
    \left\|\sup\limits_{r\in[\varepsilon,1-\varepsilon]}
    \lrabs{\frac{1}{\widetilde{N}_n} \sum\limits_{i=1}^{m_1} R_i^{\top} \lrp{\sum\limits_{j=1}^{\lrfl{nr} - \lrfl{n\varepsilon}} R_{j+m}}}\right\|_2 
&\lesssim& \frac{1}{\widetilde{N}_n} \sum\limits_{h=0}^{d} \lrp{\rho^{2m_2}\|\Gamma^{(2)}\|_F^4 + 2^{d-h} \|\Gamma^{(2)}\|_F^2}^{1/2} \\
&\lesssim& \frac{1}{n\|\Gamma^{(2)}\|_F} \lrp{\log(n)\rho^{m_2}\|\Gamma^{(2)}\|_F^2 + 2^{d/2} \|\Gamma^{(2)}\|_F} \\
&\lesssim& \frac{\log(n)}{n} \rho^{m_2}\|\Gamma^{(2)}\|_F + n^{-1/2} \\ 
&\rightarrow& 0
\EEqn
as long as $\rho^{m_2}\|\Gamma^{(2)}\|_F = o\lrp{\frac{\log(n)}{n}}$.

Similarly, the statement of $\left\|\sup\limits_{r\in[\varepsilon,1-\varepsilon]} \lrabs{\frac{1}{\widetilde{N}_n} \sum\limits_{i=1}^{m_1} R_{n+1-i}^{\top} \lrp{\sum\limits_{j=1}^{\lrfl{nr} - \lrfl{n\varepsilon}} R_{j+m}}}\right\|_2$ can be proven, for which we spare the details. 
\end{Proof}

\subsection{Lemmas for Limiting Distributions under DGP3}\label{Appdix:lemma_null_factor}

\begin{lemma}\label{Lemma:factor-aux-1-1}
Define the process $\{W_{n,1}(r):\ \varepsilon\le r\le 1-\varepsilon\}$ as $W_{n,1}(r) = \sum\limits_{i=1}^{m_1} \sum\limits_{j=1}^{\lrfl{nr} - \lrfl{n\varepsilon}} (F_{i} - F_{n+1-i})^{\top} \Lambda^{\top} \Lambda F_{j+m}$. Under Assumption \ref{Assumpt:NullDist_Wn_factor}\ref{Assumpt:NullDist_Wn_factor-1},\ref{Assumpt:NullDist_Wn_factor-3}, it holds under the null that
\begin{equation*}
    \frac{W_{n,1}(r)}{n\|\Lambda^{\top}\Lambda\|}
\leadsto b^{\top}(\varepsilon,\eta) ((\Omega^{(3)})^{1/2})^{\top} L_0 ((\Omega^{(3)})^{1/2}) \lrp{B_s(r)-B_s(\varepsilon)}
\quad \mbox{in } D[\varepsilon,1-\varepsilon],
\end{equation*}
where $b(\varepsilon,\eta) = B_s(\varepsilon-\eta) - B_s(1) + B_s(1-\varepsilon+\eta)$ and $\{B_s(r)\}_{0\le r\le 1}$ denotes a standard $s$-dimensional Brownian motion.
\end{lemma}
\begin{Proof}
Under Assumption \ref{Assumpt:NullDist_Wn_factor}\ref{Assumpt:NullDist_Wn_factor-1}, we have that
\begin{equation*}
    \frac{1}{\sqrt{n}} \sum\limits_{i=1}^{m_1} \lrp{F_i - F_{n+1-i}}
\rightarrow (\Omega^{(3)})^{1/2} \lrp{B_s(\varepsilon-\eta) - B_s(1) + B_s(1-\varepsilon+\eta)}
\end{equation*}
and 
\begin{equation*}
    \frac{1}{\sqrt{n}} \sum\limits_{j=1}^{\lrfl{nr} - \lrfl{n\varepsilon}} F_{j+m}
\leadsto (\Omega^{(3)})^{1/2} \lrp{B_s(r) - B_s(\varepsilon)}
\quad \mbox{in } D^s[\varepsilon,1-\varepsilon]
\end{equation*}
as $n\rightarrow\infty$. It follows that in $D[\varepsilon,1-\varepsilon]$ space we have that
\BEqn
    \frac{1}{n \|\Lambda^{\top} \Lambda\|_F} W_{n,1}(r) 
&=& \lrp{\frac{1}{\sqrt{n}} \sum\limits_{i=1}^{m_1} \lrp{F_i - F_{n+1-i}}}^{\top} 
    \frac{\Lambda^{\top} \Lambda}{\|\Lambda^{\top} \Lambda\|_F} 
    \lrp{\frac{1}{\sqrt{n}} \sum\limits_{j=1}^{\lrfl{nr} - \lrfl{n\varepsilon}} F_{j+m}} \\
&\leadsto& \lrp{B_s(\varepsilon-\eta) - B_s(1) + B_s(1-\varepsilon+\eta)}^{\top} ((\Omega^{(3)})^{1/2})^{\top} L_0 (\Omega^{(3)})^{1/2} \lrp{B_s(r) - B_s(\varepsilon)}
\EEqn
as $n,p \rightarrow \infty$, which leads to the desired result with $b(\varepsilon,\eta) = B_s(\varepsilon-\eta) - B_s(1) + B_s(1-\varepsilon+\eta)$.
\end{Proof}

\begin{lemma}\label{Lemma:factor-aux-1-2}
Define the process $\{W_{n,2}(r):\ \varepsilon\le r\le 1-\varepsilon\}$ as $W_{n,2}(r) = \sum\limits_{i=1}^{m_1} \sum\limits_{j=1}^{\lrfl{nr} - \lrfl{n\varepsilon}} (Z_{i} - Z_{n+1-i})^{\top} Z_{j+m}$. Under Assumption \ref{Assumpt:NullDist_Wn_lp}, if $\rho^{m_2/4}\|\Gamma^{(3)}\|_F = o\lrp{\frac{n}{\log(n)}}$, then it holds under the null that
\begin{equation*}
    \frac{W_{n,2}(r)}{\sqrt{2n\lrfl{n(\varepsilon-\eta)}} \|A^{(0)} \Gamma^{(3)} (A^{(0)})^{\top}\|_F}
\leadsto \tilde{B}(r)-\tilde{B}(\varepsilon)
\quad \mbox{in }D[\varepsilon, 1-\varepsilon],
\end{equation*}
where $\{\tilde{B}(r)\}_{0\le r\le 1}$ denotes the standard one-dimensional Brownian motion that is independent from $\{B_s(r)\}_{0\le r\le 1}$ defined as Lemma \ref{Lemma:factor-aux-1-1}.
\end{lemma}
\begin{Proof}
Recall that 
\begin{equation*}
    W_{n,2}(r) = \sum\limits_{i=1}^{m_1} \sum\limits_{j=1}^{\lrfl{nr} - \lrfl{n\varepsilon}} (Z_i - Z_{n+1-i})^{\top} Z_{j+m},
\end{equation*}
and $\{Z_t\}_{t=1}^{n}$ is a linear process. Then the proposed statement directly follows from Proposition \ref{Prop:NullDist_Wn_lp}.
\end{Proof}

\begin{lemma}\label{Lemma:factor-aux-2}
Under Assumption \ref{Assumpt:NullDist_Wn_factor}\ref{Assumpt:NullDist_Wn_factor-2}, it holds that
\begin{equation*}
    \tr\lrp{\Lambda^{\top} \Sigma \Lambda} 
  = o\lrp{\max\{\|\Lambda^{\top} \Lambda\|_F^2, \|\Gamma^{(3)}\|_F^2\}}.    
\end{equation*}
\end{lemma}
\begin{Proof}
Note that for any integers $n,m,\ell$ and $A\in\br^{m\times n}$, $B\in\br^{n\times\ell}$, we have $\|AB\| \le \|A\|\cdot\|B\|$. Additionally, for any $\Lambda\in\br^{p\times s}$, we have
\begin{equation*}
    \|\Lambda\|^2
  = \sigma_{\text{max}}^2(\Lambda)
  = \lambda_{\text{max}}(\Lambda^{\top} \Lambda)
  = \sqrt{\lambda_{\text{max}}^2(\Lambda^{\top} \Lambda)}
  = \sqrt{\lambda_{\text{max}}(\Lambda^{\top} \Lambda \Lambda^{\top} \Lambda)}
  = \sigma(\Lambda^{\top} \Lambda)
  = \|\Lambda^{\top} \Lambda\|,
\end{equation*}
where $\sigma_{\text{max}}(\cdot)$ denotes the maximal singular value and $\lambda_{\text{max}}(\cdot)$ denotes the maximal eigenvalue. Furthermore, recall that $\Lambda\in\br^{p \times s}$ with $s\ll p$, thus $\text{rank}(\Lambda) \le s$, and it follows that $\|\Lambda^{\top} \Lambda\|_F \sim \|\Lambda^{\top} \Lambda\|$ since $\|\Lambda^{\top} \Lambda\| = \sigma_{\text{max}}(\Lambda^{\top} \Lambda)$ and $\|\Lambda^{\top} \Lambda\|_F = \sqrt{\sum\limits_{i=1}^{s} \sigma_i^2(\Lambda^{\top} \Lambda)} \le \sqrt{s} \sigma_{\text{max}}(\Lambda^{\top} \Lambda)$. Therefore, we have that
\BEqn
    \tr\lrp{\Lambda^{\top} \Sigma \Lambda}
&=& \sum\limits_{i=1}^{s} \lambda_i(\Lambda^{\top} \Sigma \Lambda) 
\le s \lambda_{\text{max}}(\Lambda^{\top} \Sigma \Lambda)
= s \sqrt{\lambda_{\text{max}}(\Lambda^{\top} \Sigma \Lambda \Lambda^{\top} \Sigma \Lambda)}
= s \sigma_{\text{max}}(\Lambda^{\top} \Sigma \Lambda) \\
&=& s \|\Lambda^{\top} \Sigma \Lambda\|
\le s \|\Lambda\|^2 \|\Sigma\| 
= s \|\Lambda^{\top} \Lambda\| \cdot \|\Sigma\| \\
&\sim&\|\Lambda^{\top} \Lambda\|_F \cdot \|\Sigma\|
= o\lrp{\|\Lambda^{\top} \Lambda\|_F \|\Gamma^{(3)}\|_F}
= o\lrp{\max\{\|\Lambda^{\top} \Lambda\|_F^2, \|\Gamma^{(3)}\|_F^2\}},
\EEqn
where the second to the last step follows from Assumption \ref{Assumpt:NullDist_Wn_factor}\ref{Assumpt:NullDist_Wn_factor-2}.
\end{Proof}

\begin{lemma}\label{Lemma:factor-aux-1-3}
Define the process $\{W_{n,3}(r)\}_{\varepsilon\le r\le 1-\varepsilon}$ as $W_{n,3}(r) = \sum\limits_{i=1}^{m_1} \sum\limits_{j=1}^{\lrfl{nr} - \lrfl{n\varepsilon}} Z_{j+m}^{\top} \Lambda(F_{i} - F_{n+1-i})$, and define the process $\{W_{n,4}(r)\}_{\varepsilon\le r\le 1-\varepsilon}$ as $W_{n,4}(r) = \sum\limits_{i=1}^{m_1} \sum\limits_{j=1}^{\lrfl{nr} - \lrfl{n\varepsilon}} (Z_{i} - Z_{n+1-i})^{\top} \Lambda F_{j+m}$. Under Assumption \ref{Assumpt:NullDist_Wn_factor}\ref{Assumpt:NullDist_Wn_factor-1}, \ref{Assumpt:NullDist_Wn_factor-4}, it holds under the null that
\begin{equation*}
   \frac{W_n^{(3)}(r_1)+W_n^{(4)}(r_1)}{n \max\{\|\Lambda\|^2,\|\Gamma^{(3)}\|_F\}} \leadsto 0
   \quad \mbox{in }D[\varepsilon,1-\varepsilon].
\end{equation*}
\end{lemma}
\begin{Proof}
It follows from the definition of $W_{n,3}(r)$ and $W_{n,4}(r)$ that
\BEqn
& & W_{n,3}(r) + W_{n,4}(r) \\
&=& \sum\limits_{i=1}^{m_1} \sum\limits_{j=1}^{\lrfl{nr} - \lrfl{n\varepsilon}} Z_{j+m}^{\top} \Lambda (F_i - F_{n+1-i})
 +  \sum\limits_{i=1}^{m_1} \sum\limits_{j=1}^{\lrfl{nr} - \lrfl{n\varepsilon}} (Z_i - Z_{n+1-i})^{\top} \Lambda F_{j+m} \\
&=& \sum\limits_{i,j\in\{1,\cdots,m_1,m+1,\cdots,\lrfl{nr}\}} Z_j^{\top} \Lambda F_i 
 -  \sum\limits_{i,j=1}^{m_1} Z_j^{\top} \Lambda F_i
 -  \sum\limits_{i,j=m+1}^{\lrfl{nr}} Z_j^{\top} \Lambda F_i \\
& & - \lrp{  \sum\limits_{i,j\in\{m+1,\cdots,\lrfl{nr},n+1-m_1,\cdots,n\}} Z_j^{\top} \Lambda F_i
           - \sum\limits_{i,j=n+1-m_1}^{n} Z_j^{\top} \Lambda F_i
           - \sum\limits_{i,j=m+1}^{\lrfl{nr}} Z_j^{\top} \Lambda F_i } \\
&=& \sum\limits_{i=1}^{m_1} \sum\limits_{j=1}^{\lrfl{nr} - \lrfl{n\varepsilon}} Z_{j+m}^{\top} \Lambda (F_i - F_{n+1-i})
 +  \sum\limits_{i=1}^{m_1} \sum\limits_{j=1}^{\lrfl{nr} - \lrfl{n\varepsilon}} (Z_i - Z_{n+1-i})^{\top} \Lambda F_{j+m} \\
&=& \sum\limits_{i,j\in\{1,\cdots,m_1,m+1,\cdots,\lrfl{nr}\}} Z_j^{\top} \Lambda F_i 
 -  \sum\limits_{i,j\in\{m+1,\cdots,\lrfl{nr},n+1-m_1,\cdots,n\}} Z_j^{\top} \Lambda F_i \\
& & -  \sum\limits_{i,j=1}^{m_1} Z_j^{\top} \Lambda F_i 
 +   \sum\limits_{i,j=n+1-m_1}^{n} Z_j^{\top} \Lambda F_i.
\EEqn
Under Assumption \ref{Assumpt:NullDist_Wn_factor}\ref{Assumpt:NullDist_Wn_factor-1}, we have that
\begin{equation*}
    \frac{1}{\sqrt{n}} \sum\limits_{i=1}^{\lrfl{nr}} \bin{\Lambda^{\top} Z_i}{F_i} 
= \sqrt{n} \bin{\frac{1}{n} \sum\limits_{i=1}^{\lrfl{nr}} \Lambda^{\top} Z_i}{\frac{1}{n} \sum\limits_{i=1}^{\lrfl{nr}} F_i}
\leadsto \dmat{(\Lambda^{\top} \Sigma^{(3)} \Lambda)^{1/2}}{(\Omega^{(3)})^{1/2}} B_{2s}(r) 
=^d \bin{x(r)}{y(r)},
\end{equation*}
where $x(r) \sim (\Lambda^{\top} \Sigma^{(3)} \Lambda)^{1/2} \tilde{B}_s(r)$ and $y(r) \sim (\Omega^{(3)})^{1/2} B_s(r)$, and $\{B_s(r):\ \varepsilon \le r \le 1-\varepsilon\} \indept \{\tilde{B}_s(r):\ \varepsilon \le r \le 1-\varepsilon\}$ represent two standard $s$-dimensional Brownian motions. Let $f(x,y) = x^{\top}y$, then by using CMT, we obtain that 
\begin{equation*}
    \frac{1}{n} \sum\limits_{i=1}^{\lrfl{nr}} \sum\limits_{j=1}^{\lrfl{nr}} Z_j^{\top} \Lambda F_i
  = f\lrp{\frac{1}{\sqrt{n}} \sum\limits_{i=1}^{\lrfl{nr}} \Lambda^{\top} Z_i,\ \frac{1}{\sqrt{n}} \sum\limits_{i=1}^{\lrfl{nr}} F_i}
  \leadsto x^{\top}(r) y(r)
\end{equation*}
as $n\rightarrow\infty$. Consequently, as $n\rightarrow\infty$, we further have that
\BEqn
& & \frac{1}{n} \sum\limits_{i,j=1}^{m_1} Z_j^{\top} \Lambda F_i 
\stra{d} x^{\top}(\varepsilon-\eta) y(\varepsilon-\eta), \\
& & \frac{1}{n} \sum\limits_{i,j=n+1-m_1}^{n} Z_j^{\top} \Lambda F_i 
\stra{d} (x(1) - x(1-\varepsilon+\eta))^{\top} (y(1) - y(1-\varepsilon+\eta)), \\
& & \frac{1}{n} \sum\limits_{i,j\in\{1,\cdots,m_1,m+1,\cdots,\lrfl{nr}\}} Z_j^{\top} \Lambda F_i 
\leadsto (x(r) - x(\varepsilon) + x(\varepsilon-\eta))^{\top} (y(r) - y(\varepsilon) - y(\varepsilon-\eta)),
\EEqn
and
\begin{equation*}
    \frac{1}{n} \sum\limits_{i,j\in\{m+1,\cdots,\lrfl{nr},n+1-m_1,\cdots,n\}} Z_j^{\top} \Lambda F_i
\leadsto (x(r) - x(\varepsilon) + x(1) - x(1-\varepsilon+\eta))^{\top}
         (y(r) - y(\varepsilon) + y(1) - y(1-\varepsilon+\eta)).
\end{equation*}

Then in $D[\varepsilon,1-\varepsilon]$ space, we have that
\BEqn
& & \frac{1}{n \sqrt{\tr\lrp{\Lambda^{\top} \Sigma^{(3)} \Lambda}}} 
    \lrp{W_{n,3}(r) + W_{n,4}(r)} \\
&\leadsto&
    (\tilde{B}_s(r) - \tilde{B}_s(\varepsilon) + \tilde{B}_s(\varepsilon-\eta))^{\top} L_1 (\Omega^{(3)})^{1/2} (B_s(r) - B_s(\varepsilon) + B_s(\varepsilon-\eta)) \\
& & - (\tilde{B}_s(r) - \tilde{B}(\varepsilon) + \tilde{B}_s(1) - \tilde{B}_s(1-\varepsilon+\eta))^{\top} L_1 (\Omega^{(3)})^{1/2} (B_s(r) - B_s(\varepsilon) + B_s(1) - B_s(1-\varepsilon+\eta)) \\    
& & - \tilde{B}_s^{\top}(\varepsilon-\eta) L_1 (\Omega^{(3)})^{1/2} B_s(\varepsilon-\eta) \\
& & + (\tilde{B}_s(1) - \tilde{B}_s(1-\varepsilon+\eta))^{\top} L_1 (\Omega^{(3)})^{1/2} (B_s(1) - B_s(1-\varepsilon+\eta)) \\
&=^d& (\tilde{B}_s(r) - \tilde{B}_s(\varepsilon))^{\top} L_1 (\Omega^{(3)})^{1/2} (B_s(\varepsilon-\eta) - B_s(1) + B_s(1-\varepsilon+\eta)) \\
& & + (\tilde{B}_s(\varepsilon-\eta) - \tilde{B}_s(1) + \tilde{B}_s(1-\varepsilon+\eta))^{\top} L_1 (\Omega^{(3)})^{1/2} (B_s(r) - B_s(\varepsilon)).
\EEqn

Note that by Definition \ref{Def:Model_factor} and Assumption \ref{Assumpt:NullDist_Wn_factor}, $\Omega^{(3)},L_0\in\br^{s\times s}$ are both independent of $p$ and can be viewed as constant matrices. Additionally, it follows from Lemma \ref{Lemma:factor-aux-2} that 
\begin{equation*}
    \sqrt{\tr\lrp{\Lambda^{\top} \Sigma^{(3)} \Lambda}}
  = o\lrp{\max\{\|\Lambda\|^2,\|\Gamma^{(3)}\|_F\}}.
\end{equation*}
Therefore, we obtain that 
\begin{equation*}
    \frac{W_{n,3}(r)+W_{n,4}(r)}{n \max\{\|\Lambda\|^2,\|\Gamma^{(3)}\|_F\}}
  = \frac{\sqrt{\tr\lrp{\Lambda^{\top} \Sigma^{(3)} \Lambda}}}{\max\{\|\Lambda\|^2,\|\Gamma^{(3)}\|_F\}} \cdot \frac{W_{n,3}(r)+W_{n,4}(r)}{n \sqrt{\tr\lrp{\Lambda^{\top} \Sigma^{(3)} \Lambda}}} \\
  \leadsto 0,
\end{equation*}
which completes the proof.
\end{Proof}

\subsection{Lemmas for Theorem \ref{Thm:MultiplePower}}\label{Appdix:lemma_multiple}

\begin{lemma}\label{Lemma:G-DGP1}
Suppose that $\{\tilde{X}_t\}_{t=1}^{n}$ is a stationary sequence as defined in Definition \ref{Def:Model_Fixedp}. Assume that Assumption \ref{Assumpt:multiple} holds and Assumption \ref{Assumpt:MultiplePower} is satisfied with $N_n = n$, then it holds that 
\begin{enumerate}[label=(\roman*)]
    \item \label{Lemma:Gf-DGP1}
    if Assumption \ref{Assumpt:MultiplePower}\ref{Assumpt:MultiplePower-1} holds, then we have that
    \begin{equation*}
        \lrabs{T_n^f(1,\lrfl{n\xi_i}-m,\lrfl{n\xi_{i+1}}-m) \lrp{V_n^f(1,\lrfl{n\xi_i}-m,\lrfl{n\xi_{i+1}}-m)}^{-1/2}}
    \stra{p} \infty.
    \end{equation*}
    
    \item \label{Lemma:Gb-DGP1}
    if Assumption \ref{Assumpt:MultiplePower}\ref{Assumpt:MultiplePower-2} holds, then we have that
    \begin{equation*}
        \lrabs{T_n^b(\lrfl{n\xi_{i-1}}-m,\lrfl{n\xi_i}-m,N) \lrp{V_n^b(\lrfl{n\xi_{i-1}}-m,\lrfl{n\xi_i}-m,N)}^{-1/2}}
    \stra{p} \infty.
    \end{equation*}
\end{enumerate}
\end{lemma}

\begin{Proof}
If $\{\tilde{X}_t\}_{t=1}^{n} \in \br^{p}$ is a stationary sequence defined as Definition \ref{Def:Model_Fixedp}, with $N_n = n$, we have that $\frac{1}{\sqrt{n}} \sum\limits_{t=1}^{\lrfl{nr}} \tilde{X}_t \leadsto B_p(r)$ in $D[\varepsilon,1-\varepsilon]$, where $\{B_p(r)\}_{0\le r\le 1}$ is a standard Brownian motion in $\br^p$.

\begin{enumerate}[label=(\roman*)]
    \item We first consider the case when Assumption \ref{Assumpt:MultiplePower}\ref{Assumpt:MultiplePower-1} is satisfied with $N_n=n$. 
    
    We have shown in Lemma \ref{Lemma:Expression_Gf} that
    \BEqn
    & & T_n^f(1,\lrfl{n\xi_i}-m,\lrfl{n\xi_{i+1}}-m) \lrp{V_n^f(1,\lrfl{n\xi_i}-m,\lrfl{n\xi_{i+1}}-m)}^{-1/2} \\
    &=& \lrp{\xi_{i+1}-\varepsilon+o_p(1)} \tilde{T}_n^f(\xi_i) \lrp{\tilde{V}_n^f(\xi_i)}^{-1/2},
    \EEqn
    where
    \BEqn
        \tilde{T}_n^f(\xi_i)
    &=& \frac{1}{N_n} \lrp{\tilde{W}_n(\xi_i) - \lrp{\frac{\xi_i-\varepsilon}{\xi_{i+1}-\varepsilon}+o_p(1)} \tilde{W}_n(\xi_{i+1})} \\
    & & - \frac{n}{N_n} \sum\limits_{u=1}^{m_1} \lrp{\tilde{X}_u - \tilde{X}_{n+1-u}}^{\top} \lrp{\sum\limits_{t=1}^{i} \delta_t \lrp{\frac{(\xi_{i+1}-\xi_i)(\xi_t-\varepsilon)}{\xi_{i+1}-\varepsilon}+o_p(1)}} \\
    & & - \frac{m_1}{N_n} \delta^{\top} \lrp{\sum\limits_{j=1}^{\lrfl{n\xi_i}-m} \tilde{X}_{j+m} - \lrp{\frac{\xi_i-\varepsilon}{\xi_{i+1}-\varepsilon}+o_p(1)}  \sum\limits_{j=1}^{\lrfl{n\xi_{i+1}}-m} \tilde{X}_{j+m}} \\
    & & + \frac{m_1 n}{N_n} \sum\limits_{t=0}^{i} \delta^{\top} \delta_t \lrp{\frac{(\xi_{i+1}-\xi_i)(\xi_t-\varepsilon)}{\xi_{i+1}-\varepsilon}+o_p(1)},
    \EEqn
    and
    \BEqn
        \tilde{V}_n^f(\xi_i)
    &=& \sum\limits_{h=1}^{i} \int_{\xi_{h-1}}^{\xi_h} 
        \left(
            \frac{1}{N_n} \lrp{\tilde{W}_n(s) - \lrp{\frac{s-\varepsilon}{\xi_i-\varepsilon}+o_p(1)} \tilde{W}_n(\xi_i)} 
        \right. \\
    & & \hspace{3em}    
        - \frac{n}{N_n} \sum\limits_{u=1}^{m_1} \lrp{\tilde{X}_u - \tilde{X}_{n+1-u}}^{\top} 
        \left(
            \bone\{h>1\} \sum\limits_{t=1}^{h-1} \delta_t \lrp{\frac{(\xi_i-s)(\xi_t-\varepsilon)}{\xi_i-\varepsilon}+o_p(1)}
        \right. \\
    & & \hspace{16em}
        \left.
            + \bone\{h<i\} \sum\limits_{t=h}^{i-1} \delta_t \lrp{\frac{(\xi_i-\xi_t)(s-\varepsilon)}{\xi_i-\varepsilon}+o_p(1)}
        \right) \\
    & & \hspace{3em}    
        - \frac{m_1}{N_n} \delta^{\top} \lrp{\sum\limits_{j=1}^{\lrfl{ns}-m} \tilde{X}_{j+m} - \lrp{\frac{s-\varepsilon}{\xi_i-\varepsilon}+o_p(1)} \sum\limits_{j=1}^{\lrfl{n\xi_i}-m} \tilde{X}_{j+m}} \\ 
    & & \hspace{3em}    
        + \frac{m_1 n}{N_n}
        \left(
            \bone\{h>1\} \sum\limits_{t=1}^{h-1} \delta^{\top} \delta_t \lrp{\frac{(\xi_i-s)(\xi_t-\varepsilon)}{\xi_i-\varepsilon}+o_p(1)} 
        \right. \\
    & & \hspace{7em}      
        \left.
        \left.
            + \bone\{h<i\} \sum\limits_{t=h}^{i-1} \delta^{\top} \delta_t \lrp{\frac{(\xi_i-\xi_t)(s-\varepsilon)}{\xi_i-\varepsilon}+o_p(1)}
        \right)
        \right)^2 ds \\
    & & + \int_{\xi_i}^{\xi_{i+1}}
        \left(
            \frac{1}{N_n} \lrp{\tilde{W}_n(\xi_{i+1}) - \tilde{W}_n(s) - \lrp{\frac{\xi_{i+1}-s}{\xi_{i+1}-\xi_i}+o_p(1)} \lrp{\tilde{W}_n(\xi_{i+1})-\tilde{W}_n(\xi_i)}}
        \right. \\
    & & \hspace{5em}    
        \left.
            - \frac{m_1}{N_n} \delta^{\top} \lrp{\sum\limits_{j=\lrfl{ns}-m+1}^{\lrfl{n\xi_{i+1}}-m} \tilde{X}_{j+m} - \lrp{\frac{\xi_{i+1}-s}{\xi_{i+1}-\xi_i}+o_p(1)} \sum\limits_{j=\lrfl{n\xi_i}-m+1}^{\lrfl{n\xi_{i+1}}-m} \tilde{X}_{j+m}}
        \right)^2 ds \\
    & & + o_p(1).
    \EEqn
    
    To show the desired result, we aim to show that
    \begin{equation*}
        \lrabs{\lrp{\frac{n^2 \delta^{\top}\delta_i}{N_n}}^{-1} \cdot \tilde{T}_n^f(\xi_i)}
    \stra{p} \frac{(\varepsilon-\eta)(\xi_{i+1}-\xi_i)(\xi_i-\varepsilon)}{\xi_{i+1}-\varepsilon},
    \end{equation*}
    and
    \begin{equation*}
        \lrabs{\lrp{\frac{n^2 \delta^{\top}\delta_i}{N_n}}^{-2} \cdot \tilde{V}_n^f(\xi_i)}
    \stra{p} 0.
    \end{equation*}
    
    Under Assumption \ref{Assumpt:MultiplePower}\ref{Assumpt:MultiplePower-1} with $N_n=n$, we have that $n\lrabs{\delta^{\top}\delta_i} \rightarrow \infty$, $\|\delta\|_2^2 = O(\lrabs{\delta^{\top}\delta_i})$, $\max\limits_{1\le j\le i} \|\delta_j\|_2^2 = O(\lrabs{\delta^{\top}\delta_i})$ as well as $\max\limits_{1\le j<i}\lrabs{\delta^{\top}\delta_j} = o(\lrabs{\delta^{\top}\delta_i})$.
    
    Using these assumptions, we obtain that
    \begin{equation*}
        \lrabs{\frac{n^2 \delta^{\top}\delta_i}{N_n}}^{-1/2} \cdot \frac{1}{\sqrt{n}} \sum\limits_{t=1}^{\lrfl{nr}} \tilde{X}_t
    = \frac{1}{\sqrt{n\lrabs{\delta^{\top}\delta_i}}} \cdot \frac{1}{\sqrt{n}} \sum\limits_{t=1}^{\lrfl{nr}} \tilde{X}_t
    \leadsto 0
    \quad\mbox{in }D[\varepsilon,1-\varepsilon],
    \end{equation*}
    and it follows that
    \begin{equation*}
        \lrabs{\frac{n^2 \delta^{\top}\delta_i}{N_n}}^{-1} \cdot \frac{1}{n} \tilde{W}_n(r) 
    \leadsto 0 \cdot b^{\top}(\varepsilon,\eta) \lrp{B_p(r)-B_p(\varepsilon)}
    =^d 0
    \quad\mbox{in }D[\varepsilon,1-\varepsilon],
    \end{equation*}
    where $b(\varepsilon,\eta) = B_p(\varepsilon-\eta) - B_p(1) + B_p(1-\varepsilon+\eta)$. 
    
    It follows from direct calculation that with $N_n=n$,
    \BEqn
    & & \lrabs{\frac{n^2 \delta^{\top}\delta_i}{N_n}}^{-1} \cdot \tilde{T}_n^f(\xi_i) \\
    &=& \frac{1}{n\lrabs{\delta^{\top}\delta_i}} \cdot \frac{1}{n} \lrp{\tilde{W}_n(\xi_i) - \lrp{\frac{\xi_i-\varepsilon}{\xi_{i+1}-\varepsilon}+o_p(1)} \tilde{W}_n(\xi_{i+1})} \\
    & & - \frac{1}{n\lrabs{\delta^{\top}\delta_i}} \cdot \sum\limits_{u=1}^{m_1} \lrp{\tilde{X}_u - \tilde{X}_{n+1-u}}^{\top} \lrp{\sum\limits_{t=1}^{i} \delta_t \lrp{\frac{(\xi_{i+1}-\xi_i)(\xi_t-\varepsilon)}{\xi_{i+1}-\varepsilon}+o_p(1)}} \\
    & & - \frac{1}{n\lrabs{\delta^{\top}\delta_i}} \cdot \frac{m_1}{n} \delta^{\top} \lrp{\sum\limits_{j=1}^{\lrfl{n\xi_i}-m} \tilde{X}_{j+m} - \lrp{\frac{\xi_i-\varepsilon}{\xi_{i+1}-\varepsilon}+o_p(1)} \sum\limits_{j=1}^{\lrfl{n\xi_{i+1}}-m} \tilde{X}_{j+m}} \\
    & & + \frac{1}{n\lrabs{\delta^{\top}\delta_i}} \cdot m_1 \sum\limits_{t=0}^{i} \delta^{\top} \delta_t \lrp{\frac{(\xi_{i+1}-\xi_i)(\xi_t-\varepsilon)}{\xi_{i+1}-\varepsilon}+o_p(1)} \\
    &=& \frac{1}{n\lrabs{\delta^{\top}\delta_i}} \cdot \frac{1}{N_n} \lrp{\tilde{W}_n(\xi_i) - \lrp{\frac{\xi_i-\varepsilon}{\xi_{i+1}-\varepsilon}+o_p(1)} \tilde{W}_n(\xi_{i+1})} \\
    & & - \frac{1}{\sqrt{n\lrabs{\delta^{\top}\delta_i}}} \cdot \lrp{\frac{1}{\sqrt{n}} \sum\limits_{u=1}^{m_1} \lrp{\tilde{X}_u - \tilde{X}_{n+1-u}}}^{\top} \lrp{\sum\limits_{t=1}^{i} \frac{\delta_t}{\sqrt{\lrabs{\delta^{\top}\delta_i}}} \lrp{\frac{(\xi_{i+1}-\xi_i)(\xi_t-\varepsilon)}{\xi_{i+1}-\varepsilon}+o_p(1)}} \\
    & & - \frac{\varepsilon-\eta}{\sqrt{n\lrabs{\delta^{\top}\delta_i}}} \cdot \lrp{\frac{\delta}{\sqrt{\lrabs{\delta^{\top}\delta_i}}}}^{\top} \frac{1}{\sqrt{n}} \lrp{\sum\limits_{j=1}^{\lrfl{n\xi_i}-m} \tilde{X}_{j+m} - \lrp{\frac{\xi_i-\varepsilon}{\xi_{i+1}-\varepsilon}+o_p(1)} \sum\limits_{j=1}^{\lrfl{n\xi_{i+1}}-m} \tilde{X}_{j+m}} \\
    & & + (\varepsilon-\eta) \sum\limits_{t=0}^{i-1} \frac{\delta^{\top}\delta_t}{\lrabs{\delta^{\top}\delta_i}} \lrp{\frac{(\xi_{i+1}-\xi_i)(\xi_t-\varepsilon)}{\xi_{i+1}-\varepsilon}+o_p(1)} \\
    & & + (\varepsilon-\eta) \frac{\delta^{\top}\delta_i}{\lrabs{\delta^{\top}\delta_i}} \lrp{\frac{(\xi_{i+1}-\xi_i)(\xi_i-\varepsilon)}{\xi_{i+1}-\varepsilon}+o_p(1)} \\
    &\stra{p}& \sgn(\delta^{\top}\delta_i) \frac{(\varepsilon-\eta)(\xi_{i+1}-\xi_i)(\xi_i-\varepsilon)}{\xi_{i+1}-\varepsilon}.
    \EEqn
    
    Similarly, it follows from Continuous Mapping Theorem that
    \BEqn
    & & \lrabs{\frac{n^2 \delta^{\top}\delta_i}{N_n}}^{-2} \cdot \tilde{V}_n^f(\xi_i) \\
    &=& \sum\limits_{h=1}^{i} \int_{\xi_{h-1}}^{\xi_h} 
        \left(
            \frac{1}{n\lrabs{\delta^{\top}\delta_i}} \cdot \frac{1}{n} \lrp{\tilde{W}_n(s) - \lrp{\frac{s-\varepsilon}{\xi_i-\varepsilon}+o_p(1)} \tilde{W}_n(\xi_i)} 
        \right. \\
    & & \hspace{3em}    
        - \frac{\bone\{h>1\}}{\sqrt{n\lrabs{\delta^{\top}\delta_i}}} \cdot \lrp{\frac{1}{\sqrt{n}} \sum\limits_{u=1}^{m_1} \lrp{\tilde{X}_u - \tilde{X}_{n+1-u}}}^{\top} \lrp{\sum\limits_{t=1}^{h-1} \frac{\delta_t}{\lrabs{\delta^{\top}\delta_i}} \lrp{\frac{(\xi_i-s)(\xi_t-\varepsilon)}{\xi_i-\varepsilon}+o_p(1)}} \\
    & & \hspace{3em} 
        - \frac{\bone\{h<i\}}{\sqrt{n\lrabs{\delta^{\top}\delta_i}}} \cdot \lrp{\frac{1}{\sqrt{n}} \sum\limits_{u=1}^{m_1} \lrp{\tilde{X}_u - \tilde{X}_{n+1-u}}}^{\top} \lrp{ \sum\limits_{t=h}^{i-1} \frac{\delta_t}{\lrabs{\delta^{\top}\delta_i}} \lrp{\frac{(\xi_i-\xi_t)(s-\varepsilon)}{\xi_i-\varepsilon}+o_p(1)}} \\
    & & \hspace{3em}    
        - \frac{\varepsilon-\eta}{\sqrt{n\lrabs{\delta^{\top}\delta_i}}} \lrp{\frac{\delta}{\sqrt{\lrabs{\delta^{\top}\delta_i}}}}^{\top} \frac{1}{\sqrt{n}} \lrp{\sum\limits_{j=1}^{\lrfl{ns}-m} \tilde{X}_{j+m} - \lrp{\frac{s-\varepsilon}{\xi_i-\varepsilon}+o_p(1)} \sum\limits_{j=1}^{\lrfl{n\xi_i}-m} \tilde{X}_{j+m}} \\ 
    & & \hspace{3em}    
        + (\varepsilon-\eta) \bone\{h>1\} \sum\limits_{t=1}^{h-1} \frac{\delta^{\top}\delta_t}{\lrabs{\delta^{\top}\delta_i}} \lrp{\frac{(\xi_i-s)(\xi_t-\varepsilon)}{\xi_i-\varepsilon}+o_p(1)} \\
    & & \hspace{3em}      
        \left.
        + (\varepsilon-\eta) \bone\{h<i\} \sum\limits_{t=h}^{i-1} \frac{\delta^{\top}\delta_t}{\lrabs{\delta^{\top}\delta_i}} \lrp{\frac{(\xi_i-\xi_t)(s-\varepsilon)}{\xi_i-\varepsilon}+o_p(1)}
        \right)^2 ds \\
    &+& \int_{\xi_i}^{\xi_{i+1}}
        \left(
            \frac{1}{n\lrabs{\delta^{\top}\delta_i}} \cdot \frac{1}{n} \lrp{\tilde{W}_n(\xi_{i+1}) - \tilde{W}_n(s) - \lrp{\frac{\xi_{i+1}-s}{\xi_{i+1}-\xi_i}+o_p(1)} \lrp{\tilde{W}_n(\xi_{i+1})-\tilde{W}_n(\xi_i)}}
        \right. \\
    & & \left.
            - \frac{\varepsilon-\eta}{\sqrt{n\lrabs{\delta^{\top}\delta_i}}} \lrp{\frac{\delta}{\sqrt{\lrabs{\delta^{\top}\delta_i}}}}^{\top} \frac{1}{\sqrt{n}} \lrp{\sum\limits_{j=\lrfl{ns}-m+1}^{\lrfl{n\xi_{i+1}}-m} \tilde{X}_{j+m} - \lrp{\frac{\xi_{i+1}-s}{\xi_{i+1}-\xi_i}+o_p(1)} \sum\limits_{j=\lrfl{n\xi_i}-m+1}^{\lrfl{n\xi_{i+1}}-m} \tilde{X}_{j+m}}
        \right)^2 ds \\
    &+& o_p(1) \\
    &\stra{p}& 0,
    \EEqn
    where the last step follows from the assumption that $\max\limits_{j<i} |\delta^{\top}\delta_j| = o(\lrabs{\delta^{\top}\delta_i})$ and the observation that $\tilde{V}_n^{f}(\xi_i)$ only contains $\delta$ and $\delta_j$ for $j<i$ and does not include $\delta^{\top} \delta_i$.
    
    Finally, by using Continuous Mapping Theorem again, we obtain that
    \BEqn
    & & \lrabs{T_n^f(1,\lrfl{n\xi_i}-m,\lrfl{n\xi_{i+1}}-m) \lrp{V_n^f(1,\lrfl{n\xi_i}-m,\lrfl{n\xi_{i+1}}-m)}^{-1/2}} \\
    &=& \lrp{\xi_{i+1}-\varepsilon+o_p(1)} \lrabs{\tilde{T}_n^f(\xi_i)} \lrp{\tilde{V}_n^f(\xi_i)}^{-1/2} \\
    &=& \lrp{\xi_{i+1}-\varepsilon+o_p(1)} \frac{\lrabs{\frac{n^2 \delta^{\top}\delta_i}{N_n}}^{-1} \cdot \lrabs{\tilde{T}_n^f(\xi_i)}}{\lrp{\lrabs{\frac{n^2 \delta^{\top}\delta_i}{N_n}}^{-2} \cdot \tilde{V}_n^f(\xi_i)}^{1/2}} \\
    &\stra{p}& \infty,
    \EEqn
    which completes the proof.
    
    \item Suppose that Assumption \ref{Assumpt:MultiplePower}\ref{Assumpt:MultiplePower-2} holds, then we have that $n\lrabs{\delta^{\top}\delta_i} \rightarrow \infty$, $\|\delta\|_2^2 = O(\lrabs{\delta^{\top}\delta_i})$, $\max\limits_{M\geq j\geq i} \|\delta_j\|_2^2 = O(\lrabs{\delta^{\top}\delta_i})$ and $\max\limits_{M>j\geq i}\lrabs{\delta^{\top}\delta_j} = o(\lrabs{\delta^{\top}\delta_i})$. Note that it follows from Lemma \ref{Lemma:Expression_Gb} that
    \BEqn
    & & T_n^b(\lrfl{n\xi_{i-1}}-m,\lrfl{n\xi_i}-m,N) \lrp{V_n^b(\lrfl{n\xi_{i-1}}-m,\lrfl{n\xi_i}-m,N)}^{-1/2} \\
    &=& \lrp{1-\varepsilon-\xi_{i-1}+o_p(1)} \tilde{T}_n^b(\xi_i) \lrp{\tilde{V}_n^b(\xi_i)}^{-1/2},
    \EEqn
    where
    \BEqn
        \tilde{T}_n^b(\xi_i)
    &=& \frac{1}{N_n} \lrp{\tilde{W}_n(1-\varepsilon) - \tilde{W}_n(\xi_i) - \lrp{\frac{1-\varepsilon-\xi_i}{1-\varepsilon-\xi_{i-1}}+o_p(1)} (\tilde{W}_n(1-\varepsilon) - \tilde{W}_n(\xi_{i-1}))} \\
    & & + \frac{n}{N_n} \sum\limits_{u=1}^{m_1} \lrp{\tilde{X}_u - \tilde{X}_{n+1-u}}^{\top} \lrp{\sum\limits_{t=i}^{M} \delta_t \lrp{\frac{(1-\varepsilon-\xi_t)(\xi_i-\xi_{i-1})}{1-\varepsilon-\xi_{i-1}}+o_p(1)}} \\
    & & - \frac{m_1}{N_n} \delta^{\top} \lrp{\sum\limits_{j=\lrfl{n\xi_i}-m}^{N} \tilde{X}_{j+m} -  \lrp{\frac{1-\varepsilon-\xi_i}{1-\varepsilon-\xi_{i-1}}+o_p(1)} \sum\limits_{j=\lrfl{n\xi_{i-1}}-m}^{N} \tilde{X}_{j+m}} \\
    & & - \frac{m_1 n}{N_n} \sum\limits_{t=i}^{M} \delta^{\top} \delta_t \lrp{\frac{(1-\varepsilon-\xi_t)(\xi_i-\xi_{i-1})}{1-\varepsilon-\xi_{i-1}}+o_p(1)},
    \EEqn
    and
    \BEqn
        \tilde{V}_n^b(\xi_i)
    &=& \int_{\xi_{i-1}}^{\xi_i} 
        \left(
            \frac{1}{N_n} \lrp{\tilde{W}_n(s) - \tilde{W}_n(\xi_{i-1}) - \lrp{\frac{s-\xi_{i-1}}{\xi_i-\xi_{i-1}}+o_p(1)} \lrp{\tilde{W}_n(\xi_i) - \tilde{W}_n(\xi_{i-1})}}
        \right. \\
    & & \hspace{3em}
        \left.
            - \frac{m_1}{N_n} \delta^{\top}
            \lrp{\sum\limits_{j=\lrfl{n\xi_{i-1}}-m}^{\lrfl{ns}-m} \tilde{X}_{j+m} - \lrp{\frac{s-\xi_{i-1}}{\xi_i-\xi_{i-1}}+o_p(1)} \sum\limits_{j=\lrfl{n\xi_{i-1}}-m}^{\lrfl{n\xi_i}-m-1} \tilde{X}_{j+m}}
        \right)^2 ds \\
    & & + \sum\limits_{h=i+1}^{M+1} \int_{\xi_{h-1}}^{\xi_h} 
        \left(
            \frac{1}{N_n} \lrp{\tilde{W}_n(1-\varepsilon) - \tilde{W}_n(s) - \lrp{\frac{1-\varepsilon-s}{1-\varepsilon-\xi_i}+o_p(1)} \lrp{\tilde{W}_n(1-\varepsilon)} - \tilde{W}_n(\xi_i)}
        \right. \\
    & & \hspace{5em}
        + \frac{n}{N_n} \sum\limits_{u=1}^{m_1} \lrp{\tilde{X}_u-\tilde{X}_{n+1-u}}^{\top}
        \left(
            \bone\{h>i+1\} \sum\limits_{t=i+1}^{h-1} \delta_t \lrp{\frac{(1-\varepsilon-s)(\xi_t-\xi_i)}{1-\varepsilon-\xi_i}+o_p(1)}
        \right. \\
    & & \hspace{18em}      
        \left.
            + \bone\{h\le M\} \sum\limits_{t=h}^{M} \delta_t \lrp{\frac{(1-\varepsilon-\xi_t)(s-\xi_i)}{1-\varepsilon-\xi_i}+o_p(1)}
        \right) \\
    & & \hspace{5em}
        - \frac{m_1}{N_n} \delta^{\top}
        \lrp{\sum\limits_{j=\lrfl{ns}-m}^{N} \tilde{X}_{j+m} - \lrp{\frac{1-\varepsilon-s}{1-\varepsilon-\xi_i}+o_p(1)} \sum\limits_{j=\lrfl{n\xi_i}-m}^{N} \tilde{X}_{j+m}} \\
    & & \hspace{5em}
        - \frac{m_1 n}{N_n} 
        \left(
            \bone\{h>i+1\} \sum\limits_{t=i+1}^{h-1} \delta^{\top} \delta_t \lrp{\frac{(1-\varepsilon-s)(\xi_t-\xi_i)}{1-\varepsilon-\xi_i}+o_p(1)}
        \right. \\
    & & \hspace{9em}      
        \left.
        \left.
            + \bone\{h\le M\} \sum\limits_{t=h}^{M} \delta^{\top} \delta_t \lrp{\frac{(1-\varepsilon-\xi_t)(s-\xi_i)}{1-\varepsilon-\xi_i}+o_p(1)}
        \right)
        \right)^2 ds \\
    & & + o_p(1).    
    \EEqn
    
    By using the same arguments as for the previous case, we can again show that
    \begin{equation*}
        \lrabs{\frac{n^2 \delta^{\top}\delta_i}{N_n}}^{-1} \lrabs{\tilde{T}_n^b(\xi_i)}
    \stra{p} \frac{(\varepsilon-\eta)(\xi_i-\xi_{i-1})(1-\varepsilon-\xi_i)}{1-\varepsilon-\xi_{i-1}},
    \end{equation*}
    and
    \begin{equation*}
        \lrabs{\frac{n^2 \delta^{\top}\delta_i}{N_n}}^{-2} \cdot \tilde{V}_n^b(\xi_i) 
    \stra{p} 0,
    \end{equation*}
    which implies that
    \BEqn
    & & \lrabs{T_n^b(\lrfl{n\xi_{i-1}}-m,\lrfl{n\xi_i}-m,N) \lrp{V_n^b(\lrfl{n\xi_{i-1}}-m,\lrfl{n\xi_i}-m,N)}^{-1/2}} \\
    &=& \lrp{1-\varepsilon-\xi_{i-1}+o_p(1)} \lrabs{\tilde{T}_n^b(\xi_i)} \lrp{\tilde{V}_n^b(\xi_i)}^{-1/2} \\
    &=& \lrp{1-\varepsilon-\xi_{i-1}+o_p(1)} \frac{\lrabs{\frac{n^2 \delta^{\top}\delta_i}{N_n}}^{-1} \lrabs{\tilde{T}_n^b(\xi_i)}}{\lrp{\lrabs{\frac{n^2 \delta^{\top}\delta_i}{N_n}}^{-2} \cdot \tilde{V}_n^b(\xi_i)}^{1/2}}
    \stra{p} \infty,
    \EEqn
    which arrives that the desired result.
\end{enumerate}
\end{Proof}

\begin{lemma}\label{Lemma:G-DGP2}
Suppose that $\{\tilde{X}_t\}_{t=1}^{n}$ is a linear process as defined in Definition \ref{Def:Model_lp}. Assume that Assumption \ref{Assumpt:NullDist_Wn_lp}, Assumption \ref{Assumpt:multiple} hold and Assumption \ref{Assumpt:MultiplePower} is satisfied with $N_n = \sqrt{2nm_1} \|A^{(0)} \Gamma^{(2)} (A^{(0)})^{\top}\|_F$. Assume that $\rho^{m_2/4} \|\Gamma^{(2)}\|_F = o\lrp{\frac{n}{\log(n)}}$, then it holds that 
\begin{enumerate}[label=(\roman*)]
    \item \label{Lemma:Gf-DGP2}
    if Assumption \ref{Assumpt:MultiplePower}\ref{Assumpt:MultiplePower-1} holds, then we have that
    \begin{equation*}
        \lrabs{T_n^f(1,\lrfl{n\xi_i}-m,\lrfl{n\xi_{i+1}}-m) \lrp{V_n^f(1,\lrfl{n\xi_i}-m,\lrfl{n\xi_{i+1}}-m)}^{-1/2}}
    \stra{p} \infty.
    \end{equation*}
    
    \item\label{Lemma:Gb-DGP2}
    if Assumption \ref{Assumpt:MultiplePower}\ref{Assumpt:MultiplePower-2} holds, then we have that
    \begin{equation*}
        \lrabs{T_n^b(\lrfl{n\xi_{i-1}}-m,\lrfl{n\xi_i}-m,N) \lrp{V_n^b(\lrfl{n\xi_{i-1}}-m,\lrfl{n\xi_i}-m,N)}^{-1/2}}
    \stra{p} \infty.
    \end{equation*}
\end{enumerate}
\end{lemma}

\begin{Proof}
The statements can be proved using similar arguments as used for Lemma \ref{Lemma:G-DGP1}. 
\begin{enumerate}[label=(\roman*)]
    \item Suppose that Assumption \ref{Assumpt:MultiplePower}\ref{Assumpt:MultiplePower-1} holds with $N_n = \sqrt{2nm_1} \|A^{(0)} \Gamma^{(2)} (A^{(0)})^{\top}\|_F$, then we have $\frac{n^2 \lrabs{\delta^{\top}\delta_i}}{N_n} \rightarrow \infty$ as $n\rightarrow\infty$, $\|\delta\|_2^2 = O(\lrabs{\delta^{\top}\delta_i})$, $\max\limits_{1\le j\le i} \|\delta_j\|_2^2 = O(\lrabs{\delta^{\top}\delta_i})$ and $\max\limits_{1\le j<i} \lrabs{\delta^{\top}\delta_j} = o(\lrabs{\delta^{\top}\delta_i})$ as discussed in the proof of \ref{Lemma:G-DGP1}\ref{Lemma:Gf-DGP1}.
    
    Then under Assumption \ref{Assumpt:NullDist_Wn_lp}, Assumption \ref{Assumpt:multiple} and Assumption \ref{Assumpt:MultiplePower}, we have shown in previous proofs that if $\rho^{m_2/4}\|\Gamma^{(2)}\|_F = o\lrp{\frac{n}{\log(n)}}$, it holds that
    \begin{equation*}
        \frac{1}{N_n} \tilde{W}_n(r)
    = \frac{1}{\sqrt{2nm_1} \|A^{(0)} \Gamma^{(2)} (A^{(0)})^{\top}\|_F} \tilde{W}_n(r)
    \leadsto B(r)-B(\varepsilon)
    \quad\mbox{in }D[\varepsilon,1-\varepsilon].
    \end{equation*}
    Under the assumption that $\frac{n^2 \lrabs{\delta^{\top}\delta_i}}{N_n} \rightarrow \infty$, this further implies that
    \begin{equation*}
        \lrabs{\frac{n^2 \delta^{\top}\delta_i}{N_n}}^{-1} \cdot \frac{1}{N_n} \tilde{W}_n(r)
    \leadsto 0
    \quad\mbox{in }D[\varepsilon,1-\varepsilon].
    \end{equation*}

    Additionally, by Lemma S9.8 in \cite{wang2022inference}, it holds that
    \begin{equation*}
        \sup\limits_{u\in[\varepsilon,1-\varepsilon]} 
        \lrabs{\frac{1}{\|A^{(0)} \Gamma^{(2)} (A^{(0)})^{\top}\|_F} 
        \lrp{\frac{1}{\sqrt{n}} \sum\limits_{j=1}^{\lrfl{nu}-\lrfl{n\varepsilon}} \tilde{X}_{j+m}}^{\top} \delta} 
    = o_p\lrp{\frac{\|\delta\|_2}{\|A^{(0)} \Gamma^{(2)} (A^{(0)})^{\top}\|_F^{1/2}}},
    \end{equation*}
    and 
    \begin{equation*}
        \sup\limits_{u\in[\varepsilon,1-\varepsilon]} 
        \lrabs{\frac{1}{\|A^{(0)} \Gamma^{(2)} (A^{(0)})^{\top}\|_F} 
        \lrp{\frac{1}{\sqrt{n}} \sum\limits_{j=1}^{\lrfl{nu}-\lrfl{n\varepsilon}} \tilde{X}_{j+m}}^{\top} \delta_j}
    = o_p\lrp{\frac{\|\delta_j\|_2}{\|A^{(0)} \Gamma^{(2)} (A^{(0)})^{\top}\|_F^{1/2}}},
    \end{equation*}
    for $j=1,\cdots,M$.
    
    It follows from the first equality that
    \begin{equation*}
        \sup\limits_{u\in[\varepsilon,1-\varepsilon]} 
        \lrabs{\lrp{\frac{n}{N_n}}^{1/2} \lrp{\frac{1}{\sqrt{n}} \sum\limits_{j=1}^{\lrfl{nu}-\lrfl{n\varepsilon}} \tilde{X}_{j+m}}^{\top} \frac{\delta}{\|\delta\|_2}} 
    = o_p(1).
    \end{equation*}
    Furthermore, by using the assumptions, we have that
    \BEqn
    & & \sup\limits_{u\in[\varepsilon,1-\varepsilon]} \lrabs{ \lrabs{\frac{n^2}{N_n} \delta^{\top}\delta_i}^{-1} \cdot \frac{m_1}{N_n} \sum\limits_{j=1}^{\lrfl{nu}-\lrfl{n\varepsilon}} \tilde{X}_{j+m}^{\top} \delta} \\
    &=& (\varepsilon-\eta) \lrabs{\frac{n^2}{N_n} \delta^{\top}\delta_i}^{-1/2} \cdot \frac{\|\delta\|_2}{\sqrt{\lrabs{\delta^{\top}\delta_i}}} \cdot 
    \sup\limits_{u\in[\varepsilon,1-\varepsilon]} \lrabs{\lrp{\frac{n}{N_n}}^{1/2} \lrp{\frac{1}{\sqrt{n}} \sum\limits_{j=1}^{\lrfl{nu}-\lrfl{n\varepsilon}} \tilde{X}_{j+m}}^{\top} \frac{\delta}{\|\delta\|_2}} \\
    &=& o_p(1).
    \EEqn
    Similarly, for any $j\le i$, it holds that
    \BEqn
    & & \sup\limits_{u\in[\varepsilon,1-\varepsilon]} \lrabs{ \lrabs{\frac{n^2}{N_n} \delta^{\top}\delta_i}^{-1} \cdot \frac{n}{N_n} \sum\limits_{j=1}^{\lrfl{nu}-\lrfl{n\varepsilon}} \tilde{X}_{j+m}^{\top} \delta_j} \\
    &=& \lrabs{\frac{n^2}{N_n} \delta^{\top}\delta_i}^{-1/2} \cdot \frac{\|\delta_j\|_2}{\sqrt{\lrabs{\delta^{\top}\delta_i}}} \cdot 
    \sup\limits_{u\in[\varepsilon,1-\varepsilon]} \lrabs{  \lrp{\frac{n}{N_n}}^{1/2} \lrp{\frac{1}{\sqrt{n}} \sum\limits_{j=1}^{\lrfl{nu}-\lrfl{n\varepsilon}} \tilde{X}_{j+m}}^{\top} \frac{\delta_j}{\|\delta_j\|_2}} \\
    &=& o_p(1),
    \EEqn
    where we use the assumption that $\max\limits_{j\le i} \|\delta_j\|_2^2 = O(\lrabs{\delta^{\top}\delta_i})$.
    
    By using Lemma \ref{Lemma:Expression_Gf} again, we have that
    \BEqn
    & & T_n^f(1,\lrfl{n\xi_i}-m,\lrfl{n\xi_{i+1}}-m) \lrp{V_n^f(1,\lrfl{n\xi_i}-m,\lrfl{n\xi_{i+1}}-m)}^{-1/2} \\
    &=& \lrp{\xi_{i+1}-\varepsilon+o_p(1)} \tilde{T}_n^f(\xi_i) \lrp{\tilde{V}_n^f(\xi_i)}^{-1/2},
    \EEqn
    where $\tilde{T}_n^f(\xi_i)$ and $\tilde{V}_n^f(\xi_i)$ are defined as Lemma \ref{Lemma:Expression_Gf}. Consequently, by using the assumption that $\max\limits_{j<i} \lrabs{\delta^{\top}\delta_j} = o(\delta^{\top}\delta_i)$, it holds that
    \BEqn
    & & \lrabs{\frac{n^2}{N_n} \delta^{\top}\delta_i}^{-1} \cdot \tilde{T}_n^f(\xi_i) \\
    &=& \lrabs{\frac{n^2}{N_n} \delta^{\top}\delta_i}^{-1} \cdot \frac{1}{N_n} \lrp{\tilde{W}_n(\xi_i) - \lrp{\frac{\xi_i-\varepsilon}{\xi_{i+1}-\varepsilon}+o_p(1)} \tilde{W}_n(\xi_{i+1})} \\
    & & - \lrabs{\frac{n^2}{N_n} \delta^{\top}\delta_i}^{-1} \cdot \frac{n}{N_n} \sum\limits_{u=1}^{m_1} \lrp{\tilde{X}_u - \tilde{X}_{n+1-u}}^{\top} \lrp{\sum\limits_{t=1}^{i} \delta_t \lrp{\frac{(\xi_{i+1}-\xi_i)(\xi_t-\varepsilon)}{\xi_{i+1}-\varepsilon}+o_p(1)}} \\
    & & - \lrabs{\frac{n^2}{N_n} \delta^{\top}\delta_i}^{-1} \cdot \frac{m_1}{N_n} \delta^{\top} \lrp{\sum\limits_{j=1}^{\lrfl{n\xi_i}-m} \tilde{X}_{j+m} - \lrp{\frac{\xi_i-\varepsilon}{\xi_{i+1}-\varepsilon}+o_p(1)}  \sum\limits_{j=1}^{\lrfl{n\xi_{i+1}}-m} \tilde{X}_{j+m}} \\
    & & + \lrabs{\frac{n^2}{N_n} \delta^{\top}\delta_i}^{-1} \cdot \frac{m_1 n}{N_n} \sum\limits_{t=0}^{i} \delta^{\top} \delta_t \lrp{\frac{(\xi_{i+1}-\xi_i)(\xi_t-\varepsilon)}{\xi_{i+1}-\varepsilon}+o_p(1)} \\
    &=& \lrabs{\frac{n^2}{N_n} \delta^{\top}\delta_i}^{-1} \cdot \frac{m_1 n}{N_n} \sum\limits_{t=0}^{i} \delta^{\top} \delta_t \lrp{\frac{(\xi_{i+1}-\xi_i)(\xi_t-\varepsilon)}{\xi_{i+1}-\varepsilon}+o_p(1)} + o_p(1) \\
    &=& (\varepsilon-\eta) \sum\limits_{t=0}^{i-1} \frac{\delta^{\top}\delta_t}{\lrabs{\delta^{\top}\delta_i}} \lrp{\frac{(\xi_{i+1}-\xi_i)(\xi_t-\varepsilon)}{\xi_{i+1}-\varepsilon}+o_p(1)} \\
    & & + (\varepsilon-\eta) \sgn(\delta^{\top}\delta_i) \lrp{\frac{(\xi_{i+1}-\xi_i)(\xi_i-\varepsilon)}{\xi_{i+1}-\varepsilon}+o_p(1)} + o_p(1) \\
    &\stra{p}& \sgn(\delta^{\top}\delta_i) \frac{(\varepsilon-\eta)(\xi_{i+1}-\xi_i)(\xi_i-\varepsilon)}{\xi_{i+1}-\varepsilon}.
    \EEqn
    
    We can use the same arguments to derive the limiting distribution of $\tilde{V}_n^{f}(\xi_i)$, where the expression of $\tilde{V}_n^{f}(\xi_i)$ is derived in Lemma \ref{Lemma:Expression_Gf}, that is
    \BEqn
    & & \lrabs{\frac{n^2}{N_n} \delta^{\top}\delta_i}^{-2} \cdot \tilde{V}_n^f(\xi_i) \\
    &=& \sum\limits_{h=1}^{i} \int_{\xi_{h-1}}^{\xi_h} 
        \left(
            \lrabs{\frac{n^2}{N_n} \delta^{\top}\delta_i}^{-1} \cdot \frac{1}{N_n} \lrp{\tilde{W}_n(s) - \lrp{\frac{s-\varepsilon}{\xi_i-\varepsilon}+o_p(1)} \tilde{W}_n(\xi_i)} 
        \right. \\
    & & \hspace{2em}    
        - \lrabs{\frac{n^2}{N_n} \delta^{\top}\delta_i}^{-1} \cdot \frac{n}{N_n} \sum\limits_{u=1}^{m_1} \lrp{\tilde{X}_u - \tilde{X}_{n+1-u}}^{\top} \lrp{\bone\{h>1\} \sum\limits_{t=1}^{h-1} \delta_t \lrp{\frac{(\xi_i-s)(\xi_t-\varepsilon)}{\xi_i-\varepsilon}+o_p(1)}} \\
    & & \hspace{2em}
        - \lrabs{\frac{n^2}{N_n} \delta^{\top}\delta_i}^{-1} \cdot \frac{n}{N_n} \sum\limits_{u=1}^{m_1} \lrp{\tilde{X}_u - \tilde{X}_{n+1-u}}^{\top} \lrp{\bone\{h<i\} \sum\limits_{t=h}^{i-1} \delta_t \lrp{\frac{(\xi_i-\xi_t)(s-\varepsilon)}{\xi_i-\varepsilon}+o_p(1)}} \\
    & & \hspace{2em}    
        - \lrabs{\frac{n^2}{N_n} \delta^{\top}\delta_i}^{-1} \cdot \frac{m_1}{N_n} \delta^{\top} \lrp{\sum\limits_{j=1}^{\lrfl{ns}-m} \tilde{X}_{j+m} - \lrp{\frac{s-\varepsilon}{\xi_i-\varepsilon}+o_p(1)} \sum\limits_{j=1}^{\lrfl{n\xi_i}-m} \tilde{X}_{j+m}} \\ 
    & & \hspace{2em}    
        + \lrabs{\frac{n^2}{N_n} \delta^{\top}\delta_i}^{-1} \cdot \frac{m_1 n}{N_n} \lrp{\bone\{h>1\} \sum\limits_{t=1}^{h-1} \delta^{\top} \delta_t \lrp{\frac{(\xi_i-s)(\xi_t-\varepsilon)}{\xi_i-\varepsilon}+o_p(1)}} \\
    & & \hspace{2em}      
        \left.
        + \lrabs{\frac{n^2}{N_n} \delta^{\top}\delta_i}^{-1} \cdot \frac{m_1 n}{N_n} \lrp{\bone\{h<i\} \sum\limits_{t=h}^{i-1} \delta^{\top} \delta_t \lrp{\frac{(\xi_i-\xi_t)(s-\varepsilon)}{\xi_i-\varepsilon}+o_p(1)}}
        \right)^2 ds \\
    &+& \int_{\xi_i}^{\xi_{i+1}}
        \left(
            \lrabs{\frac{n^2}{N_n} \delta^{\top}\delta_i}^{-1} \cdot \frac{1}{N_n} \lrp{\tilde{W}_n(\xi_{i+1}) - \tilde{W}_n(s) - \lrp{\frac{\xi_{i+1}-s}{\xi_{i+1}-\xi_i}+o_p(1)} \lrp{\tilde{W}_n(\xi_{i+1})-\tilde{W}_n(\xi_i)}}
        \right. \\
    & & \hspace{2em}    
        \left.
            - \lrabs{\frac{n^2}{N_n} \delta^{\top}\delta_i}^{-1} \cdot \frac{m_1}{N_n} \delta^{\top} \lrp{\sum\limits_{j=\lrfl{ns}-m+1}^{\lrfl{n\xi_{i+1}}-m} \tilde{X}_{j+m} - \lrp{\frac{\xi_{i+1}-s}{\xi_{i+1}-\xi_i}+o_p(1)} \sum\limits_{j=\lrfl{n\xi_i}-m+1}^{\lrfl{n\xi_{i+1}}-m} \tilde{X}_{j+m}}
        \right)^2 ds \\
    & & + o_p(1) \\
    &=& o_p(1),
    \EEqn
    where the last steps follows from the assumption $\max\limits_{j<i} \lrabs{\delta^{\top}\delta_j} = o\lrp{\lrabs{\delta^{\top}\delta_i}}$.
    
    Therefore, we apply the Continuous Mapping Theorem again to obtain that
    \BEqn
    & & \lrabs{T_n^f(1,\lrfl{n\xi_i}-m,\lrfl{n\xi_{i+1}}-m) \lrp{V_n^f(1,\lrfl{n\xi_i}-m,\lrfl{n\xi_{i+1}}-m)}^{-1/2}} \\
    &=& \lrp{\xi_{i+1}-\varepsilon+o_p(1)} \lrabs{\tilde{T}_n^f(\xi_i)} \lrp{\tilde{V}_n^f(\xi_i)}^{-1/2} \\
    &=& \lrp{\xi_{i+1}-\varepsilon+o_p(1)} \frac{\lrabs{\frac{n^2 \delta^{\top}\delta_i}{N_n}}^{-1} \cdot \lrabs{\tilde{T}_n^f(\xi_i)}}{\lrp{\lrabs{\frac{n^2 \delta^{\top}\delta_i}{N_n}}^{-2} \cdot \tilde{V}_n^f(\xi_i)}^{1/2}} 
    \stra{p} \infty,
    \EEqn
    which completes the proof.
    
    \item Suppose that Assumption \ref{Assumpt:MultiplePower}\ref{Assumpt:MultiplePower-2} holds, then we have
    $\frac{n^2 \lrabs{\delta^{\top}\delta_i}}{N_n} \rightarrow \infty$ as $n\rightarrow\infty$, $\|\delta\|_2^2 = O(\lrabs{\delta^{\top}\delta_i})$, $\max\limits_{M\geq j\geq i} \|\delta_j\|_2^2 = O(\lrabs{\delta^{\top}\delta_i})$ and $\max\limits_{M\geq j>i} \lrabs{\delta^{\top}\delta_j} = o(\lrabs{\delta^{\top}\delta_i})$. Using the similar arguments as used for Lemma \ref{Lemma:G-DGP1}\ref{Lemma:Gb-DGP1} and Lemma \ref{Lemma:G-DGP2}\ref{Lemma:Gf-DGP2}, we can show that
    By using the same arguments as for the previous case, we can again show that
    \begin{equation*}
        \lrabs{\frac{n^2 \delta^{\top}\delta_i}{N_n}}^{-1} \lrabs{\tilde{T}_n^b(\xi_i)}
    \stra{p} \frac{(\varepsilon-\eta)(\xi_i-\xi_{i-1})(1-\varepsilon-\xi_i)}{1-\varepsilon-\xi_{i-1}},
    \end{equation*}
    and
    \begin{equation*}
        \lrabs{\frac{n^2 \delta^{\top}\delta_i}{N_n}}^{-2} \cdot \tilde{V}_n^b(\xi_i) 
    \stra{p} 0,
    \end{equation*}
    for which we spare the detailed steps. By applying the Continuous Mapping Theorem again, we arrive at the desired statement, that is
    \BEqn
    & & \lrabs{T_n^b(\lrfl{n\xi_{i-1}}-m,\lrfl{n\xi_i}-m,N) \lrp{V_n^b(\lrfl{n\xi_{i-1}}-m,\lrfl{n\xi_i}-m,N)}^{-1/2}} \\
    &=& \lrp{1-\varepsilon-\xi_{i-1}+o_p(1)} \lrabs{\tilde{T}_n^b(\xi_i)} \lrp{\tilde{V}_n^b(\xi_i)}^{-1/2} \\
    &=& \lrp{1-\varepsilon-\xi_{i-1}+o_p(1)} \frac{\lrabs{\frac{n^2 \delta^{\top}\delta_i}{N_n}}^{-1} \lrabs{\tilde{T}_n^b(\xi_i)}}{\lrp{\lrabs{\frac{n^2 \delta^{\top}\delta_i}{N_n}}^{-2} \cdot \tilde{V}_n^b(\xi_i)}^{1/2}}
    \stra{p} \infty.
    \EEqn
\end{enumerate}
\end{Proof}

\begin{lemma}\label{Lemma:G-DGP3}
Suppose that $\{\tilde{X}_t\}_{t=1}^{n}$ is a generated from a factor model as defined in Definition \ref{Def:Model_factor}. Assume that Assumption \ref{Assumpt:NullDist_Wn_lp} (applied to $Z_t$), Assumption \ref{Assumpt:NullDist_Wn_factor}, Assumption \ref{Assumpt:multiple} hold and Assumption \ref{Assumpt:MultiplePower} is satisfied with $N_n = \max\{n\|\Lambda^{\top}\Lambda\|, \sqrt{2nm_1} \|A^{(0)} \Gamma^{(3)} (A^{(0)})^{\top}\|_F\}$. Assume that $\rho^{m_2/4} \|\Gamma^{(3)}\|_F = o\lrp{\frac{n}{\log(n)}}$, then it holds that 
\begin{enumerate}[label=(\roman*)]
    \item\label{Lemma:Gf-DGP3}
    if Assumption \ref{Assumpt:MultiplePower}\ref{Assumpt:MultiplePower-1} holds, then we have that
    \begin{equation*}
        \lrabs{T_n^f(1,\lrfl{n\xi_i}-m,\lrfl{n\xi_{i+1}}-m) \lrp{V_n^f(1,\lrfl{n\xi_i}-m,\lrfl{n\xi_{i+1}}-m)}^{-1/2}}
    \stra{p} \infty.
    \end{equation*}
    
    \item\label{Lemma:Gb-DGP3}
    if Assumption \ref{Assumpt:MultiplePower}\ref{Assumpt:MultiplePower-2} holds, then we have that
    \begin{equation*}
        \lrabs{T_n^b(\lrfl{n\xi_{i-1}}-m,\lrfl{n\xi_i}-m,N) \lrp{V_n^b(\lrfl{n\xi_{i-1}}-m,\lrfl{n\xi_i}-m,N)}^{-1/2}}
    \stra{p} \infty.
    \end{equation*}
\end{enumerate}
\end{lemma}

\begin{Proof}
\begin{enumerate}[label=(\roman*)]
    \item We first consider the case that Assumption \ref{Assumpt:MultiplePower}\ref{Assumpt:MultiplePower-1} holds with $N_n = \max\{n\|\Lambda^{\top}\Lambda\|,\sqrt{2n m_1}\|A^{(0)} \Gamma^{(3)} (A^{(0)})^{\top}\|_F\}$. As discussed in Lemma \ref{Lemma:G-DGP1}\ref{Lemma:Gf-DGP1} and Lemma \ref{Lemma:G-DGP2}\ref{Lemma:Gf-DGP2}, it holds that $\frac{n^2 \lrabs{\delta^{\top}\delta_i}}{N_n}$, $\|\delta\|_2^2 = O(\lrabs{\delta^{\top}\delta_i})$, $\max\limits_{1\le j\le i} \|\delta_j\|_2^2 = O(\lrabs{\delta^{\top}\delta_i})$ and $\max\limits_{1\le j<i} \lrabs{\delta^{\top}\delta_j} = o\lrp{\lrabs{\delta^{\top}\delta_i}}$. 
    
    Under Assumption \ref{Assumpt:NullDist_Wn_lp} (applied to $Z_t$), Assumption \ref{Assumpt:NullDist_Wn_factor} and Assumption \ref{Assumpt:MultiplePower}, if $\rho^{m_2/4} \|\Gamma^{(3)}\|_F = o\lrp{\frac{n}{\log(n)}}$, it is shown in previous proofs that, there always exists two deterministic constants $c_1,c_2$, such that it holds in $D[\varepsilon,1-\varepsilon]$ that
    \begin{equation*}
        \frac{1}{N_n} \tilde{W}_n(r)
    \leadsto  c_1 b^{\top}(\varepsilon,\eta) ((\Omega^{(3)})^{1/2})^{\top} L_0 (\Omega^{(3)})^{1/2} (B_s(r) - B_s(\varepsilon)) 
    + c_2 \lrp{\tilde{B}(r) - \tilde{B}(\varepsilon)},
    \end{equation*}
    where $\{B_s(r)\}_{0\le r\le 1}$, $\{\tilde{B}(r)\}_{0\le r\le 1}$ are two independent Brownian motions in $\br^s$ and $\br$ respectively and $b(\varepsilon,\eta) = B_s(\varepsilon-\eta) - B_s(1) + B_s(1-\varepsilon+\eta)$.
    
    It follows that
    \begin{equation*}
        \lrabs{\frac{n^2 \delta^{\top}\delta_i}{N_n}}^{-1} \cdot \frac{1}{N_n} \tilde{W}_n(r)
    \leadsto 0
    \quad \mbox{in } D[\varepsilon,1-\varepsilon].
    \end{equation*}
    
    Next we consider the limiting distribution of $\lrabs{\frac{n^2 \delta^{\top}\delta_i}{N_n}}^{-1} \cdot \frac{n}{N_n} \sum\limits_{j=1}^{\lrfl{nu}-m} \tilde{X}_{j+m}^{\top} \delta_t$ in $D[\varepsilon,1-\varepsilon]$ for $t=1,\cdots,i$. It follows from the definition in Definition \ref{Def:Model_factor} that
    \begin{equation*}
        \frac{n}{N_n} \sum\limits_{j=1}^{\lrfl{nu}-m} \tilde{X}_{j+m}^{\top} \delta_t
      = \frac{n}{N_n} \sum\limits_{j=1}^{\lrfl{nu}-m} F_{j+m}^{\top} \Lambda^{\top} \delta_t
      + \frac{n}{N_n} \sum\limits_{j=1}^{\lrfl{nu}-m} Z_{j+m}^{\top} \delta_t.
    \end{equation*}
    Under Assumption \ref{Assumpt:NullDist_Wn_factor}\ref{Assumpt:NullDist_Wn_factor-1}, we have that 
    \begin{equation*}
        \frac{1}{\sqrt{n}} \sum\limits_{j=1}^{\lrfl{nu}-m} F_{j+m}^{\top} \Lambda^{\top} \delta_t
    \leadsto (B_s(u)-B_s(\varepsilon))^{\top} ((\Omega^{(3)})^{1/2})^{\top} \Lambda^{\top} \delta_t
    \quad
    \mbox{in }D[\varepsilon,1-\varepsilon],
    \end{equation*}
    implying that for $t=1,\cdots,i$,
    \BEqn
    & & \lrabs{\frac{n^2 \delta^{\top}\delta_i}{N_n}}^{-1} \cdot \frac{n}{N_n} \sum\limits_{j=1}^{\lrfl{nu}-m} F_{j+m}^{\top} \Lambda^{\top} \delta_t \\
    &=& \sqrt{\frac{\|\Lambda^{\top}\Lambda\|}{n\lrabs{\delta^{\top}\delta_i}}} \cdot \lrp{\frac{1}{\sqrt{n}} \sum\limits_{j=1}^{\lrfl{nu}-m} F_{j+m}}^{\top} \lrp{\frac{\Lambda}{\sqrt{\|\Lambda^{\top}\Lambda\|}}}^{\top} \lrp{\frac{\delta_t}{\sqrt{\lrabs{\delta^{\top}\delta_i}}}} \\
    &\stra{p}& 0
    \quad\mbox{in }D[\varepsilon,1-\varepsilon],
    \EEqn
    where the last step follows from the assumption $\max\limits_{t\le i} \|\delta\|_2^2 = O(\lrabs{\delta^{\top}\delta_i})$ and the observation that
    \begin{equation*}
        \frac{\|\Lambda^{\top}\Lambda\|}{n\lrabs{\delta^{\top}\delta_i}}
      = O\lrp{\frac{N_n}{n^2 \lrabs{\delta^{\top}\delta_i}}}
      = o(1).
    \end{equation*}
    
    Furthermore, by Lemma S9.8 in \cite{wang2022inference}, we have that
    \begin{equation*}
        \sup\limits_{u\in[\varepsilon,1-\varepsilon]}
        \lrabs{\frac{1}{\|A^{(0)} \Gamma^{(3)} (A^{(0)})^{\top}\|_F} \lrp{\frac{1}{\sqrt{n}} \sum\limits_{j=1}^{\lrfl{nu}-\lrfl{n\varepsilon}} Z_{j+m}}^{\top} \delta_t} 
    = o_p\lrp{\frac{\|\delta_t\|_2}{\|A^{(0)} \Gamma^{(3)} (A^{(0)})^{\top}\|_F^{1/2}}}
    \end{equation*}
    for $t=1,\cdots,i$, which implies that
    \BEqn
    & & \sup\limits_{u\in[\varepsilon,1-\varepsilon]}
        \lrabs{\lrp{\frac{n^2 \lrabs{\delta^{\top}\delta_i}}{N_n}}^{-1} \cdot \frac{n}{N_n} \sum\limits_{j=1}^{\lrfl{nu}-\lrfl{n\varepsilon}} Z_{j+m}^{\top} \delta_t} \\
    &=& \sup\limits_{u\in[\varepsilon,1-\varepsilon]}
        \lrabs{\lrp{\frac{\|A^{(0)} \Gamma^{(3)} (A^{(0)})^{\top}\|_F}{n \lrabs{\delta^{\top}\delta_i}}}^{1/2} \cdot \frac{1}{\|A^{(0)} \Gamma^{(3)} (A^{(0)})^{\top}\|_F^{1/2}} \lrp{\frac{1}{\sqrt{n}} \sum\limits_{j=1}^{\lrfl{nu}-\lrfl{n\varepsilon}} Z_{j+m}}^{\top} \frac{\delta_t}{\|\delta_t\|_2} \cdot \frac{\|\delta_t\|_2}{\sqrt{\lrabs{\delta^{\top}\delta_i}}}} \\
    &=& O\lrp{\lrp{\frac{N_n}{n^2 \lrabs{\delta^{\top}\delta_i}}}^{1/2}} \cdot \frac{\|\delta_t\|_2}{\sqrt{\lrabs{\delta^{\top}\delta_i}}} \cdot o_p(1) \\
    &=& o_p(1),
    \EEqn
    where the second to the last step follows from the definition of $N_n$ and the last step uses the assumptions that $\frac{n^2 \lrabs{\delta^{\top}\delta_i}}{N_n} \rightarrow \infty$ and $\max\limits_{t\le i} \|\delta_t\|_2^2 = O(\lrabs{\delta^{\top}\delta_i})$.
    
    By combining the two parts of analysis, we obtain that for $t=1,\cdots,i$, it holds that
    \begin{equation*}
        \lrabs{\frac{n^2 \delta^{\top}\delta_i}{N_n}}^{-1} \cdot \frac{n}{N_n} \sum\limits_{j=1}^{\lrfl{nu}-m} \tilde{X}_{j+m}^{\top} \delta_t \\
     =  o_p(1)
    \quad\mbox{in }D[\varepsilon,1-\varepsilon]
    \end{equation*}
    Using similar arguments, we also obtain that
    \begin{equation*}
        \lrabs{\frac{n^2 \delta^{\top}\delta_i}{N_n}}^{-1} \cdot \frac{m_1}{N_n} \sum\limits_{j=1}^{\lrfl{nu}-m} \tilde{X}_{j+m}^{\top} \delta \\
     =  o_p(1)
    \quad\mbox{in }D[\varepsilon,1-\varepsilon]
    \end{equation*}
    
    Finally, by using the assumption $\max\limits_{j<i} \lrabs{\delta^{\top}\delta_j} = o\lrp{\lrabs{\delta^{\top}\delta_i}}$ and the similar arguments used to prove previous lemmas, we can show that
    \begin{equation*}
        \lrabs{\frac{n^2}{N_n} \delta^{\top}\delta_i}^{-1} \cdot \lrabs{\tilde{T}_n^f(\xi_i)}
    \stra{p} \frac{(\varepsilon-\eta)(\xi_{i+1}-\xi_i)(\xi_i-\varepsilon)}{\xi_{i+1}-\varepsilon}
    \end{equation*}
    and $\lrabs{\frac{n^2}{N_n} \delta^{\top}\delta_i}^{-2} \cdot \lrabs{\tilde{V}_n^f(\xi_i)} \stra{p} 0$. Therefore, the statement in Lemma \ref{Lemma:G-DGP3}\ref{Lemma:Gf-DGP3} is a direct result of Lemma \ref{Lemma:Expression_Gf}.
    
    \item Suppose that Assumption \ref{Assumpt:MultiplePower}\ref{Assumpt:MultiplePower-2} holds, the claimed result can be shown using similar arguments as used for Lemma \ref{Lemma:G-DGP1}\ref{Lemma:Gb-DGP1}, Lemma \ref{Lemma:G-DGP2}\ref{Lemma:Gb-DGP2} and Lemma \ref{Lemma:G-DGP3}\ref{Lemma:Gf-DGP3}, thus we skip all the details.
\end{enumerate}
\end{Proof}


\section{Auxiliary Lemmas II}\label{Appdix:AuxLemma-2}

\subsection{Auxiliary Results for Appendix \ref{Appdix:lemma_null_lp}}

\begin{lemma}\label{Lemma:LP-aux-1-1}
For any $A\in\br^{p \times q}$, we have $\lrabs{A_{ij}} \le \|A\|_2$ for any $1 \le i \le p$ and $1 \le j \le q$.
\end{lemma}
\begin{Proof}
Let $e_i\in\br^p$ be the $p$-dimensional vector whose $i$-th element is one and all other elements are zero. Then it follows from the definition of the spectral norm that
\begin{equation}
    \lrabs{A_{ij}} 
  = \lrabs{e_i^{\top} A e_j}
  \le \max\limits_{\|u\|=1} \lrag{A e_j, u}
  = \|A e_j\|_2
  \le \max\limits_{\|u\|=1} \|Au\|_2
  = \|A\|_2.
\end{equation}
\end{Proof}

\begin{lemma}\label{Lemma:LP-aux-1-2}
For any $A\in\br^{r\times p}$, $\Gamma\in\br^{p\times p}$ and $B\in\br^{p\times r}$, it holds that
\begin{equation*}
    \tr(A \Gamma B) 
\le \|A\| \cdot \|B\| \cdot \|\Gamma\|_F \cdot \sqrt{r}.
\end{equation*}
\end{lemma}
\begin{Proof}
It it trivial that 
\begin{equation*}
    \tr(A \Gamma B) 
  = \tr(A \Gamma B I_r)
  \le \|A \Gamma B\|_F \|I_r\|_F
  \le \|A\| \cdot \|B\| \cdot \|\Gamma\|_F \sqrt{r}.
\end{equation*}
\end{Proof}

\begin{lemma}\label{Lemma:LP-aux-2}
Under Assumption \ref{Assumpt:NullDist_Wn_lp}\ref{Assumpt:NullDist_Wn_Lp-1}, it holds that
\begin{equation*}
    \sum\limits_{\ell_1,\ell_2,\ell_3,\ell_4=1}^{p} 
\lrp{\sum\limits_{k_1,k_2,k_3,k_4=1}^{p} A_{\ell_1,k_1}^{(s_1)} A_{\ell_2,k_2}^{(s_2)} A_{\ell_3,k_3}^{(s_3)} A_{\ell_4,k_4}^{(s_4)}  \cum(\varepsilon_{0,k_1}, \varepsilon_{0,k_2}, \varepsilon_{0,k_3}, \varepsilon_{0,k_4})}^2
\lesssim \lrp{ \prod\limits_{i=1}^{4} \|A^{(s_i)}\|^2 } \|\Gamma^{(2)}\|_F^4.
\end{equation*}
\end{lemma}
\begin{Proof}
For any fixed $\ell_2,\ell_3,\ell_4$, let $B = (B_1,\cdots,B_p)^{\top}$ denote the $p$-dimensional vector with
\begin{equation*}
    B_{k_1} = \sum\limits_{k_2,k_3,k_4=1}^{p} A_{\ell_2,k_2}^{(s_2)} A_{\ell_3,k_3}^{(s_3)} A_{\ell_4,k_4}^{(s_4)}  \cum(\varepsilon_{0,k_1}, \varepsilon_{0,k_2}, \varepsilon_{0,k_3}, \varepsilon_{0,k_4}),
\end{equation*}
then it follows that
\begin{equation*}
    \sum\limits_{\ell_1=1}^{p} \lrp{\sum\limits_{k_1=1}^{p} A_{\ell_1,k_1}^{(s_1)} B_{k_1}}^2 \\
  = \|A^{(s_1)} B\|_2^2 
  \le \|A^{(s_1)}\|^2 \|B\|_2^2
  = \|A^{(s_1)}\|^2 \sum\limits_{k_1=1}^{p} B_{k_1}^2,
\end{equation*}
where $\|A^{(s_1)}\|$ denotes the spectral norm of $A^{(s_1)}$. By repeatedly using this technique, we obtain that
\BEqn
& & \sum\limits_{\ell_1,\ell_2,\ell_3,\ell_4=1}^{p} 
\lrp{\sum\limits_{k_1,k_2,k_3,k_4=1}^{p} A_{\ell_1,k_1}^{(s_1)} A_{\ell_2,k_2}^{(s_2)} A_{\ell_3,k_3}^{(s_3)} A_{\ell_4,k_4}^{(s_4)}  \cum(\varepsilon_{0,k_1}, \varepsilon_{0,k_2}, \varepsilon_{0,k_3}, \varepsilon_{0,k_4})}^2 \\
&\le& \|A^{(s_1)}\|^2 \sum\limits_{k_1=1}^{p} \sum\limits_{\ell_2,\ell_3,\ell_4=1}^{p} \lrp{\sum\limits_{k_2,k_3,k_4=1}^{p} A_{\ell_2,k_2}^{(s_2)} A_{\ell_3,k_3}^{(s_3)} A_{\ell_4,k_4}^{(s_4)}  \cum(\varepsilon_{0,k_1}, \varepsilon_{0,k_2}, \varepsilon_{0,k_3}, \varepsilon_{0,k_4})}^2 \\
&\le& \cdots \\
&\le& \lrp{ \prod\limits_{i=1}^{4} \|A^{(s_i)}\|^2 }\sum\limits_{k_1,k_2,k_3,k_4=1}^{p} \cum^2(\varepsilon_{0,k_1}, \varepsilon_{0,k_2}, \varepsilon_{0,k_3}, \varepsilon_{0,k_4}) \\
&\lesssim& \lrp{ \prod\limits_{i=1}^{4} \|A^{(s_i)}\|^2 } \|\Gamma^{(2)}\|_F^4,
\EEqn
which completes the proof.
\end{Proof}

\begin{lemma}\label{Lemma:LP-aux-3}
For any integers $\{m_1,\cdots,m_k\}$ satisfying0 $m_1+\cdots+m_k=8$, let 
\begin{equation*}
    \pi_8(m_1,\cdots,m_k) = \{(\tau_1,\cdots,\tau_{m_1}), \cdots, (\tau_{8-m_k+1},\cdots,\tau_8)\}
\end{equation*}
denotes a $k$-part partition of $\{1,\cdots,8\}$ with sizes $m_1,\cdots,m_k$. Let $\ell(\tau) = \ell_{\lrceil{\tau/2}}$, where $\lrceil{x}$ denotes the smallest integer that is no smaller than $x$. Define 
\BEqn
    \ca(m_1,\cdots,m_k)
&=& \sum\limits_{\pi_8(m_1,\cdots,m_k)}
    \prod\limits_{r=1}^{k}
    \bone\{i_{\tau_{m_1+\cdots+m_{r-1}+1}}=\cdots=i_{\tau_{m_1+\cdots+m_r}}\} \\
& & \hspace{2em} \times 
    \cum\lrp{ (A^{(u_{\tau_{m_1+\cdots+m_{r-1}+1}})} \varepsilon_0)_{\ell(\tau_{m_1+\cdots+m_{r-1}+1})},
          \cdots,
          (A^{(u_{\tau_{m_1+\cdots+m_r}})} \varepsilon_0)_{\ell(\tau_{m_1+\cdots+m_r})} },
\EEqn
then it holds for any $u_1,\cdots,u_8\geq0$ and $i_1,\cdots,i_8\in\bz$ that
\BEqn
    \ca
&:=& \blre{ (A^{(u_1)} \varepsilon_{i_1})^{\top}
           (A^{(u_2)} \varepsilon_{i_2})^{\top}
           (A^{(u_3)} \varepsilon_{i_3})^{\top}
           (A^{(u_4)} \varepsilon_{i_4})^{\top}
           (A^{(u_5)} \varepsilon_{i_5})^{\top}
           (A^{(u_6)} \varepsilon_{i_6})^{\top}
           (A^{(u_7)} \varepsilon_{i_7})^{\top} 
           (A^{(u_8)} \varepsilon_{i_8})^{\top}} \\
&=& \sum\limits_{\ell_1,\ell_2,\ell_3,\ell_4=1}^{p} \sum\limits_{i=1}^{7} \ca_i
\EEqn
where 
\begin{equation*}
    \ca_1 = \ca(8), \
    \ca_2 = \ca(2,6), \ 
    \ca_3 = \ca(3,5), \ 
    \ca_4 = \ca(4,4), \ 
    \ca_5 = \ca(2,3,3), \
    \ca_6 = \ca(2,2,4), \
    \ca_7 = \ca(2,2,2,2).
\end{equation*}

\end{lemma}
\begin{remark}
For example, 
\BEqn
    \ca_3
&=& \sum\limits_{\pi_8(3,5)} 
    \bone\{i_{\tau_1}=i_{\tau_2}=i_{\tau_3}\}
    \cum( (A^{(u_1)} \varepsilon_0)_{\ell(\tau_1)},
          (A^{(u_2)} \varepsilon_0)_{\ell(\tau_2)},
          (A^{(u_3)} \varepsilon_0)_{\ell(\tau_2)}) \\
& & \hspace{2em} \times
    \bone\{i_{\tau_4}=\cdots=i_{\tau_8}\}
    \cum( (A^{(u_4)} \varepsilon_0)_{\ell(\tau_4)},
          (A^{(u_5)} \varepsilon_0)_{\ell(\tau_5)},
          (A^{(u_6)} \varepsilon_0)_{\ell(\tau_6)},
          (A^{(u_7)} \varepsilon_0)_{\ell(\tau_7)},
          (A^{(u_8)} \varepsilon_0)_{\ell(\tau_8)}).     
\EEqn
\end{remark}
\begin{Proof}
It is trivial that
\BEqn
& & \blre{ (A^{(u_1)} \varepsilon_{i_1})^{\top}
           (A^{(u_2)} \varepsilon_{i_2})^{\top}
           (A^{(u_3)} \varepsilon_{i_3})^{\top}
           (A^{(u_4)} \varepsilon_{i_4})^{\top}
           (A^{(u_5)} \varepsilon_{i_5})^{\top}
           (A^{(u_6)} \varepsilon_{i_6})^{\top}
           (A^{(u_7)} \varepsilon_{i_7})^{\top} 
           (A^{(u_8)} \varepsilon_{i_8})^{\top}} \\
&=& \sum\limits_{\ell_1,\ell_2,\ell_3,\ell_4=1}^{p}
    \be\left[
           (A^{(u_1)} \varepsilon_{i_1})_{\ell_1}
           (A^{(u_2)} \varepsilon_{i_2})_{\ell_1}
           (A^{(u_3)} \varepsilon_{i_3})_{\ell_2}
           (A^{(u_4)} \varepsilon_{i_4})_{\ell_2} 
    \right. \\
& & \hspace{6em} \times
    \left.
           (A^{(u_5)} \varepsilon_{i_5})_{\ell_3}
           (A^{(u_6)} \varepsilon_{i_6})_{\ell_3}
           (A^{(u_7)} \varepsilon_{i_7})_{\ell_4}
           (A^{(u_8)} \varepsilon_{i_8})_{\ell_4}
    \right].
\EEqn
Note that the index $\ell$ associated with $A^{(u_i)}$ is $\ell_{\lrceil{i/2}}$. For each $\tau_i$, let $t_{\tau_i}$ denote the corresponding index out of $\{\ell_1,\ell_2,\ell_3,\ell_4\}$, e.g. when $\tau_1=1$ and $\tau_2=2$, $A^{(u_{\tau_1})}$ and $A^{(u_{\tau_2})}$ are both associated with $t_{\tau_1}=t_{\tau_2}=\ell_1$. The decomposition of $\ca$ directly follows from the cumulant formula. 
\end{Proof}

\begin{lemma}\label{Lemma:LP-aux-3-1}
Under Assumption \ref{Assumpt:NullDist_Wn_lp}\ref{Assumpt:NullDist_Wn_Lp-2}, it holds that
\begin{equation*}
    \lrabs{\sum\limits_{\ell_1,\ell_2,\ell_3,\ell_4=1}^{p} \ca_1}
\lesssim \bone\{i_1=\cdots=i_8\}
         \lrp{ \prod\limits_{i=1}^{8} \|A^{(u_i)}\| } \|\Gamma^{(2)}\|_F^8,
\end{equation*}
where $\ca_1$ is defined as Lemma \ref{Lemma:LP-aux-3}.
\end{lemma}
\begin{Proof}
It follows from some simple calculations that
\BEqn
& & \left| \sum\limits_{\ell_1,\ell_2,\ell_3,\ell_4=1}^{p} 
    \cum( (A^{(u_1)} \varepsilon_0)_{\ell_1},
          (A^{(u_2)} \varepsilon_0)_{\ell_1},
          (A^{(u_3)} \varepsilon_0)_{\ell_2},
          (A^{(u_4)} \varepsilon_0)_{\ell_2},
    \right. \\
& & \hspace{7em}          
    \left.
          (A^{(u_5)} \varepsilon_0)_{\ell_3},
          (A^{(u_6)} \varepsilon_0)_{\ell_3},
          (A^{(u_7)} \varepsilon_0)_{\ell_4},
          (A^{(u_8)} \varepsilon_0)_{\ell_4}) \right| \\
&\le& \sum\limits_{\substack{1\le \ell_1,\cdots,\ell_4\le p \\ 1\le k_1,\cdots,k_8\le p}}
    \left|
        A_{\ell_1,k_1}^{(u_1)} A_{\ell_1,k_2}^{(u_2)} A_{\ell_2,k_3}^{(u_3)} A_{\ell_2,k_4}^{(u_4)} A_{\ell_3,k_5}^{(u_5)} A_{\ell_3,k_6}^{(u_6)} A_{\ell_4,k_7}^{(u_7)} A_{\ell_4,k_8}^{(u_8)}
    \right. \\
& & \hspace{5em} \times    
    \left.
        \cum(\varepsilon_{0,k_1}, \varepsilon_{0,k_2}, \varepsilon_{0,k_3}, \varepsilon_{0,k_4}, \varepsilon_{0,k_5}, \varepsilon_{0,k_6}, \varepsilon_{0,k_7}, \varepsilon_{0,k_8})
    \right| \\
&=& \sum\limits_{k_1,\cdots,k_8=1}^{p} 
    \lrabs{((A^{(u_1)})^{\top} A^{(u_2)})_{k_1,k_2}} \cdot
    \lrabs{((A^{(u_3)})^{\top} A^{(u_4)})_{k_3,k_4}} \cdot
    \lrabs{((A^{(u_5)})^{\top} A^{(u_6)})_{k_5,k_6}} \cdot
    \lrabs{((A^{(u_7)})^{\top} A^{(u_8)})_{k_7,k_8}} \\
& & \hspace{3em} \times
    \lrabs{\cum(\varepsilon_{0,k_1}, \varepsilon_{0,k_2}, \varepsilon_{0,k_3}, \varepsilon_{0,k_4}, \varepsilon_{0,k_5}, \varepsilon_{0,k_6}, \varepsilon_{0,k_7}, \varepsilon_{0,k_8})} \\
&\le& \|(A^{(u_1)})^{\top} A^{(u_2)}\| \cdot
      \|(A^{(u_3)})^{\top} A^{(u_4)}\| \cdot
      \|(A^{(u_5)})^{\top} A^{(u_6)}\| \cdot
      \|(A^{(u_7)})^{\top} A^{(u_8)}\| \\
& & \times \sum\limits_{k_1,\cdots,k_8=1}^{p} 
    \lrabs{ \cum(\varepsilon_{0,k_1}, \varepsilon_{0,k_2}, \varepsilon_{0,k_3}, \varepsilon_{0,k_4}, \varepsilon_{0,k_5}, \varepsilon_{0,k_6}, \varepsilon_{0,k_7}, \varepsilon_{0,k_8})} \\
&\lesssim& \lrp{ \prod\limits_{i=1}^{8} \|A^{(u_i)}\| } \|\Gamma^{(2)}\|_F^8,
\EEqn
where the second to the last step follows from Lemma \ref{Lemma:LP-aux-1-1} and the last step uses Assumption \ref{Assumpt:NullDist_Wn_lp}\ref{Assumpt:NullDist_Wn_Lp-2}.
\end{Proof}

\begin{lemma}\label{Lemma:LP-aux-3-2}
Under Assumption \ref{Assumpt:NullDist_Wn_lp}\ref{Assumpt:NullDist_Wn_Lp-2}, it holds that
\BEqn
& & \lrabs{ \sum\limits_{\ell_1,\ell_2,\ell_3,\ell_4=1}^{p} \ca_2 } \\
&\lesssim& \sum\limits_{\pi_8(2,6)}
         \bone\{i_{\tau_1}=i_{\tau_2},\ i_{\tau_3}=\cdots=i_{\tau_8}\}
         \bigp{ \prod\limits_{i=1}^{8} \|A^{(u_i)}\| } \|\Gamma^{(2)}\|_F^7 
         \bigp{ \bone\{\ell(\tau_1)=\ell(\tau_2)\}\sqrt{p} + \bone\{\ell(\tau_1)\neq\ell(\tau_2)\} },
\EEqn
where $\ca_2$ is defined as Lemma \ref{Lemma:LP-aux-3}.
\end{lemma}
\begin{Proof}
Define
\BEqn
    \widetilde{\ca}_2 
&=& (A^{(u_{\tau_1})} \Gamma^{(2)} (A^{(u_{\tau_2})})^{\top})_{\ell(\tau_1),\ell(\tau_2)} \\
& & \times 
    \cum\lrp{(A^{(u_{\tau_3})} \varepsilon_0)_{\ell(\tau_3)},
          (A^{(u_{\tau_4})} \varepsilon_0)_{\ell(\tau_4)},
          (A^{(u_{\tau_5})} \varepsilon_0)_{\ell(\tau_5)},
          (A^{(u_{\tau_6})} \varepsilon_0)_{\ell(\tau_6)},
          (A^{(u_{\tau_7})} \varepsilon_0)_{\ell(\tau_7)},
          (A^{(u_{\tau_8})} \varepsilon_0)_{\ell(\tau_8)}},
\EEqn
then we have $\ca_2 = \sum\limits_{\pi_8(2,6)} \bone\{i_{\tau_1}=i_{\tau_2},\ i_{\tau_3}=\cdots=i_{\tau_8}\} \widetilde{\ca}_2$. To find an upper bound of $\ca_2$, it suffices to investigate the upper bound of $\widetilde{\ca}_2$.

When $\ell(\tau_1) = \ell(\tau_2)$, w.l.o.g., we have that
\BEqn
    \widetilde{\ca}_2 
&=&  (A^{(u_{\tau_1})} \Gamma^{(2)} (A^{(u_{\tau_2})})^{\top})_{\ell_1,\ell_1} \\
& & \hspace{2em} \times
    \cum\lrp{(A^{(u_{\tau_3})} \varepsilon_0)_{\ell_2},
          (A^{(u_{\tau_4})} \varepsilon_0)_{\ell_2},
          (A^{(u_{\tau_5})} \varepsilon_0)_{\ell_3},
          (A^{(u_{\tau_6})} \varepsilon_0)_{\ell_3},
          (A^{(u_{\tau_7})} \varepsilon_0)_{\ell_4},
          (A^{(u_{\tau_8})} \varepsilon_0)_{\ell_4}}.
\EEqn
By using the techniques used for Lemma \ref{Lemma:LP-aux-3-1}, we obtain that
\BEqn
& & \lrabs{ \sum\limits_{\ell_1,\ell_2,\ell_3,\ell_4=1}^{p} \widetilde{\ca}_2 } \\
&=& \lrabs{ \tr\lrp{ A^{(u_{\tau_1})} \Gamma^{(2)} (A^{(u_{\tau_2})})^{\top} }} \\
& & \hspace{1em} \times
    \lrabs{ \sum\limits_{\ell_2,\ell_3,\ell_4=1}^{p} 
    \cum\lrp{(A^{(u_{\tau_3})} \varepsilon_0)_{\ell_2},
          (A^{(u_{\tau_4})} \varepsilon_0)_{\ell_2},
          (A^{(u_{\tau_5})} \varepsilon_0)_{\ell_3},
          (A^{(u_{\tau_6})} \varepsilon_0)_{\ell_3},
          (A^{(u_{\tau_7})} \varepsilon_0)_{\ell_4},
          (A^{(u_{\tau_8})} \varepsilon_0)_{\ell_4}} } \\
&\le& \lrp{\|A^{(u_{\tau_1})}\| \cdot \|A^{(u_{\tau_2})}\| \cdot \|\Gamma^{(2)}\|_F \sqrt{p}} 
      \cdot \lrp{ \prod\limits_{i=3}^{8} \|A^{(u_{\tau_i})}\| } \|\Gamma^{(2)}\|_F^6 \\
&=& \lrp{ \prod\limits_{i=1}^{8} \|A^{(u_i)}\| } \|\Gamma^{(2)}\|_F^7 \sqrt{p},
\EEqn
where the inequality follows from Lemma \ref{Lemma:LP-aux-1-2}.

When $\ell(\tau_1) \neq \ell(\tau_2)$, w.l.o.g., we have that
\BEqn
    \widetilde{\ca}_2
&=& (A^{(u_{\tau_1})} \Gamma^{(2)} (A^{(u_{\tau_2})})^{\top})_{\ell_1,\ell_2} \\
& & \hspace{2em} \times
    \cum\lrp{(A^{(u_{\tau_3})} \varepsilon_0)_{\ell_1},
          (A^{(u_{\tau_4})} \varepsilon_0)_{\ell_2},
          (A^{(u_{\tau_5})} \varepsilon_0)_{\ell_3},
          (A^{(u_{\tau_6})} \varepsilon_0)_{\ell_3},
          (A^{(u_{\tau_7})} \varepsilon_0)_{\ell_4},
          (A^{(u_{\tau_8})} \varepsilon_0)_{\ell_4}},
\EEqn
and it follows that
\BEqn
& & \lrabs{ \sum\limits_{\ell_1,\ell_2,\ell_3,\ell_4=1}^{p} \widetilde{\ca}_2 } \\
&=& \Big{|} \sum\limits_{\ell_1,\ell_2,\ell_3,\ell_4=1}^{p} \sum\limits_{k_1,\cdots,k_6=1}^{p} 
    (A^{(u_{\tau_1})} \Gamma^{(2)} (A^{(u_{\tau_2})})^{\top})_{\ell_1,\ell_2} 
    A_{\ell_1,k_1}^{(u_{\tau_3})} A_{\ell_2,k_2}^{(u_{\tau_4})} A_{\ell_3,k_3}^{(u_{\tau_5})} A_{\ell_3,k_4}^{(u_{\tau_6})} A_{\ell_4,k_5}^{(u_{\tau_7})} A_{\ell_4,k_6}^{(u_{\tau_8})} \\
& & \hspace{9em} \times
    \cum\lrp{ \varepsilon_{0,k_1}, \varepsilon_{0,k_2}, \varepsilon_{0,k_3}, \varepsilon_{0,k_4}, \varepsilon_{0,k_5}, \varepsilon_{0,k_6}} \Big{|} \\
&=& \Big{|} \sum\limits_{k_1,\cdots,k_6=1}^{p} 
    \lrp{ (A^{(u_{\tau_3})})^{\top} A^{(u_{\tau_1})} \Gamma^{(2)} (A^{(u_{\tau_2})})^{\top} A^{(u_{\tau_4})} }_{k_1,k_2} 
    \lrp{ (A^{(u_{\tau_5})})^{\top} A^{(u_{\tau_6})} }_{k_3,k_4}
    \lrp{ (A^{(u_{\tau_7})})^{\top} A^{(u_{\tau_8})} }_{k_5,k_6}  \\
& & \hspace{5em} \times    
    \cum\lrp{ \varepsilon_{0,k_1}, \varepsilon_{0,k_2}, \varepsilon_{0,k_3}, \varepsilon_{0,k_4}, \varepsilon_{0,k_5}, \varepsilon_{0,k_6}} \Big{|} \\
&\le& \| (A^{(u_{\tau_3})})^{\top} A^{(u_{\tau_1})} \Gamma^{(2)} (A^{(u_{\tau_2})})^{\top} A^{(u_{\tau_4})} \| \
    \cdot \| (A^{(u_{\tau_5})})^{\top} A^{(u_{\tau_6})} \|
    \cdot \| (A^{(u_{\tau_7})})^{\top} A^{(u_{\tau_8})} \| \\
& & \hspace{1em} \times  
    \sum\limits_{k_1,\cdots,k_6=1}^{p} 
    \lrabs{ \cum\lrp{ \varepsilon_{0,k_1}, \varepsilon_{0,k_2}, \varepsilon_{0,k_3}, \varepsilon_{0,k_4}, \varepsilon_{0,k_5}, \varepsilon_{0,k_6}} } \\
&\lesssim& \lrp{ \prod\limits_{i=1}^{8} \|A^{(u_i)}\| } \|\Gamma^{(2)}\|_F^7.
\EEqn
By combining the results from both cases, we arrive at the proposed result.
\end{Proof}

\begin{lemma}\label{Lemma:LP-aux-3-3}
Under Assumption \ref{Assumpt:NullDist_Wn_lp}\ref{Assumpt:NullDist_Wn_Lp-2}, it holds that
\begin{equation*}
    \lrabs{ \sum\limits_{\ell_1,\ell_2,\ell_3,\ell_4=1}^{p} \ca_3 }
\lesssim \sum\limits_{\pi_8(3,5)}
         \bone\{i_{\tau_1}=\cdots=i_{\tau_3},\ i_{\tau_4}=\cdots=i_{\tau_8}\}
         \bigp{ \prod\limits_{i=1}^{8} \|A^{(u_i)}\| } \|\Gamma^{(2)}\|_F^8,
\end{equation*}
where $\ca_3$ is defined as Lemma \ref{Lemma:LP-aux-3}.
\end{lemma}
\begin{Proof}
The analysis of $\ca_3$ is similar to that of $\ca_2$. Define
\BEqn
    \widetilde{\ca}_3
&=& \cum\lrp{(A^{(u_{\tau_1})} \varepsilon_0)_{\ell(\tau_1)},
          (A^{(u_{\tau_2})} \varepsilon_0)_{\ell(\tau_2)},
          (A^{(u_{\tau_3})} \varepsilon_0)_{\ell(\tau_3)}}, \\
& & \times 
    \cum\lrp{(A^{(u_{\tau_4})} \varepsilon_0)_{\ell(\tau_4)},
          (A^{(u_{\tau_5})} \varepsilon_0)_{\ell(\tau_5)},
          (A^{(u_{\tau_6})} \varepsilon_0)_{\ell(\tau_6)},
          (A^{(u_{\tau_7})} \varepsilon_0)_{\ell(\tau_7)},
          (A^{(u_{\tau_8})} \varepsilon_0)_{\ell(\tau_8)}},
\EEqn
thus it holds that $\ca_3 = \sum\limits_{\pi_8(3,5)} \bone\{i_{\tau_1}=\cdots=i_{\tau_3},\ i_{\tau_4}=\cdots=i_{\tau_8}\} \widetilde{\ca}_3$. To find the upper bound of $\widetilde{\ca}_3$, we divide into two cases.

When $\{\ell(\tau_1),\ell(\tau_2),\ell(\tau_3)\}$ are pairwise distinct, we have that
\BEqn
& & \lrabs{ \sum\limits_{\ell_1,\ell_2,\ell_3,\ell_4=1}^{p} \widetilde{\ca}_3 } \\
&=& \Big{|} \sum\limits_{\ell_1,\ell_2,\ell_3,\ell_4=1}^{p} 
    \cum\lrp{ (A^{(u_{\tau_1})} \varepsilon_0)_{\ell_1},
          (A^{(u_{\tau_2})} \varepsilon_0)_{\ell_2},
          (A^{(u_{\tau_3})} \varepsilon_0)_{\ell_3}} \\
& & \hspace{5em} \times
          \cum\lrp{(A^{(u_{\tau_4})} \varepsilon_0)_{\ell_1},
          (A^{(u_{\tau_5})} \varepsilon_0)_{\ell_2},
          (A^{(u_{\tau_6})} \varepsilon_0)_{\ell_3},
          (A^{(u_{\tau_7})} \varepsilon_0)_{\ell_4},
          (A^{(u_{\tau_8})} \varepsilon_0)_{\ell_4}} \Big{|} \\
&=& \Big{|} \sum\limits_{\ell_1,\ell_2,\ell_3,\ell_4=1}^{p} \sum\limits_{k_1,\cdots,k_8=1}^{p}
    A^{(u_{\tau_1})}_{\ell_1,k_1} A^{(u_{\tau_2})}_{\ell_2,k_2} A^{(u_{\tau_3})}_{\ell_3,k_3} A^{(u_{\tau_4})}_{\ell_1,k_4} A^{(u_{\tau_5})}_{\ell_2,k_5} A^{(u_{\tau_6})}_{\ell_3,k_6} A^{(u_{\tau_7})}_{\ell_4,k_7} A^{(u_{\tau_8})}_{\ell_4,k_8} \\
& & \hspace{9em} \times
    \cum\lrp{ \varepsilon_{0,k_1}, \varepsilon_{0,k_2}, \varepsilon_{0,k_3}}
    \cum\lrp{ \varepsilon_{0,k_4}, \varepsilon_{0,k_5}, \varepsilon_{0,k_6}, \varepsilon_{0,k_7}, \varepsilon_{0,k_8}} \Big{|} \\
&\le& \|(A^{(u_{\tau_1})})^{\top} A^{(u_{\tau_4})}\| 
      \cdot \|(A^{(u_{\tau_2})})^{\top} A^{(u_{\tau_5})}\|
      \cdot \|(A^{(u_{\tau_3})})^{\top} A^{(u_{\tau_6})}\|
      \cdot \|(A^{(u_{\tau_7})})^{\top} A^{(u_{\tau_8})}\| \\
& & \hspace{1em} \times
    \sum\limits_{k_1,\cdots,k_3=1}^{p} \lrabs{\cum\lrp{ \varepsilon_{0,k_1}, \varepsilon_{0,k_2}, \varepsilon_{0,k_3}}} \cum\lrp{ \varepsilon_{0,k_4}, \varepsilon_{0,k_5}, \varepsilon_{0,k_6}, \sum\limits_{k_4,\cdots,k_8=1}^{p} \varepsilon_{0,k_7}, \varepsilon_{0,k_8}} \\
&\lesssim& \lrp{ \prod\limits_{i=1}^{8} \|A^{(u_i)}\| } \|\Gamma^{(2)}\|_F^8.
\EEqn

When two of $\{\ell(\tau_1),\ell(\tau_2),\ell(\tau_3)\}$ are identical, by using the similar techniques, we claim without providing detailed steps that
\BEqn
& & \lrabs{ \sum\limits_{\ell_1,\ell_2,\ell_3,\ell_4=1}^{p} \widetilde{\ca}_3 } \\
&=& \Big{|} \sum\limits_{\ell_1,\ell_2,\ell_3,\ell_4=1}^{p} 
    \cum\lrp{ (A^{(u_{\tau_1})} \varepsilon_0)_{\ell_1},
          (A^{(u_{\tau_2})} \varepsilon_0)_{\ell_1},
          (A^{(u_{\tau_3})} \varepsilon_0)_{\ell_2}} \\
& & \hspace{5em} \times
          \cum\lrp{(A^{(u_{\tau_4})} \varepsilon_0)_{\ell_2},
          (A^{(u_{\tau_5})} \varepsilon_0)_{\ell_3},
          (A^{(u_{\tau_6})} \varepsilon_0)_{\ell_3},
          (A^{(u_{\tau_7})} \varepsilon_0)_{\ell_4},
          (A^{(u_{\tau_8})} \varepsilon_0)_{\ell_4}} \Big{|} \\
&\lesssim& \lrp{ \prod\limits_{i=1}^{8} \|A^{(u_i)}\| } \|\Gamma^{(2)}\|_F^8,
\EEqn
which completes the proof.
\end{Proof}

\begin{lemma}\label{Lemma:LP-aux-3-4}
Under Assumption \ref{Assumpt:NullDist_Wn_lp}\ref{Assumpt:NullDist_Wn_Lp-1},\ref{Assumpt:NullDist_Wn_Lp-2}, it holds that
\BEqn
    \lrabs{ \sum\limits_{\ell_1,\ell_2,\ell_3,\ell_4=1}^{p} \ca_4 }
&\lesssim& \sum\limits_{\pi_8(4,4)}
         \bone\{i_{\tau_1}=\cdots=i_{\tau_4},\ i_{\tau_5}=\cdots=i_{\tau_8}\}
         \bigp{ \prod\limits_{i=1}^{8} \|A^{(u_i)}\| } \|\Gamma^{(2)}\|_F^4 \\
& & \hspace{2em} \times
    \Big{(} 
    \bone\{\ell(\tau_1),\ell(\tau_2),\ell(\tau_3),\ell(\tau_4) \mbox{ are pairwise distinct}\} \\
& & \hspace{3em}
    + \bone\{\ell(\tau_1),\ell(\tau_2),\ell(\tau_3),\ell(\tau_4) \mbox{ have duplicated value(s)}\} \|\Gamma^{(2)}\|_F^4 \Big{)},     
\EEqn
where $\ca_4$ is defined as Lemma \ref{Lemma:LP-aux-3}.
\end{lemma}
\begin{Proof}
It holds that $\ca_3 = \sum\limits_{\pi_8(3,5)} \bone\{i_{\tau_1}=\cdots=i_{\tau_3},\ i_{\tau_4}=\cdots=i_{\tau_8}\} \widetilde{\ca}_3$, where
\BEqn
    \widetilde{\ca}_4
&=& \cum\lrp{(A^{(u_{\tau_1})} \varepsilon_0)_{\ell(\tau_1)},
          (A^{(u_{\tau_2})} \varepsilon_0)_{\ell(\tau_2)},
          (A^{(u_{\tau_3})} \varepsilon_0)_{\ell(\tau_3)},
          (A^{(u_{\tau_4})} \varepsilon_0)_{\ell(\tau_4)}}, \\
& & \times 
    \cum\lrp{(A^{(u_{\tau_5})} \varepsilon_0)_{\ell(\tau_5)},
          (A^{(u_{\tau_6})} \varepsilon_0)_{\ell(\tau_6)},
          (A^{(u_{\tau_7})} \varepsilon_0)_{\ell(\tau_7)},
          (A^{(u_{\tau_8})} \varepsilon_0)_{\ell(\tau_8)}},
\EEqn
We divide into the following cases to find the upper bound of $\widetilde{\ca}_4$. 

Suppose that there is at least two out of $\{\ell(\tau_1),\ell(\tau_2),\ell(\tau_3),\ell(\tau_4)\}$ that take the same value, then $\widetilde{\ca}_4$ can be written as either
\BEqn
& & \cum\lrp{(A^{(u_{\tau_1})} \varepsilon_0)_{\ell_1},
          (A^{(u_{\tau_2})} \varepsilon_0)_{\ell_1)},
          (A^{(u_{\tau_3})} \varepsilon_0)_{\ell_2},
          (A^{(u_{\tau_4})} \varepsilon_0)_{\ell_3}} \\
& & \hspace{2em} \times          
    \cum\lrp{(A^{(u_{\tau_5})} \varepsilon_0)_{\ell_2},
          (A^{(u_{\tau_6})} \varepsilon_0)_{\ell_3},
          (A^{(u_{\tau_7})} \varepsilon_0)_{\ell_4},
          (A^{(u_{\tau_8})} \varepsilon_0)_{\ell_4}}
\EEqn
or 
\BEqn
& & \cum\lrp{(A^{(u_{\tau_1})} \varepsilon_0)_{\ell_1},
          (A^{(u_{\tau_2})} \varepsilon_0)_{\ell_1)},
          (A^{(u_{\tau_3})} \varepsilon_0)_{\ell_2},
          (A^{(u_{\tau_4})} \varepsilon_0)_{\ell_2}} \\
& & \hspace{2em} \times         
    \cum\lrp{(A^{(u_{\tau_5})} \varepsilon_0)_{\ell_3},
          (A^{(u_{\tau_6})} \varepsilon_0)_{\ell_3},
          (A^{(u_{\tau_7})} \varepsilon_0)_{\ell_4},
          (A^{(u_{\tau_8})} \varepsilon_0)_{\ell_4}}.
\EEqn
Under both cases, by applying the techniques used for previous lemmas, we can show that
\begin{equation*}
    \lrabs{ \sum\limits_{\ell_1,\ell_2,\ell_3,\ell_4=1}^{p} \widetilde{\ca}_4 } 
\lesssim \lrp{ \prod\limits_{i=1}^{8} \|A^{(u_i)}\| } \|\Gamma^{(2)}\|_F^8.
\end{equation*}

It remains to consider the case when $\{\ell(\tau_1),\ell(\tau_2),\ell(\tau_3),\ell(\tau_4)\}$ are pairwise distinct. In this case, we obtain that
\BEqn
& & \lrabs{ \sum\limits_{\ell_1,\ell_2,\ell_3,\ell_4=1}^{p} \widetilde{\ca}_4 } \\
&=& \Big{|}
    \sum\limits_{\ell_1,\ell_2,\ell_3,\ell_4=1}^{p} 
    \cum\lrp{(A^{(u_{\tau_1})} \varepsilon_0)_{\ell_1},
          (A^{(u_{\tau_2})} \varepsilon_0)_{\ell_2)},
          (A^{(u_{\tau_3})} \varepsilon_0)_{\ell_3},
          (A^{(u_{\tau_4})} \varepsilon_0)_{\ell_4}} \\
& & \hspace{5em} \times    
    \cum\lrp{(A^{(u_{\tau_5})} \varepsilon_0)_{\ell_1},
          (A^{(u_{\tau_6})} \varepsilon_0)_{\ell_2},
          (A^{(u_{\tau_7})} \varepsilon_0)_{\ell_3},
          (A^{(u_{\tau_8})} \varepsilon_0)_{\ell_4}}
    \Big{|} \\
&=& \Big{|}
    \sum\limits_{\ell_1,\ell_2,\ell_3,\ell_4=1}^{p} \sum\limits_{k_1,\cdots,k_8=1}^{p}
    A^{(u_{\tau_1})}_{\ell_1,k_1} A^{(u_{\tau_2})}_{\ell_2,k_2} A^{(u_{\tau_3})}_{\ell_3,k_3} A^{(u_{\tau_4})}_{\ell_4,k_4} A^{(u_{\tau_5})}_{\ell_1,k_5} A^{(u_{\tau_6})}_{\ell_2,k_6} A^{(u_{\tau_7})}_{\ell_3,k_7} A^{(u_{\tau_8})}_{\ell_4,k_8} \\
& & \hspace{10em} \times    
    \cum(\varepsilon_{0,k_1}, \varepsilon_{0,k_2}, \varepsilon_{0,k_3}, \varepsilon_{0,k_4})
    \cum(\varepsilon_{0,k_5}, \varepsilon_{0,k_6}, \varepsilon_{0,k_7}, \varepsilon_{0,k_8}) \Big{|} \\
&\le& \lrp{ \sum\limits_{\ell_1,\ell_2,\ell_3,\ell_4=1}^{p} 
            \lrp{\sum\limits_{k_1,k_2,k_3,k_4=1}^{p} 
            A^{(u_{\tau_1})}_{\ell_1,k_1} A^{(u_{\tau_2})}_{\ell_2,k_2} A^{(u_{\tau_3})}_{\ell_3,k_3} A^{(u_{\tau_4})}_{\ell_4,k_4} \cum(\varepsilon_{0,k_1}, \varepsilon_{0,k_2}, \varepsilon_{0,k_3}, \varepsilon_{0,k_4}) }^2 }^{1/2} \\
& & \times
    \lrp{ \sum\limits_{\ell_1,\ell_2,\ell_3,\ell_4=1}^{p} 
            \lrp{\sum\limits_{k_5,k_6,k_7,k_8=1}^{p} 
            A^{(u_{\tau_5})}_{\ell_1,k_5} A^{(u_{\tau_6})}_{\ell_2,k_6} A^{(u_{\tau_7})}_{\ell_3,k_7} A^{(u_{\tau_8})}_{\ell_4,k_8} \cum(\varepsilon_{0,k_5}, \varepsilon_{0,k_6}, \varepsilon_{0,k_7}, \varepsilon_{0,k_8}) }^2 }^{1/2} \\
&\lesssim& \lrp{ \prod\limits_{i=1}^{8} \|A^{(u_i)}\| } \|\Gamma^{(2)}\|_F^4,
\EEqn
where the second to the last inequality follows from the Cauchy-Schwarz inequality and the last inequality follows from Lemma \ref{Lemma:LP-aux-2}. By unifying the aforementioned results, we obtain the desired statement.
\end{Proof}

\begin{lemma}\label{Lemma:LP-aux-3-5}
Under Assumption \ref{Assumpt:NullDist_Wn_lp}\ref{Assumpt:NullDist_Wn_Lp-1},\ref{Assumpt:NullDist_Wn_Lp-2}, it holds that
\BEqn
    \lrabs{ \sum\limits_{\ell_1,\ell_2,\ell_3,\ell_4=1}^{p} \ca_5 } 
&\lesssim& \sum\limits_{\pi_8(4,4)}
         \bone\{i_{\tau_1}={\tau_2},\ i_{\tau_3}=\cdots=i_{\tau_5},\ i_{\tau_6}=\cdots=i_{\tau_8}\}
         \bigp{ \prod\limits_{i=1}^{8} \|A^{(u_i)}\| } \|\Gamma^{(2)}\|_F^4 \\
& & \hspace{2em} \times
    \Big{(} 
    \bone\{\ell(\tau_1)=\ell(\tau_2),\ \ell(\tau_3)\neq\ell(\tau_4)\neq\ell(\tau_5),\ \ell(\tau_6)\neq\ell(\tau_7)\neq\ell(\tau_8)\} \sqrt{p} \\
& & \hspace{3em}
    + \bone\{\ell(\tau_1)=\ell(\tau_2),\ \ell(\tau_3)=\ell(\tau_4)\neq\ell(\tau_5),\ \ell(\tau_6)=\ell(\tau_7)\neq\ell(\tau_8)\} \|\Gamma^{(2)}\|_F^3 \sqrt{p} \\  
& & \hspace{3em}
    + \bone\{\ell(\tau_1)\neq\ell(\tau_2),\ \ell(\tau_3)\neq\ell(\tau_4)\neq\ell(\tau_5),\ \ell(\tau_6)\neq\ell(\tau_7)\neq\ell(\tau_8)\} \\
& & \hspace{3em}
    + \bone\{\ell(\tau_1)\neq\ell(\tau_2),\ \ell(\tau_3)=\ell(\tau_4)\neq\ell(\tau_5),\ \ell(\tau_6)\neq\ell(\tau_7)\neq\ell(\tau_8)\} \\
& & \hspace{3em}
    + \bone\{\ell(\tau_1)\neq\ell(\tau_2),\ \ell(\tau_3)=\ell(\tau_4)\neq\ell(\tau_5),\ \ell(\tau_6)=\ell(\tau_7)\neq\ell(\tau_8)\} \|\Gamma^{(2)}\|_F^3 \Big{)},     
\EEqn
where $\ca_5$ is defined as Lemma \ref{Lemma:LP-aux-3}.
\end{lemma}
\begin{Proof}
Define
\BEqn
    \widetilde{\ca}_5
&=& \lrp{A^{(u_{\tau_1})} \Gamma^{(2)} (A^{(u_{\tau_2})})^{\top}}_{\ell(\tau_1),\ell(\tau_2)}
    \cum\lrp{(A^{(u_{\tau_3})} \varepsilon_0)_{\ell(\tau_3)},
          (A^{(u_{\tau_4})} \varepsilon_0)_{\ell(\tau_4)}, 
          (A^{(u_{\tau_5})} \varepsilon_0)_{\ell(\tau_5)}} \\
& & \times 
    \cum\lrp{(A^{(u_{\tau_6})} \varepsilon_0)_{\ell(\tau_6)},
          (A^{(u_{\tau_7})} \varepsilon_0)_{\ell(\tau_7)},
          (A^{(u_{\tau_8})} \varepsilon_0)_{\ell(\tau_8)}},
\EEqn
then we have $\ca_5 = \sum\limits_{\pi_8(2,3,3)} \bone\{i_{\tau_1}={\tau_2},\ i_{\tau_3}=\cdots=i_{\tau_5},\ i_{\tau_6}=\cdots=i_{\tau_8}\} \widetilde{\ca}_5$. 

To investigate $\widetilde{\ca}_5$, we consider two major cases, that is, $\ell(\tau_1)=\ell(\tau_2)$ and $\ell(\tau_1)\neq\ell(\tau_2)$. For the first case when $\ell(\tau_1)=\ell(\tau_2)$, if $\{\ell(\tau_3),\ell(\tau_4),\ell(\tau_5)\}$ are pairwise distinct, then $\{\ell(\tau_6),\ell(\tau_7),\ell(\tau_8)\}$ are as well, and it follows that
\BEqn
    \lrabs{ \sum\limits_{\ell_1,\ell_2,\ell_3,\ell_4=1}^{p} \widetilde{\ca}_5 }
&=& \lrabs{ \sum\limits_{\ell_1=1}^{p} \lrp{A^{(u_{\tau_1})} \Gamma^{(2)} (A^{(u_{\tau_2})})^{\top}}_{\ell_1,\ell_1} } \\
& & \times
    \Big{|}
    \sum\limits_{\ell_2,\ell_3,\ell_4=1}^{p}
    \cum\lrp{(A^{(u_{\tau_3})} \varepsilon_0)_{\ell_2},
          (A^{(u_{\tau_4})} \varepsilon_0)_{\ell_3}, 
          (A^{(u_{\tau_5})} \varepsilon_0)_{\ell_4}} \\
& & \hspace{5em} \times 
    \cum\lrp{(A^{(u_{\tau_6})} \varepsilon_0)_{\ell_2},
          (A^{(u_{\tau_7})} \varepsilon_0)_{\ell_3},
          (A^{(u_{\tau_8})} \varepsilon_0)_{\ell_4}} \Big{|} \\
&\lesssim& \bigp{ \prod\limits_{i=1}^{8} \|A^{(u_i)}\| } \|\Gamma^{(2)}\|_F^4 \sqrt{p},
\EEqn
since it follows from Lemma \ref{Lemma:LP-aux-1-2} that
\begin{equation*}
    \lrabs{ \sum\limits_{\ell_1=1}^{p} \lrp{A^{(u_{\tau_1})} \Gamma^{(2)} (A^{(u_{\tau_2})})^{\top}}_{\ell_1,\ell_1} } 
  = \lrabs{\tr\lrp{A^{(u_{\tau_1})} \Gamma^{(2)} (A^{(u_{\tau_2})})^{\top}}}
  \lesssim \|A^{(u_{\tau_1})}\| \cdot \|A^{(u_{\tau_2})}\| \cdot \|\Gamma^{(2)}\|_F \sqrt{p},
\end{equation*}
and it follows from the Cauchy-Schwarz inequality and Lemma \ref{Lemma:LP-aux-2} that
\BEqn
& & \lrabs{
    \sum\limits_{\ell_2,\ell_3,\ell_4=1}^{p}
    \cum\lrp{(A^{(u_{\tau_3})} \varepsilon_0)_{\ell_2},
          (A^{(u_{\tau_4})} \varepsilon_0)_{\ell_3}, 
          (A^{(u_{\tau_5})} \varepsilon_0)_{\ell_4}}
    \cum\lrp{(A^{(u_{\tau_6})} \varepsilon_0)_{\ell_2},
          (A^{(u_{\tau_7})} \varepsilon_0)_{\ell_3},
          (A^{(u_{\tau_8})} \varepsilon_0)_{\ell_4}} } \\
&\le& \bigp{ \prod\limits_{i=3}^{8} \|A^{(u_i)}\| } \|\Gamma^{(2)}\|_F^3.
\EEqn
Under the case when $\ell(\tau_1)=\ell(\tau_2)$, if $\{\ell(\tau_3),\ell(\tau_4),\ell(\tau_5)\}$ have duplicated values, $\{\ell(\tau_6),\ell(\tau_7),\ell(\tau_8)\}$ do as well, and consequently,
\BEqn
    \lrabs{ \sum\limits_{\ell_1,\ell_2,\ell_3,\ell_4=1}^{p} \widetilde{\ca}_5 }
&=& \lrabs{\tr\lrp{A^{(u_{\tau_1})} \Gamma^{(2)} (A^{(u_{\tau_2})})^{\top}}} \\
& & \times
    \Big{|}
    \sum\limits_{\ell_2,\ell_3,\ell_4=1}^{p}
    \cum\lrp{(A^{(u_{\tau_3})} \varepsilon_0)_{\ell_2},
          (A^{(u_{\tau_4})} \varepsilon_0)_{\ell_2}, 
          (A^{(u_{\tau_5})} \varepsilon_0)_{\ell_4}} \\
& & \hspace{5em} \times 
    \cum\lrp{(A^{(u_{\tau_6})} \varepsilon_0)_{\ell_3},
          (A^{(u_{\tau_7})} \varepsilon_0)_{\ell_3},
          (A^{(u_{\tau_8})} \varepsilon_0)_{\ell_4}} \Big{|} \\
&\lesssim& \bigp{ \prod\limits_{i=1}^{8} \|A^{(u_i)}\| } \|\Gamma^{(2)}\|_F^7 \sqrt{p},
\EEqn
since we have
\BEqn
& & \lrabs{
    \sum\limits_{\ell_2,\ell_3,\ell_4=1}^{p}
    \cum\lrp{(A^{(u_{\tau_3})} \varepsilon_0)_{\ell_2},
          (A^{(u_{\tau_4})} \varepsilon_0)_{\ell_2}, 
          (A^{(u_{\tau_5})} \varepsilon_0)_{\ell_4}}
    \cum\lrp{(A^{(u_{\tau_6})} \varepsilon_0)_{\ell_3},
          (A^{(u_{\tau_7})} \varepsilon_0)_{\ell_3},
          (A^{(u_{\tau_8})} \varepsilon_0)_{\ell_4}} } \\
&\le& \bigp{ \prod\limits_{i=3}^{8} \|A^{(u_i)}\| } \|\Gamma^{(2)}\|_F^6
\EEqn
by using the similar analytic method used for previous lemmas. 

It remains to consider the major case when $\ell(\tau_1)\neq\ell(\tau_2)$, which includes three subcases. If $\{\ell(\tau_3),\ell(\tau_4),\ell(\tau_5)\}$ and $\{\ell(\tau_6),\ell(\tau_7),\ell(\tau_8)\}$ are pairwise distinct respectively, then we have that
\BEqn
& & \lrabs{ \sum\limits_{\ell_1,\ell_2,\ell_3,\ell_4=1}^{p} \widetilde{\ca}_5 } \\
&=& \Big{|} \sum\limits_{\ell_1,\ell_2,\ell_3,\ell_4=1}^{p} \lrp{A^{(u_{\tau_1})} \Gamma^{(2)} (A^{(u_{\tau_2})})^{\top}}_{\ell_1,\ell_2}  
    \cum\lrp{(A^{(u_{\tau_3})} \varepsilon_0)_{\ell_1},
          (A^{(u_{\tau_4})} \varepsilon_0)_{\ell_3}, 
          (A^{(u_{\tau_5})} \varepsilon_0)_{\ell_4}} \\
& & \hspace{5em} \times 
    \cum\lrp{(A^{(u_{\tau_6})} \varepsilon_0)_{\ell_2},
          (A^{(u_{\tau_7})} \varepsilon_0)_{\ell_3},
          (A^{(u_{\tau_8})} \varepsilon_0)_{\ell_4}} \Big{|} \\
&\le& \lrp{ \sum\limits_{\ell_2,\ell_3,\ell_4=1}^{p} 
            \lrp{\sum\limits_{\ell_1=1}^{p} 
            \lrp{A^{(u_{\tau_1})} \Gamma^{(2)} (A^{(u_{\tau_2})})^{\top}}_{\ell_1,\ell_2}  
            \cum\lrp{(A^{(u_{\tau_3})} \varepsilon_0)_{\ell_1},
                     (A^{(u_{\tau_4})} \varepsilon_0)_{\ell_3}, 
                     (A^{(u_{\tau_5})} \varepsilon_0)_{\ell_4}} }^2 }^{1/2} \\
& & \times
    \lrp{ \sum\limits_{\ell_2,\ell_3,\ell_4=1}^{p} 
          \cum^2\lrp{(A^{(u_{\tau_6})} \varepsilon_0)_{\ell_2},
                     (A^{(u_{\tau_7})} \varepsilon_0)_{\ell_3},
                     (A^{(u_{\tau_8})} \varepsilon_0)_{\ell_4}} }^{1/2} \\
&\lesssim& \bigp{ \prod\limits_{i=1}^{8} \|A^{(u_i)}\| } \|\Gamma^{(2)}\|_F^4 \sqrt{p}.
\EEqn
Note that
\BEqn
& & \sum\limits_{\ell_2,\ell_3,\ell_4=1}^{p} 
            \lrp{\sum\limits_{\ell_1=1}^{p} 
            \lrp{A^{(u_{\tau_1})} \Gamma^{(2)} (A^{(u_{\tau_2})})^{\top}}_{\ell_1,\ell_2}  
            \cum\lrp{(A^{(u_{\tau_3})} \varepsilon_0)_{\ell_1},
                     (A^{(u_{\tau_4})} \varepsilon_0)_{\ell_3}, 
                     (A^{(u_{\tau_5})} \varepsilon_0)_{\ell_4}} }^2 \\
&=& \sum\limits_{\ell_2,\ell_3,\ell_4=1}^{p} 
            \lrp{\sum\limits_{k_1,k_2,k_3=1}^{p}
            \lrp{A^{(u_{\tau_2})} \Gamma^{(2)} (A^{(u_{\tau_1})})^{\top} A^{(u_{\tau_3})}}_{\ell_2,k_1} A^{(u_{\tau_4})}_{\ell_3,k_2} A^{(u_{\tau_5})}_{\ell_4,k_3}
            \cum\lrp{\varepsilon_{0,k_1}, \varepsilon_{0,k_2}, \varepsilon_{0,k_3}} }^2 \\
&\lesssim& \|A^{(u_{\tau_2})} \Gamma^{(2)} (A^{(u_{\tau_1})})^{\top} A^{(u_{\tau_3})}\|^2 \cdot \|A^{(u_{\tau_4})}\|^2 \cdot \|A^{(u_{\tau_5})}\|^2 \|\Gamma^{(2)}\|_F^3 \\
&\le& \bigp{ \prod\limits_{i=1}^{5} \|A^{(u_{\tau_i})}\|^2 } \|\Gamma^{(2)}\|_F^5,
\EEqn
where the second to the last step follows from Lemma \ref{Lemma:LP-aux-1-2}. Also, it follows from the similar arguments used for Lemma \ref{Lemma:LP-aux-1-2} that
\BEqn
& & \sum\limits_{\ell_2,\ell_3,\ell_4=1}^{p} 
          \cum^2\lrp{(A^{(u_{\tau_6})} \varepsilon_0)_{\ell_2},
                     (A^{(u_{\tau_7})} \varepsilon_0)_{\ell_3},
                     (A^{(u_{\tau_8})} \varepsilon_0)_{\ell_4}} \\
&=& \sum\limits_{\ell_2,\ell_3,\ell_4=1}^{p} 
          \lrp{\sum\limits_{k_1,k_2,k_3=1}^{p}
               A^{(u_{\tau_6})}_{\ell_2,k_1} A^{(u_{\tau_7})}_{\ell_3,k_2} A^{(u_{\tau_8})}_{\ell_4,k_3}
               \cum\lrp{\varepsilon_{0,k_1}, \varepsilon_{0,k_2}, \varepsilon_{0,k_3}} }^2 \\
&\lesssim& \bigp{ \prod\limits_{i=6}^{8} \|A^{(u_{\tau_i})}\|^2 } \|\Gamma^{(2)}\|_F^3. 
\EEqn
In summary, when $\{\ell(\tau_3),\ell(\tau_4),\ell(\tau_5)\}$ and $\{\ell(\tau_6),\ell(\tau_7),\ell(\tau_8)\}$ are pairwise distinct respectively, we have that
\begin{equation*}
    \lrabs{ \sum\limits_{\ell_1,\ell_2,\ell_3,\ell_4=1}^{p} \widetilde{\ca}_5 } 
\lesssim \bigp{ \prod\limits_{i=1}^{8} \|A^{(u_i)}\| } \|\Gamma^{(2)}\|_F^4. 
\end{equation*}
If $\{\ell(\tau_3),\ell(\tau_4),\ell(\tau_5)\}$ have duplicated values whereas $\{\ell(\tau_6),\ell(\tau_7),\ell(\tau_8)\}$ are pairwise distinct, we have that
\BEqn
& & \lrabs{ \sum\limits_{\ell_1,\ell_2,\ell_3,\ell_4=1}^{p} \widetilde{\ca}_5 } \\
&=& \Big{|} \sum\limits_{\ell_1,\ell_2,\ell_3,\ell_4=1}^{p} \lrp{A^{(u_{\tau_1})} \Gamma^{(2)} (A^{(u_{\tau_2})})^{\top}}_{\ell_1,\ell_2}  
    \cum\lrp{(A^{(u_{\tau_3})} \varepsilon_0)_{\ell_3},
          (A^{(u_{\tau_4})} \varepsilon_0)_{\ell_3}, 
          (A^{(u_{\tau_5})} \varepsilon_0)_{\ell_4}} \\
& & \hspace{5em} \times 
    \cum\lrp{(A^{(u_{\tau_6})} \varepsilon_0)_{\ell_1},
          (A^{(u_{\tau_7})} \varepsilon_0)_{\ell_2},
          (A^{(u_{\tau_8})} \varepsilon_0)_{\ell_4}} \Big{|} \\
&\le& \lrp{ \sum\limits_{\ell_1,\ell_2,\ell_4=1}^{p} 
            \lrp{\sum\limits_{\ell_3=1}^{p} 
            \lrp{A^{(u_{\tau_1})} \Gamma^{(2)} (A^{(u_{\tau_2})})^{\top}}_{\ell_1,\ell_2}  
            \cum\lrp{(A^{(u_{\tau_3})} \varepsilon_0)_{\ell_3},
                     (A^{(u_{\tau_4})} \varepsilon_0)_{\ell_3}, 
                     (A^{(u_{\tau_5})} \varepsilon_0)_{\ell_4}} }^2 }^{1/2} \\
& & \times
    \lrp{ \sum\limits_{\ell_1,\ell_2,\ell_4=1}^{p} 
          \cum^2\lrp{(A^{(u_{\tau_6})} \varepsilon_0)_{\ell_1},
                     (A^{(u_{\tau_7})} \varepsilon_0)_{\ell_2},
                     (A^{(u_{\tau_8})} \varepsilon_0)_{\ell_4}} }^{1/2} \\
&=& \|A^{(u_{\tau_1})} \Gamma^{(2)} (A^{(u_{\tau_2})})^{\top}\|_F 
    \lrp{\sum\limits_{\ell_4=1}^{p} 
            \lrp{\sum\limits_{\ell_3=1}^{p}
            \cum\lrp{(A^{(u_{\tau_3})} \varepsilon_0)_{\ell_3},
                     (A^{(u_{\tau_4})} \varepsilon_0)_{\ell_3}, 
                     (A^{(u_{\tau_5})} \varepsilon_0)_{\ell_4}} }^2 }^{1/2} \\
& & \times
    \lrp{ \sum\limits_{\ell_1,\ell_2,\ell_4=1}^{p} 
          \cum^2\lrp{(A^{(u_{\tau_6})} \varepsilon_0)_{\ell_1},
                     (A^{(u_{\tau_7})} \varepsilon_0)_{\ell_2},
                     (A^{(u_{\tau_8})} \varepsilon_0)_{\ell_4}} }^{1/2}.
\EEqn
Note that
\BEqn
& & \sum\limits_{\ell_4=1}^{p} 
            \lrp{\sum\limits_{\ell_3=1}^{p}
            \cum\lrp{(A^{(u_{\tau_3})} \varepsilon_0)_{\ell_3},
                     (A^{(u_{\tau_4})} \varepsilon_0)_{\ell_3}, 
                     (A^{(u_{\tau_5})} \varepsilon_0)_{\ell_4}} }^2 \\
&=& \sum\limits_{\ell_4=1}^{p} 
    \lrp{ \sum\limits_{k_1,k_2,k_3=1}^{p}
          \lrp{(A^{(u_{\tau_3})})^{\top} A^{(u_{\tau_4})}}_{k_1,k_2} A^{(u_{\tau_5})}_{\ell_4,k_3}
          \cum\lrp{\varepsilon_{0,k_1}, \varepsilon_{0,k_2}, \varepsilon_{0,k_3}} }^2  \\
&\le& \|A^{(u_{\tau_3})}\| \cdot \|A^{(u_{\tau_4})}\| 
    \sum\limits_{\ell_4=1}^{p} 
    \lrp{ \sum\limits_{k_1,k_2,k_3=1}^{p} A^{(u_{\tau_5})}_{\ell_4,k_3}
          \cum\lrp{\varepsilon_{0,k_1}, \varepsilon_{0,k_2}, \varepsilon_{0,k_3}} }^2 \\
&\lesssim& \|A^{(u_{\tau_3})}\|^2 \|A^{(u_{\tau_4})}\|^2 \|A^{(u_{\tau_5})}\|^2 \cdot \|\Gamma^{(2)}\|_F^3
\EEqn
where the last step is obtained using the similar techniques as used for Lemma \ref{Lemma:LP-aux-2}, and this further implies that
\begin{equation*}
    \lrabs{ \sum\limits_{\ell_1,\ell_2,\ell_3,\ell_4=1}^{p} \widetilde{\ca}_5 } 
\lesssim \bigp{ \prod\limits_{i=1}^{8} \|A^{(u_i)}\| } \|\Gamma^{(2)}\|_F^4. 
\end{equation*}
It remains to consider the subcase when both $\{\ell(\tau_3),\ell(\tau_4),\ell(\tau_5)\}$ and $\{\ell(\tau_6),\ell(\tau_7),\ell(\tau_8)\}$ have duplicated values. It follows that
\BEqn
& & \lrabs{ \sum\limits_{\ell_1,\ell_2,\ell_3,\ell_4=1}^{p} \widetilde{\ca}_5 } \\
&=& \Big{|} \sum\limits_{\ell_1,\ell_2,\ell_3,\ell_4=1}^{p} \lrp{A^{(u_{\tau_1})} \Gamma^{(2)} (A^{(u_{\tau_2})})^{\top}}_{\ell_1,\ell_2}  
    \cum\lrp{(A^{(u_{\tau_3})} \varepsilon_0)_{\ell_1},
          (A^{(u_{\tau_4})} \varepsilon_0)_{\ell_3}, 
          (A^{(u_{\tau_5})} \varepsilon_0)_{\ell_3}} \\
& & \hspace{5em} \times 
    \cum\lrp{(A^{(u_{\tau_6})} \varepsilon_0)_{\ell_2},
          (A^{(u_{\tau_7})} \varepsilon_0)_{\ell_4},
          (A^{(u_{\tau_8})} \varepsilon_0)_{\ell_4}} \Big{|} \\
&=& \Big{|} \sum\limits_{k_1,\cdots,k_6=1}^{p} 
    \lrp{(A^{(u_{\tau_3})})^{\top} A^{(u_{\tau_1})} \Gamma^{(2)} (A^{(u_{\tau_2})})^{\top} A^{(u_{\tau_6})}}_{k_1,k_4} 
    \lrp{(A^{(u_{\tau_4})})^{\top} A^{(u_{\tau_5})}}_{k_2,k_3}
    \lrp{(A^{(u_{\tau_7})})^{\top} A^{(u_{\tau_8})}}_{k_5,k_6} \\
& & \hspace{5em} \times 
    \cum\lrp{\varepsilon_{0,k_1}, \varepsilon_{0,k_2}, \varepsilon_{0,k_3}} 
    \cum\lrp{\varepsilon_{0,k_4}, \varepsilon_{0,k_5}, \varepsilon_{0,k_6}} \Big{|} \\
&\lesssim& \|(A^{(u_{\tau_3})})^{\top} A^{(u_{\tau_1})} \Gamma^{(2)} (A^{(u_{\tau_2})})^{\top} A^{(u_{\tau_6})}\|
    \cdot \|(A^{(u_{\tau_4})})^{\top} A^{(u_{\tau_5})}\|
    \cdot \|(A^{(u_{\tau_7})})^{\top} A^{(u_{\tau_8})}\| \\
& & \hspace{5em} \times 
    \sum\limits_{k_1,k_2,k_3=1}^{p} \lrabs{ \cum\lrp{\varepsilon_{0,k_1}, \varepsilon_{0,k_2}, \varepsilon_{0,k_3}} }
    \sum\limits_{k_4,k_5,k_6=1}^{p} \lrabs{ \cum\lrp{\varepsilon_{0,k_4}, \varepsilon_{0,k_5}, \varepsilon_{0,k_6}} } \\
&\lesssim& \bigp{ \prod\limits_{i=1}^{8} \|A^{(u_i)}\| } \|\Gamma^{(2)}\|_F^7,     
\EEqn
which completes the proof.
\end{Proof}

\begin{lemma}\label{Lemma:LP-aux-3-6}
Under Assumption \ref{Assumpt:NullDist_Wn_lp}\ref{Assumpt:NullDist_Wn_Lp-1},\ref{Assumpt:NullDist_Wn_Lp-2}, it holds that
\BEqn
    \lrabs{ \sum\limits_{\ell_1,\ell_2,\ell_3,\ell_4=1}^{p} \ca_6 }
&\lesssim& \sum\limits_{\pi_8(2,2,4)} \bone\{i_{\tau_1}={\tau_2},\ i_{\tau_3}=i_{\tau_4},\ i_{\tau_5}=\cdots=i_{\tau_8}\} 
    \bigp{ \prod\limits_{i=1}^{8} \|A^{(u_i)}\| } \|\Gamma^{(2)}\|_F^4 \\
& & \hspace{2em} \times
    \Big{(} 
    \bone\{\ell(\tau_1)=\ell(\tau_2),\ \ell(\tau_3)=\ell(\tau_4)\} \|\Gamma^{(2)}\|_F^2 p \\
& & \hspace{3em}
    + \bone\{\ell(\tau_1)=\ell(\tau_2),\ \ell(\tau_3)\neq\ell(\tau_4)\} \|\Gamma^{(2)}\|_F^2 \sqrt{p} \\  
& & \hspace{3em}
    + \bone\{\ell(\tau_1)=\ell(\tau_3)\} \|\Gamma^{(2)}\|_F^2 \\
& & \hspace{3em}
    + \bone\{\ell(\tau_1),\ell(\tau_2),\ell(\tau_3),\ell(\tau_4)\mbox{ are pairwise distinct}\} \Big{)},  
\EEqn
where $\ca_6$ is defined as Lemma \ref{Lemma:LP-aux-3}.
\end{lemma}
\begin{Proof}
Define
\BEqn
    \widetilde{\ca}_6
&=& \lrp{A^{(u_{\tau_1})} \Gamma^{(2)} (A^{(u_{\tau_2})})^{\top}}_{\ell(\tau_1),\ell(\tau_2)}
    \lrp{A^{(u_{\tau_3})} \Gamma^{(2)} (A^{(u_{\tau_4})})^{\top}}_{\ell(\tau_3),\ell(\tau_4)} \\
& & \times    
    \cum\lrp{(A^{(u_{\tau_5})} \varepsilon_0)_{\ell(\tau_5)},
          (A^{(u_{\tau_6})} \varepsilon_0)_{\ell(\tau_6)},
          (A^{(u_{\tau_7})} \varepsilon_0)_{\ell(\tau_7)},
          (A^{(u_{\tau_8})} \varepsilon_0)_{\ell(\tau_8)}},
\EEqn
then we have $\ca_6 = \sum\limits_{\pi_8(2,2,4)} \bone\{i_{\tau_1}={\tau_2},\ i_{\tau_3}=i_{\tau_4},\ i_{\tau_5}=\cdots=i_{\tau_8}\} \widetilde{\ca}_6$. The analysis of $\widetilde{\ca}_6$ is similar to that of $\widetilde{\ca}_5$ and we consider three major cases, $\ell(\tau_1)=\ell(\tau_2),\ \ell(\tau_3)=\ell(\tau_4)$, and $\ell(\tau_1)=\ell(\tau_2),\ \ell(\tau_3)\neq\ell(\tau_4)$, as well as $\ell(\tau_1)\neq\ell(\tau_2),\ \ell(\tau_3)\neq\ell(\tau_4)$.

For the first case when $\ell(\tau_1)=\ell(\tau_2),\ \ell(\tau_3)=\ell(\tau_4)$, we have that
\BEqn
& & \lrabs{ \sum\limits_{\ell_1,\ell_2,\ell_3,\ell_4=1}^{p} \widetilde{\ca}_6 } \\
&=& \Big{|} \sum\limits_{\ell_1,\ell_2,\ell_3,\ell_4=1}^{p}
    \lrp{A^{(u_{\tau_1})} \Gamma^{(2)} (A^{(u_{\tau_2})})^{\top}}_{\ell_1,\ell_1}
    \lrp{A^{(u_{\tau_3})} \Gamma^{(2)} (A^{(u_{\tau_4})})^{\top}}_{\ell_2,\ell_2} \\
& & \hspace{5em} \times   
    \cum\lrp{(A^{(u_{\tau_5})} \varepsilon_0)_{\ell_3},
          (A^{(u_{\tau_6})} \varepsilon_0)_{\ell_3},
          (A^{(u_{\tau_7})} \varepsilon_0)_{\ell_4},
          (A^{(u_{\tau_8})} \varepsilon_0)_{\ell_4}} \Big{|} \\
&=& \lrabs{ \tr\lrp{A^{(u_{\tau_1})} \Gamma^{(2)} (A^{(u_{\tau_2})})^{\top}} }
    \cdot \lrabs{ \tr\lrp{A^{(u_{\tau_3})} \Gamma^{(2)} (A^{(u_{\tau_4})})^{\top}} } \\
& & \hspace{5em} \times      
    \lrabs{\sum\limits_{\ell_3,\ell_4=1}^{p}
           \cum\lrp{(A^{(u_{\tau_5})} \varepsilon_0)_{\ell_3},
                    (A^{(u_{\tau_6})} \varepsilon_0)_{\ell_3},
                    (A^{(u_{\tau_7})} \varepsilon_0)_{\ell_4},
                    (A^{(u_{\tau_8})} \varepsilon_0)_{\ell_4}} } \\
&\lesssim&  \bigp{ \prod\limits_{i=1}^{8} \|A^{(u_i)}\| } \|\Gamma^{(2)}\|_F^6 p,
\EEqn
where the last step follows from Lemma \ref{Lemma:LP-aux-1-2}.

Under the second case when $\ell(\tau_1)=\ell(\tau_2)$ whereas $\ell(\tau_3)\neq\ell(\tau_4)$, we have that
\BEqn
& & \lrabs{ \sum\limits_{\ell_1,\ell_2,\ell_3,\ell_4=1}^{p} \widetilde{\ca}_6 } \\
&=& \Big{|} \sum\limits_{\ell_1,\ell_2,\ell_3,\ell_4=1}^{p}
    \lrp{A^{(u_{\tau_1})} \Gamma^{(2)} (A^{(u_{\tau_2})})^{\top}}_{\ell_1,\ell_1}
    \lrp{A^{(u_{\tau_3})} \Gamma^{(2)} (A^{(u_{\tau_4})})^{\top}}_{\ell_2,\ell_3} \\
& & \hspace{5em} \times   
    \cum\lrp{(A^{(u_{\tau_5})} \varepsilon_0)_{\ell_2},
          (A^{(u_{\tau_6})} \varepsilon_0)_{\ell_3},
          (A^{(u_{\tau_7})} \varepsilon_0)_{\ell_4},
          (A^{(u_{\tau_8})} \varepsilon_0)_{\ell_4}} \Big{|} \\
&=& \lrabs{ \tr\lrp{A^{(u_{\tau_1})} \Gamma^{(2)} (A^{(u_{\tau_2})})^{\top}} } \\
& & \times
    \left|
        \sum\limits_{k_1,k_2,k_3,k_4=1}^{p} 
           \lrp{(A^{(u_{\tau_5})})^{\top} A^{(u_{\tau_3})} \Gamma^{(2)} (A^{(u_{\tau_4})})^{\top} A^{(u_{\tau_6})}}_{k_1,k_2}\right. \\
& & \hspace{7em} \times
    \left.
        \lrp{(A^{(u_{\tau_7})})^{\top} A^{(u_{\tau_8})}}_{k_3,k_4} 
        \cum\lrp{\varepsilon_{0,k_1}, \varepsilon_{0,k_2}, \varepsilon_{0,k_3}, \varepsilon_{0,k_4}}
    \right| \\
&\lesssim&  \bigp{ \prod\limits_{i=1}^{8} \|A^{(u_i)}\| } \|\Gamma^{(2)}\|_F^6 \sqrt{p}.
\EEqn

For the last case when $\ell(\tau_1)\neq\ell(\tau_2)$ and $\ell(\tau_3)\neq\ell(\tau_4)$, we consider three subcases. If $\ell(\tau_1)=\ell(\tau_3)$ and $\ell(\tau_2)=\ell(\tau_4)$, we have that
\BEqn
& & \lrabs{ \sum\limits_{\ell_1,\ell_2,\ell_3,\ell_4=1}^{p} \widetilde{\ca}_6 } \\
&=& \Big{|} \sum\limits_{\ell_1,\ell_2,\ell_3,\ell_4=1}^{p}
    \lrp{A^{(u_{\tau_1})} \Gamma^{(2)} (A^{(u_{\tau_2})})^{\top}}_{\ell_1,\ell_2}
    \lrp{A^{(u_{\tau_3})} \Gamma^{(2)} (A^{(u_{\tau_4})})^{\top}}_{\ell_1,\ell_2} \\
& & \hspace{5em} \times   
    \cum\lrp{(A^{(u_{\tau_5})} \varepsilon_0)_{\ell_3},
          (A^{(u_{\tau_6})} \varepsilon_0)_{\ell_3},
          (A^{(u_{\tau_7})} \varepsilon_0)_{\ell_4},
          (A^{(u_{\tau_8})} \varepsilon_0)_{\ell_4}} \Big{|} \\
&=& \lrabs{ \tr\lrp{A^{(u_{\tau_1})} \Gamma^{(2)} (A^{(u_{\tau_2})})^{\top} A^{(u_{\tau_4})} \Gamma^{(2)} (A^{(u_{\tau_3})})^{\top}} } \\
& & \times
    \lrabs{\sum\limits_{k_1,k_2,k_3,k_4=1}^{p} 
           \lrp{(A^{(u_{\tau_5})})^{\top} A^{(u_{\tau_6})}}_{k_1,k_2}
           \lrp{(A^{(u_{\tau_7})})^{\top} A^{(u_{\tau_8})}}_{k_3,k_4} 
           \cum\lrp{\varepsilon_{0,k_1}, \varepsilon_{0,k_2}, \varepsilon_{0,k_3}, \varepsilon_{0,k_4}}} \\
&\lesssim&  \bigp{ \prod\limits_{i=1}^{8} \|A^{(u_i)}\| } \|\Gamma^{(2)}\|_F^6.
\EEqn
If $\ell(\tau_1)=\ell(\tau_3)$ whereas $\ell(\tau_2)\neq\ell(\tau_4)$, we have that
\BEqn
& & \lrabs{ \sum\limits_{\ell_1,\ell_2,\ell_3,\ell_4=1}^{p} \widetilde{\ca}_6 } \\
&=& \Big{|} \sum\limits_{\ell_1,\ell_2,\ell_3,\ell_4=1}^{p}
    \lrp{A^{(u_{\tau_1})} \Gamma^{(2)} (A^{(u_{\tau_2})})^{\top}}_{\ell_1,\ell_2}
    \lrp{A^{(u_{\tau_3})} \Gamma^{(2)} (A^{(u_{\tau_4})})^{\top}}_{\ell_1,\ell_3} \\
& & \hspace{5em} \times   
    \cum\lrp{(A^{(u_{\tau_5})} \varepsilon_0)_{\ell_2},
          (A^{(u_{\tau_6})} \varepsilon_0)_{\ell_3},
          (A^{(u_{\tau_7})} \varepsilon_0)_{\ell_4},
          (A^{(u_{\tau_8})} \varepsilon_0)_{\ell_4}} \Big{|} \\
&=& \Big{|} \sum\limits_{k_1,k_2,k_3,k_4=1}^{p} 
    \lrp{A^{(u_{\tau_5})})^{\top} A^{(u_{\tau_2})} \Gamma^{(2)} (A^{(u_{\tau_1})})^{\top} A^{(u_{\tau_3})} \Gamma^{(2)} (A^{(u_{\tau_4})})^{\top} A^{(u_{\tau_6})}}_{k_1,k_2} 
    \lrp{(A^{(u_{\tau_7})})^{\top} A^{(u_{\tau_8})}}_{k_3,k_4} \\
& & \hspace{6em} \times
    \cum\lrp{\varepsilon_{0,k_1}, \varepsilon_{0,k_2}, \varepsilon_{0,k_3}, \varepsilon_{0,k_4}} \Big{|} \\
&\lesssim&  \bigp{ \prod\limits_{i=1}^{8} \|A^{(u_i)}\| } \|\Gamma^{(2)}\|_F^6.
\EEqn
Lastly, if $\{\ell(\tau_1),\ell(\tau_2),\ell(\tau_3),\ell(\tau_4)\}$ are pairwise distinct, we have that
\BEqn
& & \lrabs{ \sum\limits_{\ell_1,\ell_2,\ell_3,\ell_4=1}^{p} \widetilde{\ca}_6 } \\
&=& \Big{|} \sum\limits_{\ell_1,\ell_2,\ell_3,\ell_4=1}^{p}
    \lrp{A^{(u_{\tau_1})} \Gamma^{(2)} (A^{(u_{\tau_2})})^{\top}}_{\ell_1,\ell_2}
    \lrp{A^{(u_{\tau_3})} \Gamma^{(2)} (A^{(u_{\tau_4})})^{\top}}_{\ell_3,\ell_4} \\
& & \hspace{5em} \times   
    \cum\lrp{(A^{(u_{\tau_5})} \varepsilon_0)_{\ell_1},
          (A^{(u_{\tau_6})} \varepsilon_0)_{\ell_2},
          (A^{(u_{\tau_7})} \varepsilon_0)_{\ell_3},
          (A^{(u_{\tau_8})} \varepsilon_0)_{\ell_4}} \Big{|} \\
&\le& \lrp{ \sum\limits_{\ell_1,\ell_2,\ell_3,\ell_4=1}^{p}
            \lrp{A^{(u_{\tau_1})} \Gamma^{(2)} (A^{(u_{\tau_2})})^{\top}}_{\ell_1,\ell_2}^2
            \lrp{A^{(u_{\tau_3})} \Gamma^{(2)} (A^{(u_{\tau_4})})^{\top}}_{\ell_3,\ell_4}^2 }^{1/2} \\
& & \times              
    \lrp{ \sum\limits_{\ell_1,\ell_2,\ell_3,\ell_4=1}^{p}
          \cum^2\lrp{(A^{(u_{\tau_5})} \varepsilon_0)_{\ell_1},
          (A^{(u_{\tau_6})} \varepsilon_0)_{\ell_2},
          (A^{(u_{\tau_7})} \varepsilon_0)_{\ell_3},
          (A^{(u_{\tau_8})} \varepsilon_0)_{\ell_4}} }^{1/2} \\
&=& \|A^{(u_{\tau_1})} \Gamma^{(2)} (A^{(u_{\tau_2})})^{\top}\|_F \cdot \|A^{(u_{\tau_3})} \Gamma^{(2)} (A^{(u_{\tau_4})})^{\top}\|_F \\
& & \times
    \lrp{ \sum\limits_{\ell_1,\ell_2,\ell_3,\ell_4=1}^{p}
          \lrp{\sum\limits_{k_1,k_2,k_3,k_4=1}^{p} 
          A^{(u_{\tau_5})}_{\ell_1,k_1} A^{(u_{\tau_6})}_{\ell_2,k_2} A^{(u_{\tau_7})}_{\ell_3,k_3} A^{(u_{\tau_8})}_{\ell_4,k_4} 
          \cum\lrp{\varepsilon_{0,k_1}, \varepsilon_{0,k_2}, \varepsilon_{0,k_3}, \varepsilon_{0,k_4}} }^2 }^{1/2} \\
&\lesssim&  \bigp{ \prod\limits_{i=1}^{8} \|A^{(u_i)}\| } \|\Gamma^{(2)}\|_F^4,
\EEqn
where the last step follows from Lemma \ref{Lemma:LP-aux-2}. By unifying the results from all previous cases, we obtain the desired statement.
\end{Proof}

\begin{lemma}\label{Lemma:LP-aux-3-7}
Under Assumption \ref{Assumpt:NullDist_Wn_lp}\ref{Assumpt:NullDist_Wn_Lp-1},\ref{Assumpt:NullDist_Wn_Lp-2}, it holds that
\BEqn
    \lrabs{ \sum\limits_{\ell_1,\ell_2,\ell_3,\ell_4=1}^{p} \ca_7 }
&\lesssim& \sum\limits_{\pi_8(2,2,4)} \bone\{i_{\tau_1}=i_{\tau_2},\ i_{\tau_3}=i_{\tau_4},\ i_{\tau_5}=i_{\tau_6},\ i_{\tau_7}=i_{\tau_8}\}
    \bigp{ \prod\limits_{i=1}^{8} \|A^{(u_i)}\| } \|\Gamma^{(2)}\|_F^4 \\
& & \hspace{2em} \times
    \Big{(} 
    \bone\{\ell(\tau_1)=\ell(\tau_2),\ \ell(\tau_3)=\ell(\tau_4),\ \ell(\tau_5)=\ell(\tau_6),\ \ell(\tau_7)=\ell(\tau_8)\} p^2 \\
& & \hspace{3em}
    + \bone\{\ell(\tau_1)=\ell(\tau_2),\ \ell(\tau_3)=\ell(\tau_4),\ \ell(\tau_5)\neq\ell(\tau_6),\ \ell(\tau_7)\neq\ell(\tau_8)\} p \\
& & \hspace{3em}
    + \bone\{\ell(\tau_1)=\ell(\tau_2),\ \ell(\tau_3)\neq\ell(\tau_4),\ \ell(\tau_5)\neq\ell(\tau_6),\ \ell(\tau_7)\neq\ell(\tau_8)\} \sqrt{p} \\
& & \hspace{3em}
    + \bone\{\ell(\tau_1)\neq\ell(\tau_2),\ \ell(\tau_3)\neq\ell(\tau_4),\ \ell(\tau_5)\neq\ell(\tau_6),\ \ell(\tau_7)\neq\ell(\tau_8)\} \Big{)},     
\EEqn
where $\ca_7$ is defined as Lemma \ref{Lemma:LP-aux-3}.
\end{lemma}
\begin{Proof}
Define
\BEqn
    \widetilde{\ca}_7
&=& \lrp{A^{(u_{\tau_1})} \Gamma^{(2)} (A^{(u_{\tau_2})})^{\top}}_{\ell(\tau_1),\ell(\tau_2)}
    \lrp{A^{(u_{\tau_3})} \Gamma^{(2)} (A^{(u_{\tau_4})})^{\top}}_{\ell(\tau_3),\ell(\tau_4)} \\
& & \times    
    \lrp{A^{(u_{\tau_5})} \Gamma^{(2)} (A^{(u_{\tau_6})})^{\top}}_{\ell(\tau_5),\ell(\tau_6)}
    \lrp{A^{(u_{\tau_7})} \Gamma^{(2)} (A^{(u_{\tau_8})})^{\top}}_{\ell(\tau_7),\ell(\tau_9)},
\EEqn
then we have $\ca_6 = \sum\limits_{\pi_8(2,2,2,2)} \bone\{i_{\tau_1}=i_{\tau_2},\ i_{\tau_3}=i_{\tau_4},\ i_{\tau_5}=i_{\tau_6},\ i_{\tau_7}=i_{\tau_8}\} \widetilde{\ca}_7$. Consider the four pairs $\{\ell(\tau_1),\ell(\tau_2)\}$, $\{\ell(\tau_3),\ell(\tau_4)\}$, $\{\ell(\tau_5),\ell(\tau_6)\}$ and $\{\ell(\tau_7),\ell(\tau_8)\}$, we consider four cases based on the number of pair(s) that admit the same value.

For the first case when all the four pairs take the same value, i.e. $\ell(\tau_1)=\ell(\tau_2),\ \ell(\tau_3)=\ell(\tau_4),\ \ell(\tau_5)=\ell(\tau_6)$ and $\ell(\tau_7)=\ell(\tau_8)$, we have that
\BEqn
    \lrabs{ \sum\limits_{\ell_1,\ell_2,\ell_3,\ell_4=1}^{p} \widetilde{\ca}_7 }
&=& \Big{|} \sum\limits_{\ell_1,\ell_2,\ell_3,\ell_4=1}^{p}
    \lrp{A^{(u_{\tau_1})} \Gamma^{(2)} (A^{(u_{\tau_2})})^{\top}}_{\ell_1,\ell_1}
    \lrp{A^{(u_{\tau_3})} \Gamma^{(2)} (A^{(u_{\tau_4})})^{\top}}_{\ell_2,\ell_2} \\
& & \hspace{6em} \times    
    \lrp{A^{(u_{\tau_5})} \Gamma^{(2)} (A^{(u_{\tau_6})})^{\top}}_{\ell_3,\ell_3}
    \lrp{A^{(u_{\tau_7})} \Gamma^{(2)} (A^{(u_{\tau_8})})^{\top}}_{\ell_4,\ell_4} \Big{|} \\
&=& \lrabs{ \tr\lrp{A^{(u_{\tau_1})} \Gamma^{(2)} (A^{(u_{\tau_2})})^{\top}} }
    \cdot \lrabs{ \tr\lrp{A^{(u_{\tau_3})} \Gamma^{(2)} (A^{(u_{\tau_4})})^{\top}} } \\
& & \hspace{2em} \times
    \cdot \lrabs{ \tr\lrp{A^{(u_{\tau_5})} \Gamma^{(2)} (A^{(u_{\tau_6})})^{\top}} }
    \cdot \lrabs{ \tr\lrp{A^{(u_{\tau_7})} \Gamma^{(2)} (A^{(u_{\tau_8})})^{\top}} } \\
&\lesssim&  \bigp{ \prod\limits_{i=1}^{8} \|A^{(u_i)}\| } \|\Gamma^{(2)}\|_F^4 p^2,
\EEqn
where the last step follows from Lemma \ref{Lemma:LP-aux-1-2}.

Now consider the case when two out of the four pairs take the same value, i.e. $\ell(\tau_1)=\ell(\tau_2),\ \ell(\tau_3)=\ell(\tau_4),\ \ell(\tau_5)=\ell(\tau_7)$ and $\ell(\tau_6)=\ell(\tau_8)$, we have that
\BEqn
    \lrabs{ \sum\limits_{\ell_1,\ell_2,\ell_3,\ell_4=1}^{p} \widetilde{\ca}_7 } 
&=& \Big{|} \sum\limits_{\ell_1,\ell_2,\ell_3,\ell_4=1}^{p}
    \lrp{A^{(u_{\tau_1})} \Gamma^{(2)} (A^{(u_{\tau_2})})^{\top}}_{\ell_1,\ell_1}
    \lrp{A^{(u_{\tau_3})} \Gamma^{(2)} (A^{(u_{\tau_4})})^{\top}}_{\ell_2,\ell_2} \\
& & \hspace{7em} \times
    \lrp{A^{(u_{\tau_5})} \Gamma^{(2)} (A^{(u_{\tau_6})})^{\top}}_{\ell_3,\ell_4}
    \lrp{A^{(u_{\tau_7})} \Gamma^{(2)} (A^{(u_{\tau_8})})^{\top}}_{\ell_3,\ell_4} \Big{|} \\
&=& \lrabs{ \tr\lrp{A^{(u_{\tau_1})} \Gamma^{(2)} (A^{(u_{\tau_2})})^{\top}} }
    \cdot \lrabs{ \tr\lrp{A^{(u_{\tau_3})} \Gamma^{(2)} (A^{(u_{\tau_4})})^{\top}} } \\
& & \hspace{2em} \times
    \cdot \lrabs{ \tr\lrp{A^{(u_{\tau_5})} \Gamma^{(2)} (A^{(u_{\tau_6})})^{\top} A^{(u_{\tau_8})} \Gamma^{(2)} (A^{(u_{\tau_7})})^{\top}} } \\
&\lesssim&  \bigp{ \prod\limits_{i=1}^{8} \|A^{(u_i)}\| } \|\Gamma^{(2)}\|_F^4 p.
\EEqn

When only one out of four pairs takes the same value, we obtain that
\BEqn
& & \lrabs{ \sum\limits_{\ell_1,\ell_2,\ell_3,\ell_4=1}^{p} \widetilde{\ca}_7 } \\
&=& \Big{|} \sum\limits_{\ell_1,\ell_2,\ell_3,\ell_4=1}^{p}
    \lrp{A^{(u_{\tau_1})} \Gamma^{(2)} (A^{(u_{\tau_2})})^{\top}}_{\ell_1,\ell_1}
    \lrp{A^{(u_{\tau_3})} \Gamma^{(2)} (A^{(u_{\tau_4})})^{\top}}_{\ell_2,\ell_3} \\
& & \hspace{7em} \times
    \lrp{A^{(u_{\tau_5})} \Gamma^{(2)} (A^{(u_{\tau_6})})^{\top}}_{\ell_3,\ell_4}
    \lrp{A^{(u_{\tau_7})} \Gamma^{(2)} (A^{(u_{\tau_8})})^{\top}}_{\ell_2,\ell_4} \Big{|} \\
&=& \lrabs{ \tr\lrp{A^{(u_{\tau_1})} \Gamma^{(2)} (A^{(u_{\tau_2})})^{\top}} }
    \cdot \lrabs{ \tr\lrp{A^{(u_{\tau_3})} \Gamma^{(2)} (A^{(u_{\tau_4})})^{\top} A^{(u_{\tau_5})} \Gamma^{(2)} (A^{(u_{\tau_6})})^{\top} A^{(u_{\tau_8})} \Gamma^{(2)} (A^{(u_{\tau_7})})^{\top}} } \\
&\lesssim&  \bigp{ \prod\limits_{i=1}^{8} \|A^{(u_i)}\| } \|\Gamma^{(2)}\|_F^4 \sqrt{p}.
\EEqn

Finally, if none of the pairs takes the same value, we either have that
\BEqn
& & \lrabs{ \sum\limits_{\ell_1,\ell_2,\ell_3,\ell_4=1}^{p} \widetilde{\ca}_7 } \\
&=& \Big{|} \sum\limits_{\ell_1,\ell_2,\ell_3,\ell_4=1}^{p}
    \lrp{A^{(u_{\tau_1})} \Gamma^{(2)} (A^{(u_{\tau_2})})^{\top}}_{\ell_1,\ell_2}
    \lrp{A^{(u_{\tau_3})} \Gamma^{(2)} (A^{(u_{\tau_4})})^{\top}}_{\ell_1,\ell_2} \\
& & \hspace{7em} \times    
    \lrp{A^{(u_{\tau_5})} \Gamma^{(2)} (A^{(u_{\tau_6})})^{\top}}_{\ell_3,\ell_4}
    \lrp{A^{(u_{\tau_7})} \Gamma^{(2)} (A^{(u_{\tau_8})})^{\top}}_{\ell_3,\ell_4} \Big{|} \\
&=& \lrabs{ \tr\lrp{A^{(u_{\tau_1})} \Gamma^{(2)} (A^{(u_{\tau_2})})^{\top} A^{(u_{\tau_4})} \Gamma^{(2)} (A^{(u_{\tau_3})})^{\top}} }
    \cdot \lrabs{ \tr\lrp{ A^{(u_{\tau_5})} \Gamma^{(2)} (A^{(u_{\tau_6})})^{\top} A^{(u_{\tau_8})} \Gamma^{(2)} (A^{(u_{\tau_7})})^{\top}} } \\
&\lesssim&  \bigp{ \prod\limits_{i=1}^{8} \|A^{(u_i)}\| } \|\Gamma^{(2)}\|_F^4,
\EEqn
or have that
\BEqn
& & \lrabs{ \sum\limits_{\ell_1,\ell_2,\ell_3,\ell_4=1}^{p} \widetilde{\ca}_7 } \\
&=& \Big{|} \sum\limits_{\ell_1,\ell_2,\ell_3,\ell_4=1}^{p}
    \lrp{A^{(u_{\tau_1})} \Gamma^{(2)} (A^{(u_{\tau_2})})^{\top}}_{\ell_1,\ell_2}
    \lrp{A^{(u_{\tau_3})} \Gamma^{(2)} (A^{(u_{\tau_4})})^{\top}}_{\ell_2,\ell_3} \\
& & \hspace{7em} \times
    \lrp{A^{(u_{\tau_5})} \Gamma^{(2)} (A^{(u_{\tau_6})})^{\top}}_{\ell_3,\ell_4}
    \lrp{A^{(u_{\tau_7})} \Gamma^{(2)} (A^{(u_{\tau_8})})^{\top}}_{\ell_1,\ell_4} \Big{|} \\
&=& \lrabs{ \tr\lrp{A^{(u_{\tau_1})} \Gamma^{(2)} (A^{(u_{\tau_2})})^{\top} A^{(u_{\tau_3})} \Gamma^{(2)} (A^{(u_{\tau_4})})^{\top} A^{(u_{\tau_5})} \Gamma^{(2)} (A^{(u_{\tau_6})})^{\top} A^{(u_{\tau_8})} \Gamma^{(2)} (A^{(u_{\tau_7})})^{\top}} } \\
&\lesssim&  \bigp{ \prod\limits_{i=1}^{8} \|A^{(u_i)}\| } \|\Gamma^{(2)}\|_F^4.
\EEqn
In summary, we obtain the proposed result.
\end{Proof}

\subsection{Auxiliary Results for Appendix \ref{Appdix:lemma_multiple}}

\begin{lemma}\label{Lemma:Expression_S}
For any $1\le \ell_1\le \ell_2\le N$, define 
\begin{equation*}
    S_{\ell_1,\ell_2} 
  = \sum\limits_{j=\ell_1}^{\ell_2} Y_j
  = \frac{1}{m_1} \sum\limits_{i=1}^{m_1} \lrp{X_i - X_{n+1-i}}^{\top} \lrp{\sum\limits_{j=\ell_1}^{\ell_2} X_{j+m}},
\end{equation*}
and
\begin{equation*}
    \tilde{S}_{\ell_1,\ell_2} 
  = \frac{1}{m_1} \sum\limits_{i=1}^{m_1} \lrp{\tilde{X}_i - \tilde{X}_{n+1-i}}^{\top} \lrp{\sum\limits_{j=\ell_1}^{\ell_2} \tilde{X}_{j+m}},
\end{equation*}
then it holds that 
\begin{equation*}
    S_{\ell_1,\ell_2}
 =  \tilde{S}_{\ell_1,\ell_2} 
 + \frac{1}{m_1} \sum\limits_{i=1}^{m_1} \lrp{\tilde{X}_i - \tilde{X}_{n+1-i}}^{\top} \lrp{\sum\limits_{j=\ell_1}^{\ell_2} \mu_{j+m}} 
 - \sum\limits_{j=\ell_1}^{\ell_2} \delta^{\top} \tilde{X}_{j+m} 
 - \sum\limits_{j=\ell_1}^{\ell_2} \delta^{\top} \mu_{j+m}.
\end{equation*}
\end{lemma}

\begin{Proof}
Under the assumption that $\varepsilon_0 \le \varepsilon < \xi_1 < \cdots < \xi_M < 1-\varepsilon < 1-\varepsilon_0$, we observe that
\begin{equation*}
    \hat\mu_1 
  = \frac{1}{m_1} \sum\limits_{i=1}^{m_1} X_i
  = \mu_{k_1} + \frac{1}{m_1}\sum\limits_{i=1}^{m_1} \tilde{X}_{i}
\end{equation*}
and
\begin{equation*}
    \hat\mu_n
  = \frac{1}{m_1} \sum\limits_{i=1}^{m_1} X_{n+1-i}
  = \mu_{k_{M+1}} + \frac{1}{m_1}\sum\limits_{i=1}^{m_1} \tilde{X}_{n+1-i},
\end{equation*}
both of which imply that
\begin{equation*}
    \frac{1}{m_1} \sum\limits_{i=1}^{m_1} \lrp{X_i - X_{n+1-i}}
  = \frac{1}{m_1} \sum\limits_{i=1}^{m_1} \lrp{\tilde{X}_i - \tilde{X}_{n+1-i}} - \delta.
\end{equation*}
Then it follows from direct calculation that
\BEqn
& & S_{\ell_1,\ell_2} \\
&=& \frac{1}{m_1} \sum\limits_{i=1}^{m_1} \lrp{X_i - X_{n+1-i}}^{\top} \lrp{\sum\limits_{j=\ell_1}^{\ell_2} X_{j+m}} \\
&=& \lrp{\frac{1}{m_1} \sum\limits_{i=1}^{m_1} \lrp{\tilde{X}_i - \tilde{X}_{n+1-i}} - \delta}^{\top} \lrp{\sum\limits_{j=\ell_1}^{\ell_2} \lrp{\tilde{X}_{j+m} + \mu_{j+m}}} \\
&=& \tilde{S}_{\ell_1,\ell_2} + \frac{1}{m_1} \sum\limits_{i=1}^{m_1} \lrp{\tilde{X}_i - \tilde{X}_{n+1-i}}^{\top} \lrp{\sum\limits_{j=\ell_1}^{\ell_2} \mu_{j+m}} - \sum\limits_{j=\ell_1}^{\ell_2} \delta^{\top} \tilde{X}_{j+m} - \sum\limits_{j=\ell_1}^{\ell_2} \delta^{\top} \mu_{j+m}.
\EEqn
for any $1\le \ell_1\le \ell_2\le N$.
\end{Proof}

\begin{lemma}\label{Lemma:Expression_W}
Define the process $\{W_n(r)\}_{\varepsilon\le r\le 1-\varepsilon}$ as $W_n(r) = m_1 S_{1,\lrfl{nr}-\lrfl{n\varepsilon}}$ and the process $\{\tilde{W}_n(r)\}_{\varepsilon\le r\le 1-\varepsilon}$ as $\tilde{W}_n(r) = m_1 \tilde{S}_{1,\lrfl{nr}-\lrfl{n\varepsilon}}$, then it holds for any $\varepsilon\le r\le 1-\varepsilon$ that
\begin{equation*}
    W_n(r)
  = \tilde{W}_n(r)
    + \sum\limits_{i=1}^{m_1} \lrp{\tilde{X}_i - \tilde{X}_{n+1-i}}^{\top} \lrp{\sum\limits_{j=1}^{\lrfl{nr}-\lrfl{n\varepsilon}} \mu_{j+m}} 
    - m_1 \sum\limits_{j=1}^{\lrfl{nr}-\lrfl{n\varepsilon}} \delta^{\top} \tilde{X}_{j+m} 
    - m_1 \sum\limits_{j=1}^{\lrfl{nr}-\lrfl{n\varepsilon}} \delta^{\top} \mu_{j+m}.
\end{equation*}
\end{lemma}

\begin{Proof}
The relationship between $W_n(r)$ and $\tilde{W}_n(r)$ is a direct result of Lemma \ref{Lemma:Expression_S}.
\end{Proof}

\begin{lemma}\label{Lemma:Expression_aux-1}
For any $\varepsilon\le r_1\le r_2\le r_3\le 1-\varepsilon$, it holds that
\BEqn
& & \sum\limits_{j=\lrfl{nr_1}-m+1}^{\lrfl{nr_2}-m} \mu_{j+m} - \frac{\lrfl{nr_2}-\lrfl{nr_1}}{\lrfl{nr_3}-\lrfl{nr_1}} \sum\limits_{j=\lrfl{nr_1}-m+1}^{\lrfl{nr_3}-m} \mu_{j+m} \\
&=& - \sum\limits_{t=0}^{M} n \delta_t \lrp{\lrp{\frac{(r_3-r_2)(\xi_t-r_1)}{r_3-r_1}+o_p(1)}\bone\{r_1\le\xi_t<r_2\} + \lrp{\frac{(r_3-\xi_t)(r_2-r_1)}{r_3-r_1}+o_p(1)} \bone\{r_2\le \xi_t<r_3\}},
\EEqn
and
\BEqn
& & \sum\limits_{j=\lrfl{nr_2}-m+1}^{\lrfl{nr_3}-m} \mu_{j+m} - \frac{\lrfl{nr_3}-\lrfl{nr_2}}{\lrfl{nr_3}-\lrfl{nr_1}} \sum\limits_{j=\lrfl{nr_1}-m+1}^{\lrfl{nr_3}-m} \mu_{j+m} \\
&=& \sum\limits_{t=0}^{M} n \delta_t \lrp{\lrp{\frac{(r_3-r_2)(\xi_t-r_1)}{r_3-r_1}+o_p(1)}\bone\{r_1\le\xi_t<r_2\} + \lrp{\frac{(r_3-\xi_t)(r_2-r_1)}{r_3-r_1}+o_p(1)} \bone\{r_2\le \xi_t<r_3\}}.
\EEqn
\end{lemma}

\begin{Proof}
Recall that $\delta_0 = \mu_{k_1}$ and $k_0=m$, then for any $m+1\le j\le n-m$, it holds that 
\begin{equation*}
    \mu_j 
  = \mu_{k_1} + \delta_1 \bone\{j>k_1\} + \delta_2 \bone\{j>k_2\} + \cdots + \delta_M \bone\{j>k_M\}
  = \sum\limits_{t=0}^{M} \delta_{t} \bone\{j>k_t\}.
\end{equation*}
Furthermore, for any $\varepsilon\le r\le 1-\varepsilon$, it holds that
\begin{equation*}
   \sum\limits_{j=1}^{\lrfl{nr}-m} \mu_{j+m} 
 = \sum\limits_{j=m+1}^{\lrfl{nr}} \mu_{j}
 = \sum\limits_{t=0}^{M} \delta_{t} \sum\limits_{j=m+1}^{\lrfl{nr}} \bone\{j>k_t\}
 = \sum\limits_{t=0}^{M} \delta_{t} (\lrfl{nr}-k_t) \bone\{\lrfl{nr}>k_t\},
\end{equation*}
which implies that for any $\varepsilon\le r_1\le r_3\le 1-\varepsilon$, we have that
\BEqn
& & \sum\limits_{j=\lrfl{nr_1}-m+1}^{\lrfl{nr_2}-m} \mu_{j+m} - \frac{\lrfl{nr_2}-\lrfl{nr_1}}{\lrfl{nr_3}-\lrfl{nr_1}} \sum\limits_{j=\lrfl{nr_1}-m+1}^{\lrfl{nr_3}-m} \mu_{j+m} \\
&=& \sum\limits_{t=0}^{M} \delta_t \lrp{\lrp{\lrfl{nr_2}-k_t} \bone\{\lrfl{nr_2}>k_t\} - \lrp{\frac{r_2-r_1}{r_3-r_1}+o_p(1)} \lrp{\lrfl{nr_3}-k_t} \bone\{\lrfl{nr_3}>k_t\}} \\
& & - \sum\limits_{t=0}^{M} \delta_t \lrp{\lrp{\lrfl{nr_1}-k_t} \bone\{\lrfl{nr_1}>k_t\} - \lrp{\frac{r_2-r_1}{r_3-r_1}+o_p(1)} \lrp{\lrfl{nr_1}-k_t} \bone\{\lrfl{nr_1}>k_t\}} \\
&=& - \sum\limits_{t=0}^{M} \delta_t \lrp{\frac{r_2-r_1}{r_3-r_1}+o_p(1)} \lrp{\lrfl{nr_3}-k_t} \bone\{\lrfl{nr_3}>k_t\} 
    + \sum\limits_{t=0}^{M} \delta_t \lrp{\lrfl{nr_2}-k_t} \bone\{\lrfl{nr_2}>k_t\} \\
& & - \sum\limits_{t=0}^{M} \delta_t \lrp{\frac{r_3-r_2}{r_3-r_1}+o_p(1)} \lrp{\lrfl{nr_1}-k_t} \bone\{\lrfl{nr_1}>k_t\} \\
&=& - \sum\limits_{t=0}^{M} n \delta_t \lrp{\frac{(r_3-\xi_t)(r_2-r_1)}{r_3-r_1}+o_p(1)} \bone\{r_3>\xi_t\}
    + \sum\limits_{t=0}^{M} n \delta_t \lrp{r_2-\xi_t+o_p(1)} \bone\{r_2>\xi_t\} \\
& & - \sum\limits_{t=0}^{M} n \delta_t \lrp{\frac{(r_3-r_2)(r_1-\xi_t)}{r_3-r_1}+o_p(1)} \bone\{r_1>\xi_t\} \\
&=& - \sum\limits_{t=0}^{M} n \delta_t \lrp{\lrp{\frac{(r_3-r_2)(\xi_t-r_1)}{r_3-r_1}+o_p(1)}\bone\{r_1\le\xi_t<r_2\} + \lrp{\frac{(r_3-\xi_t)(r_2-r_1)}{r_3-r_1}+o_p(1)} \bone\{r_2\le \xi_t<r_3\}},
\EEqn

Similarly, for any $\varepsilon\le r_1\le r_2\le r_3\le 1-\varepsilon$, it follows from simple calculation that
\BEqn
& & \sum\limits_{j=\lrfl{nr_2}-m+1}^{\lrfl{nr_3}-m} \mu_{j+m} - \frac{\lrfl{nr_3}-\lrfl{nr_2}}{\lrfl{nr_3}-\lrfl{nr_1}} \sum\limits_{j=\lrfl{nr_1}-m+1}^{\lrfl{nr_3}-m} \mu_{j+m} \\
&=& \sum\limits_{t=0}^{M} \delta_t \lrp{\lrp{\lrfl{nr_3}-k_t} \bone\{\lrfl{nr_3}>k_t\} - \lrp{\frac{r_3-r_2}{r_3-r_1}+o_p(1)} \lrp{\lrfl{nr_3}-k_t} \bone\{\lrfl{nr_3}>k_t\}} \\
& & - \sum\limits_{t=0}^{M} \delta_t \lrp{\lrp{\lrfl{nr_2}-k_t} \bone\{\lrfl{nr_2}>k_t\} - \lrp{\frac{r_3-r_2}{r_3-r_1}+o_p(1)} \lrp{\lrfl{nr_1}-k_t} \bone\{\lrfl{nr_1}>k_t\}} \\
&=& \sum\limits_{t=0}^{M} \delta_t \lrp{\frac{r_2-r_1}{r_3-r_1}+o_p(1)} \lrp{\lrfl{nr_3}-k_t} \bone\{\lrfl{nr_3}>k_t\} 
    - \sum\limits_{t=0}^{M} \delta_t \lrp{\lrfl{nr_2}-k_t} \bone\{\lrfl{nr_2}>k_t\} \\
& & + \sum\limits_{t=0}^{M} \delta_t \lrp{\frac{r_3-r_2}{r_3-r_1}+o_p(1)} \lrp{\lrfl{nr_1}-k_t} \bone\{\lrfl{nr_1}>k_t\} \\
&=& \sum\limits_{t=0}^{M} n \delta_t \lrp{\frac{(r_3-\xi_t)(r_2-r_1)}{r_3-r_1}+o_p(1)} \bone\{r_3>\xi_t\}
    - \sum\limits_{t=0}^{M} n \delta_t \lrp{r_2-\xi_t+o_p(1)} \bone\{r_2>\xi_t\} \\
& & + \sum\limits_{t=0}^{M} n \delta_t \lrp{\frac{(r_3-r_2)(r_1-\xi_t)}{r_3-r_1}+o_p(1)} \bone\{r_1>\xi_t\} \\
&=& \sum\limits_{t=0}^{M} n \delta_t \lrp{\lrp{\frac{(r_3-r_2)(\xi_t-r_1)}{r_3-r_1}+o_p(1)}\bone\{r_1\le\xi_t<r_2\} + \lrp{\frac{(r_3-\xi_t)(r_2-r_1)}{r_3-r_1}+o_p(1)} \bone\{r_2\le \xi_t<r_3\}},
\EEqn
which completes the proof.
\end{Proof}

\begin{lemma}\label{Lemma:Expression_aux-2}
For any $1\le i\le M$ and $\ell\geq1$ satisfying that $i+\ell\le M$, it holds that
\BEqn
& & \sum\limits_{j=1}^{\lrfl{n\xi_i}-m} \mu_{j+m} - \frac{\lrfl{n\xi_i}-m}{\lrfl{n\xi_{i+\ell}}-m} \sum\limits_{j=1}^{\lrfl{n\xi_{i+\ell}}-m} \mu_{j+m} \\
&=& - \sum\limits_{t=0}^{i-1} n \delta_t \lrp{\frac{(\xi_{i+\ell}-\xi_i)(\xi_t-\varepsilon)}{\xi_{i+\ell}-\varepsilon}+o_p(1)} - \sum\limits_{t=i}^{i+\ell-1} n \delta_t \lrp{\frac{(\xi_{i+\ell}-\xi_t)(\xi_i-\varepsilon)}{\xi_{i+\ell}-\varepsilon}+o_p(1)}.
\EEqn
\end{lemma}

\begin{Proof}
Let $r_1=\varepsilon$, $r_2 = \xi_i$ and $r_3 = \xi_{i+\ell}$, then it directly follows from Lemma \ref{Lemma:Expression_aux-1} that
\BEqn
& & \sum\limits_{j=1}^{\lrfl{n\xi_i}-m} \mu_{j+m} - \frac{\lrfl{n\xi_i}-m}{\lrfl{n\xi_{i+\ell}}-m} \sum\limits_{j=1}^{\lrfl{n\xi_{i+\ell}}-m} \mu_{j+m} \\
&=& - \sum\limits_{t=0}^{M} n \delta_t \lrp{\lrp{\frac{(\xi_{i+\ell}-\xi_i)(\xi_t-\varepsilon)}{\xi_{i+\ell}-\varepsilon}+o_p(1)} \bone\{\xi_i>\xi_t\} + \lrp{\frac{(\xi_{i+\ell}-\xi_t)(\xi_i-\varepsilon)}{\xi_{i+\ell}-\varepsilon}+o_p(1)} \bone\{\xi_i\le \xi_t<\xi_{i+\ell}\}} \\
&=& - \sum\limits_{t=0}^{i-1} n \delta_t \lrp{\frac{(\xi_{i+\ell}-\xi_i)(\xi_t-\varepsilon)}{\xi_{i+\ell}-\varepsilon}+o_p(1)} - \sum\limits_{t=i}^{i+\ell-1} n \delta_t \lrp{\frac{(\xi_{i+\ell}-\xi_t)(\xi_i-\varepsilon)}{\xi_{i+\ell}-\varepsilon}+o_p(1)}.
\EEqn
\end{Proof}

\begin{lemma}\label{Lemma:Expression_aux-3}
For any $1\le i\le M$ and $\ell\geq1$ satisfying that $i+\ell\le M$, it holds that
\BEqn
& & \sum\limits_{j=\lrfl{n\xi_{i+\ell}}-m}^{N} \mu_{j+m} - \frac{N-\lrfl{n\xi_{i+\ell}}+m+1}{N-\lrfl{n\xi_i}+m+1} \sum\limits_{j=\lrfl{n\xi_i}-m}^{N} \mu_{j+m} \\
&=& \sum\limits_{t=i}^{i+\ell-1} n \delta_t \lrp{\frac{(1-\varepsilon-\xi_{i+\ell})(\xi_t-\xi_i)}{1-\varepsilon-\xi_i}+o_p(1)} + \sum\limits_{t=i+\ell}^{M} n \delta_t \lrp{\frac{(1-\varepsilon-\xi_t)(\xi_{i+\ell}-\xi_i)}{1-\varepsilon-\xi_i}+o_p(1)}.
\EEqn
\end{lemma}

\begin{Proof}
Let $r_1=\xi_i$, $r_2=\xi_{i+\ell}$ and $r_3=1-\varepsilon$, then it follows from the established results in Lemma \ref{Lemma:Expression_aux-1} that
\BEqn
& & \sum\limits_{j=\lrfl{n\xi_{i+\ell}}-m}^{N} \mu_{j+m} - \frac{N-\lrfl{n\xi_{i+\ell}}+m+1}{N-\lrfl{n\xi_i}+m+1} \sum\limits_{j=\lrfl{n\xi_i}-m}^{N} \mu_{j+m} \\
&=& \sum\limits_{t=0}^{M} n \delta_t \lrp{\frac{(1-\varepsilon-\xi_{i+\ell})(\xi_t-\xi_i)}{1-\varepsilon-\xi_i}+o_p(1)}\bone\{\xi_i\le\xi_t<\xi_{i+\ell}\} \\
& & + \sum\limits_{t=0}^{M} n \delta_t \lrp{\frac{(1-\varepsilon-\xi_t)(\xi_{i+\ell}-\xi_i)}{1-\varepsilon-\xi_i}+o_p(1)} \bone\{\xi_{i+\ell}\le \xi_t<1-\varepsilon\} \\
&=& \sum\limits_{t=i}^{i+\ell-1} n \delta_t \lrp{\frac{(1-\varepsilon-\xi_{i+\ell})(\xi_t-\xi_i)}{1-\varepsilon-\xi_i}+o_p(1)} + \sum\limits_{t=i+\ell}^{M} n \delta_t \lrp{\frac{(1-\varepsilon-\xi_t)(\xi_{i+\ell}-\xi_i)}{1-\varepsilon-\xi_i}+o_p(1)},
\EEqn
which completes the proof.
\end{Proof}

\begin{lemma}\label{Lemma:Expression_Tf}
For any $1\le i\le M$ and $\ell\geq1$ satisfying that $i+\ell\le M$, it holds that 
\BEqn
& & \frac{m_1 \sqrt{\lrfl{n\xi_{i+\ell}}-m}}{N_n} T_n^f(1,\lrfl{n\xi_i}-m,\lrfl{n\xi_{i+\ell}}-m) \\
&=& \frac{1}{N_n} \lrp{\tilde{W}_n(\xi_i) - \lrp{\frac{\xi_i-\varepsilon}{\xi_{i+\ell}-\varepsilon}+o_p(1)} \tilde{W}_n(\xi_{i+\ell})} \\
& & - \frac{n}{N_n} \sum\limits_{u=1}^{m_1} \lrp{\tilde{X}_u - \tilde{X}_{n+1-u}}^{\top}
    \left(
      \sum\limits_{t=0}^{i-1} \delta_t \lrp{\frac{(\xi_{i+\ell}-\xi_i)(\xi_t-\varepsilon)}{\xi_{i+\ell}-\varepsilon}+o_p(1)} 
    \right. \\
& & \hspace{13em}    
    \left.
    + \sum\limits_{t=i}^{i+\ell-1} \delta_t \lrp{\frac{(\xi_{i+\ell}-\xi_t)(\xi_i-\varepsilon)}{\xi_{i+\ell}-\varepsilon}+o_p(1)}
    \right) \\
& & - \frac{m_1}{N_n} \delta^{\top} \lrp{\sum\limits_{j=1}^{\lrfl{n\xi_i}-m} \tilde{X}_{j+m} - \lrp{\frac{\xi_i-\varepsilon}{\xi_{i+\ell}-\varepsilon}+o_p(1)}  \sum\limits_{j=1}^{\lrfl{n\xi_{i+\ell}}-m} \tilde{X}_{j+m}} \\
& & + \frac{m_1 n}{N_n} \lrp{\sum\limits_{t=0}^{i-1} \delta^{\top} \delta_t \lrp{\frac{(\xi_{i+\ell}-\xi_i)(\xi_t-\varepsilon)}{\xi_{i+\ell}-\varepsilon}+o_p(1)} + \sum\limits_{t=i}^{i+\ell-1} \delta^{\top} \delta_t \lrp{\frac{(\xi_{i+\ell}-\xi_t)(\xi_i-\varepsilon)}{\xi_{i+\ell}-\varepsilon}+o_p(1)}}.
\EEqn
\end{lemma}

\begin{Proof}
It follows from the definition of $T_n^f(1,\ell_1,\ell_2)$ and Lemma \ref{Lemma:Expression_aux-2} that, when $\ell_1 = \lrfl{n\xi_i}-m$ and $\ell_2 = \lrfl{n\xi_{i+\ell}}-m$, we have that
\BEqn
& & \frac{m_1 \sqrt{\lrfl{n\xi_{i+\ell}}-m}}{N_n} T_n^f(1,\lrfl{n\xi_i}-m,\lrfl{n\xi_{i+\ell}}-m) \\
&=& \frac{1}{N_n} \lrp{W_n(\xi_i) - \frac{\lrfl{n\xi_i}-m}{\lrfl{n\xi_{i+\ell}}-m} W_n(\xi_{i+\ell})} \\
&=& \frac{1}{N_n} \lrp{\tilde{W}_n(\xi_i) - \frac{\lrfl{n\xi_i}-m}{\lrfl{n\xi_{i+\ell}}-m} \tilde{W}_n(\xi_{i+\ell})} \\
& & + \frac{1}{N_n} \sum\limits_{u=1}^{m_1} \lrp{\tilde{X}_u - \tilde{X}_{n+1-u}}^{\top} \lrp{\sum\limits_{j=1}^{\lrfl{n\xi_i}-m} \mu_{j+m} - \frac{\lrfl{n\xi_i}-m}{\lrfl{n\xi_{i+\ell}}-m} \sum\limits_{j=1}^{\lrfl{n\xi_{i+\ell}}-m} \mu_{j+m}} \\
& & - \frac{m_1}{N_n} \delta^{\top} \lrp{\sum\limits_{j=1}^{\lrfl{n\xi_i}-m} \tilde{X}_{j+m} - \frac{\lrfl{n\xi_i}-m}{\lrfl{n\xi_{i+\ell}}-m} \sum\limits_{j=1}^{\lrfl{n\xi_{i+\ell}}-m} \tilde{X}_{j+m}} \\
& & - \frac{m_1}{N_n} \delta^{\top} \lrp{\sum\limits_{j=1}^{\lrfl{n\xi_i}-m} \mu_{j+m} - \frac{\lrfl{n\xi_i}-m}{\lrfl{n\xi_{i+\ell}}-m} \sum\limits_{j=1}^{\lrfl{n\xi_{i+\ell}}-m} \mu_{j+m}} \\
&=& \frac{1}{N_n} \lrp{\tilde{W}_n(\xi_i) - \lrp{\frac{\xi_i-\varepsilon}{\xi_{i+\ell}-\varepsilon}+o_p(1)} \tilde{W}_n(\xi_{i+\ell})} \\
& & - \frac{n}{N_n} \sum\limits_{u=1}^{m_1} \lrp{\tilde{X}_u - \tilde{X}_{n+1-u}}^{\top}
\left(
  \sum\limits_{t=0}^{i-1} \delta_t \lrp{\frac{(\xi_{i+\ell}-\xi_i)(\xi_t-\varepsilon)}{\xi_{i+\ell}-\varepsilon}+o_p(1)}
\right. \\
& & \hspace{13em}    
    \left. 
      + \sum\limits_{t=i}^{i+\ell-1} \delta_t   \lrp{\frac{(\xi_{i+\ell}-\xi_t)(\xi_i-\varepsilon)}{\xi_{i+\ell}-\varepsilon}+o_p(1)}
    \right) \\
& & - \frac{m_1}{N_n} \delta^{\top} \lrp{\sum\limits_{j=1}^{\lrfl{n\xi_i}-m} \tilde{X}_{j+m} - \lrp{\frac{\xi_i-\varepsilon}{\xi_{i+\ell}-\varepsilon}+o_p(1)}  \sum\limits_{j=1}^{\lrfl{n\xi_{i+\ell}}-m} \tilde{X}_{j+m}} \\
& & + \frac{m_1 n}{N_n} \lrp{\sum\limits_{t=0}^{i-1} \delta^{\top} \delta_t \lrp{\frac{(\xi_{i+\ell}-\xi_i)(\xi_t-\varepsilon)}{\xi_{i+\ell}-\varepsilon}+o_p(1)} + \sum\limits_{t=i}^{i+\ell-1} \delta^{\top} \delta_t \lrp{\frac{(\xi_{i+\ell}-\xi_t)(\xi_i-\varepsilon)}{\xi_{i+\ell}-\varepsilon}+o_p(1)}}.
\EEqn
\end{Proof}

\begin{lemma}\label{Lemma:Expression_Tb}
For any $1\le i\le M$ and $\ell\geq1$ satisfying that $i+\ell\le M$, it holds that 
\BEqn
& & \frac{m_1 \sqrt{N-\lrfl{n\xi_i}+m+1}}{N_n} T_n^b(\lrfl{n\xi_i}-m,\lrfl{n\xi_{i+\ell}}-m,N) \\
&=& \frac{1}{N_n} \lrp{\tilde{W}_n(1-\varepsilon) - \tilde{W}_n(\xi_{i+\ell}) - \lrp{\frac{1-\varepsilon-\xi_{i+\ell}}{1-\varepsilon-\xi_i}+o_p(1)} (\tilde{W}_n(1-\varepsilon) - \tilde{W}_n(\xi_i))} \\
& & + \frac{n}{N_n} \sum\limits_{u=1}^{m_1} \lrp{\tilde{X}_u - \tilde{X}_{n+1-u}}^{\top} \\
& & \hspace{2em}
    \times
    \lrp{\sum\limits_{t=i}^{i+\ell-1} \delta_t \lrp{\frac{(1-\varepsilon-\xi_{i+\ell})(\xi_t-\xi_i)}{1-\varepsilon-\xi_i}+o_p(1)} + \sum\limits_{t=i+\ell}^{M} \delta_t \lrp{\frac{(1-\varepsilon-\xi_t)(\xi_{i+\ell}-\xi_i)}{1-\varepsilon-\xi_i}+o_p(1)}} \\
& & - \frac{m_1}{N_n} \delta^{\top} \lrp{\sum\limits_{j=\lrfl{n\xi_{i+\ell}}-m}^{N} \tilde{X}_{j+m} -  \lrp{\frac{1-\varepsilon-\xi_{i+\ell}}{1-\varepsilon-\xi_i}+o_p(1)} \sum\limits_{j=\lrfl{n\xi_i}-m}^{N} \tilde{X}_{j+m}} \\
& & - \frac{m_1 n}{N_n} \lrp{\sum\limits_{t=i}^{i+\ell-1} \delta^{\top} \delta_t \lrp{\frac{(1-\varepsilon-\xi_{i+\ell})(\xi_t-\xi_i)}{1-\varepsilon-\xi_i}+o_p(1)} + \sum\limits_{t=i+\ell}^{M} \delta^{\top} \delta_t \lrp{\frac{(1-\varepsilon-\xi_t)(\xi_{i+\ell}-\xi_i)}{1-\varepsilon-\xi_i}+o_p(1)}}.
\EEqn
\end{lemma}

\begin{Proof}
Let $\ell_1=\lrfl{n\xi_i}-m$, $\ell_2=\lrfl{n\xi_{i+\ell}}-m$, then it follows from the expression of $T_n^b(\ell_1,\ell_2,N)$ and the resutls of Lemma \ref{Lemma:Expression_aux-3} that
\BEqn
& & \frac{m_1 \sqrt{N-\lrfl{n\xi_i}+m+1}}{N_n} T_n^{b}(\lrfl{n\xi_i}-m,\lrfl{n\xi_{i+\ell}}-m,N) \\
&=& \frac{1}{N_n} \lrp{W_n(1-\varepsilon) - W_n(\xi_{i+\ell}) - \frac{N-\lrfl{n\xi_{i+\ell}}+m+1}{N-\lrfl{n\xi_i}+m+1} (W_n(1-\varepsilon) - W_n(\xi_i))} \\
&=& \frac{1}{N_n} \lrp{\tilde{W}_n(1-\varepsilon) - \tilde{W}_n(\xi_{i+\ell}) - \frac{N-\lrfl{n\xi_{i+\ell}}+m+1}{N-\lrfl{n\xi_i}+m+1} (\tilde{W}_n(1-\varepsilon) - \tilde{W}_n(\xi_i))} \\
& & + \frac{1}{N_n} \sum\limits_{u=1}^{m_1} \lrp{\tilde{X}_u - \tilde{X}_{n+1-u}}^{\top} \lrp{\sum\limits_{j=\lrfl{n\xi_{i+\ell}}-m}^{N} \mu_{j+m} - \frac{N-\lrfl{n\xi_{i+\ell}}+m+1}{N-\lrfl{n\xi_i}+m+1} \sum\limits_{j=\lrfl{n\xi_i}-m}^{N} \mu_{j+m}} \\
& & - \frac{m_1}{N_n} \delta^{\top} \lrp{\sum\limits_{j=\lrfl{n\xi_{i+\ell}}-m}^{N} \tilde{X}_{j+m} - \frac{N-\lrfl{n\xi_{i+\ell}}+m+1}{N-\lrfl{n\xi_i}+m+1} \sum\limits_{j=\lrfl{n\xi_i}-m}^{N} \tilde{X}_{j+m}} \\
& & - \frac{m_1}{N_n} \delta^{\top} \lrp{\sum\limits_{j=\lrfl{n\xi_{i+\ell}}-m}^{N} \mu_{j+m} -  \frac{N-\lrfl{n\xi_{i+\ell}}+m+1}{N-\lrfl{n\xi_i}+m+1} \sum\limits_{j=\lrfl{n\xi_i}-m}^{N} \mu_{j+m}} \\
&=& \frac{1}{N_n} \lrp{\tilde{W}_n(1-\varepsilon) - \tilde{W}_n(\xi_{i+\ell}) - \lrp{\frac{1-\varepsilon-\xi_{i+\ell}}{1-\varepsilon-\xi_i}+o_p(1)} (\tilde{W}_n(1-\varepsilon) - \tilde{W}_n(\xi_i))} \\
& & + \frac{n}{N_n} \sum\limits_{u=1}^{m_1} \lrp{\tilde{X}_u - \tilde{X}_{n+1-u}}^{\top} \\
& & \hspace{2em}
    \times
    \lrp{\sum\limits_{t=i}^{i+\ell-1} \delta_t \lrp{\frac{(1-\varepsilon-\xi_{i+\ell})(\xi_t-\xi_i)}{1-\varepsilon-\xi_i}+o_p(1)} + \sum\limits_{t=i+\ell}^{M} \delta_t \lrp{\frac{(1-\varepsilon-\xi_t)(\xi_{i+\ell}-\xi_i)}{1-\varepsilon-\xi_i}+o_p(1)}} \\
& & - \frac{m_1}{N_n} \delta^{\top} \lrp{\sum\limits_{j=\lrfl{n\xi_{i+\ell}}-m}^{N} \tilde{X}_{j+m} -  \lrp{\frac{1-\varepsilon-\xi_{i+\ell}}{1-\varepsilon-\xi_i}+o_p(1)} \sum\limits_{j=\lrfl{n\xi_i}-m}^{N} \tilde{X}_{j+m}} \\
& & - \frac{m_1 n}{N_n} \lrp{\sum\limits_{t=i}^{i+\ell-1} \delta^{\top} \delta_t \lrp{\frac{(1-\varepsilon-\xi_{i+\ell})(\xi_t-\xi_i)}{1-\varepsilon-\xi_i}+o_p(1)} + \sum\limits_{t=i+\ell}^{M} \delta^{\top} \delta_t \lrp{\frac{(1-\varepsilon-\xi_t)(\xi_{i+\ell}-\xi_i)}{1-\varepsilon-\xi_i}+o_p(1)}},
\EEqn
which arrives at the desired result.
\end{Proof}

\begin{lemma}\label{Lemma:Expression_Vf}
For any $1\le i\le M$ and $\ell\geq1$ satisfying that $i+\ell\le M$, it holds that 
\BEqn
& & \frac{m_1^2 (\lrfl{n\xi_{i+\ell}}-m)}{N_n^2} V_n^f(1,\lrfl{n\xi_i}-m,\lrfl{n\xi_{i+\ell}}-m) \\
&=& \lrp{\frac{1}{\xi_{i+\ell}-\varepsilon}+o_p(1)}
    \sum\limits_{h=1}^{i} \int_{\xi_{h-1}}^{\xi_h} 
    \left(
      \frac{1}{N_n} \lrp{\tilde{W}_n(s) - \lrp{\frac{s-\varepsilon}{\xi_i-\varepsilon}+o_p(1)} \tilde{W}_n(\xi_i)} 
    \right. \\
& & \hspace{11em}    
      - \frac{n}{N_n} \sum\limits_{u=1}^{m_1} \lrp{\tilde{X}_u - \tilde{X}_{n+1-u}}^{\top} 
      \left(
        \sum\limits_{t=0}^{h-1} \delta_t \lrp{\frac{(\xi_i-s)(\xi_t-\varepsilon)}{\xi_i-\varepsilon}+o_p(1)}
      \right. \\
& & \hspace{24em}
      \left.
        + \sum\limits_{t=h}^{i-1} \delta_t \lrp{\frac{(\xi_i-\xi_t)(s-\varepsilon)}{\xi_i-\varepsilon}+o_p(1)}
      \right) \\
& & \hspace{11em}    
      - \frac{m_1}{N_n} \delta^{\top} \lrp{\sum\limits_{j=1}^{\lrfl{ns}-m} \tilde{X}_{j+m} - \lrp{\frac{s-\varepsilon}{\xi_i-\varepsilon}+o_p(1)} \sum\limits_{j=1}^{\lrfl{n\xi_i}-m} \tilde{X}_{j+m}}
    \\ 
& & \hspace{11em}    
      + \frac{m_1 n}{N_n}
      \left(
        \sum\limits_{t=0}^{h-1} \delta^{\top} \delta_t \lrp{\frac{(\xi_i-s)(\xi_t-\varepsilon)}{\xi_i-\varepsilon}+o_p(1)} 
      \right. \\
& & \hspace{15em}      
    \left.
      \left.
        + \sum\limits_{t=h}^{i-1} \delta^{\top} \delta_t \lrp{\frac{(\xi_i-\xi_t)(s-\varepsilon)}{\xi_i-\varepsilon}+o_p(1)}
      \right)
    \right)^2 ds \\
& & + \lrp{\frac{1}{\xi_{i+\ell}-\varepsilon}+o_p(1)}    
    \sum\limits_{h=i+1}^{i+\ell} \int_{\xi_{h-1}}^{\xi_h}
    \left(
      \frac{1}{N_n} \lrp{\tilde{W}_n(\xi_{i+\ell}) - \tilde{W}_n(s) - \lrp{\frac{\xi_{i+\ell}-s}{\xi_{i+\ell}-\xi_i}+o_p(1)} \lrp{\tilde{W}_n(\xi_{i+\ell})-\tilde{W}_n(\xi_i)}}
    \right. \\
& & \hspace{11em}    
      + \frac{n}{N_n} \sum\limits_{u=1}^{m_1} \lrp{\tilde{X}_u - \tilde{X}_{n+1-u}}^{\top} 
      \left(
        \sum\limits_{t=i}^{h-1} \delta_t \lrp{\frac{(\xi_{i+\ell}-s)(\xi_t-\xi_i)}{\xi_{i+\ell}-\xi_i}+o_p(1)}
      \right. \\
& & \hspace{24em}
      \left.
      + \sum\limits_{t=h}^{i+\ell-1} \delta_t \lrp{\frac{(\xi_{i+\ell}-\xi_t)(s-\xi_i)}{\xi_{i+\ell}-\xi_i}+o_p(1)}
      \right) \\
& & \hspace{11em}    
      - \frac{m_1}{N_n} \delta^{\top} \lrp{\sum\limits_{j=\lrfl{ns}-m+1}^{\lrfl{n\xi_{i+\ell}}-m} \tilde{X}_{j+m} - \lrp{\frac{\xi_{i+\ell}-s}{\xi_{i+\ell}-\xi_i}+o_p(1)} \sum\limits_{j=\lrfl{n\xi_i}-m+1}^{\lrfl{n\xi_{i+\ell}}-m} \tilde{X}_{j+m}} \\
    \\
& & \hspace{11em}
      - \frac{m_1 n}{N_n}
      \left(
        \sum\limits_{t=i}^{h-1} \delta^{\top} \delta_t \lrp{\frac{(\xi_{i+\ell}-s)(\xi_t-\xi_i)}{\xi_{i+\ell}-\xi_i}+o_p(1)} 
      \right. \\
& & \hspace{15em}
    \left.
      \left.
        + \sum\limits_{t=h}^{i+\ell-1} \delta^{\top} \delta_t \lrp{\frac{(\xi_{i+\ell}-\xi_t)(s-\xi_i)}{\xi_{i+\ell}-\xi_i}+o_p(1)}
      \right)
    \right)^2 ds \\
& & + o_p(1).    
\EEqn
\end{lemma}

\begin{Proof}
For any $\varepsilon\le r_1\le r_2\le 1-\varepsilon$, with $\ell_1=\lrfl{nr_1}-m$ and $\ell_2=\lrfl{nr_2}-m$, we observe that
\BEqn
& & \frac{m_1^2 \ell_2}{N_n^2} V_n^f(1,\ell_1,\ell_2) \\
&=& \frac{n}{(\lrfl{nr_2}-m) N_n^2} \int_{\varepsilon}^{r_1} \lrp{W_n(s) - \frac{\lrfl{ns}-m}{\lrfl{nr_1}-m} W_n(r_1)}^2 ds \\
& & + \frac{n}{(\lrfl{nr_2}-m) N_n^2} \int_{r_1}^{r_2} \lrp{W_n(r_2) - W_n(s) - \frac{\lrfl{nr_2}-\lrfl{ns}}{\lrfl{nr_2}-\lrfl{nr_1}} \lrp{W_n(r_2)-W_n(r_1)}}^2 ds + o_p(1),
\EEqn
where
\BEqn
& & \frac{1}{N_n} \lrp{W_n(s) - \frac{\lrfl{ns}-m}{\lrfl{nr_1}-m} W_n(r_1)} \\
&=& \frac{1}{N_n} \lrp{\tilde{W}_n(s) - \frac{\lrfl{ns}-m}{\lrfl{nr_1}-m} \tilde{W}_n(r_1)} \\
& & + \frac{1}{N_n} \sum\limits_{i=1}^{m_1} \lrp{\tilde{X}_i - \tilde{X}_{n+1-i}}^{\top} \lrp{\sum\limits_{j=1}^{\lrfl{ns}-m} \mu_{j+m} - \frac{\lrfl{ns}-m}{\lrfl{nr_1}-m} \sum\limits_{j=1}^{\lrfl{nr_1}-m} \mu_{j+m}} \\
& & - \frac{m_1}{N_n} \delta^{\top} \lrp{\sum\limits_{j=1}^{\lrfl{ns}-m} \tilde{X}_{j+m} - \frac{\lrfl{ns}-m}{\lrfl{nr_1}-m} \sum\limits_{j=1}^{\lrfl{nr_1}-m} \tilde{X}_{j+m}} \\
& & - \frac{m_1}{N_n} \delta^{\top} \lrp{\sum\limits_{j=1}^{\lrfl{ns}-m} \mu_{j+m} - \frac{\lrfl{ns}-m}{\lrfl{nr_1}-m} \sum\limits_{j=1}^{\lrfl{nr_1}-m} \mu_{j+m}},
\EEqn
and
\BEqn
& & \frac{1}{N_n} \lrp{W_n(r_2) - W_n(s) - \frac{\lrfl{nr_2}-\lrfl{ns}}{\lrfl{nr_2}-\lrfl{nr_1}} \lrp{W_n(r_2)-W_n(r_1)}} \\
&=& \frac{1}{N_n} \lrp{\tilde{W}_n(r_2) - \tilde{W}_n(s) - \frac{\lrfl{nr_2}-\lrfl{ns}}{\lrfl{nr_2}-\lrfl{nr_1}} \lrp{\tilde{W}_n(r_2)-\tilde{W}_n(r_1)}} \\
& & + \frac{1}{N_n} \sum\limits_{i=1}^{m_1} \lrp{\tilde{X}_i - \tilde{X}_{n+1-i}}^{\top} \lrp{\sum\limits_{j=\lrfl{ns}-m+1}^{\lrfl{nr_2}-m} \mu_{j+m} - \frac{\lrfl{nr_2}-\lrfl{ns}}{\lrfl{nr_2}-\lrfl{nr_1}} \sum\limits_{j=\lrfl{nr_1}-m+1}^{\lrfl{nr_2}-m} \mu_{j+m}} \\
& & - \frac{m_1}{N_n} \delta^{\top} \lrp{\sum\limits_{j=\lrfl{ns}-m+1}^{\lrfl{nr_2}-m} \tilde{X}_{j+m} - \frac{\lrfl{nr_2}-\lrfl{ns}}{\lrfl{nr_2}-\lrfl{nr_1}} \sum\limits_{j=\lrfl{nr_1}-m+1}^{\lrfl{nr_2}-m} \tilde{X}_{j+m}} \\
& & - \frac{m_1}{N_n} \delta^{\top} \lrp{\sum\limits_{j=\lrfl{ns}-m+1}^{\lrfl{nr_2}-m} \mu_{j+m} - \frac{\lrfl{nr_2}-\lrfl{ns}}{\lrfl{nr_2}-\lrfl{nr_1}} \sum\limits_{j=\lrfl{nr_1}-m+1}^{\lrfl{nr_2}-m} \mu_{j+m}}.
\EEqn

Now consider the case when $r_1=\xi_i$ and $r_2=\xi_{i+\ell}$, it follows from Lemma \ref{Lemma:Expression_aux-1} that
\BEqn
& & \sum\limits_{j=1}^{\lrfl{ns}-m} \mu_{j+m} - \frac{\lrfl{ns}-m}{\lrfl{n\xi_i}-m} \sum\limits_{j=1}^{\lrfl{n\xi_i}-m} \mu_{j+m} \\
&=& - \sum\limits_{t=0}^{M} n \delta_t \lrp{\lrp{\frac{(\xi_i-s)(\xi_t-\varepsilon)}{\xi_i-\varepsilon}+o_p(1)} \bone\{s>\xi_t\} + \lrp{\frac{(\xi_i-\xi_t)(s-\varepsilon)}{\xi_i-\varepsilon}+o_p(1)} \bone\{s\le \xi_t<\xi_i\}},
\EEqn
and consequently, 
\BEqn
& & \frac{1}{N_n^2} \int_{\varepsilon}^{\xi_i} \lrp{W_n(s) - \frac{\lrfl{ns}-m}{\lrfl{n\xi_i}-m} W_n(\xi_i)}^2 ds \\
&=& \frac{1}{N_n^2} \sum\limits_{h=1}^{i} \int_{\xi_{h-1}}^{\xi_h} \lrp{W_n(s) - \frac{\lrfl{ns}-m}{\lrfl{n\xi_i}-m} W_n(\xi_i)}^2 ds \\
&=& \sum\limits_{h=1}^{i} \int_{\xi_{h-1}}^{\xi_h} 
    \left(
      \frac{1}{N_n} \lrp{\tilde{W}_n(s) - \lrp{\frac{s-\varepsilon}{\xi_i-\varepsilon}+o_p(1)} \tilde{W}_n(\xi_i)} 
    \right. \\
& & \hspace{5em}    
      - \frac{n}{N_n} \sum\limits_{u=1}^{m_1} \lrp{\tilde{X}_u - \tilde{X}_{n+1-u}}^{\top} 
      \left(
        \sum\limits_{t=0}^{M} \delta_t \lrp{\frac{(\xi_i-s)(\xi_t-\varepsilon)}{\xi_i-\varepsilon}+o_p(1)} \bone\{s>\xi_t\} 
      \right. \\
& & \hspace{18em}      
      \left.
        + \sum\limits_{t=0}^{M} \delta_t \lrp{\frac{(\xi_i-\xi_t)(s-\varepsilon)}{\xi_i-\varepsilon}+o_p(1)} \bone\{s\le \xi_t<\xi_i\}
      \right) \\
& & \hspace{5em}    
      - \frac{m_1}{N_n} \delta^{\top} \lrp{\sum\limits_{j=1}^{\lrfl{ns}-m} \tilde{X}_{j+m} - \lrp{\frac{s-\varepsilon}{\xi_i-\varepsilon}+o_p(1)} \sum\limits_{j=1}^{\lrfl{n\xi_i}-m} \tilde{X}_{j+m}}
    \\ 
& & \hspace{5em}    
      + \frac{m_1 n}{N_n}
      \left(
        \sum\limits_{t=0}^{M} \delta^{\top} \delta_t \lrp{\frac{(\xi_i-s)(\xi_t-\varepsilon)}{\xi_i-\varepsilon}+o_p(1)} \bone\{s>\xi_t\} 
      \right. \\
& & \hspace{9em}    
    \left.
      \left.
        + \sum\limits_{t=0}^{M} \delta^{\top} \delta_t \lrp{\frac{(\xi_i-\xi_t)(s-\varepsilon)}{\xi_i-\varepsilon}+o_p(1)} \bone\{s\le \xi_t<\xi_i\}
      \right)
    \right)^2 ds \\   
&=& \sum\limits_{h=1}^{i} \int_{\xi_{h-1}}^{\xi_h} 
    \left(
      \frac{1}{N_n} \lrp{\tilde{W}_n(s) - \lrp{\frac{s-\varepsilon}{\xi_i-\varepsilon}+o_p(1)} \tilde{W}_n(\xi_i)} 
    \right. \\
& & \hspace{1em}    
      - \frac{n}{N_n} \sum\limits_{u=1}^{m_1} \lrp{\tilde{X}_u - \tilde{X}_{n+1-u}}^{\top} 
      \lrp{
        \sum\limits_{t=0}^{h-1} \delta_t \lrp{\frac{(\xi_i-s)(\xi_t-\varepsilon)}{\xi_i-\varepsilon}+o_p(1)}
        + \sum\limits_{t=h}^{i-1} \delta_t \lrp{\frac{(\xi_i-\xi_t)(s-\varepsilon)}{\xi_i-\varepsilon}+o_p(1)}} \\
& & \hspace{1em}    
      - \frac{m_1}{N_n} \delta^{\top} \lrp{\sum\limits_{j=1}^{\lrfl{ns}-m} \tilde{X}_{j+m} - \lrp{\frac{s-\varepsilon}{\xi_i-\varepsilon}+o_p(1)} \sum\limits_{j=1}^{\lrfl{n\xi_i}-m} \tilde{X}_{j+m}}
    \\ 
& & \hspace{1em}    
    \left.
      + \frac{m_1 n}{N_n}
      \lrp{
        \sum\limits_{t=0}^{h-1} \delta^{\top} \delta_t \lrp{\frac{(\xi_i-s)(\xi_t-\varepsilon)}{\xi_i-\varepsilon}+o_p(1)}
        + \sum\limits_{t=h}^{i-1} \delta^{\top} \delta_t \lrp{\frac{(\xi_i-\xi_t)(s-\varepsilon)}{\xi_i-\varepsilon}+o_p(1)}}
    \right)^2 ds.
\EEqn

Similarly, from Lemma \ref{Lemma:Expression_aux-1} we can also show that
\BEqn
& & \sum\limits_{j=\lrfl{ns}-m+1}^{\lrfl{n\xi_{i+\ell}}-m} \mu_{j+m} - \frac{\lrfl{n\xi_{i+\ell}}-\lrfl{ns}}{\lrfl{n\xi_{i+\ell}}-\lrfl{n\xi_i}} \sum\limits_{j=\lrfl{n\xi_i}-m+1}^{\lrfl{n\xi_{i+\ell}}-m} \mu_{j+m} \\
&=& \sum\limits_{t=0}^{M} n \delta_t \lrp{\lrp{\frac{(\xi_{i+\ell}-s)(\xi_t-\xi_i)}{\xi_{i+\ell}-\xi_i}+o_p(1)}\bone\{\xi_i\le\xi_t<s\} + \lrp{\frac{(\xi_{i+\ell}-\xi_t)(s-\xi_i)}{\xi_{i+\ell}-\xi_i}+o_p(1)} \bone\{s\le \xi_t<\xi_{i+\ell}\}},
\EEqn
and it follows that
\BEqn
& & \frac{1}{N_n^2} \int_{\xi_i}^{\xi_{i+\ell}} \lrp{W_n(\xi_{i+\ell}) - W_n(s) - \frac{\lrfl{n\xi_{i+\ell}}-\lrfl{ns}}{\lrfl{n\xi_{i+\ell}}-\lrfl{n\xi_i}} \lrp{W_n(\xi_{i+\ell})-W_n(\xi_i)}}^2 ds \\
&=& \sum\limits_{h=i+1}^{i+\ell} \int_{\xi_{h-1}}^{\xi_h}
    \left(
      \frac{1}{N_n} \lrp{\tilde{W}_n(\xi_{i+\ell}) - \tilde{W}_n(s) - \lrp{\frac{\xi_{i+\ell}-s}{\xi_{i+\ell}-\xi_i}+o_p(1)} \lrp{\tilde{W}_n(\xi_{i+\ell})-\tilde{W}_n(\xi_i)}}
    \right. \\
& & \hspace{2em}    
      + \frac{n}{N_n} \sum\limits_{u=1}^{m_1} \lrp{\tilde{X}_u - \tilde{X}_{n+1-u}}^{\top} 
      \left(
        \sum\limits_{t=i}^{h-1} \delta_t \lrp{\frac{(\xi_{i+\ell}-s)(\xi_t-\xi_i)}{\xi_{i+\ell}-\xi_i}+o_p(1)}
      \right. \\
& & \hspace{15em}
      \left.
      + \sum\limits_{t=h}^{i+\ell-1} \delta_t \lrp{\frac{(\xi_{i+\ell}-\xi_t)(s-\xi_i)}{\xi_{i+\ell}-\xi_i}+o_p(1)}
      \right) \\
& & \hspace{2em}    
      - \frac{m_1}{N_n} \delta^{\top} \lrp{\sum\limits_{j=\lrfl{ns}-m+1}^{\lrfl{n\xi_{i+\ell}}-m} \tilde{X}_{j+m} - \lrp{\frac{\xi_{i+\ell}-s}{\xi_{i+\ell}-\xi_i}+o_p(1)} \sum\limits_{j=\lrfl{n\xi_i}-m+1}^{\lrfl{n\xi_{i+\ell}}-m} \tilde{X}_{j+m}} \\
    \\
& & \hspace{2em}
    \left.
      - \frac{m_1 n}{N_n} \lrp{\sum\limits_{t=i}^{h-1} \delta^{\top} \delta_t \lrp{\frac{(\xi_{i+\ell}-s)(\xi_t-\xi_i)}{\xi_{i+\ell}-\xi_i}+o_p(1)} + \sum\limits_{t=h}^{i+\ell-1} \delta^{\top} \delta_t \lrp{\frac{(\xi_{i+\ell}-\xi_t)(s-\xi_i)}{\xi_{i+\ell}-\xi_i}+o_p(1)}}
    \right)^2 ds
\EEqn

In summary, we obtain that
\BEqn
& & \frac{m_1^2 (\lrfl{n\xi_{i+\ell}}-m)}{N_n^2} V_n^f(1,\lrfl{n\xi_i}-m,\lrfl{n\xi_{i+\ell}}-m) \\
&=& \lrp{\frac{1}{\xi_{i+\ell}-\varepsilon}+o_p(1)}
    \sum\limits_{h=1}^{i} \int_{\xi_{h-1}}^{\xi_h} 
    \left(
      \frac{1}{N_n} \lrp{\tilde{W}_n(s) - \lrp{\frac{s-\varepsilon}{\xi_i-\varepsilon}+o_p(1)} \tilde{W}_n(\xi_i)} 
    \right. \\
& & \hspace{11em}    
      - \frac{n}{N_n} \sum\limits_{u=1}^{m_1} \lrp{\tilde{X}_u - \tilde{X}_{n+1-u}}^{\top} 
      \left(
        \sum\limits_{t=0}^{h-1} \delta_t \lrp{\frac{(\xi_i-s)(\xi_t-\varepsilon)}{\xi_i-\varepsilon}+o_p(1)}
      \right. \\
& & \hspace{24em}
      \left.
        + \sum\limits_{t=h}^{i-1} \delta_t \lrp{\frac{(\xi_i-\xi_t)(s-\varepsilon)}{\xi_i-\varepsilon}+o_p(1)}
      \right) \\
& & \hspace{11em}    
      - \frac{m_1}{N_n} \delta^{\top} \lrp{\sum\limits_{j=1}^{\lrfl{ns}-m} \tilde{X}_{j+m} - \lrp{\frac{s-\varepsilon}{\xi_i-\varepsilon}+o_p(1)} \sum\limits_{j=1}^{\lrfl{n\xi_i}-m} \tilde{X}_{j+m}}
    \\ 
& & \hspace{11em}    
      + \frac{m_1 n}{N_n}
      \left(
        \sum\limits_{t=0}^{h-1} \delta^{\top} \delta_t \lrp{\frac{(\xi_i-s)(\xi_t-\varepsilon)}{\xi_i-\varepsilon}+o_p(1)} 
      \right. \\
& & \hspace{15em}      
    \left.
      \left.
        + \sum\limits_{t=h}^{i-1} \delta^{\top} \delta_t \lrp{\frac{(\xi_i-\xi_t)(s-\varepsilon)}{\xi_i-\varepsilon}+o_p(1)}
      \right)
    \right)^2 ds \\
&+& \lrp{\frac{1}{\xi_{i+\ell}-\varepsilon}+o_p(1)}    
    \sum\limits_{h=i+1}^{i+\ell} \int_{\xi_{h-1}}^{\xi_h}
    \left(
      \frac{1}{N_n} \lrp{\tilde{W}_n(\xi_{i+\ell}) - \tilde{W}_n(s) - \lrp{\frac{\xi_{i+\ell}-s}{\xi_{i+\ell}-\xi_i}+o_p(1)} \lrp{\tilde{W}_n(\xi_{i+\ell})-\tilde{W}_n(\xi_i)}}
    \right. \\
& & \hspace{11em}    
      + \frac{n}{N_n} \sum\limits_{u=1}^{m_1} \lrp{\tilde{X}_u - \tilde{X}_{n+1-u}}^{\top} 
      \left(
        \sum\limits_{t=i}^{h-1} \delta_t \lrp{\frac{(\xi_{i+\ell}-s)(\xi_t-\xi_i)}{\xi_{i+\ell}-\xi_i}+o_p(1)}
      \right. \\
& & \hspace{24em}
      \left.
      + \sum\limits_{t=h}^{i+\ell-1} \delta_t \lrp{\frac{(\xi_{i+\ell}-\xi_t)(s-\xi_i)}{\xi_{i+\ell}-\xi_i}+o_p(1)}
      \right) \\
& & \hspace{11em}    
      - \frac{m_1}{N_n} \delta^{\top} \lrp{\sum\limits_{j=\lrfl{ns}-m+1}^{\lrfl{n\xi_{i+\ell}}-m} \tilde{X}_{j+m} - \lrp{\frac{\xi_{i+\ell}-s}{\xi_{i+\ell}-\xi_i}+o_p(1)} \sum\limits_{j=\lrfl{n\xi_i}-m+1}^{\lrfl{n\xi_{i+\ell}}-m} \tilde{X}_{j+m}} \\
    \\
& & \hspace{11em}
      - \frac{m_1 n}{N_n}
      \left(
        \sum\limits_{t=i}^{h-1} \delta^{\top} \delta_t \lrp{\frac{(\xi_{i+\ell}-s)(\xi_t-\xi_i)}{\xi_{i+\ell}-\xi_i}+o_p(1)} 
      \right. \\
& & \hspace{15em}
    \left.
      \left.
        + \sum\limits_{t=h}^{i+\ell-1} \delta^{\top} \delta_t \lrp{\frac{(\xi_{i+\ell}-\xi_t)(s-\xi_i)}{\xi_{i+\ell}-\xi_i}+o_p(1)}
      \right)
    \right)^2 ds \\
& & + o_p(1),
\EEqn
which completes the proof.
\end{Proof}

\begin{lemma}\label{Lemma:Expression_Vb}
For any $1\le i\le M$ and $\ell\geq1$ satisfying that $i+\ell\le M$, it holds that 
\BEqn
& & \frac{m_1^2 (N-\lrfl{n\xi_i}+m+1)}{N_n^2} V_n^b(\lrfl{n\xi_i}-m,\lrfl{n\xi_{i+\ell}}-m,N) \\
&=& \lrp{\frac{1}{1-\varepsilon-\xi_i}+o_p(1)} \\
& & \times
    \sum\limits_{h=i+1}^{i+\ell} \int_{\xi_{h-1}}^{\xi_h} 
    \left(
      \frac{1}{N_n} \lrp{\tilde{W}_n(s) - \tilde{W}_n(\xi_i) - \lrp{\frac{s-\xi_i}{\xi_{i+\ell}-\xi_i}+o_p(1)} \lrp{\tilde{W}_n(\xi_{i+\ell}) - \tilde{W}_n(\xi_i)}}
    \right. \\
& & \hspace{7em}
      - \frac{n}{N_n} \sum\limits_{u=1}^{m_1} \lrp{\tilde{X}_u-\tilde{X}_{n+1-u}}^{\top}
      \left(
        \sum\limits_{t=i}^{h-1} \delta_t \lrp{\frac{(\xi_{i+\ell}-s)(\xi_t-\xi_i)}{\xi_{i+\ell}-\xi_i}+o_p(1)}
      \right. \\
& & \hspace{20em}      
      \left.
        + \sum\limits_{t=h}^{i+\ell-1} \delta_t \lrp{\frac{(\xi_{i+\ell}-\xi_t)(s-\xi_i)}{\xi_{i+\ell}-\xi_i}+o_p(1)}
      \right) \\
& & \hspace{7em}
      - \frac{m_1}{N_n} \delta^{\top}
      \lrp{\sum\limits_{j=\lrfl{n\xi_i}-m}^{\lrfl{ns}-m} \tilde{X}_{j+m} - \lrp{\frac{s-\xi_i}{\xi_{i+\ell}-\xi_i}+o_p(1)} \sum\limits_{j=\lrfl{n\xi_i}-m}^{\lrfl{n\xi_{i+\ell}}-m-1} \tilde{X}_{j+m}} \\
& & \hspace{7em}
      + \frac{m_1 n}{N_n} 
      \left(
        \sum\limits_{t=i}^{h-1} \delta^{\top} \delta_t \lrp{\frac{(\xi_{i+\ell}-s)(\xi_t-\xi_i)}{\xi_{i+\ell}-\xi_i}+o_p(1)} 
      \right. \\
& & \hspace{11em}      
    \left.
      \left.
        + \sum\limits_{t=h}^{i+\ell-1} \delta^{\top} \delta_t \lrp{\frac{(\xi_{i+\ell}-\xi_t)(s-\xi_i)}{\xi_{i+\ell}-\xi_i}+o_p(1)} 
      \right)
    \right)^2 ds \\
& & + \lrp{\frac{1}{1-\varepsilon-\xi_i}+o_p(1)} \\
& & \times
    \sum\limits_{h=i+\ell+1}^{M+1} \int_{\xi_{h-1}}^{\xi_h} 
    \left(
      \frac{1}{N_n} \lrp{\tilde{W}_n(1-\varepsilon) - \tilde{W}_n(s) - \lrp{\frac{1-\varepsilon-s}{1-\varepsilon-\xi_{i+\ell}}+o_p(1)} \lrp{\tilde{W}_n(1-\varepsilon)} - \tilde{W}_n(\xi_{i+\ell})}
    \right. \\
& & \hspace{8em}
      + \frac{n}{N_n} \sum\limits_{u=1}^{m_1} \lrp{\tilde{X}_u-\tilde{X}_{n+1-u}}^{\top}
      \left(
        \sum\limits_{t=i+\ell}^{h-1} \delta_t \lrp{\frac{(1-\varepsilon-s)(\xi_t-\xi_{i+\ell})}{1-\varepsilon-\xi_{i+\ell}}+o_p(1)}
      \right. \\
& & \hspace{21em}      
      \left.
        + \sum\limits_{t=h}^{M} \delta_t \lrp{\frac{(1-\varepsilon-\xi_t)(s-\xi_{i+\ell})}{1-\varepsilon-\xi_{i+\ell}}+o_p(1)}
      \right) \\
& & \hspace{8em}
      - \frac{m_1}{N_n} \delta^{\top}
      \lrp{\sum\limits_{j=\lrfl{ns}-m}^{N} \tilde{X}_{j+m} - \lrp{\frac{1-\varepsilon-s}{1-\varepsilon-\xi_{i+\ell}}+o_p(1)} \sum\limits_{j=\lrfl{n\xi_{i+\ell}}-m}^{N} \tilde{X}_{j+m}} \\
& & \hspace{8em}
      - \frac{m_1 n}{N_n} 
      \left(
        \sum\limits_{t=i+\ell}^{h-1} \delta^{\top} \delta_t \lrp{\frac{(1-\varepsilon-s)(\xi_t-\xi_{i+\ell})}{1-\varepsilon-\xi_{i+\ell}}+o_p(1)}
      \right. \\
& & \hspace{12em}      
    \left.
      \left.
        + \sum\limits_{t=h}^{M} \delta^{\top} \delta_t \lrp{\frac{(1-\varepsilon-\xi_t)(s-\xi_{i+\ell})}{1-\varepsilon-\xi_{i+\ell}}+o_p(1)}
      \right)
    \right)^2 ds \\
& & + o_p(1).    
\EEqn
\end{lemma}

\begin{Proof}
From the definition of $V_n^b(j_1,j_2,j_3)$, we have that
\BEqn
& & \frac{m_1^2 (N-\lrfl{n\xi_i}+m+1)}{N_n^2} V_n^b(\lrfl{n\xi_i}-m,\lrfl{n\xi_{i+\ell}}-m,N) \\
&=& \frac{n}{(N-\lrfl{n\xi_i}+m+1)N_n^2} \int_{\xi_i}^{\xi_{i+\ell}} \lrp{W_n(s) - W_n(\xi_i) - \frac{\lrfl{ns}-\lrfl{n\xi_i}+1}{\lrfl{n\xi_{i+\ell}}-\lrfl{n\xi_i}} \lrp{W_n(\xi_{i+\ell}) - W_n(\xi_i)}}^2 ds \\
& & + \frac{n}{(N-\lrfl{n\xi_i}+m+1)N_n^2} 
      \int_{\xi_{i+\ell}}^{1-\varepsilon} 
      \Big( 
          W_n(1-\varepsilon) - W_n(s) \\
& & \hspace{14em}
    \left.      
          - \frac{N-\lrfl{ns}+m+1}{N-\lrfl{n\xi_{i+\ell}}+m+1} \lrp{W_n(1-\varepsilon) - W_n(\xi_{i+\ell})}
    \right)^2 ds.
\EEqn

Again, by using the results of Lemma \ref{Lemma:Expression_aux-1}, we have that
\BEqn
& & \sum\limits_{j=\lrfl{n\xi_i}-m}^{\lrfl{ns}-m} \mu_{j+m} - \frac{\lrfl{ns}-\lrfl{n\xi_i}+1}{\lrfl{n\xi_{i+\ell}}-\lrfl{n\xi_i}} \sum\limits_{j=\lrfl{n\xi_i}-m}^{\lrfl{n\xi_{i+\ell}}-m-1} \mu_{j+m} \\
&=& - \sum\limits_{t=0}^{M} n \delta_t
\left( 
  \lrp{\frac{(\xi_{i+\ell}-s)(\xi_t-\xi_i)}{\xi_{i+\ell}-\xi_i}+o_p(1)}\bone\{\xi_i\le\xi_t<s\} 
\right. \\
& & \hspace{5em} 
\left.
  + \lrp{\frac{(\xi_{i+\ell}-\xi_t)(s-\xi_i)}{\xi_{i+\ell}-\xi_i}+o_p(1)} \bone\{s\le \xi_t<\xi_{i+\ell}\}
\right),
\EEqn
and
\BEqn
& & \sum\limits_{j=\lrfl{ns}-m}^{N} \mu_{j+m} - \frac{N-\lrfl{ns}+m+1}{N-\lrfl{n\xi_{i+\ell}}+m+1} \sum\limits_{j=\lrfl{n\xi_{i+\ell}}-m}^{N} \mu_{j+m} \\
&=& \sum\limits_{t=0}^{M} n \delta_t
    \left(
      \lrp{\frac{(1-\varepsilon-s)(\xi_t-\xi_{i+\ell})}{1-\varepsilon-\xi_{i+\ell}}+o_p(1)}\bone\{\xi_{i+\ell}\le\xi_t<s\}
    \right. \\
& & \hspace{4em}
    \left.
      + \lrp{\frac{(1-\varepsilon-\xi_t)(s-\xi_{i+\ell})}{1-\varepsilon-\xi_{i+\ell}}+o_p(1)} \bone\{s\le \xi_t<1-\varepsilon\}
    \right).
\EEqn

Furthermore, by applying Lemma \ref{Lemma:Expression_W}, we obtain that
\BEqn
& & \frac{1}{N_n^2} \int_{\xi_i}^{\xi_{i+\ell}} \lrp{W_n(s) - W_n(\xi_i) - \frac{\lrfl{ns}-\lrfl{n\xi_i}+1}{\lrfl{n\xi_{i+\ell}}-\lrfl{n\xi_i}} \lrp{W_n(\xi_{i+\ell}) - W_n(\xi_i)}}^2 ds \\
&=& \sum\limits_{h=i+1}^{i+\ell} \int_{\xi_{h-1}}^{\xi_h} 
    \left(
      \frac{1}{N_n} \lrp{\tilde{W}_n(s) - \tilde{W}_n(\xi_i) - \lrp{\frac{s-\xi_i}{\xi_{i+\ell}-\xi_i}+o_p(1)} \lrp{\tilde{W}_n(\xi_{i+\ell}) - \tilde{W}_n(\xi_i)}}
    \right. \\
& & \hspace{6em}
      - \frac{n}{N_n} \sum\limits_{u=1}^{m_1} \lrp{\tilde{X}_u-\tilde{X}_{n+1-u}}^{\top}
      \left(
        \sum\limits_{t=0}^{M} \delta_t \lrp{\frac{(\xi_{i+\ell}-s)(\xi_t-\xi_i)}{\xi_{i+\ell}-\xi_i}+o_p(1)}\bone\{\xi_i\le\xi_t<s\} 
      \right. \\
& & \hspace{16em}      
      \left.
        + \sum\limits_{t=0}^{M} \delta_t \lrp{\frac{(\xi_{i+\ell}-\xi_t)(s-\xi_i)}{\xi_{i+\ell}-\xi_i}+o_p(1)} \bone\{s\le \xi_t<\xi_{i+\ell}\}
      \right) \\
& & \hspace{6em}
      - \frac{m_1}{N_n} \delta^{\top}
      \lrp{\sum\limits_{j=\lrfl{n\xi_i}-m}^{\lrfl{ns}-m} \tilde{X}_{j+m} - \lrp{\frac{s-\xi_i}{\xi_{i+\ell}-\xi_i}+o_p(1)} \sum\limits_{j=\lrfl{n\xi_i}-m}^{\lrfl{n\xi_{i+\ell}}-m-1} \tilde{X}_{j+m}} \\
& & \hspace{6em}
      + \frac{m_1 n}{N_n} 
      \left(
        \sum\limits_{t=0}^{M} \delta^{\top} \delta_t \lrp{\frac{(\xi_{i+\ell}-s)(\xi_t-\xi_i)}{\xi_{i+\ell}-\xi_i}+o_p(1)}\bone\{\xi_i\le\xi_t<s\} 
      \right. \\
& & \hspace{10em}      
    \left.
      \left.
        + \sum\limits_{t=0}^{M} \delta^{\top} \delta_t \lrp{\frac{(\xi_{i+\ell}-\xi_t)(s-\xi_i)}{\xi_{i+\ell}-\xi_i}+o_p(1)} \bone\{s\le \xi_t<\xi_{i+\ell}\}
      \right)
    \right)^2 ds \\
&=& \sum\limits_{h=i+1}^{i+\ell} \int_{\xi_{h-1}}^{\xi_h} 
    \left(
      \frac{1}{N_n} \lrp{\tilde{W}_n(s) - \tilde{W}_n(\xi_i) - \lrp{\frac{s-\xi_i}{\xi_{i+\ell}-\xi_i}+o_p(1)} \lrp{\tilde{W}_n(\xi_{i+\ell}) - \tilde{W}_n(\xi_i)}}
    \right. \\
& & \hspace{6em}
      - \frac{n}{N_n} \sum\limits_{u=1}^{m_1} \lrp{\tilde{X}_u-\tilde{X}_{n+1-u}}^{\top}
      \left(
        \sum\limits_{t=i}^{h-1} \delta_t \lrp{\frac{(\xi_{i+\ell}-s)(\xi_t-\xi_i)}{\xi_{i+\ell}-\xi_i}+o_p(1)}
      \right. \\
& & \hspace{19em}      
      \left.
        + \sum\limits_{t=h}^{i+\ell-1} \delta_t \lrp{\frac{(\xi_{i+\ell}-\xi_t)(s-\xi_i)}{\xi_{i+\ell}-\xi_i}+o_p(1)}
      \right) \\
& & \hspace{6em}
      - \frac{m_1}{N_n} \delta^{\top}
      \lrp{\sum\limits_{j=\lrfl{n\xi_i}-m}^{\lrfl{ns}-m} \tilde{X}_{j+m} - \lrp{\frac{s-\xi_i}{\xi_{i+\ell}-\xi_i}+o_p(1)} \sum\limits_{j=\lrfl{n\xi_i}-m}^{\lrfl{n\xi_{i+\ell}}-m-1} \tilde{X}_{j+m}} \\
& & \hspace{6em}
      + \frac{m_1 n}{N_n} 
      \left(
        \sum\limits_{t=i}^{h-1} \delta^{\top} \delta_t \lrp{\frac{(\xi_{i+\ell}-s)(\xi_t-\xi_i)}{\xi_{i+\ell}-\xi_i}+o_p(1)} 
      \right. \\
& & \hspace{10em}      
    \left.
      \left.
        + \sum\limits_{t=h}^{i+\ell-1} \delta^{\top} \delta_t \lrp{\frac{(\xi_{i+\ell}-\xi_t)(s-\xi_i)}{\xi_{i+\ell}-\xi_i}+o_p(1)} 
      \right)
    \right)^2 ds,
\EEqn
and similarly,
\BEqn
& & \frac{1}{N_n^2} \int_{\xi_{i+\ell}}^{1-\varepsilon} \lrp{W_n(1-\varepsilon) - W_n(s) - \frac{N-\lrfl{ns}+m+1}{N-\lrfl{n\xi_{i+\ell}}+m+1} \lrp{W_n(1-\varepsilon) - W_n(\xi_{i+\ell})}}^2 ds \\
&=& \sum\limits_{h=i+\ell+1}^{M+1} \int_{\xi_{h-1}}^{\xi_h} 
    \left(
      \frac{1}{N_n} \lrp{\tilde{W}_n(1-\varepsilon) - \tilde{W}_n(s) - \lrp{\frac{1-\varepsilon-s}{1-\varepsilon-\xi_{i+\ell}}+o_p(1)} \lrp{\tilde{W}_n(1-\varepsilon)} - \tilde{W}_n(\xi_{i+\ell})}
    \right. \\
& & \hspace{7em}
      + \frac{n}{N_n} \sum\limits_{u=1}^{m_1} \lrp{\tilde{X}_u-\tilde{X}_{n+1-u}}^{\top}
      \left(
        \sum\limits_{t=i+\ell}^{h-1} \delta_t \lrp{\frac{(1-\varepsilon-s)(\xi_t-\xi_{i+\ell})}{1-\varepsilon-\xi_{i+\ell}}+o_p(1)}
      \right. \\
& & \hspace{20em}      
      \left.
        + \sum\limits_{t=h}^{M} \delta_t \lrp{\frac{(1-\varepsilon-\xi_t)(s-\xi_{i+\ell})}{1-\varepsilon-\xi_{i+\ell}}+o_p(1)}
      \right) \\
& & \hspace{7em}
      - \frac{m_1}{N_n} \delta^{\top}
      \lrp{\sum\limits_{j=\lrfl{ns}-m}^{N} \tilde{X}_{j+m} - \lrp{\frac{1-\varepsilon-s}{1-\varepsilon-\xi_{i+\ell}}+o_p(1)} \sum\limits_{j=\lrfl{n\xi_{i+\ell}}-m}^{N} \tilde{X}_{j+m}} \\
& & \hspace{7em}
      - \frac{m_1 n}{N_n} 
      \left(
        \sum\limits_{t=i+\ell}^{h-1} \delta^{\top} \delta_t \lrp{\frac{(1-\varepsilon-s)(\xi_t-\xi_{i+\ell})}{1-\varepsilon-\xi_{i+\ell}}+o_p(1)}
      \right. \\
& & \hspace{11em}      
    \left.
      \left.
        + \sum\limits_{t=h}^{M} \delta^{\top} \delta_t \lrp{\frac{(1-\varepsilon-\xi_t)(s-\xi_{i+\ell})}{1-\varepsilon-\xi_{i+\ell}}+o_p(1)}
      \right)
    \right)^2 ds.
\EEqn
Finally, by combining two individual parts, we obtain the desired result.
\end{Proof}

\begin{lemma}\label{Lemma:Expression_Gf}
For any $1\le i\le M-1$, it holds that
\BEqn
& & T_n^f(1,\lrfl{n\xi_i}-m,\lrfl{n\xi_{i+1}}-m) \lrp{V_n^f(1,\lrfl{n\xi_i}-m,\lrfl{n\xi_{i+1}}-m)}^{-1/2} \\
&=& \lrp{\xi_{i+1}-\varepsilon+o_p(1)} \tilde{T}_n^f(\xi_i) \lrp{\tilde{V}_n^f(\xi_i)}^{-1/2},
\EEqn
where
\BEqn
    \tilde{T}_n^f(\xi_i)
&=& \frac{1}{N_n} \lrp{\tilde{W}_n(\xi_i) - \lrp{\frac{\xi_i-\varepsilon}{\xi_{i+1}-\varepsilon}+o_p(1)} \tilde{W}_n(\xi_{i+1})} \\
& & - \frac{n}{N_n} \sum\limits_{u=1}^{m_1} \lrp{\tilde{X}_u - \tilde{X}_{n+1-u}}^{\top} \lrp{\sum\limits_{t=1}^{i} \delta_t \lrp{\frac{(\xi_{i+1}-\xi_i)(\xi_t-\varepsilon)}{\xi_{i+1}-\varepsilon}+o_p(1)}} \\
& & - \frac{m_1}{N_n} \delta^{\top} \lrp{\sum\limits_{j=1}^{\lrfl{n\xi_i}-m} \tilde{X}_{j+m} - \lrp{\frac{\xi_i-\varepsilon}{\xi_{i+1}-\varepsilon}+o_p(1)}  \sum\limits_{j=1}^{\lrfl{n\xi_{i+1}}-m} \tilde{X}_{j+m}} \\
& & + \frac{m_1 n}{N_n} \sum\limits_{t=0}^{i} \delta^{\top} \delta_t \lrp{\frac{(\xi_{i+1}-\xi_i)(\xi_t-\varepsilon)}{\xi_{i+1}-\varepsilon}+o_p(1)},
\EEqn
and
\BEqn
    \tilde{V}_n^f(\xi_i)
&=& \sum\limits_{h=1}^{i} \int_{\xi_{h-1}}^{\xi_h} 
    \left(
      \frac{1}{N_n} \lrp{\tilde{W}_n(s) - \lrp{\frac{s-\varepsilon}{\xi_i-\varepsilon}+o_p(1)} \tilde{W}_n(\xi_i)} 
    \right. \\
& & \hspace{5em}    
      - \frac{n}{N_n} \sum\limits_{u=1}^{m_1} \lrp{\tilde{X}_u - \tilde{X}_{n+1-u}}^{\top} 
      \left(
        \bone\{h>1\} \sum\limits_{t=1}^{h-1} \delta_t \lrp{\frac{(\xi_i-s)(\xi_t-\varepsilon)}{\xi_i-\varepsilon}+o_p(1)}
      \right. \\
& & \hspace{18em}
      \left.
        + \bone\{h<i\} \sum\limits_{t=h}^{i-1} \delta_t \lrp{\frac{(\xi_i-\xi_t)(s-\varepsilon)}{\xi_i-\varepsilon}+o_p(1)}
      \right) \\
& & \hspace{5em}    
      - \frac{m_1}{N_n} \delta^{\top} \lrp{\sum\limits_{j=1}^{\lrfl{ns}-m} \tilde{X}_{j+m} - \lrp{\frac{s-\varepsilon}{\xi_i-\varepsilon}+o_p(1)} \sum\limits_{j=1}^{\lrfl{n\xi_i}-m} \tilde{X}_{j+m}}
    \\ 
& & \hspace{5em}    
      + \frac{m_1 n}{N_n}
      \left(
        \bone\{h>1\} \sum\limits_{t=1}^{h-1} \delta^{\top} \delta_t \lrp{\frac{(\xi_i-s)(\xi_t-\varepsilon)}{\xi_i-\varepsilon}+o_p(1)} 
      \right. \\
& & \hspace{9em}      
    \left.
      \left.
        + \bone\{h<i\} \sum\limits_{t=h}^{i-1} \delta^{\top} \delta_t \lrp{\frac{(\xi_i-\xi_t)(s-\varepsilon)}{\xi_i-\varepsilon}+o_p(1)}
      \right)
    \right)^2 ds \\
& & + \int_{\xi_i}^{\xi_{i+1}}
    \left(
      \frac{1}{N_n} \lrp{\tilde{W}_n(\xi_{i+1}) - \tilde{W}_n(s) - \lrp{\frac{\xi_{i+1}-s}{\xi_{i+1}-\xi_i}+o_p(1)} \lrp{\tilde{W}_n(\xi_{i+1})-\tilde{W}_n(\xi_i)}}
    \right. \\
& & \hspace{5em}    
    \left.
      - \frac{m_1}{N_n} \delta^{\top} \lrp{\sum\limits_{j=\lrfl{ns}-m+1}^{\lrfl{n\xi_{i+1}}-m} \tilde{X}_{j+m} - \lrp{\frac{\xi_{i+1}-s}{\xi_{i+1}-\xi_i}+o_p(1)} \sum\limits_{j=\lrfl{n\xi_i}-m+1}^{\lrfl{n\xi_{i+1}}-m} \tilde{X}_{j+m}}
    \right)^2 ds \\
& & + o_p(1).
\EEqn
\end{lemma}

\begin{Proof}
It follow from Lemma \ref{Lemma:Expression_Tf} that
\BEqn
& & \frac{m_1 \sqrt{\lrfl{n\xi_{i+1}}-m}}{N_n} T_n^f(1,\lrfl{n\xi_i}-m,\lrfl{n\xi_{i+1}}-m) \\
&=& \frac{1}{N_n} \lrp{\tilde{W}_n(\xi_i) - \lrp{\frac{\xi_i-\varepsilon}{\xi_{i+1}-\varepsilon}+o_p(1)} \tilde{W}_n(\xi_{i+1})} \\
& & - \frac{n}{N_n} \sum\limits_{u=1}^{m_1} \lrp{\tilde{X}_u - \tilde{X}_{n+1-u}}^{\top} \lrp{\sum\limits_{t=0}^{i-1} \delta_t \lrp{\frac{(\xi_{i+1}-\xi_i)(\xi_t-\varepsilon)}{\xi_{i+1}-\varepsilon}+o_p(1)} + \delta_i \lrp{\frac{(\xi_{i+1}-\xi_i)(\xi_i-\varepsilon)}{\xi_{i+1}-\varepsilon}+o_p(1)}} \\
& & - \frac{m_1}{N_n} \delta^{\top} \lrp{\sum\limits_{j=1}^{\lrfl{n\xi_i}-m} \tilde{X}_{j+m} - \lrp{\frac{\xi_i-\varepsilon}{\xi_{i+1}-\varepsilon}+o_p(1)}  \sum\limits_{j=1}^{\lrfl{n\xi_{i+1}}-m} \tilde{X}_{j+m}} \\
& & + \frac{m_1 n}{N_n} \lrp{\sum\limits_{t=0}^{i-1} \delta^{\top} \delta_t \lrp{\frac{(\xi_{i+1}-\xi_i)(\xi_t-\varepsilon)}{\xi_{i+1}-\varepsilon}+o_p(1)} + \delta^{\top} \delta_i \lrp{\frac{(\xi_{i+1}-\xi_i)(\xi_i-\varepsilon)}{\xi_{i+1}-\varepsilon}+o_p(1)}} \\
&=& \frac{1}{N_n} \lrp{\tilde{W}_n(\xi_i) - \lrp{\frac{\xi_i-\varepsilon}{\xi_{i+1}-\varepsilon}+o_p(1)} \tilde{W}_n(\xi_{i+1})} \\
& & - \frac{n}{N_n} \sum\limits_{u=1}^{m_1} \lrp{\tilde{X}_u - \tilde{X}_{n+1-u}}^{\top} \lrp{\sum\limits_{t=0}^{i} \delta_t \lrp{\frac{(\xi_{i+1}-\xi_i)(\xi_t-\varepsilon)}{\xi_{i+1}-\varepsilon}+o_p(1)}} \\
& & - \frac{m_1}{N_n} \delta^{\top} \lrp{\sum\limits_{j=1}^{\lrfl{n\xi_i}-m} \tilde{X}_{j+m} - \lrp{\frac{\xi_i-\varepsilon}{\xi_{i+1}-\varepsilon}+o_p(1)}  \sum\limits_{j=1}^{\lrfl{n\xi_{i+1}}-m} \tilde{X}_{j+m}} \\
& & + \frac{m_1 n}{N_n} \sum\limits_{t=0}^{i} \delta^{\top} \delta_t \lrp{\frac{(\xi_{i+1}-\xi_i)(\xi_t-\varepsilon)}{\xi_{i+1}-\varepsilon}+o_p(1)} \\
&=& \frac{1}{N_n} \lrp{\tilde{W}_n(\xi_i) - \lrp{\frac{\xi_i-\varepsilon}{\xi_{i+1}-\varepsilon}+o_p(1)} \tilde{W}_n(\xi_{i+1})} \\
& & - \frac{n}{N_n} \sum\limits_{u=1}^{m_1} \lrp{\tilde{X}_u - \tilde{X}_{n+1-u}}^{\top} \lrp{\sum\limits_{t=1}^{i} \delta_t \lrp{\frac{(\xi_{i+1}-\xi_i)(\xi_t-\varepsilon)}{\xi_{i+1}-\varepsilon}+o_p(1)}} \\
& & - \frac{m_1}{N_n} \delta^{\top} \lrp{\sum\limits_{j=1}^{\lrfl{n\xi_i}-m} \tilde{X}_{j+m} - \lrp{\frac{\xi_i-\varepsilon}{\xi_{i+1}-\varepsilon}+o_p(1)}  \sum\limits_{j=1}^{\lrfl{n\xi_{i+1}}-m} \tilde{X}_{j+m}} \\
& & + \frac{m_1 n}{N_n} \sum\limits_{t=1}^{i} \delta^{\top} \delta_t \lrp{\frac{(\xi_{i+1}-\xi_i)(\xi_t-\varepsilon)}{\xi_{i+1}-\varepsilon}+o_p(1)} \\
&=& \tilde{T}_n^{f}(\xi_i),
\EEqn
where the second to the last equality uses the observation that $\xi_0 = \varepsilon$ and drops the term with $t=0$ from the summation.

Additionally, from Lemma \ref{Lemma:Expression_Vf}, we obtain that
\BEqn
& & \frac{m_1^2 (\lrfl{n\xi_{i+1}}-m)}{N_n^2} V_n^f(1,\lrfl{n\xi_i}-m,\lrfl{n\xi_{i+1}}-m) \\
&=& \lrp{\frac{1}{\xi_{i+1}-\varepsilon}+o_p(1)}
    \sum\limits_{h=1}^{i} \int_{\xi_{h-1}}^{\xi_h} 
    \left(
      \frac{1}{N_n} \lrp{\tilde{W}_n(s) - \lrp{\frac{s-\varepsilon}{\xi_i-\varepsilon}+o_p(1)} \tilde{W}_n(\xi_i)} 
    \right. \\
& & \hspace{11em}    
      - \frac{n}{N_n} \sum\limits_{u=1}^{m_1} \lrp{\tilde{X}_u - \tilde{X}_{n+1-u}}^{\top} 
      \left(
        \bone\{h>1\} \sum\limits_{t=1}^{h-1} \delta_t \lrp{\frac{(\xi_i-s)(\xi_t-\varepsilon)}{\xi_i-\varepsilon}+o_p(1)}
      \right. \\
& & \hspace{24em}
      \left.
        + \bone\{h<i\} \sum\limits_{t=h}^{i-1} \delta_t \lrp{\frac{(\xi_i-\xi_t)(s-\varepsilon)}{\xi_i-\varepsilon}+o_p(1)}
      \right) \\
& & \hspace{11em}    
      - \frac{m_1}{N_n} \delta^{\top} \lrp{\sum\limits_{j=1}^{\lrfl{ns}-m} \tilde{X}_{j+m} - \lrp{\frac{s-\varepsilon}{\xi_i-\varepsilon}+o_p(1)} \sum\limits_{j=1}^{\lrfl{n\xi_i}-m} \tilde{X}_{j+m}}
    \\ 
& & \hspace{11em}    
      + \frac{m_1 n}{N_n}
      \left(
        \bone\{h>1\} \sum\limits_{t=1}^{h-1} \delta^{\top} \delta_t \lrp{\frac{(\xi_i-s)(\xi_t-\varepsilon)}{\xi_i-\varepsilon}+o_p(1)} 
      \right. \\
& & \hspace{15em}      
    \left.
      \left.
        + \bone\{h<i\} \sum\limits_{t=h}^{i-1} \delta^{\top} \delta_t \lrp{\frac{(\xi_i-\xi_t)(s-\varepsilon)}{\xi_i-\varepsilon}+o_p(1)}
      \right)
    \right)^2 ds \\
& & + \lrp{\frac{1}{\xi_{i+1}-\varepsilon}+o_p(1)}    
    \int_{\xi_i}^{\xi_{i+1}}
    \left(
      \frac{1}{N_n} \lrp{\tilde{W}_n(\xi_{i+1}) - \tilde{W}_n(s) - \lrp{\frac{\xi_{i+1}-s}{\xi_{i+1}-\xi_i}+o_p(1)} \lrp{\tilde{W}_n(\xi_{i+1})-\tilde{W}_n(\xi_i)}}
    \right. \\
& & \hspace{10em}    
    \left.
      - \frac{m_1}{N_n} \delta^{\top} \lrp{\sum\limits_{j=\lrfl{ns}-m+1}^{\lrfl{n\xi_{i+1}}-m} \tilde{X}_{j+m} - \lrp{\frac{\xi_{i+1}-s}{\xi_{i+1}-\xi_i}+o_p(1)} \sum\limits_{j=\lrfl{n\xi_i}-m+1}^{\lrfl{n\xi_{i+1}}-m} \tilde{X}_{j+m}}
    \right)^2 ds \\
& & + o_p(1) \\
&=& \lrp{\frac{1}{\xi_{i+1}-\varepsilon}+o_p(1)} \tilde{V}_n^{f}(\xi_i).
\EEqn

Finally, by noting that
\BEqn
    \frac{T_n^f(1,\lrfl{n\xi_i}-m,\lrfl{n\xi_{i+1}}-m)}{\lrp{V_n^f(1,\lrfl{n\xi_i}-m,\lrfl{n\xi_{i+1}}-m)}^{1/2}} 
&=& \frac{\displaystyle\frac{m_1 \sqrt{\lrfl{n\xi_{i+1}}-m}}{N_n} T_n^f(1,\lrfl{n\xi_i}-m,\lrfl{n\xi_{i+1}}-m)}{\displaystyle\lrp{\frac{m_1^2 (\lrfl{n\xi_{i+1}}-m)}{N_n^2} V_n^f(1,\lrfl{n\xi_i}-m,\lrfl{n\xi_{i+1}}-m)}^{1/2}} \\
&=& \lrp{\xi_{i+1}-\varepsilon+o_p(1)} \tilde{T}_n^f(\xi_i) \lrp{\tilde{V}_n^f(\xi_i)}^{1/2},
\EEqn
we thus arrive at the claimed result.
\end{Proof}

\begin{lemma}\label{Lemma:Expression_Gb}
For any $1\le i\le M$, it holds that
\BEqn
& & T_n^b(\lrfl{n\xi_{i-1}}-m,\lrfl{n\xi_i}-m,N) \lrp{V_n^b(\lrfl{n\xi_{i-1}}-m,\lrfl{n\xi_i}-m,N)}^{-1/2} \\
&=& \lrp{1-\varepsilon-\xi_{i-1}+o_p(1)} \tilde{T}_n^b(\xi_i) \lrp{\tilde{V}_n^b(\xi_i)}^{-1/2},
\EEqn
where
\BEqn
    \tilde{T}_n^b(\xi_i)
&=& \frac{1}{N_n} \lrp{\tilde{W}_n(1-\varepsilon) - \tilde{W}_n(\xi_i) - \lrp{\frac{1-\varepsilon-\xi_i}{1-\varepsilon-\xi_{i-1}}+o_p(1)} (\tilde{W}_n(1-\varepsilon) - \tilde{W}_n(\xi_{i-1}))} \\
& & + \frac{n}{N_n} \sum\limits_{u=1}^{m_1} \lrp{\tilde{X}_u - \tilde{X}_{n+1-u}}^{\top} \lrp{\sum\limits_{t=i}^{M} \delta_t \lrp{\frac{(1-\varepsilon-\xi_t)(\xi_i-\xi_{i-1})}{1-\varepsilon-\xi_{i-1}}+o_p(1)}} \\
& & - \frac{m_1}{N_n} \delta^{\top} \lrp{\sum\limits_{j=\lrfl{n\xi_i}-m}^{N} \tilde{X}_{j+m} -  \lrp{\frac{1-\varepsilon-\xi_i}{1-\varepsilon-\xi_{i-1}}+o_p(1)} \sum\limits_{j=\lrfl{n\xi_{i-1}}-m}^{N} \tilde{X}_{j+m}} \\
& & - \frac{m_1 n}{N_n} \sum\limits_{t=i}^{M} \delta^{\top} \delta_t \lrp{\frac{(1-\varepsilon-\xi_t)(\xi_i-\xi_{i-1})}{1-\varepsilon-\xi_{i-1}}+o_p(1)},
\EEqn
and
\BEqn
    \tilde{V}_n^b(\xi_i)
&=& \int_{\xi_{i-1}}^{\xi_i} 
    \left(
      \frac{1}{N_n} \lrp{\tilde{W}_n(s) - \tilde{W}_n(\xi_{i-1}) - \lrp{\frac{s-\xi_{i-1}}{\xi_i-\xi_{i-1}}+o_p(1)} \lrp{\tilde{W}_n(\xi_i) - \tilde{W}_n(\xi_{i-1})}}
    \right. \\
& & \hspace{3em}
    \left.
      - \frac{m_1}{N_n} \delta^{\top}
      \lrp{\sum\limits_{j=\lrfl{n\xi_{i-1}}-m}^{\lrfl{ns}-m} \tilde{X}_{j+m} - \lrp{\frac{s-\xi_{i-1}}{\xi_i-\xi_{i-1}}+o_p(1)} \sum\limits_{j=\lrfl{n\xi_{i-1}}-m}^{\lrfl{n\xi_i}-m-1} \tilde{X}_{j+m}}
    \right)^2 ds \\
& & + \sum\limits_{h=i+1}^{M+1} \int_{\xi_{h-1}}^{\xi_h} 
    \left(
      \frac{1}{N_n} \lrp{\tilde{W}_n(1-\varepsilon) - \tilde{W}_n(s) - \lrp{\frac{1-\varepsilon-s}{1-\varepsilon-\xi_i}+o_p(1)} \lrp{\tilde{W}_n(1-\varepsilon)} - \tilde{W}_n(\xi_i)}
    \right. \\
& & \hspace{4em}
      + \frac{n}{N_n} \sum\limits_{u=1}^{m_1} \lrp{\tilde{X}_u-\tilde{X}_{n+1-u}}^{\top}
      \left(
        \bone\{h>i+1\} \sum\limits_{t=i+1}^{h-1} \delta_t \lrp{\frac{(1-\varepsilon-s)(\xi_t-\xi_i)}{1-\varepsilon-\xi_i}+o_p(1)}
      \right. \\
& & \hspace{17em}      
      \left.
        + \bone\{h\le M\} \sum\limits_{t=h}^{M} \delta_t \lrp{\frac{(1-\varepsilon-\xi_t)(s-\xi_i)}{1-\varepsilon-\xi_i}+o_p(1)}
      \right) \\
& & \hspace{4em}
      - \frac{m_1}{N_n} \delta^{\top}
      \lrp{\sum\limits_{j=\lrfl{ns}-m}^{N} \tilde{X}_{j+m} - \lrp{\frac{1-\varepsilon-s}{1-\varepsilon-\xi_i}+o_p(1)} \sum\limits_{j=\lrfl{n\xi_i}-m}^{N} \tilde{X}_{j+m}} \\
& & \hspace{4em}
      - \frac{m_1 n}{N_n} 
      \left(
        \bone\{h>i+1\} \sum\limits_{t=i+1}^{h-1} \delta^{\top} \delta_t \lrp{\frac{(1-\varepsilon-s)(\xi_t-\xi_i)}{1-\varepsilon-\xi_i}+o_p(1)}
      \right. \\
& & \hspace{8em}      
    \left.
      \left.
        + \bone\{h\le M\} \sum\limits_{t=h}^{M} \delta^{\top} \delta_t \lrp{\frac{(1-\varepsilon-\xi_t)(s-\xi_i)}{1-\varepsilon-\xi_i}+o_p(1)}
      \right)
    \right)^2 ds \\
& & + o_p(1).    
\EEqn
\end{lemma}

\begin{Proof}
By using the results of Lemma \ref{Lemma:Expression_Tb}, we observe that
\BEqn
& & \frac{m_1 \sqrt{N-\lrfl{n\xi_{i-1}}+m+1}}{N_n} T_n^b(\lrfl{n\xi_{i-1}}-m,\lrfl{n\xi_i}-m,N) \\
&=& \frac{1}{N_n} \lrp{\tilde{W}_n(1-\varepsilon) - \tilde{W}_n(\xi_i) - \lrp{\frac{1-\varepsilon-\xi_i}{1-\varepsilon-\xi_{i-1}}+o_p(1)} (\tilde{W}_n(1-\varepsilon) - \tilde{W}_n(\xi_{i-1}))} \\
& & + \frac{n}{N_n} \sum\limits_{u=1}^{m_1} \lrp{\tilde{X}_u - \tilde{X}_{n+1-u}}^{\top} \lrp{\sum\limits_{t=i}^{M} \delta_t \lrp{\frac{(1-\varepsilon-\xi_t)(\xi_i-\xi_{i-1})}{1-\varepsilon-\xi_{i-1}}+o_p(1)}} \\
& & - \frac{m_1}{N_n} \delta^{\top} \lrp{\sum\limits_{j=\lrfl{n\xi_i}-m}^{N} \tilde{X}_{j+m} -  \lrp{\frac{1-\varepsilon-\xi_i}{1-\varepsilon-\xi_{i-1}}+o_p(1)} \sum\limits_{j=\lrfl{n\xi_{i-1}}-m}^{N} \tilde{X}_{j+m}} \\
& & - \frac{m_1 n}{N_n} \sum\limits_{t=i}^{M} \delta^{\top} \delta_t \lrp{\frac{(1-\varepsilon-\xi_t)(\xi_i-\xi_{i-1})}{1-\varepsilon-\xi_{i-1}}+o_p(1)} \\
&=& \tilde{T}_n^b(\xi_i).
\EEqn

Furthermore, it follows from Lemma \ref{Lemma:Expression_Vb} that
\BEqn
& & \frac{m_1^2 (N-\lrfl{n\xi_{i-1}}+m+1)}{N_n^2} V_n^b(\lrfl{n\xi_{i-1}}-m,\lrfl{n\xi_i}-m,N) \\
&=& \lrp{\frac{1}{1-\varepsilon-\xi_{i-1}}+o_p(1)} \\
& & \hspace{1em}
    \times
    \int_{\xi_{i-1}}^{\xi_i} 
    \left(
      \frac{1}{N_n} \lrp{\tilde{W}_n(s) - \tilde{W}_n(\xi_{i-1}) - \lrp{\frac{s-\xi_{i-1}}{\xi_i-\xi_{i-1}}+o_p(1)} \lrp{\tilde{W}_n(\xi_i) - \tilde{W}_n(\xi_{i-1})}}
    \right. \\
& & \hspace{5em}
    \left.
      - \frac{m_1}{N_n} \delta^{\top}
      \lrp{\sum\limits_{j=\lrfl{n\xi_{i-1}}-m}^{\lrfl{ns}-m} \tilde{X}_{j+m} - \lrp{\frac{s-\xi_{i-1}}{\xi_i-\xi_{i-1}}+o_p(1)} \sum\limits_{j=\lrfl{n\xi_{i-1}}-m}^{\lrfl{n\xi_i}-m-1} \tilde{X}_{j+m}}
    \right)^2 ds \\
& & + \lrp{\frac{1}{1-\varepsilon-\xi_{i-1}}+o_p(1)} \\
& & \hspace{2em}
    \times
    \sum\limits_{h=i+1}^{M+1} \int_{\xi_{h-1}}^{\xi_h} 
    \left(
      \frac{1}{N_n} \lrp{\tilde{W}_n(1-\varepsilon) - \tilde{W}_n(s) - \lrp{\frac{1-\varepsilon-s}{1-\varepsilon-\xi_i}+o_p(1)} \lrp{\tilde{W}_n(1-\varepsilon)} - \tilde{W}_n(\xi_i)}
    \right. \\
& & \hspace{6em}
      + \frac{n}{N_n} \sum\limits_{u=1}^{m_1} \lrp{\tilde{X}_u-\tilde{X}_{n+1-u}}^{\top}
      \left(
        \bone\{h>i+1\} \sum\limits_{t=i+1}^{h-1} \delta_t \lrp{\frac{(1-\varepsilon-s)(\xi_t-\xi_i)}{1-\varepsilon-\xi_i}+o_p(1)}
      \right. \\
& & \hspace{19em}      
      \left.
        + \bone\{h\le M\} \sum\limits_{t=h}^{M} \delta_t \lrp{\frac{(1-\varepsilon-\xi_t)(s-\xi_i)}{1-\varepsilon-\xi_i}+o_p(1)}
      \right) \\
& & \hspace{6em}
      - \frac{m_1}{N_n} \delta^{\top}
      \lrp{\sum\limits_{j=\lrfl{ns}-m}^{N} \tilde{X}_{j+m} - \lrp{\frac{1-\varepsilon-s}{1-\varepsilon-\xi_i}+o_p(1)} \sum\limits_{j=\lrfl{n\xi_i}-m}^{N} \tilde{X}_{j+m}} \\
& & \hspace{6em}
      - \frac{m_1 n}{N_n} 
      \left(
        \bone\{h>i+1\} \sum\limits_{t=i+1}^{h-1} \delta^{\top} \delta_t \lrp{\frac{(1-\varepsilon-s)(\xi_t-\xi_i)}{1-\varepsilon-\xi_i}+o_p(1)}
      \right. \\
& & \hspace{10em}      
    \left.
      \left.
        + \bone\{h\le M\} \sum\limits_{t=h}^{M} \delta^{\top} \delta_t \lrp{\frac{(1-\varepsilon-\xi_t)(s-\xi_i)}{1-\varepsilon-\xi_i}+o_p(1)}
      \right)
    \right)^2 ds \\
& & + o_p(1) \\
&=& \lrp{\frac{1}{1-\varepsilon-\xi_{i-1}}+o_p(1)} \tilde{V}_n^b(\xi_i).
\EEqn

Therefore, we have that
\BEqn
& & T_n^b(\lrfl{n\xi_{i-1}}-m,\lrfl{n\xi_i}-m,N) \lrp{V_n^b(\lrfl{n\xi_{i-1}}-m,\lrfl{n\xi_i}-m,N)}^{-1/2} \\
&=& \frac{\displaystyle \frac{m_1 \sqrt{N-\lrfl{n\xi_{i-1}}+m+1}}{N_n} T_n^b(\lrfl{n\xi_{i-1}}-m,\lrfl{n\xi_i}-m,N)}{\lrp{\displaystyle \frac{m_1^2 (N-\lrfl{n\xi_{i-1}}+m+1)}{N_n^2} V_n^b(\lrfl{n\xi_{i-1}}-m,\lrfl{n\xi_i}-m,N)}^{1/2}} \\
&=& (1-\varepsilon-\xi_{i-1}+o_p(1)) \tilde{T}_n^b(\xi_i) \lrp{V_n^b(\xi_i)}^{1/2}.
\EEqn
\end{Proof}

\end{appendices}

\clearpage
\bibliographystyle{agsm}
\bibliography{arXiv_V2.bib}

\end{document}